\documentclass[superscriptaddress, nofootinbib, longbibliography
]{revtex4-1}

\usepackage{amsmath}	
\usepackage{amssymb}	
\usepackage{graphicx}
\usepackage{bm}

\usepackage{color}

\usepackage[colorlinks=true, citecolor=midblue, linkcolor=midblue, urlcolor=midblue]{hyperref}

\usepackage{setspace}

\usepackage{datetime}

\usepackage{fancyhdr}

\usepackage{lastpage}

\usepackage{titlesec}
\usepackage{ulem}               

\usepackage{letltxmacro}		

\numberwithin{equation}{section}

\titleformat{\section}
       {\centering\normalfont\fontsize{14}{17}\bfseries}{\thesection}{1em}{}

\makeatletter
	\renewcommand{\fnum@figure}{{\bf Figure \thefigure}}
	\renewcommand{\fnum@table}{{\bf Table \thetable}}
\makeatother

\makeatletter
\let\oldr@@t\r@@t
\def\r@@t#1#2{%
\setbox0=\hbox{$\oldr@@t#1{#2\,}$}\dimen0=\ht0
\advance\dimen0-0.2\ht0
\setbox2=\hbox{\vrule height\ht0 depth -\dimen0}%
{\box0\lower0.4pt\box2}}
\LetLtxMacro{\oldsqrt}{\sqrt}
\renewcommand*{\sqrt}[2][\ ]{\oldsqrt[#1]{#2}}
\makeatother

\def\aap{Astron.\ Astrophys.}
\def\apj{Astrophys.\ J.}

\def\apjl{Astrophys.\ J.\ Lett.}
\def\apjs{Astrophys.\ J.\ Suppl.\ S.}
\def\mnras{Mon.\ Not.\ R.\ Astron.\ Soc.}
\def\araa{Annu.\  Rev.\ Astron.\ Astr.}

\def\jcap{J.\ Cosmol.\ Astropart.\ Phys.}

\def\pasj{Pub.\ Astron.\ Soc.\ Jap.}

\def\prd{Phys.\ Rev.\ D}

\def\nat{Nature}

\def\sovast{Sov.\ Astron.}
\def\prl{Phys.\ Rev.\ Lett.}
\def\prd{Phys.\ Rev.\ D}

\definecolor{grey}{rgb}{0.4,0.4,0.4}
\definecolor{dullmagenta}{rgb}{0.4,0,0.4}
\definecolor{darkblue}{rgb}{0,0,0.4}
\definecolor{midblue}{rgb}{0,0,0.7}
\definecolor{midred}{rgb}{0.5,0,0}
\definecolor{orange}{rgb}{1,0.5,0}
\definecolor{lightbrown}{rgb}{0.75,0.5,0.25}
\definecolor{tan}{cmyk}{0.14,0.42,0.56,0}
\definecolor{djunglegreen}{cmyk}{0.99,0,0.52,0}
\definecolor{lightgreen}{rgb}{0,1,0}
\definecolor{olivegreen}{cmyk}{0.64,0,0.95,0.40}
\definecolor{midgreen}{rgb}{0.0,0.675,0.0}
\definecolor{darkgreen}{rgb}{0,0.5,0}
\definecolor{pink}{rgb}{1,0.078,0.57}

\newcommand{\q}{\quad}

\newcommand{\vs}{\vspace}
\newcommand{\hs}{\hspace}
\renewcommand{\.}{\hspace{0.5mm}}

\newcommand{\la}{\ensuremath{\leftarrow}}

\newcommand{\Brm}{\ensuremath{\mathrm{B}}}
\newcommand{\Crm}{\ensuremath{\mathrm{C}}}
\newcommand{\Drm}{\ensuremath{\mathrm{D}}}
\newcommand{\Erm}{\ensuremath{\mathrm{E}}}

\newcommand{\Hrm}{\ensuremath{\mathrm{H}}}

\newcommand{\Krm}{\ensuremath{\mathrm{K}}}
\newcommand{\Lrm}{\ensuremath{\mathrm{L}}}
\newcommand{\Mrm}{\ensuremath{\mathrm{M}}}

\newcommand{\Rrm}{\ensuremath{\mathrm{R}}}
\newcommand{\Srm}{\ensuremath{\mathrm{S}}}
\newcommand{\Trm}{\ensuremath{\mathrm{T}}}

\newcommand{\Xrm}{\ensuremath{\mathrm{X}}}

\newcommand{\arm}{\ensuremath{\mathrm{a}}}
\newcommand{\brm}{\ensuremath{\mathrm{b}}}
\newcommand{\crm}{\ensuremath{\mathrm{c}}}
\newcommand{\drm}{\ensuremath{\mathrm{d}}}

\newcommand{\grm}{\ensuremath{\mathrm{g}}}
\newcommand{\hrm}{\ensuremath{\mathrm{h}}}
\newcommand{\irm}{\ensuremath{\mathrm{i}}}

\newcommand{\mrm}{\ensuremath{\mathrm{m}}}

\newcommand{\orm}{\ensuremath{\mathrm{o}}}
\newcommand{\prm}{\ensuremath{\mathrm{p}}}

\newcommand{\srm}{\ensuremath{\mathrm{s}}}

\newcommand{\vrm}{\ensuremath{\mathrm{v}}}

\newcommand{\Mcal}{\ensuremath{\mathcal{M}}}

\newcommand{\Ocal}{\ensuremath{\mathcal{O}}}

\newcommand{\Rcal}{\ensuremath{\mathcal{R}}}

\renewcommand{\d}{\ensuremath{\mathrm{d}}}

\newcommand{\MeV}{\ensuremath{\mathrm{MeV}}}
\newcommand{\GeV}{\ensuremath{\mathrm{GeV}}}
\newcommand{\TeV}{\ensuremath{\mathrm{TeV}}}

\newcommand{\eg}{e.g.}
\newcommand{\ie}{i.e.}

\newcommand{\cf}{c.f.}
\newcommand{\be}{\begin{equation}}
\newcommand{\ee}{\end{equation}}
\newcommand{\ba}{\begin{eqnarray}}
\newcommand{\ea}{\end{eqnarray}}

\smallskipamount
\settimeformat{ampmtime}

\def\ga{\mathrel{\raise.3ex\hbox{$>$\kern-.75em\lower1ex\hbox{$\sim$}}}}
\def\la{\mathrel{\raise.3ex\hbox{$<$\kern-.75em\lower1ex\hbox{$\sim$}}}}

\def\Msun{M_\odot}

\def\fPBH{f_{\rm PBH}}
\def\MPBH{M_{\rm PBH}}

\def\dCP{\delta_{\rm CP}}

\newcommand{\sv}{\langle\sigma v \rangle}

\def\be{\begin{equation}}
\def\ee{\end{equation}}
\def\bea{\begin{eqnarray}}
\def\eea{\end{eqnarray}}

\begin{document}

\title{\mbox{\!\!\!\! Observational Evidence for Primordial Black Holes: A Positivist Perspective}}

\author{B.~J.~Carr}
\email{b.j.carr@qmul.ac.uk}
\affiliation{School of Physics and Astronomy,
	Queen Mary University of London
	}

\author{S.~Clesse}
\email{sebastien.clesse@ulb.ac.be}
\affiliation{Service de Physique Th{\'e}orique,
	University of Brussels (ULB)
	}

\author{J.~Garc{\'i}a-Bellido}
\email{juan.garciabellido@uam.es}
\affiliation{Instituto de F{\'i}sica Te{\'o}rica,
	Universidad Auton{\'o}ma de Madrid
	}

\author{M.~R.~S.~Hawkins}
\email{mrsh@roe.ac.uk}
\affiliation{Royal Observatory Edinburgh
	}

\author{F.~K{\"u}hnel}
\email{fkuehnel@mpp.mpg.de}
\affiliation{Max Planck Institute for Physics,
	$\vphantom{_{_{_{_{_{_{_{}}}}}}}}$
	}

\date{\formatdate{\day}{\month}{\year}, \currenttime$\vphantom{_{_{_{_{_{_{_{}}}}}}}}$}

\begin{abstract}
We review numerous arguments for primordial black holes (PBHs) based on observational evidence from a variety of lensing, dynamical, accretion and gravitational-wave effects. This represents a shift from the usual emphasis on PBH constraints and provides what we term a positivist perspective. Microlensing observations of stars and quasars suggest that PBHs of around $1\.\Msun$ could provide much of the dark matter in galactic halos, this being allowed by the Large Magellanic Cloud microlensing observations if the PBHs have an extended mass function. More generally, providing the mass and dark matter fraction of the PBHs is large enough, the associated Poisson fluctuations could generate the first bound objects at a much earlier epoch than in the standard cosmological scenario. This simultaneously explains the recent detection of high-redshift dwarf galaxies, puzzling correlations of the source-subtracted infrared and X-ray cosmic backgrounds, the size and the mass-to-light ratios of ultra-faint-dwarf galaxies, the dynamical heating of the Galactic disk, and the binary coalescences observed by LIGO/Virgo/KAGRA in a mass range not usually associated with stellar remnants. Even if PBHs provide only a small fraction of the dark matter, they could explain various other observational conundra, and sufficiently large ones could seed the supermassive black holes in galactic nuclei or even early galaxies themselves. We argue that PBHs would naturally have formed around the electroweak, quantum chromodynamics and electron-positron annihilation epochs, when the sound-speed inevitably dips. This leads to an extended PBH mass function with a number of distinct bumps, the most prominent one being at around $1\.\Msun$, and this would allow PBHs to explain many of the observations in a unified way.
\end{abstract}

\maketitle
\newpage
\begin{spacing}{1.15}
	\tableofcontents
\end{spacing}
\newpage

\section{Introduction}
\label{sec:Introduction}

\noindent It is over 50 years since Hawking proposed the formation of PBHs~\cite{Hawking:1971ei}. Although his specific model turned out to be wrong (because it assumed the black holes were electrically charged and therefore able to capture charged particles to form neutral ``atoms''), this might be regarded as the first {\it positive} paper on the topic. Zeldovich and Novikov~\cite{1967SvA....10..602Z} had also speculated about this possibility in 1967 but concluded on the basis of a simple Newtonian argument that PBHs were unlikely to have formed because they would have grown through accretion to masses of order $10^{15}\.\Msun$ and the existence of such behemoths was excluded. However, in 1974 their accretion argument was disproved by a more precise relativistic analysis and it was shown that PBHs would not grow much at all~\cite{Carr:1974nx}.

PBHs would be expected to have the cosmological horizon mass at formation, which is well below the mass $\sim 1\.\Msun$ of the smallest astrophysical black holes for those forming before the quantum chromodynamics (QCD) epoch at $10^{-5}\,$s. This prompted Hawking to consider the quantum effects of smaller black holes and led to his discovery in 1974 that they emit particles like a black-body with temperature $T \propto M^{-1}$ and evaporate completely on a timescale $\tau \propto M^{3}$~\cite{Hawking:1974rv}. (We note that the study of PBHs prompted the study of black hole quantum effects and not vice versa.) PBHs of around $10^{15}\,$g would be completing their evaporation at the present epoch and those smaller than this would no longer exist. Even though Hawking's prediction has still not been experimentally confirmed, this might be regarded as one of the most important theoretical developments of the late 20th century since it unifies general relativity, quantum theory and thermodynamics. Since only PBHs could be small enough for this effect to be important, this shows that it has been useful to think about them even if they never actually formed.

In the aftermath of Hawking's discovery, there was a flurry of papers about the cosmological consequences of evaporating PBHs (such as their contribution to cosmic rays or gamma-ray bursts) but this did not lead to conclusive evidence for such objects. Later attention turned to the consequences of PBHs larger than $10^{15}\,$g, for which quantum effects are unimportant, but there was no compelling evidence even in this context, so the topic remained a minority interest for many years. It was clear that PBHs were a unique probe of the early Universe, even their non-existence giving valuable information (\eg~about the form of the density fluctuations on small scales and inflationary scenarios), but few people thought they might actually have formed. For this reason the number of publications in the area was initially very small.

However, in recent years the situation has changed radically and PBH publications have soared to several hundred per year. There are many reasons for this but perhaps the main ones are:
	(1)	growing interest in the possibility that PBHs 
		larger than $10^{15}\,$g could provide the dark 
		matter in galactic halos; 
	(2)	the LIGO/Virgo/KAGRA (LVK) detection of 
		gravitational waves from coalescing binary 
		black holes, at least some of which might be 
		primordial; 
	(3)	the possibility that very large PBHs could 
		provide seeds for the supermassive black holes 
		(SMBHs)	which reside in galactic nuclei.
None of these arguments is decisive but they are sufficiently suggestive to have excited both the cosmological and particle physics communities.

As regards (1), the suggestion that PBHs could provide the dark matter goes back to the earliest days of PBH research~\cite{1975Natur.253..251C}, as does the realisation that this would have important consequences for galaxy formation~\cite{Meszaros:1975ef}. This interest was intensified in the 1990s, when the {\it Massive Compact Halo Object} (MACHO) microlensing survey suggested the presence of compact objects with mass around $0.5\.\Msun$ and about $20\%$ of the dark matter~\cite{MACHO:1996qam}. The most likely candidates would be PBHs and those in this mass range might naturally form at the QCD transition~\cite{Jedamzik:1998hc}. Although later microlensing data reduced the dark matter fraction~\cite{Alcock:2000ph} and indeed excluded the mass range $10^{-7}\.\Msun$ to $10\.\Msun$~\cite{Tisserand:2006zx} from providing $100\%$, we will argue that PBHs are still a viable explanation for the MACHO events if they have an extended mass function.

As regards (2), although the mainstream view is that all the LVK detections derive from astrophysical black holes, the sources are larger than initially expected~\cite{Kovetz:2017rvv} and include mass gaps where stellar remnants should not be found. Also the observations suggest smaller spins than would be expected for astrophysical holes~\cite{Garcia-Bellido:2017fdg}. The PBH proposal is still controversial but future observations of the mass, spin and redshift distribution of the objects should clarify the issue soon. It should be stressed that the PBHs invoked in this context do not necessarily provide most of the dark matter.

As regards (3), the mainstream view is that the SMBHs in galactic nuclei form from dynamical processes {\it after} galaxies, but this proposal is becoming increasingly challenged by the high mass and redshift of some SMBHs~\cite{Dolgov:2023ijt}. It is unclear that such objects can form quickly enough in the standard model, so this has led to the suggestion that the SMBHs{\,---\,}or at least their seeds{\,---\,}could form {\it before} galaxies. Such seeds would only have a tiny fraction of the dark matter density and would probably undergo a huge amount of accretion after decoupling.

Although these three arguments have attracted most attention, several other cosmological conundra might be explained by PBHs. For example, provided the PBHs have an appreciable density, the formation of some galaxies at times earlier than predicted by the standard cold dark matter (CDM) model can be explained through their Poisson fluctuations. There are also other problems with the standard scenario which Silk claims can be resolved if intermediate-mass PBHs are ubiquitous in early dwarf galaxies, with their early feedback providing a unified explanation for many dwarf galaxy anomalies~\cite{Silk:2017yai}. In this context, it should be stressed that non-evaporating PBHs may still be of great interest even if they provide only a small fraction of the dark matter. For example, the PBHs seeding SMBHs in galactic nuclei and the PBHs required to explain planetary-mass microlensing events would only have a tiny cosmological density.

\begin{figure}[t]
	\centering
	\vs{-9mm}
	\includegraphics[width=0.90\textwidth]{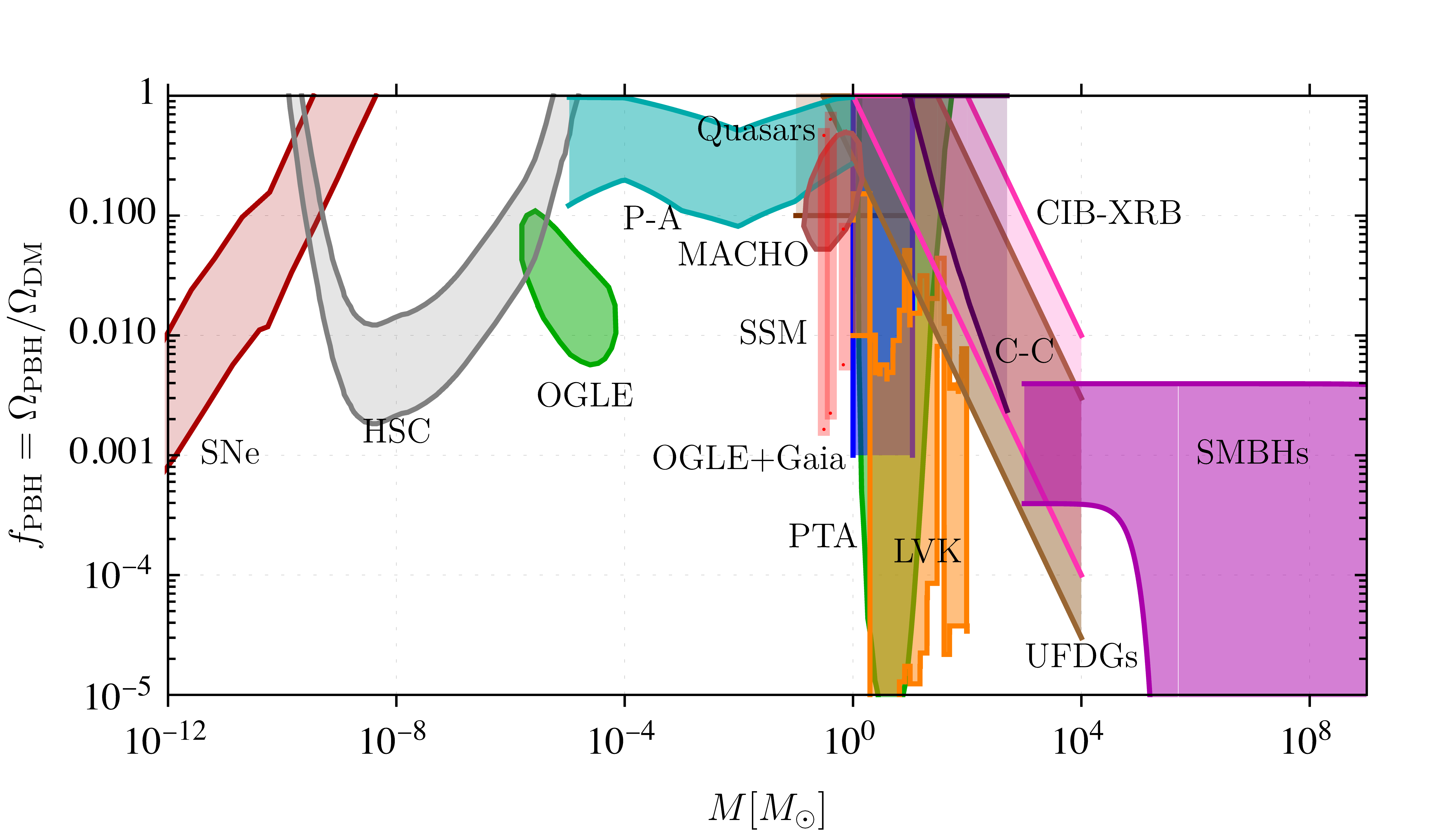}
	\caption{
		Summary of positive indications of
		PBHs in terms of $\fPBH$ values required 
		by or consistent with the claimed detections.
		These come from PBH-attributed 
		signals from 
			supernov{\ae} (SNe), 
			various microlensing surveys 
				({\it Gaia}, HSC, OGLE, MACHO), 
			POINT-AGAPE pixel-lensing (P-A), 
			gravitational waves (LVK), 
			ultra-faint dwarf galaxies (UFDGs), 
			supermassive black holes (SMBHs), 
			core/cusp (C-C)
			profiles for inner galactic halos, and
			correlations of the source-subtracted 
			cosmic infrared and X-ray backgrounds 
			(CIB-XRB).
		The method used to calculate each region 
		is described in Section~\ref{sec:Comparing-Evidence-with-Thermal--History-Model}.
		\vs{-1.5mm}
		}
	\label{fig:PositiveEvidence}
\end{figure}

The possible evidence for PBHs is summarised in Fig.~\ref{fig:PositiveEvidence} in terms of $\fPBH( M )$, the fraction of the dark matter in PBHs with mass around $M$. The derivation of this figure is the main purpose of this review, although a detailed justification for each region is postponed until later. This figure supports what we term a ``positivist" approach to the topic because of the association with the philosophy movement known as ``logical positivism". This was born nearly a century ago and stressed the importance of positive evidence as opposed to theoretical speculation~\cite{Passmore1967-PASLP-6}. The movement is no longer popular in philosophical circles but the term seems appropriate in the context of this review. For comparison, Darwin's ideas about natural selection were developed through detailed and careful study of a variety of disparate biological systems.

Besides the positive evidence for PBHs, there are numerous constraints on their number density. The limits on $\fPBH( M )$ have recently been reviewed in Ref.~\cite{Carr:2020gox}. If the PBH mass function is monochromatic, these limits are often thought to imply that $10^{17}\,\text{--}\,10^{23}$\,g is the only mass range where $\fPBH \sim 1$ is possible. However, it must be stressed that all limits have caveats and other mass ranges are possible if the PBH mass function is extended. Specifically, we will argue that the mass range $1\,\text{--}\,10^{2}\.\Msun$ is still possible. On the other hand, some PBH advocates prefer the lower mass range, so there is a sociological parallel with the split between the search for light and heavy particle candidates. However, whereas advocates of low-mass PBHs rely on the absence of limits, stellar-mass advocates rely on the existence of evidence.

Although this review focusses on the positive evidence for PBHs, it should be stressed that none of the evidence is conclusive because there is always some uncertainty associated with either the observations or the theoretical interpretations. Even when the observations are definite, we cannot exclude some other explanation (\eg~one involving some other dark matter candidate). For example, there is clear evidence for microlensing and gravitational wave events but still no consensus that these are associated with PBHs. Therefore, as with the constraints, one can associate varying degrees of confidence with each item of evidence and the confidence level is in part associated with the possibility of other explanations. Nevertheless, we would claim that the attraction of PBHs (at least with an extended mass function) is that they can provide a {\it unified} explanation of all the evidence.

This review does not include a detailed discussion of the numerous formation mechanisms for PBHs, although the credibility of the evidence is clearly influenced by the plausibility of the mechanisms. The high density of the early Universe does not itself guarantee PBH formation. One also requires density inhomogeneities of some form, so that a small fraction of regions are dense enough to collapse. The most natural source of such inhomogeneities is inflation and there is a huge literature on this topic, starting with Refs.~\cite{Carr:1993aq, Ivanov:1994pa, 1996PhRvD..54.6040G, 1996NuPhB.472..377R}. However, there are also other mechanisms, including a reduction of pressure due to early matter-domination~\cite{Khlopov:1980mg}, quark confinement~\cite{Dvali:2021byy}, the collapse of cosmic strings~\cite{1989PhLB..231..237H, Polnarev:1988dh} or vacuum bubbles~\cite{Deng:2017uwc} or a scalar field~\cite{2018PhRvD..98h3513C}, and collisions of domain walls~\cite{2001JETP...92..921R} or bubbles of broken symmetry~\cite{Hawking:1982ga}. Nor will we cover the numerical studies of PBH formation~\cite{1978SvA....22..129N, 2013CQGra..30n5009M, 2009CQGra..26w5001M, 2019PhRvD.100l3524M, 2021PhRvD.103f3538M}. For a more comprehensive review of these issues, see Refs.~\cite{Khlopov:2008qy, Escriva:2022duf}.
\newpage

Despite our neglect of these topics, we will highlight one particular formation mechanism. This is because the concentration of evidence at the solar-mass scale suggests a possible association with the QCD epoch~\cite{Jedamzik:1996mr, 2018JCAP...08..041B} and a particular unified scenario in this context is described in Ref.~\cite{Carr:2019kxo}. This is based on the idea that the thermal history of the Universe naturally led to dips in the sound-speed and therefore enhanced PBH formation at scales corresponding to the electroweak phase transition ($10^{-6}\.\Msun$), the QCD transition ($1\.\Msun$), the pion-plateau ($10\.\Msun$) and $e^{+}e^{-}$ annihilation ($10^{6}\.\Msun$). This scenario allows an appreciable fraction of the dark matter to be PBHs formed at the QCD peak. Indeed, this fraction would naturally be around the value inferred by the MACHO project if the {\it total} PBH density is comparable to that of the dark matter. Towards the end we discuss our preferred model in some detail. In particular, Fig.~\ref{fig:PositiveEvidence} is reproduced, together with a curve representing the prediction of our unifying PBH scenario, showing how they are remarkably concordant.

The plan of this paper is as follows. We start with a discussion of the inevitable Poisson clustering of PBHs in Section~\ref{sec:Clustering-of-PBHs} because this is an important aspect of the positive evidence and also affects the constraints. We discuss lensing evidence in Section~\ref{sec:Lensing-Evidence}, dynamical/accretion evidence in Section~\ref{sec:Dynamical-and-Accretion-Evidence} and gravitational-wave evidence in Section~\ref{sec:Gravitational--Wave-Evidence}. Various other problems with the standard $\Lambda$CDM model which could easily be resolved by PBHs are covered in Section~\ref{sec:Potential-Solutions-to-Cosmic-Problems}. We consider some more general issues related to the PBH scenario in Section~\ref{sec:Evidence-from-Dark-Matter-and-Unified-Model}, including a discussion of our favoured model for their production at the QCD epoch, a possible resolution of the fine-tuning problem and mixed dark matter scenarios. We draw some conclusions in Section~\ref{sec:Conclusions} and append a table of acronyms.
\newpage

\section{Clustering of PBHs}
\label{sec:Clustering-of-PBHs}

\noindent A major difference between dark matter particles and PBHs comes from the larger mass of the latter. This implies significant Poisson fluctuations in the PBH spatial distribution and eventually leads to bound PBH clusters. This was first pointed out by M{\'e}sz{\'a}ros~\cite{Meszaros:1975ef} and later many other authors studied this effect~\cite{1977A&A....56..377C, 1983ApJ...268....1C, 1983ApJ...275..405F, Chisholm:2005vm, Chisholm:2011kn}. It is crucial for sufficiently massive and abundant PBHs and has several significant consequences:
	(1) Some of the PBH constraints{\,---\,}including 
		the microlensing 
		ones~\cite{DeLuca:2020jug, Gorton:2022fyb, 
		Petac:2022rio}{\,---\,}may be affected;
	(2) The clustering influences the PBH merger rate 
		and thereby provides a connection between 
		the dark matter and gravitational-wave (GW) 
		observations;
	(3) It implies the formation of non-linear 
		structures at higher redshifts than in the 
		standard formation history, with implications 
		for cosmic backgrounds and 
		the observations of high-redshift galaxies;
	(4) It may explain the minimum size and large 
		mass-to-light ratio of ultra-faint dwarf 
		galaxies (UFDGs) and the identification of a 
		subset of PBH clusters with these means that 
		we can regard clustering as providing positive 
		evidence for PBHs.
The Coulomb effect of an individual PBH, which might be considered a special case of the Poisson effect, is also important. Because some of the observational evidence for PBHs depends on their clustering properties, we begin with a discussion of this topic. We first recall the heuristic treatment of the Poisson and Coulomb effects in Ref.~\cite{Carr:2018rid}. We then present a more detailed account of the formation and dynamical evolution of PBH clusters, including some extra effects which were not fully incorporated in earlier treatments. In this section the PBH mass is denoted by $m$ rather than $M$ and $f( m )$ is the fraction of the dark matter density in those of mass $m$ for an extended mass function.

\subsection{Seed and Poisson Effects}
\label{sec:Seed-and-Poisson-Effects}

PBHs provide a source of fluctuations for objects of larger mass in two ways: 
	(1) via the seed effect, associated with the 
		Coulomb attraction of a {\it single} black 
		hole; 
	(2) via the Poisson effect, associated with the 
		$\sqrt{N}$ fluctuation in the number of black 
		holes.
If the PBHs have a single mass $m$, the initial fluctuation in the matter density on a scale $M$ is~\cite{Carr:2018rid}
\begin{equation}
	\delta_{i}
		\approx
				\begin{cases}
					m / M
						& ( {\rm seed} )
					\\[0.5mm]
					( \fPBH \,m / M )^{1/2}
						& ( {\rm Poisson} )
					\, ,
				\end{cases}
				\label{eq:initial}
\end{equation}
where $\fPBH$ is the fraction of the dark matter in the PBHs, assumed to be constant during the clustering process. If $\fPBH \sim 1$, the Poisson effect dominates for all $M$; if $\fPBH \ll 1$, the Poisson effect dominates for $M > m / \fPBH$ and the seed effect for $M < m / \fPBH$. Indeed, the mass bound by a single seed can never exceed $m / \fPBH$ because of competition from other seeds.

There is always a mass $M_{\rm CDM}$ below which the PBH fluctuations dominate the CDM fluctuations, so this produces extra power on small scales. However, the CDM fluctuations fall off slower than both the Poisson and seed fluctuations with increasing $M$ and so generally dominate for sufficiently large $M$. For this reason, the structure of the Universe is unchanged in $\Lambda$CDM models on sufficiently large scales but it can be radically different on small-scales.
\newpage

Both types of PBH fluctuations are frozen during the radiation-dominated era~\cite{1974A&A....37..225M} but grow as $( 1 + z )^{-1}$ from the start of the matter-dominated era. Since matter-radiation equality corresponds to a redshift $z_{\rm eq} \approx 4000$ and an overdense region binds when $\delta \approx 1$, one can estimate the mass of regions which become gravitationally bound at redshift $z_{\crm}$ as
\vs{-1mm}
\begin{equation}
	M_{\crm}
		\approx
				\begin{cases}
					4000\;m\.( 1 + z_{\crm} )^{-1}
						& ( {\rm seed} )
						\\[1mm]
					10^{7} \fPBH\,m\.
					( 1 + z_{\crm} )^{-2}
						& ( {\rm Poisson} )
					\, .
				\end{cases}
				\label{eq:bind}
\end{equation}
However, there is an important difference between the two effects. In the Poisson case, most of the Universe goes into bound regions of mass $M_{\crm}$ at redshift $z_{\crm}$. For instance, if $\fPBH \sim 1$ and $m\sim 1~\.\Msun$, the Universe is already inhomogeneous at $z \sim 100$ with the formation of PBH clusters of around $10^{3}\.\Msun$. A similar result is obtained for any combination for which $m\mspace{1.5mu}\fPBH \sim 1\.\Msun$. In the seed case, the fraction of the Universe in bound regions at redshift $z_{\crm}$ is approximately $\fPBH\,z_{\rm eq} / z_{\crm}$, which is initially small for $\fPBH \ll 1$ and only reaches $1$ at $z_{\crm} \approx \fPBH\,z_{\rm eq}$. Thereafter, competition between the seeds will limit the mass of each seed-bound region to at most $M \sim m / \fPBH$. This is just the value of $M$ above which the Poisson effect dominates. Each PBH is surrounded by a local bound region of this mass, so one has a combination of the seed and Poisson effect. Indeed, the regions of mass $m/\fPBH$ can themselves be regarded as sourcing Poisson fluctuations: putting $\fPBH \rightarrow 1$ and $m \rightarrow m / \fPBH$ in Eq.~\eqref{eq:initial}, and assuming growth starts at $z \approx \fPBH\,z_{\rm eq}$, gives the same bound mass as the original Poisson expression~\eqref{eq:bind}. The dependences of $\delta_{\irm}$ on $M$ and $M_{\crm}$ on $z_{\crm}$ are shown in Fig.~\ref{fig:seedvsposson}.

For an extended PBH mass function, the relationship between the seed and Poisson effect is more complicated~\cite{Carr:2018rid}. For simplicity, we consider a power-law mass function with $\d n / \drm m \propto m^{-\alpha}$ with upper and lower cut-offs at $m_{\rm max}$ and $m_{\rm min}$, respectively. This implies the mass fraction of the Universe in PBHs of mass $m$ is $f( m ) \propto m^{2}\.\d n / \d m \propto m^{- \alpha + 2}$. The total dark matter fraction $\fPBH$ is therefore dominated by the $m_{\rm min}$ PBHs for $\alpha > 2$. The dominant fluctuation on a scale $M$ comes from the seed effect for $M < m_{\rm min} / \fPBH$, so Eq.~\eqref{eq:bind} implies that the mass binding at redshift $z_{\crm}$ is
\begin{equation}
\label{eq:boundmass}
	M_{\crm} \equiv M( z_{\crm} )
		\sim
				4000\;m_{\rm min}\.
				( 1 + z_{\crm} )^{-1}
	\quad {\rm for} \quad
	M_{\crm} 
		<
				m_{\rm min}/\fPBH
				\,.
\end{equation}
The Poisson effect dominates for $M > m_{\rm min} / \fPBH$ but the relevant value of $m$ for given $M$ depends on the form of the function $f( m )\.m$. If this decreases with increasing $m$ (corresponding to $\alpha > 3$), the holes of mass $m_{\rm min}$ dominate the Poisson effect and Eq.~\eqref{eq:bind} implies that the mass binding at redshift $z_{\crm}$ is
\vs{-1mm}
\begin{equation}
	M_{\crm}
		\sim
				10^{7}\mspace{1mu}\fPBH\,m_{\rm min}\.
				( 1 + z_{\crm} )^{-2}
	\quad {\rm for} \quad
	M_{\crm}
		>
				m_{\rm min} / \fPBH
				\,.
\end{equation}
The expressions for $M_{\crm}$ cross at $z_{\crm} = z_{\rm eq}\,\fPBH$, which is when the Poisson effect takes over.

If $f( m )\.m$ is an increasing function of $m$ (corresponding to $\alpha < 3$), the largest $m$ dominates the Poisson fluctuation. If the mass of the largest PBH in the region is less than $m_{\rm max}$, \ie~if $M < m_{\rm max} / f_{\rm max}$ where $f_{\rm max} \equiv f( m_{\rm max} )$, then the Poisson scenario effectively reduces to the seed scenario with an $M$-dependent seed mass 
\begin{equation}
	m_{\rm seed}
		\sim
				\big(
					M \fPBH\,m_{\rm min}^{\alpha - 2}
				\big)^{1 / ( \alpha - 1 )}
				\, .
				\label{eq:seed}
\end{equation}
\newpage

\noindent The mass binding at redshift $z_{\crm}$ is therefore
\begin{equation}
	M_{\crm}
		\sim
				\fPBH^{1/( \alpha - 2 )}\.m_{\rm min}
				\big(
					1 + z_{\crm} / 4000
				\big)^{(\alpha - 1)/(2 - \alpha)} 
	\quad
	{\rm for}
	\quad
	m_{\rm min} / \fPBH
		<
				M_{\crm}
		<
				m_{\rm max} / f_{\rm max}
				\, .
\ee 
However, if $M$ is larger than $m_{\rm max} / f_{\rm max}$, it will contain many PBHs of mass $m_{\rm max}$ and the associated Poisson fluctuation will bind a mass 
\begin{equation}
	M_{\crm}
		\sim
				10^{7}\,f_{\rm max}\.m_{\rm max}\.
				( 1 + z_{\crm} )^{-2}
		\sim
				10^{7}\,\fPBH\,
				m_{\rm min}^{\alpha - 2}\,
				m_{\rm max}^{3 - \alpha}\.
				\big(
					1 + z_{\crm}
				\big)^{-2}
	\quad {\rm for} \quad
	M_{\crm}
		>
				m_{\rm max} / f_{\rm max}
				\, .
\end{equation}
There is therefore a complicated transition between the seed and Poisson effects in the extended case, this depending sensitively on the shape of the PBH mass function. One has $M_{\crm} \propto ( 1 + z_{\crm} )^{-1}$ at large $z_{\crm}$, $( 1 + z_{\crm} )^{-2}$ at low $z_{\crm}$ and an intermediate regime with $M_{\crm} \propto ( 1 + z_{\crm} )^{- ( \alpha - 1 ) / ( \alpha - 2 )}$ for $\alpha < 3$. Numerical simulations are required for a more detailed analysis. Note that a more complicated mass function may still be approximated by a power-law over some mass range.

\begin{figure*}[t]
	\vs{-3mm}
	\includegraphics[width=0.45\textwidth]{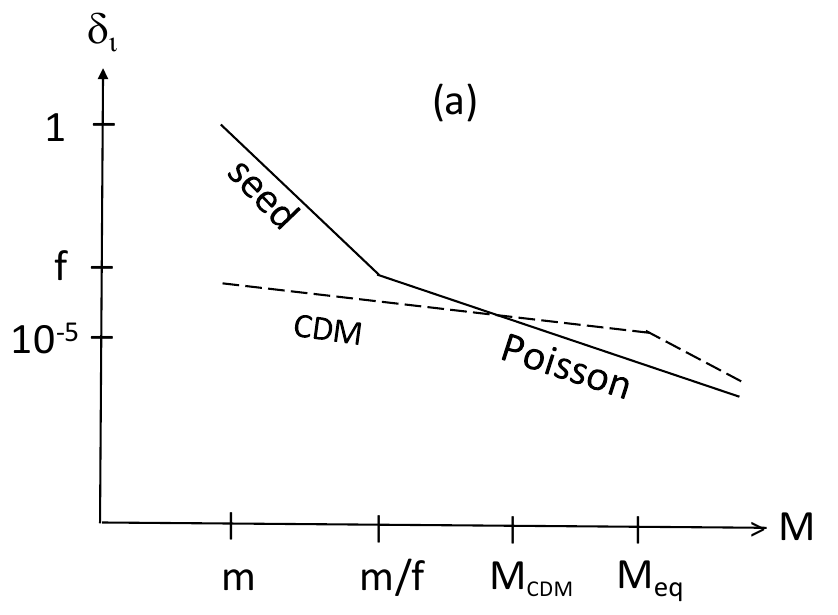}
	\includegraphics[width=0.45\textwidth]{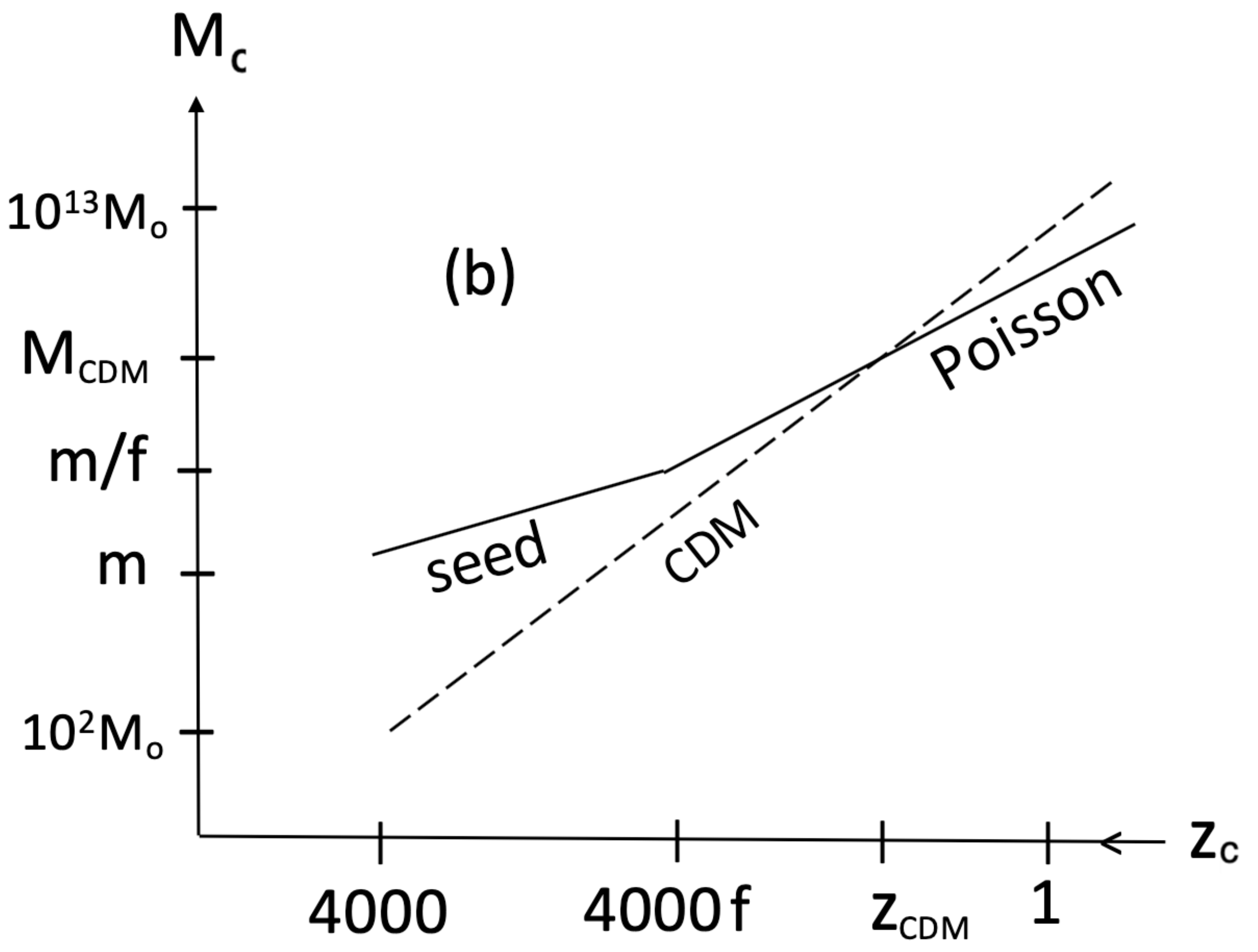}
	\caption{
		Relative importance of the seed and 
		Poisson effects as compared to the standard CDM 
		scenario in terms of the initial density 
		fluctuation $\delta_{\irm}$ as a function of 
		mass (left) and the binding mass $M$ as a 
		function of redshift $z_{\crm}$ (right).
		From Ref.~\cite{Carr:2018rid}.
		\vs{-3mm}
		}
	\label{fig:seedvsposson}
\end{figure*}

\subsection{Evolution of Poisson Fluctuations and Formation of PBH Clusters}
\label{sec:Evolution-of-Poisson-Fluctuations-and-Formation-of-PBH-Clusters}

We now provide a more accurate description of PBH clustering induced by Poisson fluctuations. Going beyond the simple analytical estimates of the previous section is essential in assessing the positive observational evidence linked to the dynamics of these clusters and calculating the accretion onto the PBHs they contain. We consider the case of a monochromatic mass function. Since the Poisson fluctuation scales like $m\mspace{1.5mu}\fPBH$, we can generalise to any PBH mass function by replacing $m\mspace{1.5mu}\fPBH$ with $\int{\d \ln m}\;m\.f( m )$.

The formation of Poisson-induced PBH clusters has been studied using $N$-body simulations~\cite{Inman:2019wvr, Raidal:2018bbj} and analytically using the Press--Schechter formalism and the theory of spherical collapse~\cite{Afshordi:2003zb}. As an illustrative example, Fig.~\ref{fig:Nbody} shows the density distribution at redshift $z = 100$ obtained by Ref.~\cite{Inman:2019wvr} for monochromatic PBHs with $m = 30\.\Msun$ and $\fPBH = 10^{-5}$ and $10^{-1}$. These results also apply for any combination with $m\mspace{1.5mu}\fPBH/\Msun = 0.0003$ and $3$. The difference with the standard cosmological scenario, in which the Universe is still homogeneous at such a high redshift, is striking.
\newpage

\begin{figure*}
	\includegraphics[width=0.4\textwidth]{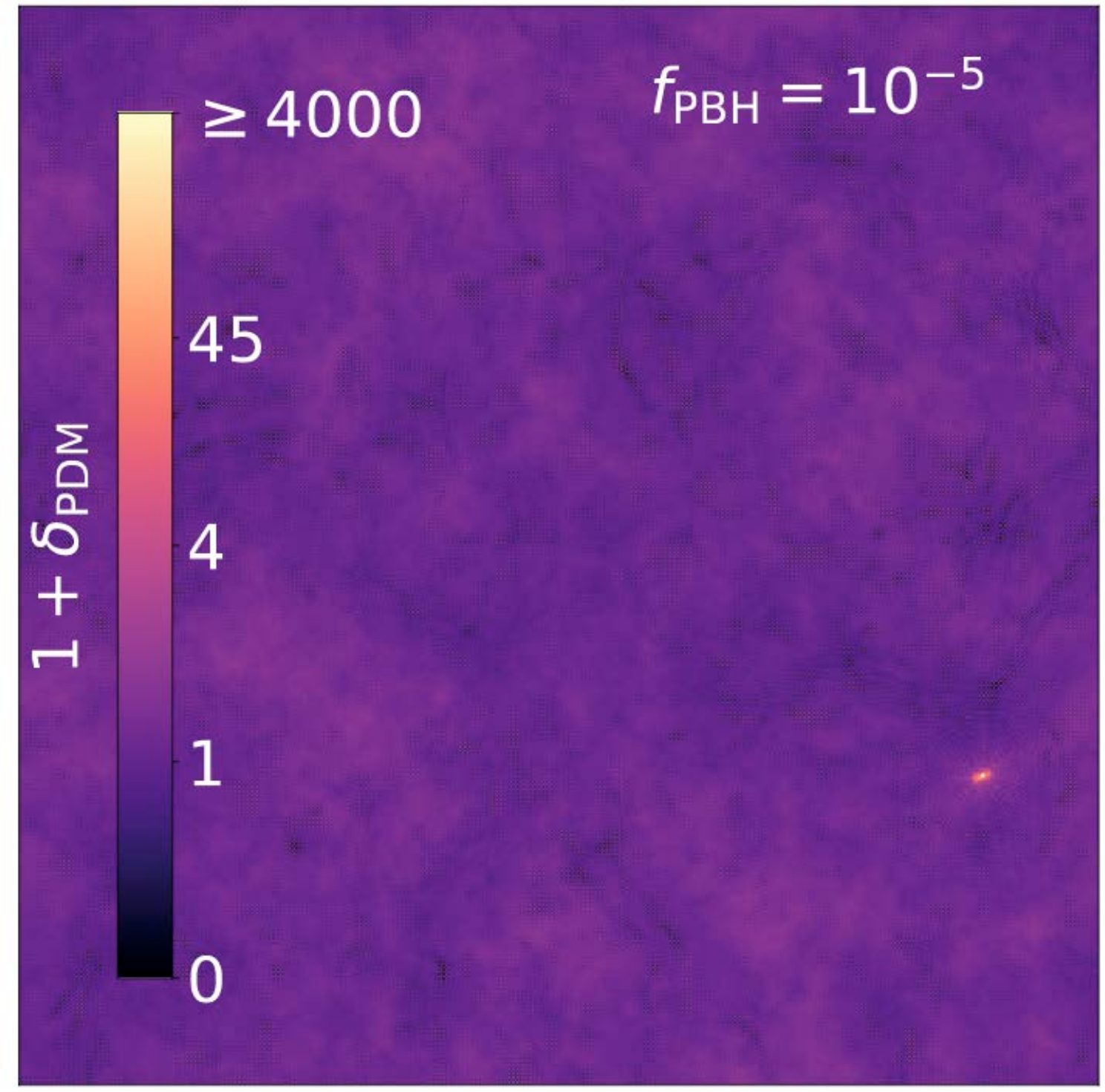}
	\includegraphics[width=0.4\textwidth]{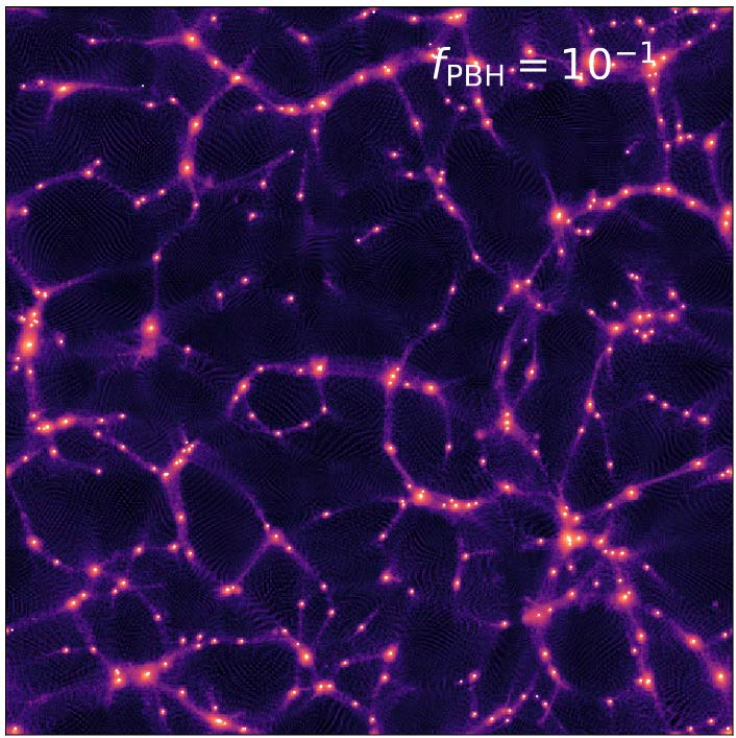}
	\caption{Expected dark matter density distribution 
		over a scale of $2\.h^{-1}\,{\rm kpc}$ at 
		redshift $z = 99$ obtained from $N$-body 
		simulations of Ref.~\cite{Inman:2019wvr} for 
		$m = 30\.\Msun$ and
			$\fPBH = 10^{-5}$ ({\it left}\hs{0.2mm})
		or
			$\fPBH = 0.1$ ({\it right}\hs{0.2mm}).
		}
	\label{fig:Nbody}
\end{figure*}

As pointed out in Ref.~\cite{Kashlinsky:2016sdv}, the contribution from Poisson fluctuations to the matter power spectrum becomes dominant on small scales and for stellar-mass PBHs leads to the high-redshift gravitational collapse of almost all small-scale perturbations into clusters with masses up to $10^{6}\,\text{--}\,10^{7}\.\Msun$. The Poisson contribution in the matter power spectrum is scale-invariant and its amplitude at redshift $z$ is well approximated by~\cite{Afshordi:2003zb}
\begin{equation}
	P_{\rm Poisson}
		\simeq
				2 \times 10^{-3}\,
				\frac{ \fPBH }{ g( z )^{2} }
				\mspace{-1.5mu}
				\left(
					\frac{ m }{ 3\mspace{1mu}\Msun }
				\right)
				{\rm Mpc}^{3}
				\, ,
\end{equation}
where $g( z )$ is the growth factor for isocurvature fluctuations; this increases linearly with the scale factor in the matter-dominated era, when clusters are formed. Later, the growth function takes a more complex but still analytic form.

For example, if we consider the standard $\Lambda$CDM model with $m = 3\.\Msun$ and $\fPBH = 1$, the Poisson term dominates for comoving wavemodes $k \gtrsim 100\,{\rm Mpc}^{-1}$ at $z \approx 20$. The current (dimensionless) density perturbation $\Delta( k )$, related to the power spectrum through $\Delta( k )^{2} = P( k )\.k^{3} / ( 2 \pi^{2} )$, is shown in Fig.~\ref{fig:deltas} (neglecting non-linear effects). For comparison, we also show the expected spectrum for the standard $\Lambda$CDM scenario and for a primordial spectrum with a sharp enhancement at $k_{\rm trans} = 10^{3}\,{\rm Mpc^{-1}}$ to generate PBHs at small scales. This effect modifies the spectrum below the scale at which the Poisson effect becomes dominant, but even for lower values of $m$ or $\fPBH$ one may have to take into account the effect of the enhanced primordial power spectrum on the PBH cluster formation, which is sometimes overlooked.

One can associate a cluster mass scale with each fluctuation wavelength $\lambda = 2\pi / k$:
\begin{equation}
\label{eq:Mlambda}
	M_{\crm} ( \lambda )
		=
				( 4 \pi / 3 )\.\lambda^{3}\.
				\rho_{\rm m}^{0}\.
				( 1 + \delta_{\crm} )
		\simeq
				1.15 \times 10^{12}\.
				( \lambda / {\rm Mpc } )^{3}\.
				( 1 + \delta_{\crm} )\.\Msun
				\, ,
\end{equation}
where $\rho_{\rm m}^{0}$ is the current matter density. Poisson-induced fluctuations decouple from the expansion and form a bound PBH cluster when they become larger than the overdensity threshold $\delta_{\crm} \simeq 1.686\vphantom{1_{_{_{_{_{_{_{_{_{_{_{_{_{_{_{_{_{_{_{_{1}}}}}}}}}}}}}}}}}}}}}$, which implies a redshift of collapse given by
\begin{equation}
	z_{\crm} + 1
		\simeq
				3.7 \times 10^{-3}\.k^{-3/2}
				\left(
					\frac{ m\mspace{1.5mu}\fPBH }{\.\Msun }
				\right)^{\!-1/2}
		\simeq
				39 \times\!
				\left[
					\frac{ 10^{6}\.m\mspace{1.5mu}\fPBH }
					{ M_{\crm} }
				\right]^{1/2}
\end{equation}
in the matter-dominated era. One recognises the same dependence on $m$ and $M_{\crm}$ as in Eq.~\eqref{eq:bind}, but the formation redshift is higher, which demonstrates the need to go beyond those estimations. When a bound cluster is formed, the theory of spherical collapse predicts that its density is $178$ times the background density, giving a cluster radius at formation:
\begin{equation}
	r_{\crm}
		\simeq 
				135\,{\rm pc}
				\left(
					\frac{ m\mspace{1.5mu}\fPBH}{\.\Msun }
				\right)^{\!-1/2} 
				\left(
					\frac{ M_{\crm} }
					{ 10^{6}\mspace{1mu}\Msun }
				\right)^{\!5/6}
				\, .
				\label{eq:rM}
\end{equation}
According to the Press--Schechter formalism~\cite{Bond:1990iw, Bower:1991kf, Lacey:1993iv, Lacey:1994su}, the fraction of fluctuations that collapse into halos with a mass $M_{\crm}$ is given by
\begin{equation}
	F_{\crm}( M_{\crm},\.z_{\crm} )
		=
				{\rm erfc}\!
				\left[
					\frac{ \delta_{\crm} }
					{ \sqrt{2}\.
					\sigma( M_{\crm}, z_{\crm} ) }
				\right]
				,
\end{equation}
where 
\begin{equation}
	\sigma^{2}( M_{\crm}, z_{\crm} )
		=
				\int\!\drm \ln k\;
				\frac{ k^{3} }{ 2 \pi^{2} }\.
				P( k, z_{\crm} )\,W( k )
				\, .
\end{equation}
Here $W( k )$ a window function, commonly taken to be a top-hat at $k( M_{\crm} )$ with width $\ln k ( M_{\crm} )$. Figure~\ref{fig:deltas} shows the expected value of $F_{\crm}$ for the previous example ($m = 2.6\.\Msun$, $\fPBH = 1$) and indicates that $F_{\crm}( M_{\crm} < 10^{6}\.\Msun )$ is very close to unity. This means that almost all fluctuations below this scale collapse to clusters, so most PBHs end up in such clusters at some point. The maximum mass of PBH clusters for this model is therefore around $10^{6}\,\text{--}\,10^{7}\.\Msun$.

\begin{figure*}[t]
	\includegraphics[width=0.85\textwidth]{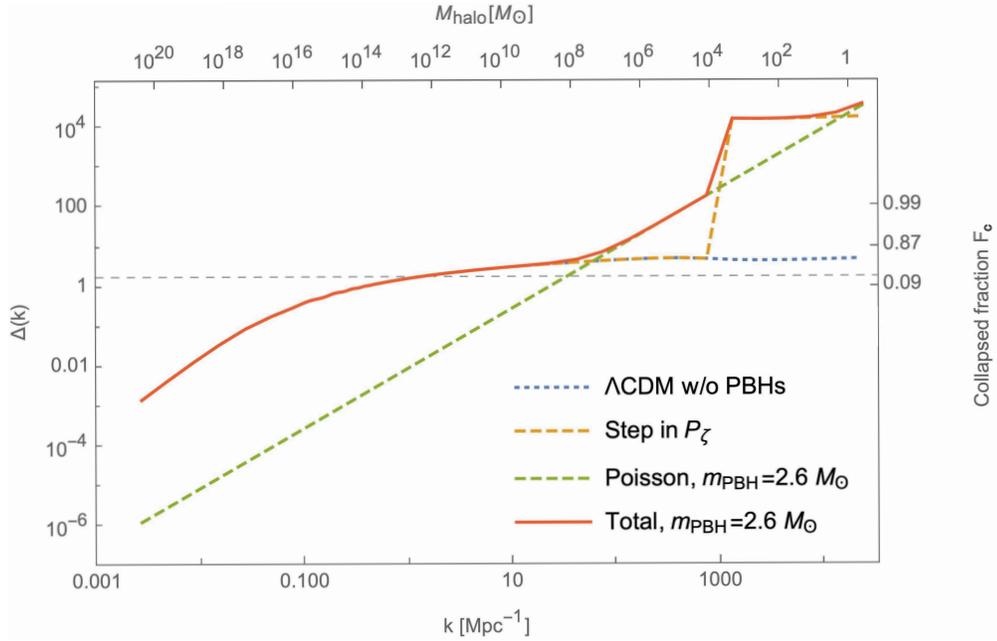}
	\caption{Dimensionless (linear) density contrast 
		$\Delta( k )$ today in the standard 
		$\Lambda$CDM model without PBHs (dotted blue 
		line) and with them for $\fPBH = 1$ at 
		$m = 2.6\.\Msun$ (solid red line), 
		including the inevitable Poisson term in the 
		matter power spectrum (dashed green line). For 
		comparison, we also show the possible power 
		spectrum enhancement leading to PBH formation 
		for a transition scale at 
		$k_{\rm trans} = 10^{3}\,{\rm Mpc^{-1}}$ 
		(dashed orange line). The horizontal dashed 
		line represents the critical threshold for halo 
		formation $\delta_{\rm clust} \simeq 1.686$.
		The upper $x$-axis is the halo mass implied by
		Eq.~\eqref{eq:Mlambda}.	The right $y$-axis 
		shows the halo collapsed fraction 
		$F_{\crm}( M_{\crm} )$; this is almost $1$ for 
		$k \gtrsim 100\,{\rm Mpc}^{-1}$, meaning that 
		all PBHs are regrouped in clusters with masses 
		of $10^{6}\,\text{--}\,10^{7}\.\Msun$. Figure 
		adapted from Ref.~\cite{Clesse:2020ghq}.
		\vs{3mm}} 
	\label{fig:deltas}
\end{figure*}

\subsection{Relaxation, Dynamical Heating and Evaporation of PBH Clusters}
\label{sec:Relaxation,-Dynamical-Heating-and-Evaporation-of-PBH-Clusters}

When Poisson-induced PBH clusters have formed, it takes some time for the PBH velocities and the cluster size to relax towards the values expected from the virial theorem. The relaxation time for a halo of $N = \fPBH\.M_{\crm}/{m}$ objects is 
\begin{equation}
\label{eq:reltime}
	t_{\rm rel}
		\simeq
				\left[
					\frac{ 0.14\.N }{ \ln( 0.14\.N ) }
				\right]\!
				\sqrt{\frac{ r_{\crm}^{3} }
				{ G M_{\crm} }}\.
		= 
				2.1 \times 10^{6}\,{\rm yr}\,
				\frac{ (M_{\crm}/\Msun)^{1/2} }
				{ (m/\Msun)
				\ln\!
				\big[
					0.14\.M_{\crm} / m
				\big] }
				\left(
					\frac{ r_{\crm}}{\rm pc}
				\right)^{\!3/2}
				\, ,
\end{equation}
where $r_{\crm}$ is the initial halo median radius, estimated in the previous subsection. After relaxation the clusters continue to expand due to dynamical heating. This process is taken into account, for instance, to derive limits on $\fPBH$ from the minimum size of UDFGs or their globular star clusters.$\vphantom{_{_{_{_{_{_{_{_{_{_{_{_{_{_{_{_{_{_{_{_{_{_{_{_{_{_{_{_{_{_{_{_{_{_{_{_{_{_{_{_{_{_{_{_{_{_{_{1}}}}}}}}}}}}}}}}}}}}}}}}}}}}}}}}}}}}}}}}}}}}}}}}$ The cluster radius increases according to
\begin{equation}
	\frac{ \drm\.r_{\crm} }{ \drm t }
		=
				\frac{ 4\.\sqrt{2} \pi\.G \fPBH\.m
				\ln\mspace{1mu}( \fPBH\mspace{1mu}
				M_{\crm} / 2\.m ) }
				{ 2\.\beta\.v_{\rm vir}\.r_{\crm} }
				\, ,
\end{equation}
where $v_{\rm vir} = \sqrt{G M_{\crm} / 2\.r_{\crm}}$ is the halo virial velocity and $\beta \approx 10$ is a parameter depending on the halo profile. By integrating this equation, one can relate the time elapsed since cluster formation to the current cluster radius $r_{\crm}$: 
\begin{equation}
	t_{\rm dyn}
		\simeq
				\beta\,t_{\rm rel}
				\left[
					1
					-
					\left(
						\frac{ r_{\rm c,\mspace{1mu} i} }
						{ r_{\crm} }
					\right)^{\!3/2}
				\right]
	\quad\Rightarrow\quad
		r_{\crm}( t_{0} )
			\simeq
					400\,\beta^{-2/3}
					\left(
						\frac{M_{\crm}}{\Msun} 
					\right)^{\!-1/3}\!
					\left(
						\frac{m}{\Msun}
					\right)^{\!2/3}
					\ln\!
					\left[
						\frac{M_{\crm}}{2\.m}
					\right]
					.
					\label{eq:dyntime}
\end{equation}
Here $r_{\rm c,\mspace{1mu}i}$ is the initial cluster radius, given by Eq.~\eqref{eq:rM}, and the second expression gives the current cluster radius, obtained by putting $t_{\rm dyn} = t_{0} \approx 14\,$Gyr. 

Eventually, if the PBH clusters are initially very compact and dense, they will completely evaporate through dynamical heating. This happens within the evaporation time, given approximately by $t_{\rm ev} \sim 100\.t_{\rm rel}$. As a result, for a given cluster mass $M_{\rm c,}$ one can deduce a minimum radius $r_{\crm}$ from the requirement that the cluster has not totally evaporated. The time scales $t_{\rm rel}$, $t_{\rm dyn}$, $t_{\rm ev}$ have the same dependence on $M_{\crm}$, $f_{\rm PBH}$ and $m$ but with different numerical factors. They have sometimes been confused. For example, in the context of the microlensing limits, the evaporation time was considered in Ref.~\cite{Petac:2022rio} in order to find the minimum cluster mass, whereas the dynamical heating time should have been used. Another point that has often been overlooked is that the initial size of the PBH clusters can be large enough for heating over the age of the Universe to be inefficient. This applies above the line specified by Eq.~\eqref{eq:dyntime}.

This discussion assumes that the evolution of a cluster can be considered in isolation. However, in the Poisson situation, a cluster always becomes part of a bigger cluster as long as $F_{\crm} \approx 1$. So as it expands, it becomes subsumed into progressively larger structures and loses its integrity. This means the values of $r_{\crm}$ and $M_{\crm}$ for surviving clusters at the present epoch are related by Eq.~\eqref{eq:rM}, so long as the expansion time $t_{\rm dyn}$ is more than the age of the Universe. The expansion process stops above the mass for which the lines given by Eqs.~\eqref{eq:rM} and \eqref{eq:dyntime} cross, so this introduces a minimum mass ($M_{\crm} \propto m\.\fPBH^{3/7}$) and radius ($r_{\crm} \propto m^{1/3}\.\fPBH^{-1/7}$) for surviving clusters. These are of order $10^{5}\.\Msun$ and $r_{\crm} \approx 20\,$pc for $f_{\rm PBH}\.m / \Msun \sim \Ocal( 1 )$.

As discussed below, clusters larger than this minimum mass may still lose their integrity as a result of collisions within a larger cluster. However, this can lead to their disruption only if the velocity dispersion $\sigma_{\crm}$ within clusters increases with their mass. So long as the Poisson effect dominates, Eq.~\eqref{eq:rM} implies $\sigma_{\crm} \sim M_{\crm}^{1/12}$, which increases only very weakly, so it is marginal whether the hierarchy of clustering can be destroyed. However, on scales larger than about $10^{8}\.\Msun$, the CDM fluctuations take over (see Fig.~\ref{fig:deltas}) and in this case $\sigma_{\crm} \sim M_{\crm}^{1/3}$. This means that collisional disruption may occur, at least in the central region of the larger bound region. In particular, we show below that collisions in our own halo may destroy clusters within a sufficiently small Galactocentric radius.

In order to illustrate these different processes and infer the PBH cluster mass-size relation today, Fig.~\ref{fig:rcvsmc} shows as broken black lines the initial cluster size, from Eq.~\eqref{eq:rM}, and the region where dynamical expansion occurs within the age of the Universe, from Eq.~\eqref{eq:dyntime}, for the cases $\fPBH\.m / \Msun = 0.3$, $3$ (relevant for the thermal-history model discussed in Section~\ref{sec:Thermal--History-Scenario}) and $30$. The current relationship is given by the solid black lines with the minimum depending on $f_{\rm PBH}$ and $m$. The increase of size with mass roughly follows the distribution of observed UFDGs (whose most likely half-light radius and mass are represented by crosses), which indicates that they form from Poisson rather than CDM fluctuations. The figure also indicates the cluster masses for which $F_{\crm}( M_{\crm} ) = 0.9$ and $0.99$ (vertical blue lines), showing that almost all Poisson fluctuations have collapsed into such clusters on the relevant mass scale.

\begin{figure*}[t]
	\includegraphics[width=0.9\textwidth]{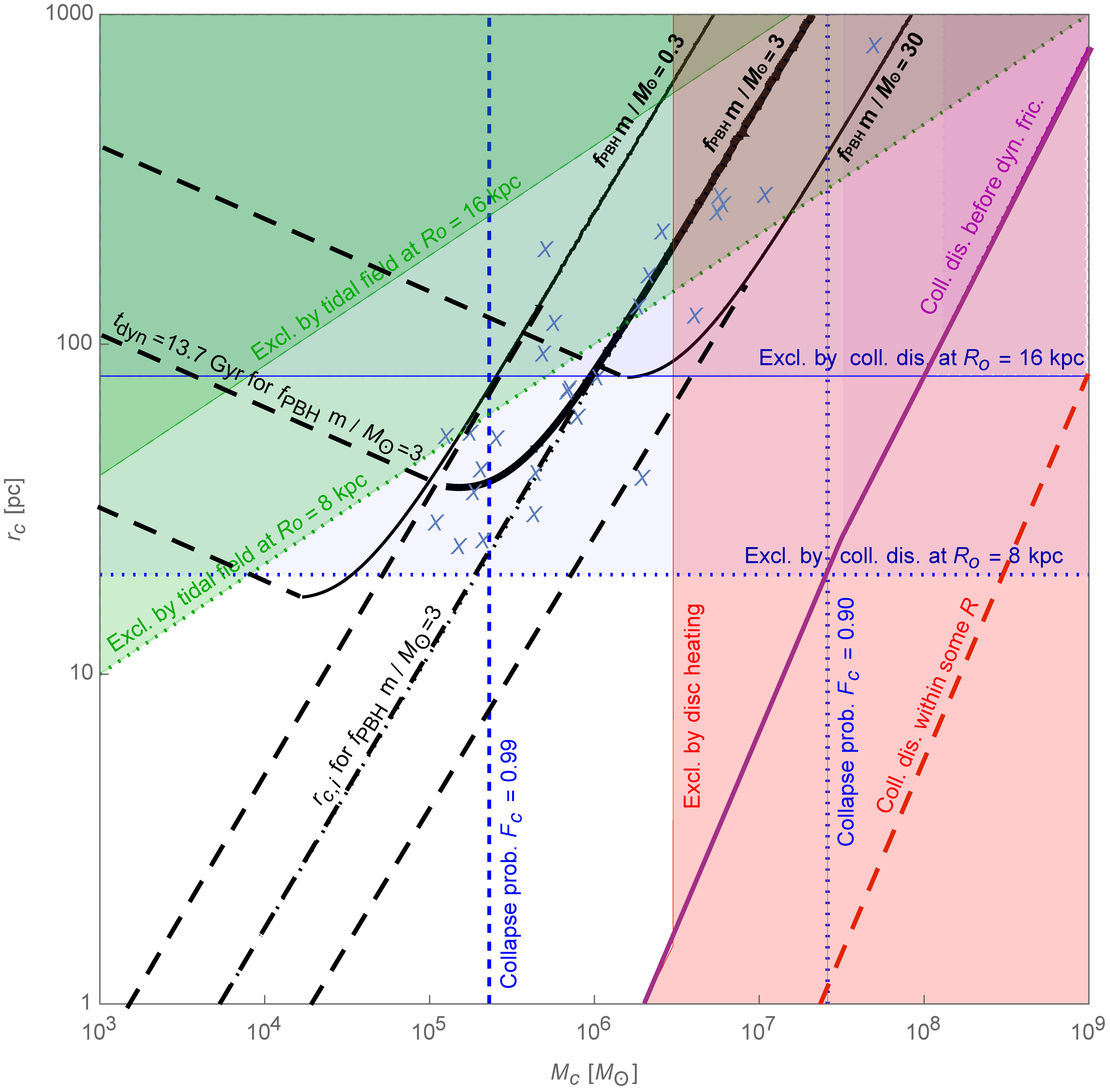}
	\caption{Regions in the $( M_{\crm},\.r_{\crm} )$ 
		plane of PBH clusters induced by Poisson 
		fluctuations that are excluded or favoured by 
		(1)	Galactic disc heating (red region), 
		(2)	disruption by the Galactic tidal field 
			(green region) for a Galactocentric 
			distance of $8$\,kpc (dotted line) and 
			$16$\,kpc (solid line), 
		(3)	cluster collisions (blue lines), 
		(4)	dynamical friction (purple line), 
		(5)	dynamical heating (upper dashed black 
		    lines).
		The expected size of clusters in a scenario 
		with $\fPBH\.m = 3\.\Msun$ is 
		indicated by the thick solid black line that 
		takes into account the dilution of sub-clusters 
		inside more massive host clusters. For 
		comparison, the cases $\fPBH\.m = 0.3\.\Msun$ 
		and $30\.\Msun$ are also represented (thin 
		black lines). The lower dashed black lines show 
		the initial cluster size from 
		Eq.~\eqref{eq:rM}. The vertical blue lines 
		indicate the cluster mass for which there is a 
		probability of collapse into halos 
		$F_{\crm} = 0.99$ and $F_{\crm} = 0.90$ (dashed 
		and dotted lines, respectively) for Poisson 
		fluctuations in the Press--Schechter formalism. 
		The crosses indicate actual observations of 
		UFDGs, with $r_{\crm} \approx r_{1/2}$ 
		and $M_{\crm}$ computed from the virial 
		theorem, for the best-fit values of the 
		half-light-radius $r_{1/2}$ and velocity 
		dispersion.
		}
\label{fig:rcvsmc}
\end{figure*}
{\color{white}.}

\subsection{Dynamical Friction}
\label{sec:Dynamical-Friction}

Encounters between either SMBHs or clusters in our halo and lower mass objects (such as the spheroid stars) will steal energy from them so that they drift towards the Galactic centre. One can show that the clusters will reach the nucleus within the age of the Milky Way$\vphantom{_{_{_{_{_{_{_{_{_{_{_{_{_{_{_{_{_{_{_{_{_{_{_{_{_{_{_{_{_{_{_{_{_{_{_{_{_{_{_{_{_{_{_{_{_{_{_{_{_{}}}}}}}}}}}}}}}}}}}}}}}}}}}}}}}}}}}}}}}}}}}}}}}}}}$ ($t_{\grm} \approx 1.5 \times 10^{10}\,$yr) from inside a Galactocentric radius 
\begin{equation}
	R_{\rm df}
		\approx
				\begin{cases}
					1.3\,{\rm kpc}
					\left(
						\frac{ M_{\crm} }
						{ 10^{6}\mspace{1mu}\Msun }
					\right)^{\!10/21}
						& ( R_{\rm df} < R_{1} )
					\\[2.0mm]
					2.2\,{\rm kpc}
					\left(
						\frac{ M_{\crm} }
						{ 10^{6}\mspace{1mu}\Msun }
					\right)^{\!2/3}
						& ( R_{1} < R_{\rm df} < 
							R_{2} )
					\\[2.0mm]
					1.4\,{\rm kpc}
					\left(
						\frac{ M_{\crm} }
						{ 10^{6}\mspace{1mu}\Msun }
					\right)^{\!1/2}
						& ( R_{\rm df} > R_{2} ) 
					\, ,
				\end{cases}
				\label{eq:DF}
\end{equation}
where $R_{1} \approx 800\,$pc is the core radius of the spheroid stars and $R_{2}$ is the radius beyond which the halo density dominates the spheroid density. The first two expressions were derived by Carr \& Lacey~\cite{1987ApJ...316...23C} but this neglects the dynamical friction effects of the halo objects themselves if they have an extended mass function. The third expression allows for this and comes from an analysis by Carr \& Sakellariadou~\cite{Carr:1997cn}. In the SMBH case, the mass dragged into the Galactic nucleus will exceed the observational upper limit of $4 \times 10^{6}\.\Msun$~\cite{Gillessen:2008qv} unless one imposes an upper limit on the SMBH mass of around $10^{4}\.( R_{\crm} / 2 {\rm\,kpc} )^{2}\.\Msun$. This assumes that the halo density is constant within some core radius $R_{\crm}$ and falls off like $R^{-2}$ outside this. However, this conclusion can be avoided for halo clusters since, as discussed below, they may be disrupted by collisions before dynamical friction can drag them into the nucleus.

\subsection{Collisional and Tidal Disruption}
\label{Collisional-and-Tidal-Disruption}

The proposal that galactic halos comprise SMBHs is precluded by dynamical and accretion constraints. However, it is still possible that halos are made of supermassive clusters of black holes of more modest mass. Indeed, many years ago Carr \& Lacey~\cite{1987ApJ...316...23C} suggested that the dark matter could be clusters of around $10^{6}\.\Msun$ comprising massive compact halo objects (MACHOs), this being the Jeans mass at decoupling. The accretion luminosity is then reduced by the number of objects per cluster, since the Bondi accretion rate and hence luminosity scale as the square of the black hole mass (see Section~\ref{sec:PBH-Accretion:-Pregalactic-Era}), and the dynamical limits can be circumvented if the clusters are disrupted by collisions before the dynamical effects become operative. Although Carr \& Lacey did not discuss the nature of the MACHOs or assume any particular mass for them, it is natural to assume that they are PBHs if they provide the dark matter because other MACHO candidates would have to derive from baryons and could only comprise $20\%$ of the dark matter. A simplified discussion of the dynamical constraints on the mass and radius of such clusters follows below and the results are summarised in Fig.~\ref{fig:rcvsmc}. Most of the constraints are for the Milky Way halo but the arguments could also be applied to other galaxies.

Collisions between clusters will lead to their disruption if the velocity dispersion of the halo objects $\sigma_{\hrm}$ exceeds their internal dispersion $\sigma_{\crm}$, as shown by the broken red line in Fig.~\ref{fig:rcvsmc}. Outside the halo core ($R > R_{\crm}$), this requires $r_{\crm} > 0.05\,( M_{\crm}/10^{6}\.\Msun )\,$pc. The time scale for collisional disruption is constant within the core and increases further outward, so clusters will only be disrupted if the time scale at $R_{\crm}$ is less than the lifetime of the Milky Way. This requires $r_{\crm} > 0.9\,f_{\hrm}^{-1} ( R_{\crm} / 2\,{\rm kpc} )^{2} \,{\rm pc}$ and clusters will be disrupted within a Galactocentric distance $2.1\.f_{\hrm}^{1/2}\,( r_{\crm} / {\rm pc} )^{1/2}\,$kpc, where $f_{\hrm}$ is the fraction of the halo mass in the clusters. Clusters avoid dynamical friction providing this distance exceeds the value $R_{\rm df}$ given by Eq.~\eqref{eq:DF}. An upper limit on $r_{\crm}$ comes from requiring that the clusters survive at our own Galactocentric distance,
\vs{-1mm}
\begin{equation}
	r_{\crm}
		<
				20 \.f_{\hrm}^{-1}
				( R_{\orm} / 8\,\rm{kpc} )^{2}
				\,{\rm pc}\,.
\end{equation}
We have represented this line in Fig.~\ref{fig:rcvsmc}, together with the limit obtained at a Galactocentric distance of $16$\,kpc, and interestingly this scale corresponds to the minimum size of PBH clusters if $\fPBH\,m /\Msun \sim \Ocal( 1 )$. It is therefore expected that in the inner part of the Milky Way halo, PBH clusters in this case would have been disrupted by collisions. They nevertheless remain stable at the larger Galactocentric distance that is typical of UFDGs.

Clusters will also be destroyed by the Galactic tidal field at $R_{\orm}$ unless
\be 
	r_{\crm}
		<
				100\.( M_{\crm} / 10^{6}\.\Msun )^{1/3}
				( R_{\orm} / 8\,\rm{kpc} )^{2} 
				\,{\rm pc}\,.
\end{equation}
As for collisional disruption, we also show this limit in Fig.~\ref{fig:rcvsmc} for a Galactocentric distance of $16\,$kpc. Again, the predictions for PBH clusters cross these regions, implying that they should have been destroyed by the tidal field in the inner part of halo. But at larger Galactocentric distances, corresponding to UFDGs, they should remain stable. On the other hand, clusters will be destroyed by collisions within the Galactocentric radius at which dynamical friction would drag them into the nucleus, given by Eq.~\eqref{eq:DF}, only if
\vs{-2mm}
\begin{equation}
	r_{\crm}
		> 
				\begin{cases}
					0.5\.f_{\hrm}^{-1}\!
					\left(
						\frac{ M_{\crm} }
						{ 10^{6}\mspace{1mu}\Msun }
					\right)^{\!4/3}
					{\rm pc}
						& ( M_{\crm} < M_{2} )
					\\[2.0mm]
					0.8 \.f_{\hrm}^{-1}\!
					\left(
						\frac{ M_{\crm} }
						{ 10^{6}\mspace{1mu}\Msun }
					\right)
					{\rm pc}
						& ( M_{\crm} > M_{2} )
					\, ,
				\end{cases}
\end{equation}
where $M_{2}$ is the mass for which Eq.~\eqref{eq:DF} gives $R_{\rm df} = R_{2}$. This condition is necessary in order to obviate the upper limit on their mass which would apply for SMBHs and corresponds to the region above the purple line in Fig.~\ref{fig:rcvsmc}.

\subsection{PBH Clustering from Non-Gaussian Effects during Inflation}
\label{sec:PBH-Clustering-from-Non--Gaussian-Effects-during-Inflation}

Non-Gaussian corrections to the matter density distribution may provide an important additional component to the initial PBH clustering during the radiation-dominated era. This arises because quantum diffusion during inflation creates high exponential tails in the probability distribution function for the density contrast~\cite{Ezquiaga:2019ftu}. These non-Gaussian effects increase the probability of finding another PBH near a given one, with successive waves of enhanced curvature fluctuations inevitably generating dense clusters. Their size and mass depend very strongly on the form of the non-Gaussian tails and is still under investigation. In particular, plateau-type potentials like those of Critical Higgs Inflation~\cite{Ezquiaga:2017fvi, Garcia-Bellido:2017mdw} naturally generate such tails, and significant clustering is also expected in the thermal-history scenario~\cite{Carr:2019kxo}. These non-Gaussian exponential tails also have important consequences for the early formation of massive structures like galaxies and clusters at high redshift~\cite{Ezquiaga:2022qpw}. In fact, the {\it James Webb SpaceTelescope} (JWST) has detected massive galaxies at $13\,\text{--}\,16$, while the {\it Dark Energy Spectroscopic Instrument} (DESI) and other telescopes found SMBHs at the centres of galaxies up to $z \sim 9$, whereas according to the $\Lambda$CDM model there should be none. These exponential tails not only affect the large-scale structure evolution during the matter-dominated era, but also early PBH formation and clustering. Although primordial non-Gaussianity at CMB scales is usually characterised by a local $f_{\rm NL}$ parameter, the non-Gaussian exponential tails arising from fundamental quantum diffusion during inflation~\cite{Ezquiaga:2019ftu} cannot be described in terms of perturbative parameters like $f_{\rm NL}$ or $g_{\rm NL}$~\cite{Ezquiaga:2022qpw}. Indeed, they are always positively squeezed with a large kurtosis and higher-order cumulants, enhancing the probability of collapse to PBHs.

The effect of large exponential tails in the distribution of density contrasts on the evolution of PBH clusters comprising some or all of the dark matter has been explored using $N$-body simulations of large-sale structure, these also including gas and other components~\cite{Tkachev:2020uin, 2021Univ....7...18T}. However, Ref.~\cite{Ali-Haimoud:2018dau} claims that one does not expect clustering on very small scales beyond what is expected from a Poisson distribution. Moreover, the usual analysis of the formation of binaries during the radiation-dominated era~\cite{Nakamura:1997sm} assumes Poisson statistics, with binary formation occurring due to (rare) three-body encounters. Nevertheless, the additional clustering effect discussed here modifies both the formation of binaries and their disruption due to energetic close encounters. This prevents the early development of a stochastic GW background.
\newpage

\section{Lensing Evidence}
\label{sec:Lensing-Evidence}

\noindent This section discusses the evidence for PBHs from gravitational lensing. We distinguish between four effects, all of which have been observed and are relevant to the detection of PBHs: 
	(1) multiple images of quasars produced by massive
		objects such as galaxies (macrolensing); 
	(2) brightness variations in isolated quasars 
		induced by much less massive objects, 
		typically in the stellar mass range 
		(microlensing), including microlensing by an 
		intergalactic population of compact bodies, 
		or individual stars or halo objects within 
		galaxies; 
	(3) microlensing by such objects of the images of a 
		macrolensed quasar; 
	(4) microlensing of individual nearby stars by 
		other compact bodies.

The idea that a cosmological distribution of compact bodies might be detected by observing the microlensing of distant point-like light sources such as quasars or stars has a long history. In an early paper~\cite{1973ApJ...185..397P}, Press \& Gunn proposed that a cosmological distribution of compact bodies could be detected from the amplification of quasar light due to microlensing, an idea which was put on a more practical basis~\cite{1979Natur.282..561C} by Chang \& Refsdal. A similar idea was suggested by Gott~\cite{1981ApJ...243..140G}, who proposed that this method could determine whether the dark halos of galaxies were composed of compact bodies. The first practical suggestion as to how such bodies might be detected was put forward by Paczy{\'n}ski~\cite{1986ApJ...304....1P}. His idea was to determine whether the dark halo of the Milky Way was composed of stellar-mass bodies by monitoring several million stars in the Magellanic Clouds to look for the distinctive signature of microlensing events. This was put to the test by the MACHO collaboration~\cite{1996ApJ...461...84A, 2000ApJ...541..270A} and the experiment was a resounding success. We discuss the results in more detail below.

Attempts to detect a cosmological distribution of compact bodies from the microlensing of quasars were more problematic. The difficulty was that should compact bodies make up a sizeable fraction of dark matter, a web of caustics would be formed, leading to the erratic variation of the quasar light as it traversed the complex amplification pattern. The resulting light curve would not have an easily identifiable shape and would be hard to distinguish from an intrinsic variation in the quasar accretion disc. There are, nonetheless, unmistakable features associated with caustic crossings which have been observed and cannot plausibly be associated with intrinsic variations~\cite{Hawkins:1998fk}. We discuss some typical examples below. Despite these difficulties, Hawkins~\cite{1993Natur.366..242H} presented the first evidence for a cosmological distribution of PBHs from the microlensing of quasar light curves. Further lines of evidence were later summarised~\cite{Hawkins:2011qz} in a more comprehensive case for PBHs as dark matter.

The difficulty of distinguishing microlensing events from intrinsic variations can largely be removed by focussing on multiply-lensed quasar systems. In these systems intrinsic variations in the quasar light will show up in each image, with a small time lag corresponding to differences in light travel time to the observer. However, it is well known that the individual quasar images can vary independently, often by large amounts. This is generally accepted to be the result of microlensing by compact bodies along the line of sight to each image, which might suggest a cosmological distribution of lenses. However, since a massive galaxy will inevitably lie along the line of sight to a gravitationally lensed quasar system, it is also possible that stars in the galaxy act as the lenses. The problem with this idea is that there are a number of systems where the microlensed quasar images lie well beyond the stellar population of the lensing galaxy~\cite{Hawkins:2020zie}, so the stellar lenses must either reside in the dark halo or more generally along the line of sight to the quasar images. The most striking example is a cluster lens~\cite{Hawkins:2020rqu} which we discuss below. If such a population of compact bodies exists, it must make up a significant fraction of the dark matter and should be detectable by microlensing along more general lines of sight to quasars. This possibility was investigated in early work by Schneider~\cite{1993A&A...279....1S} and has more recently been updated by Hawkins~\cite{2022MNRAS.tmp..849H} to reveal a cosmological population of compact bodies associated with the dark matter distribution, most plausibly identified as PBHs.

There are a number of other puzzling features of quasar light curves which seem to point towards microlensing variations. These include lack of correlation between emission line and continuum variation, lack of evidence for time dilation in samples of quasar light curves, statistical symmetry of the variations, and colour effects which seem inconsistent with accretion-disc instabilities~\cite{Hawkins:2011qz}. For the remainder of this section we shall focus on the most conclusive evidence for microlensing by a cosmological distribution of compact bodies. Should this be confirmed, then at present the only plausible candidates would appear to be PBHs~\cite{Hawkins:2020zie}.

\subsection{Galactic Halo}
\label{sec:Galactic-Halo}

\begin{figure}[t]
	\centering
	\vs{-2mm}
	\includegraphics[width=0.475\textwidth]{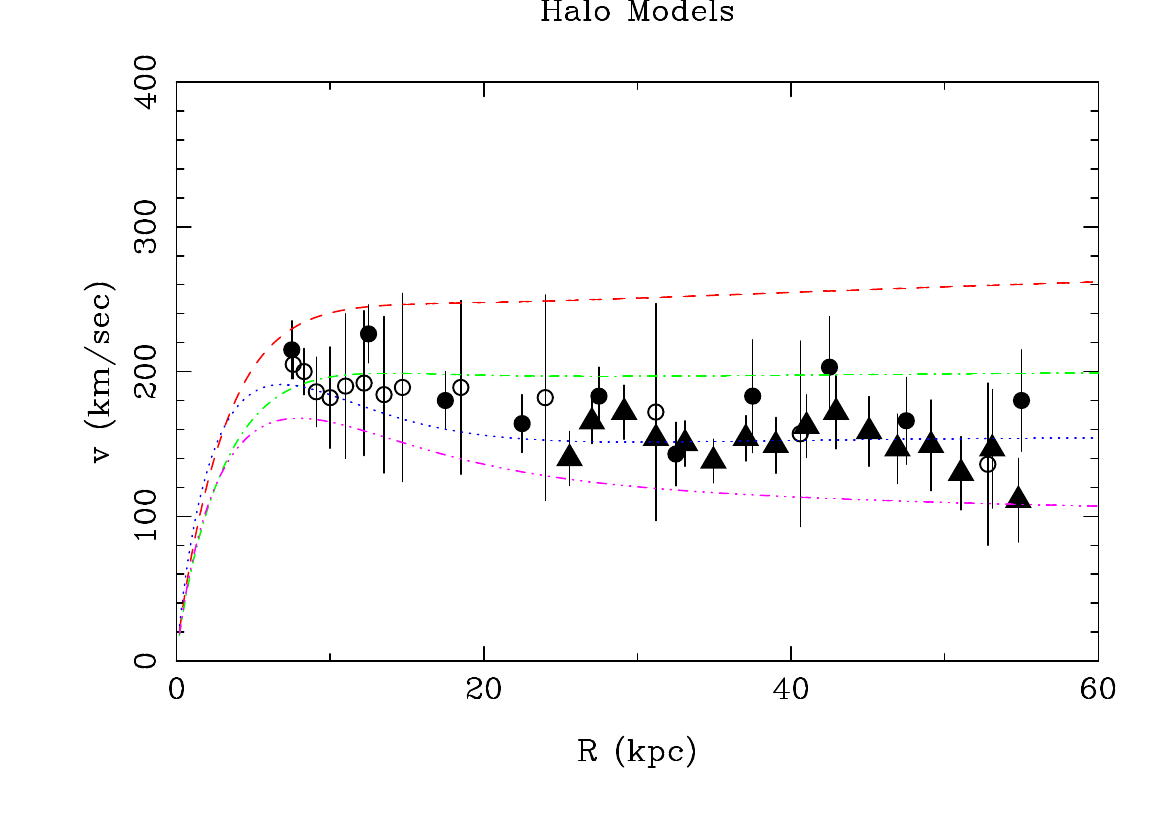}
	\includegraphics[width=0.475\textwidth]{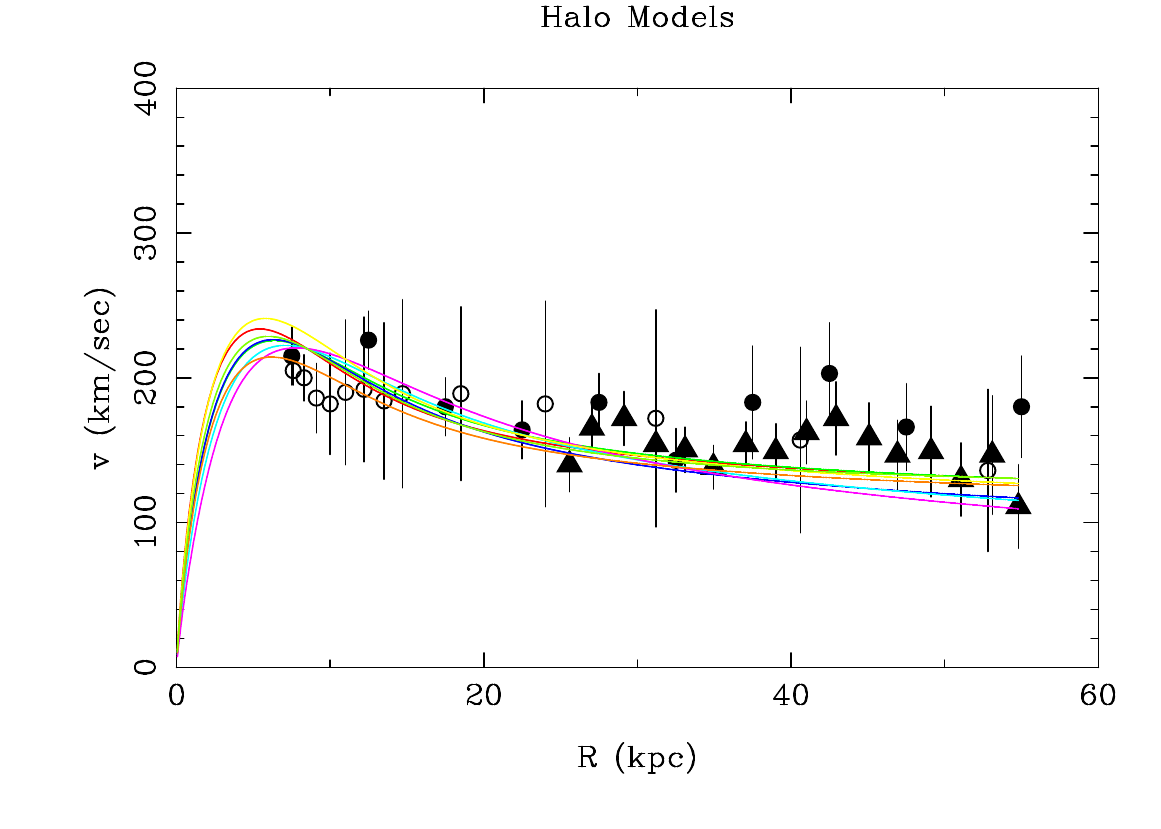}
	\includegraphics[width=0.94\textwidth]{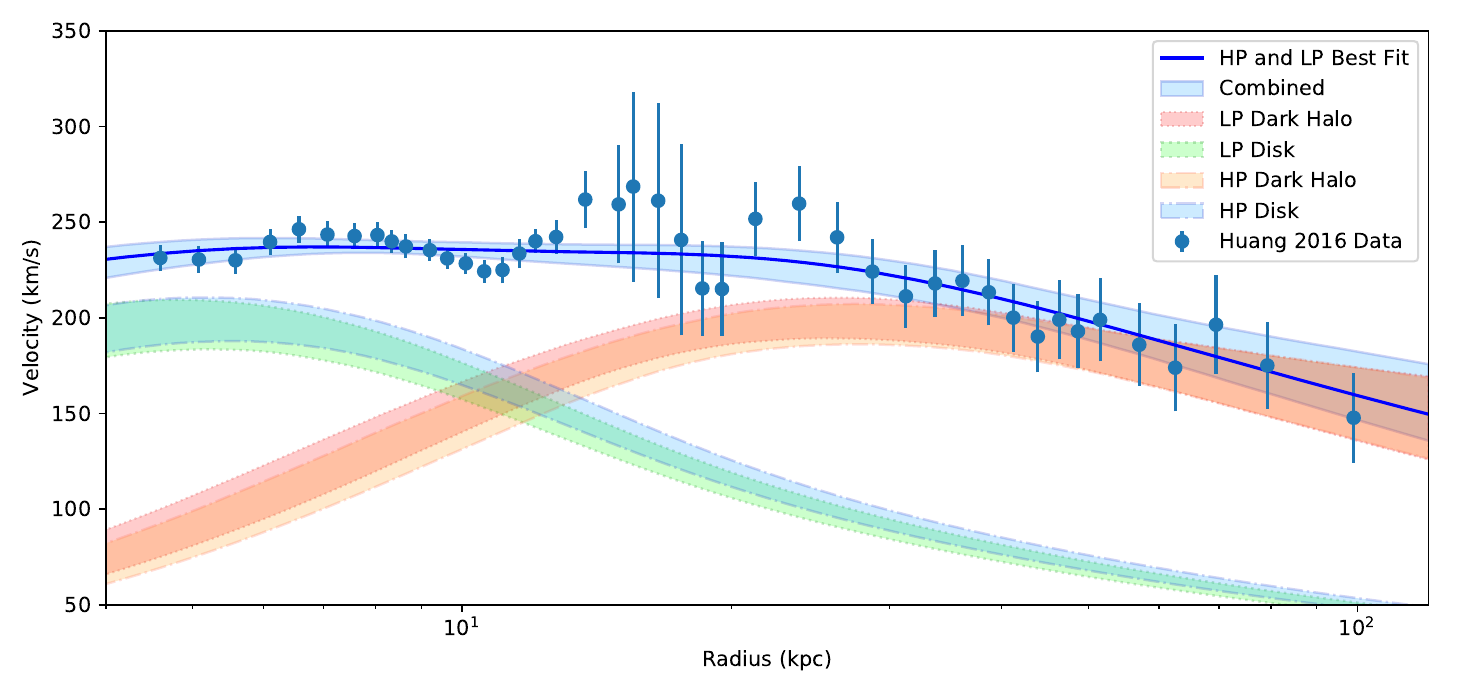}
	\vs{-1mm}
	\caption{
		{\it Upper panel:}
			Rotation curves for models of the Milky Way 
			out to $60\,$kpc from
			Ref.~\cite{Hawkins:2015uja}.
			The observed rotation curves are shown as 
			filled circles~\cite{SDSS:2008nmx}, open 
			circles~\cite{2013PASJ...65..118S} 
			and filled 
			triangles~\cite{2014ApJ...785...63B}.
			In the {\it top-left panel}, the red 
			(dashed), green (dot-dash),	blue (dotted) 
			and purple (dash-dot-dot-dot) curves are 
			for	Models B, S, F and E, respectively, 
			from Ref.~\cite{1996ApJ...461...84A}.
			In the {\it top-right} panel the coloured 
			curves are for power-law models from 
			Ref.~\cite{1994MNRAS.267L..21E}, 
			consistent with the optical depth 
			measurements of 
			Ref.~\cite{2000ApJ...541..270A}.
		{\it Lower panel:}
			More recent data for the rotation curve,
			extending to $100\,$kpc, from 
			Ref.~\cite{Calcino:2018mwh}. This shows the 
			best-fit plots and $2\sigma$ uncertainties 
			(shaded) for the power-law model with fixed 
			bulge model included in the combined 
			constraints. The difference between the 
			combined rotation curve for the low prior 
			(LP) and high prior (HP) cases is not 
			discernible at the resolution of this plot, 
			so a combined shaded region is shown.
			The HP case makes a slightly more massive 
			disk and therefore a slightly lighter dark 
			halo in the inner regions.
		}
	\label{fig:fig1}
\end{figure}

The proposal of Paczy{\'n}ski~\cite{1986ApJ...304....1P} to search for compact objects in the halo of the Milky Way by looking for microlensing of stars in the Magellanic Clouds was based on the fact that the half-crossing time of the magnification curve depends on three quantities: the mass of the deflector $M_{\Drm}$, the reduced distance $1 / d = 1 / d_{\rm OD} + 1 / d_{\rm DS}$ (where $d_{\rm OD}$ and $d_{\rm DS}$ are the observer-deflector and deflector-source distances, respectively) and the velocity $v$ of the lens across the line of sight. This gives
\begin{equation}
	\Delta t_{1/2}
		=
				\frac{ \sqrt{4\.G M_{\Drm}
				\mspace{1mu}d} }{ v }
		\approx
				0.1\,{\rm yr}
				\left(
					\frac{ M_{\Drm} }{\.\Msun }
				\right)^{\!1/2}
				\left(
					\frac{ d }{ 10\,{\rm kpc} }
				\right)^{\!1/2}
				\left(
					\frac{ v }{ 200\,{\rm km/s} }
				\right)^{\!-1}
				\, .
\end{equation}

The results of the survey for compact bodies in the Milky Way by the MACHO collaboration~\cite{1996ApJ...461...84A, 2000ApJ...541..270A} have been widely and perhaps simplistically interpreted as ruling out the possibility that stellar-mass compact bodies make up the dark matter. However, we shall argue that the experiment puts no useful constraints on this proposal and, on the contrary, provides the first positive detections of stellar-mass PBHs. The number of microlensing events detected was around $15$, depending on the details of the selection criteria. This is far larger than can be attributed to known stellar populations in the Milky Way or Magellanic Clouds~\cite{2000ApJ...541..270A}. The population of compact bodies making up the lenses was an obvious dark matter candidate, but the MACHO collaboration stopped short of making such a claim because they concluded on the basis of their preferred `heavy' halo model that there were too few compact bodies to account for the Milky Way rotation curve. Indeed, they concluded that such objects could make up at most $40\%$ of the halo. This limit was, among other things, based on the assumption of a narrow mass range for the lenses making up the dark matter. However, if the lenses have an extended mass function, as predicted for PBHs~\cite{Carr:2019kxo}, it is to be expected that the dark matter fraction at around $1\.\Msun$ will be significantly less than $1$.

When the first MACHO results were published, little was known about the rotation curve of the Milky Way and a variety of models were considered, as illustrated in the top left panel of Fig.~\ref{fig:fig1}. However, most of these models were rejected as being unrealistic and it is now clear that their Model F (blue dotted line) fits recent rotation curve measures well and is consistent with their observed microlensing optical depth. The collaboration assumed an essentially flat rotation curve, in line with the consensus at the time on the statistics of the rotation curves of nearby galaxies. This implied a massive halo and considerably more microlensing events than were observed, so compact bodies were rejected as candidates to account for all of the halo dark matter~\cite{2000ApJ...541..270A}, although the possibility of a partial contribution was left open. Since then, much work has been done on the dynamics of the Milky Way and it is clear that the rotation curve is not flat but declines, as illustrated by the top panels of Fig.~\ref{fig:fig1}, which are taken from Ref.~\cite{Hawkins:2015uja}. The observations shown there are from Refs.~\cite{SDSS:2008nmx, 2013PASJ...65..118S, 2014ApJ...785...63B} and can easily be fitted by halo models~\cite{1994MNRAS.267L..21E} with a microlensing optical depth consistent with the MACHO observations. More recently, the declining rotation curve has been confirmed by Ref.~\cite{Calcino:2018mwh}, shown in the lower panel of Fig.~\ref{fig:fig1}, and Refs.~\cite{Ou:2023adg,Jiao:2023aci}. However, there is still inertia in the literature in accepting the reality of the new measures of halo kinematics. For example, in a recent paper~\cite{2022A&A...664A.106B} the old superseded `S' model for the halo is still used without any reference or challenge to the papers which show it to be inconsistent with recent observations of the Milky Way rotation curve. This inevitably appears to support earlier constraints on the population of PBHs in the Milky Way halo. Further doubt has been cast on the validity of the MACHO constraints by considering the effects of a broad mass function, as mentioned above.

The microlensing rate measured by MACHO was not confirmed by the EROS~\cite{Tisserand:2006zx} and OGLE~\cite{Wyrzykowski:2011tr} collaborations. Comparison of the results of these three groups is complicated, as EROS and OGLE failed to observe the excess of microlensing events detected by MACHO. The statistical incompatibility of the MACHO and EROS results has been the subject of extensive discussion~\cite{Tisserand:2006zx, Moniez:2010zt}. One of the most obvious differences between the MACHO and EROS experiments was their respective use of faint and bright star samples. Only two of the $17$ MACHO microlensing candidates were bright enough to be included in the EROS Bright-Stars sample~\cite{Tisserand:2006zx} but these were not detected in the standard EROS analysis for well-understood reasons. Although there are advantages in the restriction of the EROS analysis to bright stars, it would have resulted in problems associated with the resolution of stellar discs. In this sense, the MACHO detections can be seen as a lower limit to the total population of compact bodies.

This raises the question of the reliability of the estimates of detection efficiency, which are essential to setting limits on the frequency of microlensing events. There are a number of problems with these estimates, as discussed in detail by Hawkins~\cite{Hawkins:2015uja}. These include gaps in the data due to weather and mechanical failure, the consequences of variable seeing on overlapping images in the dense star fields of the Magellanic Clouds, the elimination of true variable stars, and the distortion of light curves due to the presence of binary or planetary companions. A related issue is the question of self-lensing by stars in the Magellanic Clouds rather than the Galactic halo. The MACHO fields covered only very dense regions of the LMC, which may increase the probability of self-lensing by stars there~\cite{Sahu:1994pj}. This effect is hard to quantify, given lack of detailed knowledge of the structure of the Magellanic Clouds, but subsequent analysis~\cite{Mancini:2004pb} has shown that self-lensing can only have a small effect on the observed optical depth to microlensing. A more detailed comparison between MACHO, EROS and OGLE events is beyond this review and their compatibility has recently been discussed elsewhere~\cite{2022A&A...664A.106B}.

To summarise, the large excess of events associated with solar-mass lenses detected by the MACHO project cannot be attributed to the stellar population of the Milky Way and at present the only plausible candidates are PBHs~\cite{Hawkins:2020zie}. This possibility was considered by the MACHO collaboration~\cite{2001ApJ...550L.169A}, but rejected on the basis of their constraints. However, both MACHO and EROS only show microlensing constraints for monochromatic mass functions, while extended mass functions{\,---\,}further discussed in Section~\ref{sec:Thermal--History-Scenario}{\,---\,}allow compatibility of the present microlensing constraints with a halo made entirely of PBHs~\cite{Calcino:2018mwh}. The reason is that the new Milky Way halo profile is no longer understood to be spherical, but triaxial, and the integrated optical depth due to dark matter in the form of PBHs along the line of sight between Earth and the LMC has been reduced by almost an order of magnitude. This, together with the observed MACHO microlensing events, implies that around $40\%$ of the Milky Way halo could be in the form of stellar mass PBHs~\cite{Hawkins:2020zie}. An extended mass function then allows for a peak at around a few solar masses, with smaller contributions at the few percent level, at smaller and larger masses. This would be inhomogeneously populated (\eg~in the Galactic Center, dwarf spheroidals etc.), giving $100\%$ of the dark matter in the form of PBHs.
\newpage

\subsection{Galactic Bulge}
\label{sec:Galactic-Bulge}

Another approach for the detection of PBHs in the Milky Way has involved the monitoring of large numbers of stars in the Galactic bulge by the OGLE collaboration~\cite{2016MNRAS.458.3012W}. These observations are particularly focussed on the detection of white dwarf, neutron star and black hole lensing events, using {\it Gaia} parallax measurements and brightness changes to estimate the mass of the lenses. This approach has led to the detection of a small sample of `dark' lenses, where the contribution of light from the lensing body is negligible~\cite{2020A&A...636A..20W}. This sample is of particular interest as it appears to bridge the mass range between $2\.\Msun$ and $5\.\Msun$, where there is a gap in the distribution of X-ray binaries~\cite{2010ApJ...725.1918O}. This gap may be explained by `natal kicks', where the tangential velocity of the remnant is given a boost at birth, thus distorting the measurement of the lens mass. However, a more significant possibility~\cite{2020A&A...636A..20W} is that the mass gap is populated by PBHs which would not be part of X-ray binary systems.

\begin{figure}[t]
	\centering
	\includegraphics[width=0.77\textwidth]{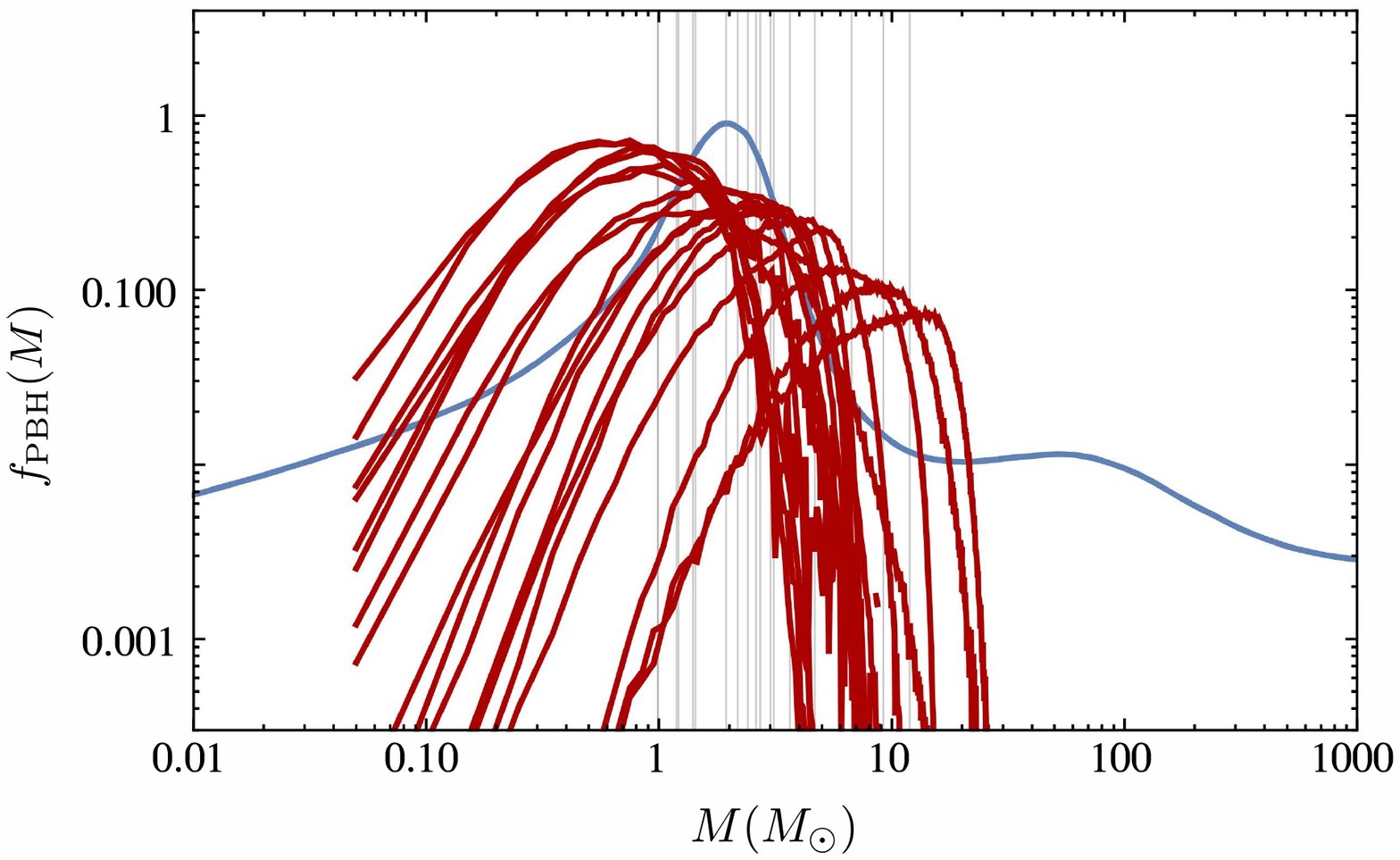}
	\caption{
		Mass distribution of OGLE/{\it Gaia} events 
		(red curves) overlaid onto the model of 	
		Ref.~\cite{Carr:2019kxo}. The mode of the 
		distribution for each microlensing event 
		is shown by grey vertical lines.
		Figure taken from 
		Ref.~\cite{Garcia-Bellido:2019tvz}.
		\vs{-2mm}
		}
	\label{fig:figX}
\end{figure}

\begin{figure}[t]
	\centering
	\includegraphics[width=0.45\textwidth]{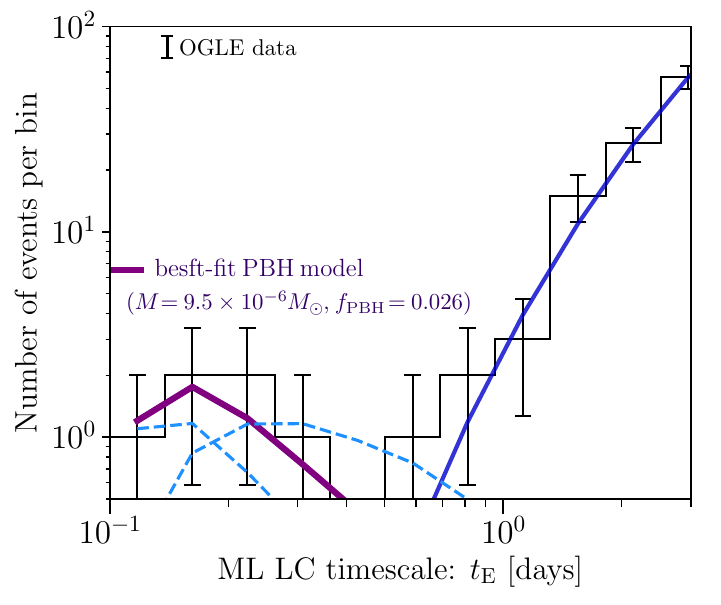}
	\includegraphics[width=0.45\textwidth]{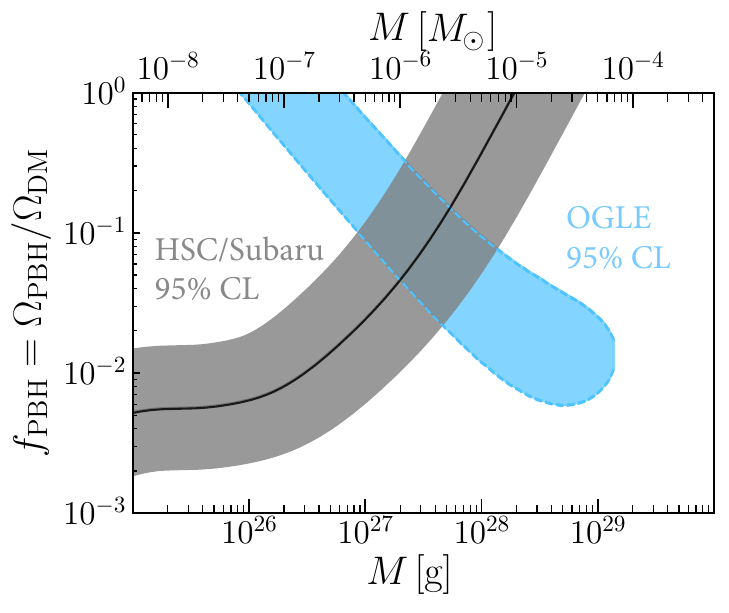}
	\caption{
		{\it Left panel:}
			Number of observed microlensing events as a 
			function of lightcurve (LC) timescale in 
			days from 
			Ref.~\cite{Niikura:2019kqi}. The 
			bold-purple solid line indicates the 
			best-fit PBH model (with 
			$M = 9.5 \times 10^{-6}\.\Msun$ and 
			$\fPBH = 0.026$) for explaining the 
			distribution of six ultrashort-timescale 
			OGLE events.
		{\it Right panel:}\hs{-0.5mm}
			$95\%$ confidence level (CL) region for
			PBH\hs{-0.25mm} abundance (blue-shaded 
			region), assuming that the six observed 
			ultrashort-timescale events in the OGLE 
			data are due to PBHs with a monochromatic 
			mass function. Also shown (grey band) is 
			the $95\%$ CL prediction for the abundance 
			inferred from the HSC/Subaru 
			observations.
		From Refs.~\cite{Niikura:2019kqi} 
		and~\cite{2019NatAs...3..524N}.
		}
	\label{fig:OGLEPLAN}
\end{figure}

An unexpected indication of the existence of Earth-mass PBHs has come from a microlensing survey for unbound planetary mass bodies~\cite{2017Natur.548..183M}. Although the survey was primarily focussed on Jupiter-mass planets with an expected microlensing timescale of $1\,\text{--}\,2$ days, a small number of very short events with timescales less than half a day were also detected. This implies a lens with around an Earth mass and raises the question of the identity of such objects. This was addressed by Niikura {\it et al.}~\cite{Niikura:2019kqi}, who used an accurate model for the stellar population of the Galactic bulge and disc, including their remnants, to account for the observed microlensing events in this mass range. The most obvious explanation for the six Earth-mass outliers is that they are detached or free floating planets, resulting from gravitational interactions between stellar systems. However, there is no continuity of detections between these outliers and events associated with low-mass stars or brown dwarfs. Figure~\ref{fig:figX} indicates the mass distribution of the OGLE/{\it Gaia} events. Figure~\ref{fig:OGLEPLAN} shows their distribution as a function of timescale (left panel) and also the values of $\fPBH( M )$ required to explain the data for a monochromatic mass function (blue band of right panel).

\subsection{Andromeda Galaxy}
\label{sec:Andromeda-Galaxy}

The search for the microlensing of stars in Andromeda is more challenging because the microlensed star is unresolved and just one of many stars contributing to each CCD pixel. The pixel-lensing technique, which involves subtracting a reference image from a target image taken at a different epoch, must therefore be used. This has been attempted by several groups. In particular, the POINT-AGAPE collaboration detected six microlensing events~\cite{CalchiNovati:2005cd}. They argued that this was more than could be expected from self-lensing alone and concluded that $20\%$ of the halo mass in the direction of Andromeda could be PBHs in the range $0.5\,\text{--}\,1\.\Msun$. This supports our conclusion above. In addition, they discussed a likely binary microlensing candidate with caustic crossing, whose location supported the conclusion that they were detecting a MACHO signal in the direction of Andromeda.

\begin{figure}[t]
	\centering
	\includegraphics[width=0.65\textwidth]{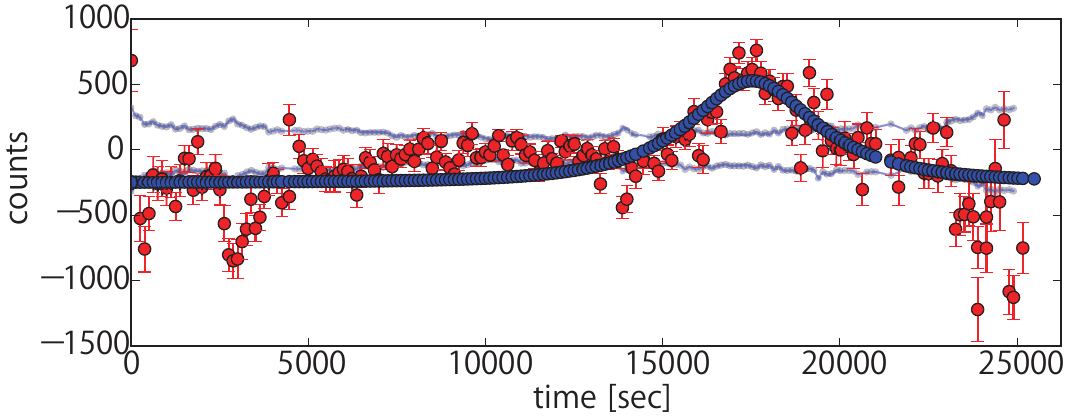}
	\caption{
		Best-fit microlensing model (blue curve) for 
		the light curve of the single pixel-lensing 
		event reported by Niikura {\it et al.}
		~\cite{Niikura:2017zjd}. Error bars denote 
		photometric errors in the brightness 
		measurement in the image difference at each 
		epoch. From Ref.~\cite{Niikura:2017zjd}.
		}
	\label{fig:Subaru/HSC}
\end{figure}

\begin{figure}[t]
	\centering
	\includegraphics[width=0.65\textwidth]{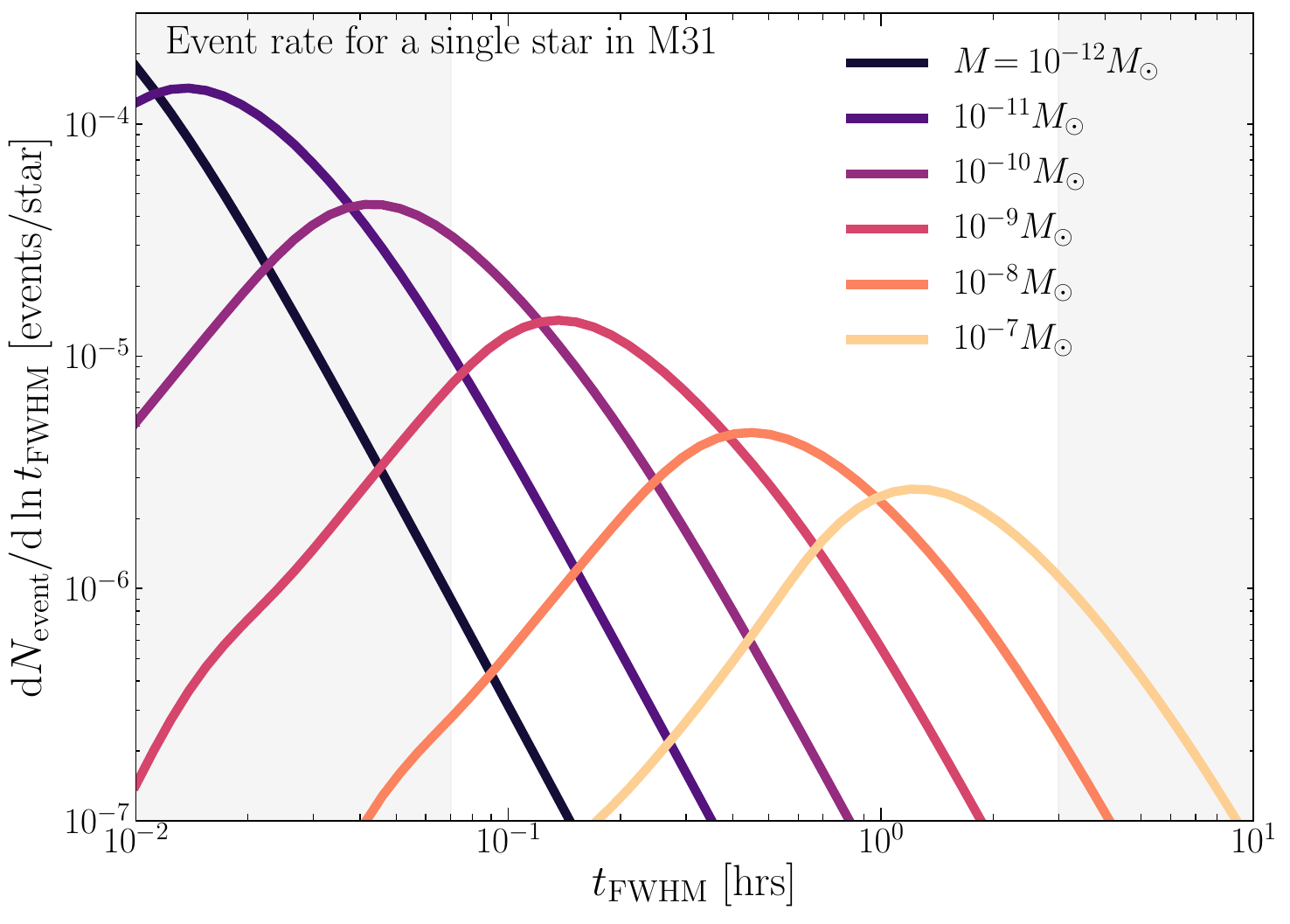}
	\caption{
		The expected differential number of 
		microlensing events per logarithmic interval of 
		the full-width-at-half-maximum timescale 
		$t_{\rm FWHM}$ for a single star in Andromeda.
		Solid lines correspond to a monochromatic PBH 
		mass function with $\fPBH = 1$.	The unshaded 
		band is where the utilised data has the highest 
		sensitivity to the lightcurves 
		($t_{\rm FWHM} \simeq 0.07\,\text{--}\,3~{\rm hours}$).
		From Ref.~\cite{Niikura:2017zjd}.
		}
	\label{fig:eventrate}
\end{figure}

Niikura {\it et al.}~\cite{Niikura:2017zjd} carried out a dense-cadence, seven-hour observation of Andromeda with the Subaru Hyper Suprime-Camera (HSC) and then used the pixel-lensing technique to search for microlensing of stars by PBHs lying in the halo of the Milky Way or Andromeda. The pointing was towards the central region of Andromeda and they reported the observation of a single microlensing event by a compact body with mass in the range $10^{-11}\,\text{--}\,10^{-5}\.\Msun$. A number of $15,571$ candidate variable stars were extracted from the difference images, and this was subsequently reduced to $66$ by selecting only those with a best-fit reduced $\chi$-squared value and a light curve which is symmetrical around the peak. Further visual inspection left a single but solid candidate which passed all applied criteria and its light-curve is shown in Fig.~\ref{fig:Subaru/HSC}. Assuming that the number of microlensing events follows a Poisson distribution, and given the $10^{8}$ monitored stars, Niikura {\it et al.}~estimated that about a thousand microlensing would have been observed during the seven-hour observation period if the PBHs constituted all the dark matter. Here, it is assumed that the PBHs are not subject to any additional (\ie~non-Poissonian) clustering and have a monochromatic mass spectrum. Figure~\ref{fig:eventrate} shows the expected event rate for various PBH masses as a function of the full-width-at-half-maximum timescale $t_{\rm FWHM}$. Translating this into estimates for $\fPBH$ yields the grey band in the right panel of Fig.~\ref{fig:OGLEPLAN}.

\subsection{Caustic Crossings of Quasars}
\label{sec:Caustic-Crossings-of-Quasars}

\begin{figure}[t]
	\centering
	\includegraphics[width=1\textwidth]{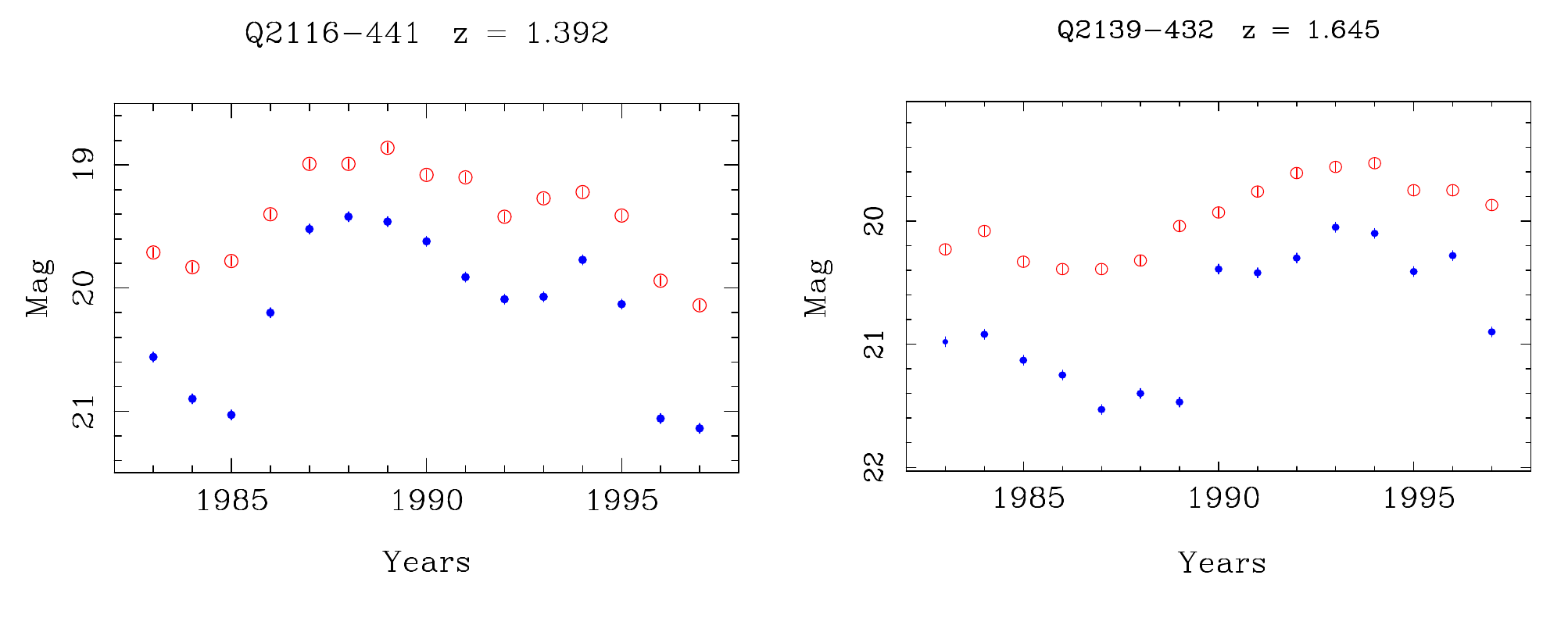}
	\vs{-9mm}
	\caption{
		Light curves for two quasars showing all the 
		characteristics expected of caustic crossing 
		events. Filled and open circles are blue and 
		red passband measures, respectively. Error bars 
		are based on measured photometric errors.
		From Ref.~\cite{Hawkins:1998fk}.
		}
	\label{fig:fig2}
\end{figure}

If solar-mass PBHs make up a significant fraction of dark matter, then the detection of microlensing in the light curves of distant compact sources such as quasars becomes an interesting possibility. The frequency of such events is a function of the cosmological density of the lenses and the redshift of the quasar~\cite{1992ApJ...393....3F}, while the structure of the resulting light curve depends upon the relative sizes of the quasar accretion disc and the Einstein radius of the lenses, as well as the optical depth to microlensing $\tau$. From simulations it is clear that it will be hard to unambiguously identify most microlensing events in quasar light curves. A large source size will smear out the characteristic features and make them hard to distinguish from intrinsic changes in quasar luminosity. Some interesting candidates for isolated microlensing events have recently been proposed~\cite{10.1093/mnras/staa972} but such featureless light curves are hard to pin down as the result of microlensing, as opposed to intrinsic variations in the quasar luminosity.

The most distinctive features of microlensing light-curves are caustic crossings. For low optical depth ($\tau \lesssim 0.3$), microlensing events tend to be isolated, with no readily identifiable features. However, as $\tau$ increases, caustic features emerge in the amplification pattern, with characteristic shapes and structures. These are well illustrated by simulations~\cite{1986A&A...166...36K}, where the effect of increasing source size can also be seen. Although the identification of caustic crossings in quasar light curves would be an unambiguous signature of a component of dark matter in the form of PBHs, the difficulty of disentangling microlensing events from the distorting effects of intrinsic variations means that uncontaminated detections of caustic crossings will be quite rare. However, in a large enough sample of light curves these difficulties can be largely overcome~\cite{Hawkins:1998fk}.
\newpage

The requirements for detecting caustic crossings in quasar light-curves are quite demanding. For a population of solar-mass lenses, the typical time for a compact source to cross a caustic is $10$ to $20$ years. To define the structure of the resulting light curve, it must be sampled at least once a year with an accuracy of $\sim 0.05$ mags, ideally for around $20$ years. Given that for a cosmological distribution of lenses well-defined caustic crossings are quite rare, a large sample of light curves will be required for a reasonable chance to find an unambiguous detection of a caustic crossing. These requirements are largely satisfied in the large-scale quasar monitoring programme undertaken by the UK $1.2\,$m Schmidt telescope from 1975 to 2002 in ESO/SERC Field $287$. The survey comprises yearly sampled light curves covering up to 25 years for over a thousand quasars~\cite{Madgwick:2003bd}, and has proved to be an excellent database for the detection of caustic-crossing events.

In Fig.~\ref{fig:fig2} we show two examples of caustic crossing events from the Field 287 survey, where the blue and red symbols are for blue and red passband magnitudes, respectively. The two light curves were selected as they clearly show the characteristic distinguishing features of caustic crossing events. A well-known feature of the gravitational lensing of a point source is that changes in amplitude are achromatic. The question of whether intrinsic quasar variations are achromatic is complicated. Low-luminosity active galactic nuclei are known to become bluer as they brighten~\cite{Peterson:2002nh}, whereas more luminous quasars can show little or no colour change as they vary~\cite{2018ApJ...862..123M}. This means that for point sources, colour changes are not a useful way of distinguishing between microlensing and intrinsic variations. A more interesting situation arises where the source is resolved, or larger than the Einstein radii of the lenses, and there is a radial colour gradient, as might be expected in a quasar accretion disc. In this case the blue and red passband light curves can have very different and characteristic structures. The colour gradient effectively results in the source of blue light being more compact than the red source. When a quasar crosses a caustic in the amplification pattern of the lenses, the blue light curve can show the unresolved features of the caustic, while in the red light curve the features are smoothed out from the partial resolution of the red source. An additional effect is that as the quasar accretion disc traverses the caustic pattern, it will start to amplify the red light before the blue, and will continue to do so for longer. The light curves in Fig.~\ref{fig:fig2} clearly show all these features. The blue curves show the characteristic cusps associated with caustic crossings, which are smoothed out in the red curves. Amplification of the quasar light first appears in the red curves, followed by sharper and larger rises and falls in the blue. The red light then continues to decline to the base level.

The features in the two light curves in Fig.~\ref{fig:fig2} are commonly seen in the Field 287 sample, although in some cases apparently distorted by intrinsic variations. It is not easy to see how accretion-disc instabilities can produce these symmetrical feature, as instabilities originating at the centre should start simultaneously in the red and blue light curves, and persist longer in the red as the disc cools with increasing radius. This asymmetry can be seen in individual light curves~\cite{Hawkins:1998fk}, and statistically in the form of time lags~\cite{2018ApJ...862..123M}. The identification of caustic crossings in quasar light curves implies a cosmological distribution of lenses. This rules out stars as lens candidates, which would have negligible probability of forming a universal cosmic web. Given the large optical depth to microlensing required to produce caustics, the lenses must make up a large part of the dark matter, and in this case the only plausible candidates are PBHs~\cite{Hawkins:2020zie}.

Another example of caustic crossing has come from Kelly {\it et al.}~\cite{Kelly:2017fps}, who have reported the magnification of an individual star by the galaxy cluster MACSJ1149.5+2223 at $z = 1.49$. They claim that dark matter subhalos or massive compact objects may account for this and that PBHs of $30\.\Msun$ with $3\%$ of the dark matter could potentially explain some features of the observations. Diego {\it et al.}~\cite{Diego:2017drh} have used simulations of ray-tracing to study this case; they allow for both stars and PBHs, and also include the effects of the intracluster medium. They point out that this phenomenon should be widespread and that HST monitoring of such arcs could yield the mass spectrum of the compact objects. Venumadhav {\it et al.}~\cite{Venumadhav:2017pps} have studied this in more detail and find that the exquisite sensitivity of caustic-crossing events to the granularity of the lens-mass distribution makes them ideal probes of dark matter components, such as PBHs.
\newpage

\subsection{Multiply Lensed Quasars}
\label{sec:Multiply-Lensed-Quasars}

Perhaps the most convincing detections of stellar-mass PBHs come from the analysis of microlensing events in the light curves of multiply lensed quasar systems. The first detection of such a gravitational lens~\cite{1979Natur.279..381W} showed the quasar Q0957+561 split into two separate images by a massive galaxy along the line of sight. A photometric monitoring programme revealed that small brightness changes in one image were repeated in the second image around a year later. This confirmed the identification of the system as a gravitational lens, but also revealed that the images varied in brightness independently of each other~\cite{1991AJ....101..813S}. This result has been universally interpreted as the result of microlensing by a population of solar-mass bodies along the line of sight to the quasar images. The different light trajectories from each of the quasar images to the observer result in independent amplification patterns, and hence light curves, for each image. Somewhat surprisingly, it was widely assumed from the outset~\cite{1990AJ....100.1771S, 1991MNRAS.251..698F, 1993ApJ...404..455K} that the microlenses were in the lensing galaxy, and it was only some time later~\cite{Hawkins:2011qz, Hawkins:2020zie, Hawkins:2020rqu} that evidence was presented that the microlenses were part of a cosmological distribution of stellar-mass compact bodies making up the dark matter. In this case, they would be situated in the halo of the lensing galaxy, or in other galaxy halos, along the line of sight to the quasar images.

Since the identification of microlensing features in lensed quasar systems, there have been a number of attempts to demonstrate that these can be attributed to stars in the halo of the lensing galaxy and to put limits on any population of non-stellar compact bodies~\cite{2009ApJ...706.1451M, 2012ApJ...744..111P}. Using a large sample of lensed quasar systems, a maximum likelihood analysis is used to estimate the most likely percentage of compact bodies in the galaxy halos, consistent with the observed microlensing statistics of the quasar images. The results of this type of analysis generally imply consistency of the observed microlensing with a halo population of compact bodies of about $10\%$, similar to the expected stellar component.

The problem with this approach is that the sample of gravitational lenses is dominated by systems where due to the configuration of the quasar and lens redshift and alignment, and the mass and compactness of the lensing galaxy, the quasar images lie deeply embedded in the stellar component of lensing galaxy. In these cases, there is no question that the observed microlensing can be attributed to the stellar population. There are however a few cases where the microlensed quasar images lie well outside the stellar distribution of the lensing galaxy, where the probability of microlensing by stars is negligible, and in this case the observed microlensing must be seen as the detection of a large population of non-stellar compact bodies. This point is in fact implicit in the separate analysis of the few wide-separation systems in the sample of Pooley {\it et al.}~\cite{2012ApJ...744..111P}, but not commented upon by them. The issue is that in their sample of wide-separation lens systems, there is a much higher likelihood than in the sample as a whole that the dark matter component of the lensing galaxies is made up of compact objects~\cite{Hawkins:2020zie}.

In order to clarify this issue, a more direct approach was taken by Hawkins~\cite{Hawkins:2020zie} where, instead of treating the stellar population as a free parameter, the surface mass density in stars was measured from observations of the lensing galaxy, thus enabling a direct estimation of the probability of microlensing by the stellar population. The optical depth to microlensing $\tau$ is given by:
\begin{equation}
	\tau
		=
				\kappa_{*} / ( 1 - \kappa_{\crm} )
				\, ,
				\label{eq:eqn1}
\end{equation}
where $\kappa_{*}$ is the convergence due to the stars (the surface mass density in units of some appropriate critical density) and $\kappa_{\crm}$ is the convergence due to smoothly distributed dark matter, which for most lens systems is close to $0.5$ in the vicinity of the quasar images. Precise values can be obtained from mass modelling of the system. For large values of $\tau$ a network of caustics is formed, resulting in a non-linear amplification pattern for the source, but for low values, $\tau$ is effectively the probability that the quasar image will be significantly microlensed.

\begin{figure}[t]
	\centering
	\vs{-2mm}
	\includegraphics[width=0.72\textwidth]{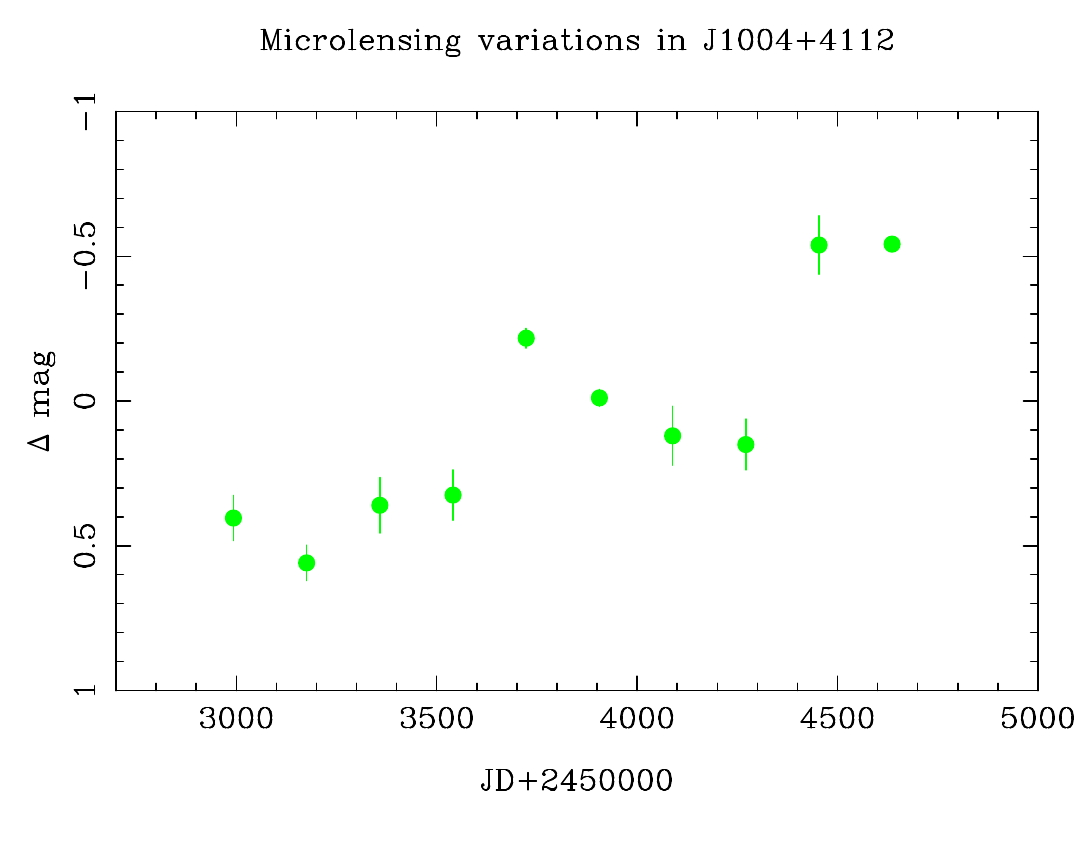}
	\vs{-3mm}
	\caption{
		Difference light curve for images A and C of 
		the gravitationally lensed quasar SDSS 
		J1004+4112. The differential variations of the 
		images are attributed to microlensing.
		From Ref.~\cite{Hawkins:2020rqu}.
		\vs{3mm}
		}
	\label{fig:fig3}
\end{figure}

In order to test the possibility that in wide separation systems the observed microlensing can be attributed to the stellar population of the lensing galaxy, a sample of four of the widest separation lens systems was selected where the quasar images appeared to lie well clear of the distribution of star light~\cite{Hawkins:2020zie}. As for most gravitationally lensed quasars, the images in this sample were strongly microlensed, in the sense that photometric variations in the individual images were not correlated.

The first step was to measure the surface brightness profile of the lensing galaxy from deep {\it Hubble Space Telescope} (HST) frames in the F160W infrared passband. The surface brightness was then converted to surface mass density using a mass-to-light ratio based on stars in globular clusters, where the mass can be measured directly from velocity dispersions. Working in the infrared made this step more straightforward than in optical bands, as the mass-to-light ratio for the F160W band is only weakly dependent on stellar type. The resulting values of $\kappa_{*}$ were then combined with values for $\kappa_{\crm}$ to derive $\tau$ from Eq.~\eqref{eq:eqn1} and hence the probability of microlensing at the positions of the quasar images. The result of this analysis~\cite{Hawkins:2020zie} is that any individual quasar image has a probability of $\sim 0.05$ of being microlensed by the stellar population of the lensing galaxy. This probability is confirmed by computer simulations where non-linear effects and the amplitude of the light curves are taken into account. The simulations also show that the combined probability of the observed microlensing is of order $10^{-4}$~\cite{Hawkins:2020zie}.
\newpage

Although for most quasar gravitational lens systems, the lens is a massive early type galaxy, where the image separation is typically less than $3\,$arcsec, there is one well known example (SDSS J1004+4112) of a quadruple system where the lens is a massive cluster, with a central dominant galaxy together with a number of smaller galaxies. The separation of the quasar images is $16\,$arcsec, putting them well clear of any detectable starlight, but the most striking feature is that the images are strongly microlensed. Figure~\ref{fig:fig3} shows the difference light curve for two of the quasar images. The amplitude of variation is more than a magnitude, which is very large for any known quasar system, and raises the question of the identity of the lenses.

\begin{figure}[t]
	\centering
	\vs{-4mm}
	\includegraphics[width=0.72\textwidth]{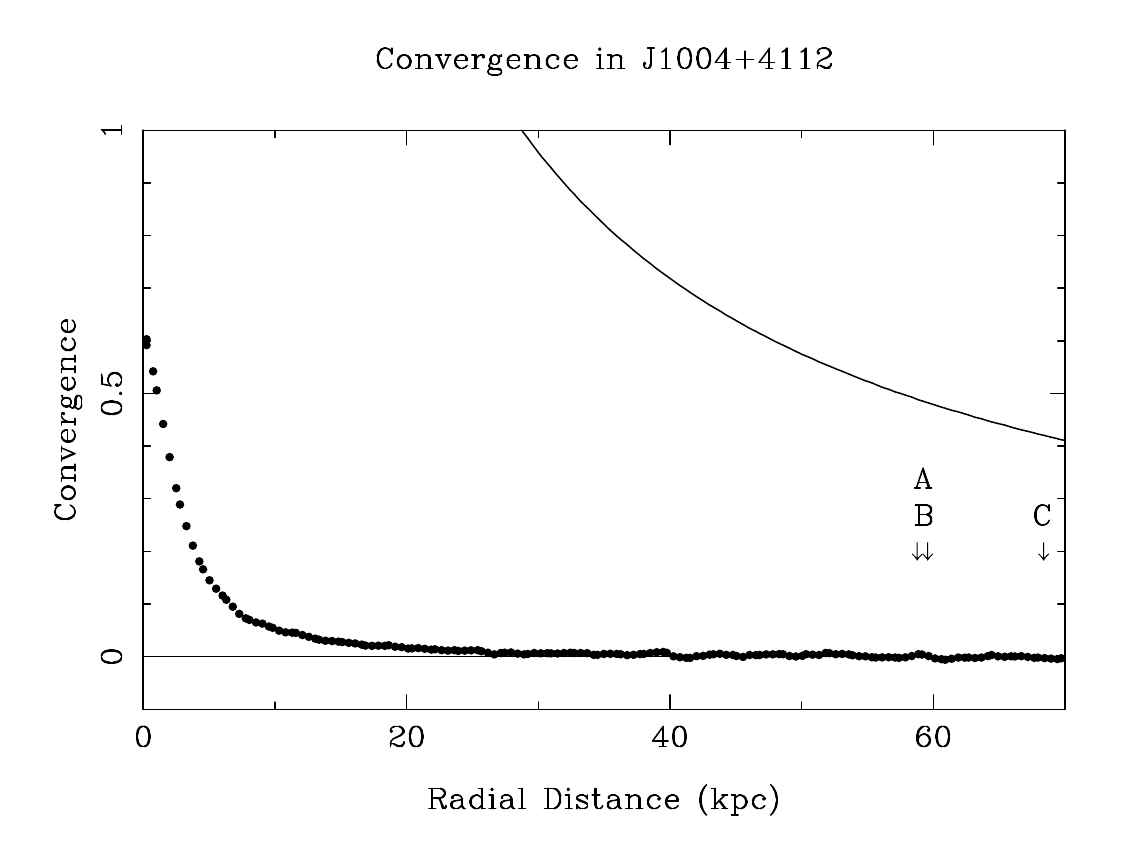}
	\vs{-2mm}
	\caption{
		Convergence as a function of radial distance 
		for the lensing cluster J1004+4112. The solid 
		line shows the total convergence $\kappa$ 
		(\ie~the optical depth $\tau$) from 
		the lensing model and the filled circles show 
		convergence $\kappa_{*}$ for the stellar 
		population of the lensing cluster. The arrows 
		indicate the positions of the quasar images.
		From Ref.~\cite{Hawkins:2020rqu}.
		}
	\label{fig:fig4}
\end{figure}

\begin{figure}[t]
	\centering
	\vs{0.5mm}
	\includegraphics[width=0.72\textwidth]{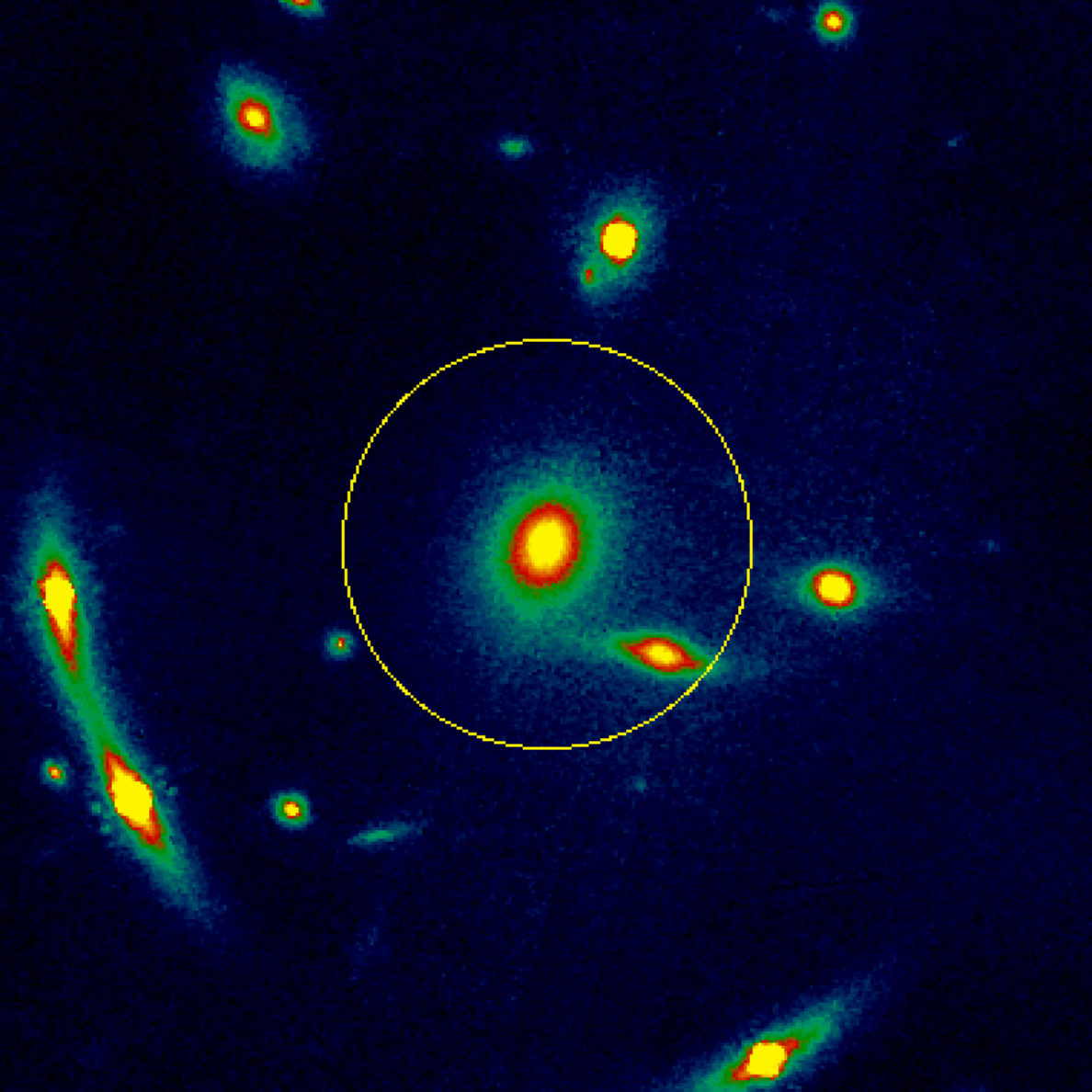}
	\vs{1mm}
	\caption{
		Deep HST frame of the area around the lensed 
		quasar images of the cluster J1004+4112, in the 
		photometric passband F160W, close to the $H$ 
		band. The frame is approximately $15\,$arcsec 
		on a side. The yellow circle defines the point 
		at which starlight from the central cluster 
		galaxy is lost in the background noise at 
		$25\,$kpc. From Ref.~\cite{Hawkins:2020rqu}.
		\vs{2mm}
		}
\label{fig:fig5}
\end{figure}

Figure~\ref{fig:fig4} shows the relation between $\kappa_{*}$ or $\tau$ and the radial distance from the cluster centre, constructed as described above for galaxy lenses. It can be seen that the optical depth to microlensing from stars is effectively negligible beyond $25\,$kpc from the cluster centre. The overall configuration is illustrated in Fig.~\ref{fig:fig5}, showing the cluster galaxies from an HST frame in the F160W passband, close to the infrared $H$-band. The quasar images are visible as part of arcs centred on the cluster. The yellow circle has a radius of $25\,$kpc and shows the point at which convergence from starlight becomes negligible, as illustrated in Fig.~\ref{fig:fig5}. The whole analysis is described in detail in Ref.~\cite{Hawkins:2020rqu}, including the investigation of the possibility of the chance superposition of a small cluster galaxy exactly in the position of one of the quasar images. The overall conclusion of this work is that the probability that the observed microlensing of the quasar images can be attributed to stars is less than $10^{-4}$.

The only way to account for the observed microlensing of the quasar images in SDSS J1004+4112 is to invoke a cosmological distribution of solar-mass compact bodies. For a number of reasons summarised in Ref.~\cite{Hawkins:2020zie}, these bodies can only be PBHs making up a significant fraction of the dark matter. This would appear to be a clear detection of solar-mass PBHs.
\newpage

\begin{figure}[t]
	\vs{-5mm}
	\includegraphics[width=0.8\textwidth]{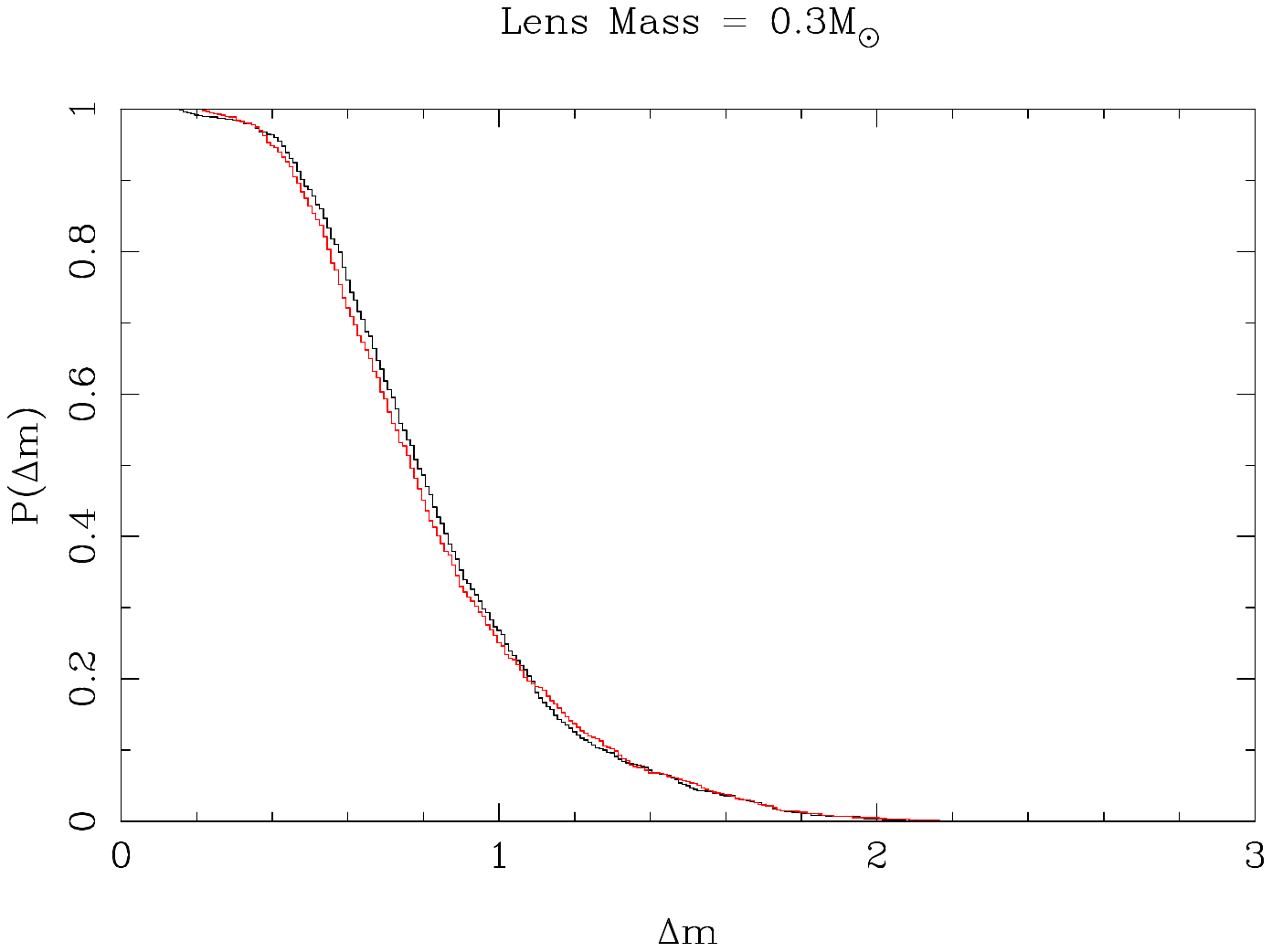}
	\vs{-4mm}
	\caption{
		Cumulative probability $P( \Delta m )$ that a 
		source varies by more than $\Delta m$ 
		magnitudes for the sample of 1033 quasars 
		from the Field 287 survey for lens mass 
		$0.3\.\Msun$. The black curve represents the 
		data, and the red curve shows the cumulative 
		probability for a combination of lognormal 
		intrinsic variation and microlensing 
		amplification, assuming a quasar disc radius of
		$4$ light-days. From 
		Ref.~\cite{2022MNRAS.tmp..849H}.
		\vs{2mm}
		}
	\label{fig:fig6b}
\end{figure}

\subsection{Cosmological Distribution of Primordial Black Holes}
\label{sec:Cosmological-Distribution-of-Primordial-Black-Holes}

The detection of a significant population of PBHs in the halo of the Milky Way and more distant galaxies and clusters suggests that these objects might also betray their presence by microlensing the light from quasars with a cosmological distribution. The detection of caustic crossings discussed in subsection~\ref{sec:Caustic-Crossings-of-Quasars} supports this possibility, but the idea of putting limits on dark matter by comparing the expected microlensing amplifications with observed quasar amplitudes was first suggested by Schneider~\cite{1993A&A...279....1S}, some $30\,$years ago. He found that assuming a point source for the quasar light, the expected variation due to microlensing far exceeded the observed amplitudes for plausible values of lens mass and lens density $\Omega_{\Lrm}$. Subsequent work~\cite{2003A&A...399...23Z} showed that for more realistic values of the quasar disc size, the simulated microlensing amplitudes put no meaningful constraint on $\Omega_{\Lrm}$. The main drawback with this approach is the unknown contribution of intrinsic variations to the microlensing amplifications.
\newpage

In a recent paper~\cite{2022MNRAS.tmp..849H}, the problem of modelling intrinsic quasar variations has been addressed by using luminous quasars where the accretion disc is too large to be significantly microlensed as a template for changes in quasar luminosity. The idea was to simulate the intrinsic variations in luminosity for a large sample of quasars and to apply the amplifications expected for a dark matter component of stellar-mass compact bodies, and then to compare the resulting distribution in amplitudes with the observed distribution from the quasar sample. The optical depth to microlensing was calculated from the redshift of each quasar in a standard $\Lambda$CDM cosmology. Other input parameters, such as the size of the quasar accretion disc and the mass of the lenses, were not optimised but taken from independent and unrelated measurements in the literature. The quasar sample for comparison with the observations comprised 1033 with blue-band lightcurves covering $26\,$years from the UK $1.2\,$m Schmidt telescope~\cite{10.1046/j.1365-8711.2003.06828.x}, and covering a wide range of luminosity and redshift.

Figure~\ref{fig:fig6b} shows the result of repeating Schneider's original test with the new large sample and modelling the intrinsic changes in quasar brightness. The black line shows the cumulative probability of quasar amplitudes from the observed quasar light curves. The red line shows a similar curve for the combination of intrinsic luminosity changes amplified by the expected effects of microlensing due to the optical depth corresponding to each quasar redshift. The parameters of the simulations as described above are not fitted to the data but independently measured, with the result that to understand the distribution of quasar amplitudes it is necessary to include the microlensing effects of a cosmologically-distributed population of stellar-mass compact bodies. These bodies must make up a significant fraction of the dark matter and are most plausibly identified as PBHs.
\newpage

\subsection{Galaxy-Galaxy Strong Lensing}
\label{sec:Galaxy--Galaxy-Strong-Lensing}

\begin{figure}[t]
	\centering
	\vs{-2mm}
	\includegraphics[width = 0.77\textwidth]{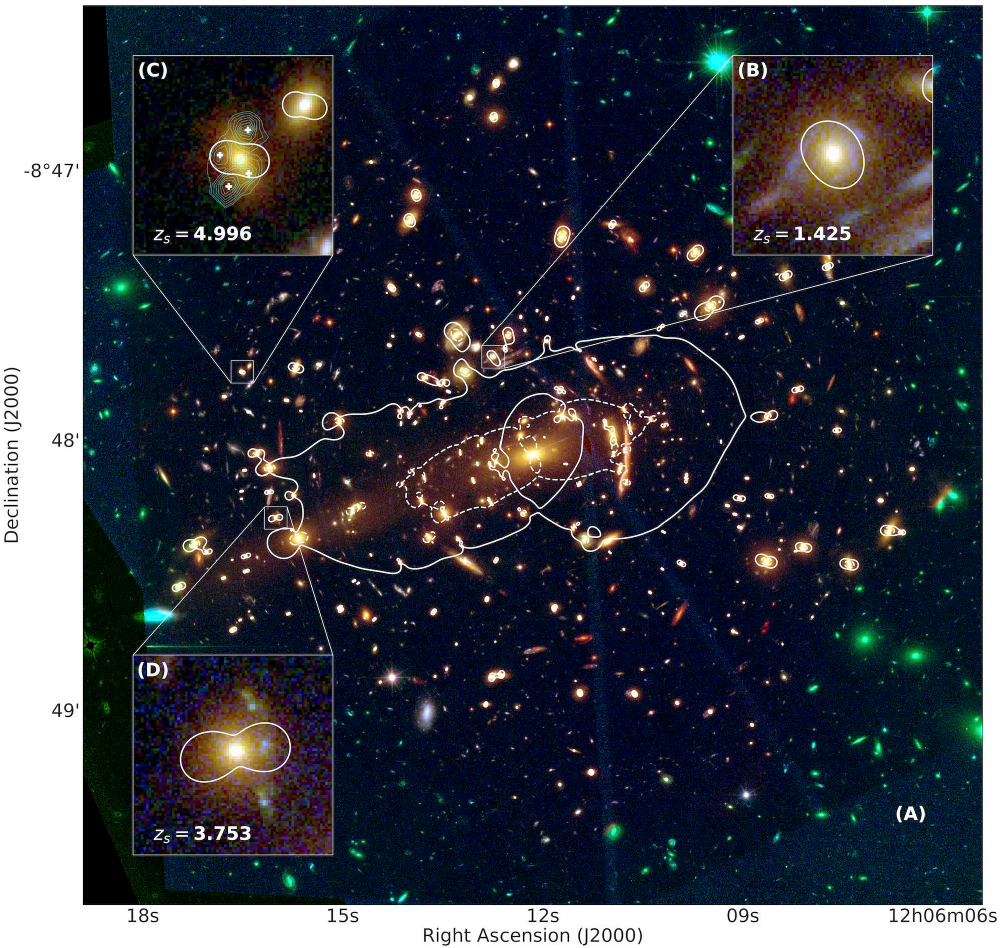}
	\caption{
		Composite HST image of the central region of 
		the galaxy cluster MACSJ1206, where the 
		critical lines of the cluster at source 
		redshifts of $1$ and $7$ are indicated with 
		dashed and solid lines, respectively. The 
		panels (B), (C), and (D) zoom into three 
		galaxy-galaxy strong lensing events.
		From Ref.~\cite{Meneghetti:2020yif}.
		}
	\label{fig:HST-GGSL}
\end{figure}

\begin{figure}[t]
	\centering
	\includegraphics[width = 0.72\textwidth]{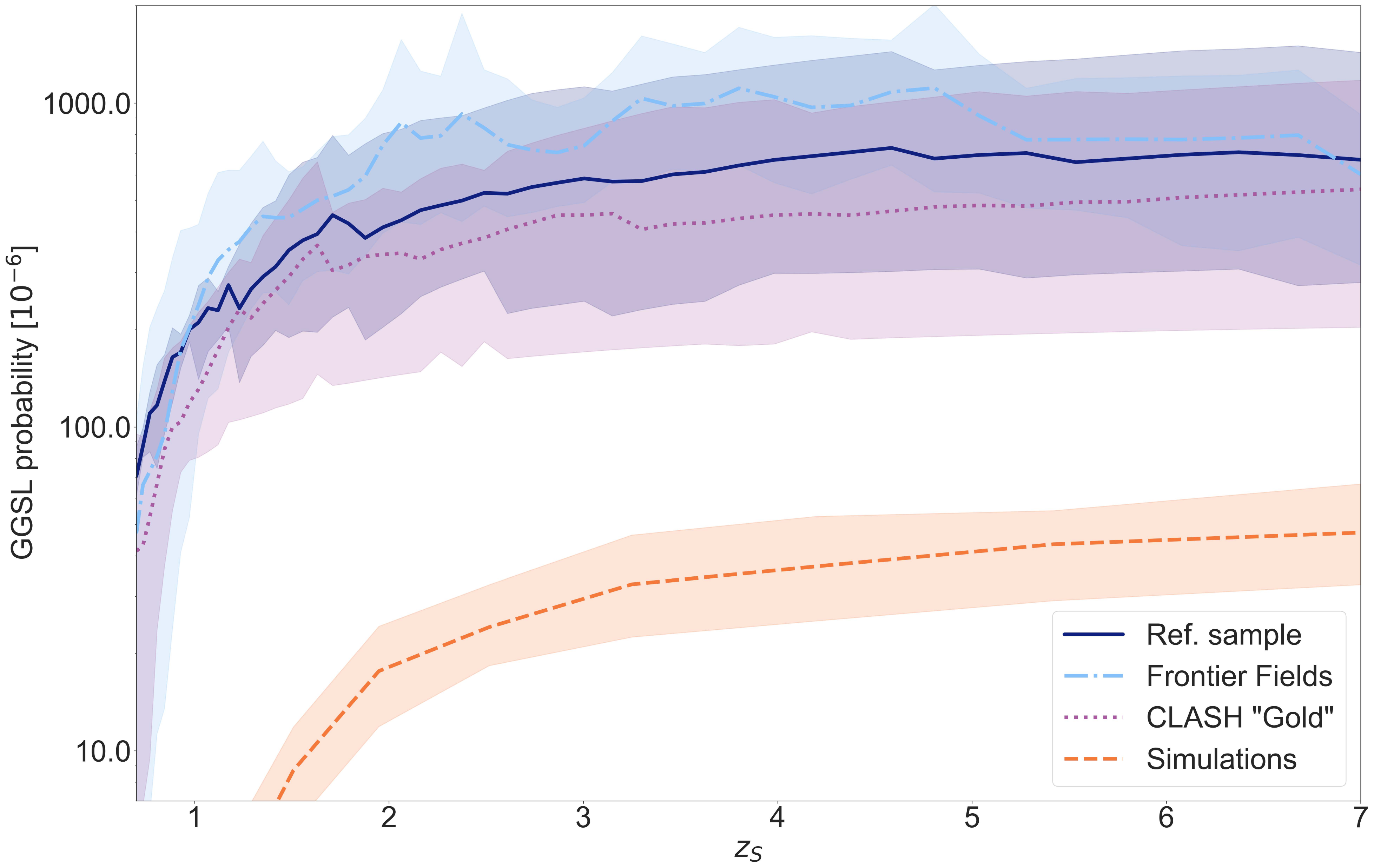}
	\caption{
		Mean observed galaxy-galaxy strong lensing 
		(GGSL) probability (solid blue line) based on 
		three reference samples (coloured bands show 
		the $99.9\%$ CL) versus median simulation 
		results (orange line). From 
		Ref.~\cite{Meneghetti:2020yif}.
		}
	\label{fig:HST-Number-Of-Clusters-Comparison}
\end{figure}

Meneghetti {\it et al.}~\cite{Meneghetti:2020yif} have recently investigated the probability of strong lensing events produced by dark matter substructure using HST data for eleven galaxy clusters. As an example, Fig.~\ref{fig:HST-GGSL} shows a colour-composite image of the galaxy cluster MACSJ1206, together with several of its critical lines, which indicate the boundaries of the inner regions of the cluster near which the most dramatically distorted gravitational arcs occur. This implies significantly more substructure than expected. Indeed, all studied galaxy clusters show an order of magnitude more substructures than predicted by standard CDM simulations (\cf~Fig.~\ref{fig:HST-Number-Of-Clusters-Comparison}). Either systematic issues with simulations or incorrect assumptions about the properties of dark matter could explain these findings.

In contrast to the solutions to other well-known issues, like the ``missing satellite''~\cite{1999ApJ...524L..19M, Klypin:1999uc}, ``cusp-core''~\cite{Flores:1994gz}, and ``too-big-to-fail'' problems~\cite{Boylan-Kolchin:2009ztl, Boylan-Kolchin:2011qkt}, or discrepancies with planes of satellite galaxies~\cite{Muller:2018hks}, which require that observed small satellite galaxies are fewer in number and less dense than expected, the excess in dark matter clumps discussed above requires that subhalos are {\it more} concentrated than predicted by simulations. Strikingly, the clustering of PBHs induced by an extended mass function of the form imprinted by the thermal history of the Universe~\cite{Carr:2019kxo} {\it predicts} the required excess of compact dark substructures. Furthermore, this scenario can address a related issue recently pointed out by Safarzadeh \& Loeb~\cite{2021arXiv210703478S}. They found that several Milky Way satellites, in particular {\it Horologium I} and {\it Tucana II}, are too dense, leading to the conclusion that the formation masses and redshifts of CDM halos are incompatible with their being satellites. These ``too-dense-to-be-satellite systems'' can easily be accounted for by clustered PBHs with an extended mass function.

\subsection{Identity of Lenses}
\label{sec:Identity-of-Lenses}

If dark matter is indeed made up of compact bodies, then there are several strong constraints on the form it can take. Combining measurements of light element abundances with the photon density from the CMB gives a fairly robust limit of $\Omega_{\brm} < 0.03$~\cite{1993ApJS...85..219S} which effectively rules out baryonic objects as dark matter candidates. This of course includes stars and planets as well as later products of stellar evolution such as white dwarfs, neutron stars and stellar black holes. A second constraint comes from the timescale of microlensing events, which constitute much of the evidence for compact bodies as dark matter. Both the MACHO events in the Milky Way halo and the microlensing events observed in multiple quasar systems of the type discussed here point to a lens mass of around a solar mass, with an uncertainty of at least a factor of $10$ either way. This is nonetheless sufficient to rule out a wide range of candidates, which we discuss below. A final constraint is compactness. Again, since much of the evidence for a cosmological distribution of compact bodies comes from microlensing, dark matter bodies must be compact enough to act as lenses in the configuration in which they are observed.

In many respects, stars are the most obvious candidates for the microlenses, and in a review of baryonic dark matter, Carr~\cite{Carr:1994ci} discusses the plausibility of various stellar populations acting as lenses. This is relevant to the microlensing observations described in this paper, as a major issue is the possibility that a significant fraction of the lenses might be stars. In her review of the nature of dark matter, Trimble~\cite{Trimble:1987ee} lists quark nuggets, cosmic strings and PBHs as non-baryonic non-particle candidates. Quark nuggets have some virtue as dark matter candidates~\cite{Alam:1997ij}, and the basic idea has been generalised in a number of ways~\cite{Jacobs:2014yca}, but as their mass is limited to around $10^{-8}\.\Msun$, they cannot be the objects detected in the microlensing experiments. This low-mass constraint also applies to other classes of compact dark matter objects such as {\it N}-MACHOs~\cite{Dvali:2019ewm}. Another recent paper~\cite{Fujikura:2021omw} has investigated the possibility that boson stars might contribute to the dark matter in the Milky Way halo, but the problem here is that such objects are predicted to be insufficiently compact to microlens solar-mass stars. The possibility that cosmic strings might betray their presence by the microlensing of distant quasars has been investigated in some detail by Kuijken {\it et al}.~\cite{2008MNRAS.384..161K}. Their motivation is to establish a method for detecting the presence of cosmic strings, and they estimate that a typical microlensing event would last about $20$ years. This fits in well with other microlensing signatures, but the problem they face is that such events would be extremely rare. The optical depth to microlensing for cosmic strings derived in Ref.~\cite{2008MNRAS.384..161K} is of the order $\tau \sim 10^{-8}$ which completely rules them out as the source of the microlensing seen in multiple quasar systems.

An interesting idea proposed by Walker \& Wardle~\cite{1998ApJ...498L.125W} suggesting that extreme radio scattering events might be caused by cold self-gravitating gas clouds has been investigated by Rafikov \& Draine~\cite{2001ApJ...547..207R} with a view to establishing whether such clouds could be detected by microlensing searches. The clouds are predicted to have masses of around $10^{-3}\.\Msun$ and radii of $\sim 10\,$AU, and would make up a significant fraction of the mass in galaxy halos. Reference~\cite{2001ApJ...547..207R} concluded that although strongly constrained by MACHO microlensing searches, such clouds could not be definitively ruled out as the dark matter component of the Milky Way halo. However, as an explanation for the microlensing seen in multiple quasar systems they can be ruled out on the basis of their low proposed mass, and also the fact that they are essentially baryonic and so cannot make up the dark matter on cosmological scales.

The steady improvement in numerical simulations of the formation of large-scale structure from primordial perturbations culminating in the discovery of a universal profile for dark matter halos~\cite{1996ApJ...462..563N} has raised the question of the nature of their internal structure on small scales. An ambitious programme to solve this problem was undertaken by the {\it Aquarius Project}~\cite{2008MNRAS.391.1685S} which focussed on simulating the evolution of individual galaxy halos with unprecedented resolution. The two main results of this and other similar simulations were that galaxy halos have cuspy dark matter profiles, and that there should be a very large population of dark matter subhalos. In their largest simulation they resolve nearly $300,000$ gravitationally bound subhalos within the virialised region of the halo. At first sight these subhalos might seem attractive candidates for microlensing quasars, but their structure as described in Ref.~\cite{2008MNRAS.391.1685S} imply that they would not be compact enough to act as microlenses for the accretion discs of distant quasars, even if subhalos as small as a solar mass can form. The objects which have received the most attention as candidates for dark matter in the form of compact bodies are PBHs. They satisfy the basic constraints mentioned above in that they are non-baryonic, very compact and can in principle form over a wide range of masses, including those relevant to quasar microlensing, and as things stand they appear to be the only plausible candidates for dark matter in the form of compact bodies.
\newpage

\section{Dynamical and Accretion Evidence}
\label{sec:Dynamical-and-Accretion-Evidence}

\noindent In this section we consider a number of constraints and signatures related to dynamical and accretion effects. Numerous dynamical effects were discussed in Ref.~\cite{Carr:1997cn}, these being associated with the heating of galactic disks~\cite{1985ApJ...299..633L}, effects on dwarf galaxies~\cite{2017PhRvL.119d1102K}, triggering of white dwarf explosions~\cite{2015PhRvD..92f3007G}, disruption of wide binaries~\cite{1985ApJ...290...15B, 1987ApJ...312..367W, Monroy-Rodriguez:2014ula} and globular clusters~\cite{1993ApJ...413L..93M} and star clusters~\cite{2016ApJ...824L..31B}, capture of PBHs by white dwarfs and neutron stars~\cite{Capela:2014qea, 2014JCAP...06..026P}, dynamical friction effect of spheroid stars~\cite{Carr:1997cn}, tidal distortions in galaxies~\cite{1969Natur.224..891V} and the CMB dipole anisotropy~\cite{Carr:1997cn}. Many of them involve the heating or destruction of astronomical systems by the passage of nearby PBHs. If the PBHs have density $\rho$ and velocity dispersion $V$, while the systems have mass $M_{\crm}$, radius $R_{\crm}$, velocity dispersion $V_{\crm}$ and survival time $t_{\Lrm}$, then the constraint has the form~\cite{Carr:1997cn}
\begin{align}
	\label{eq:carsaklim}
	\fPBH( M )
		< 
				\begin{cases}
					M_{\crm}\.V / 
					( G\.M \rho\.t_{\Lrm}\.R_{\crm}) 
						& \big[ 
							M < M_{\crm}\.
							( V / V_{\crm} )
						 \big]
						\\[1mm]
					M_{\crm} / ( \rho\.V_{\crm}\.
					t_{\Lrm}\.R_{\crm}^{2} )
						&
						\big[
							M_{\crm}\.(V / V_{\crm} ) < 
							M < M_{\crm}\.
							( V / V_{\crm} )^{3}
						\big]
						\\[1mm]
					M\.V_{\crm}^{2} / 
					\big(
						\rho\.
						R_{\crm}^{2}\.V^{3}\.t_{\Lrm}
					\big)\.
					\exp\!
					\big[
						( M / M_{\crm} )
						( V_{\crm} / V )^{3}
					\big]
						&
						\big[
							M > M_{\crm}\.
							( V / V_{\crm} )^{3}
						\big] 
					\, .
				\end{cases}
\end{align}
The three limits correspond to disruption by multiple encounters, one-off encounters and non-impulsive encounters, respectively. These are all potential signatures of PBHs but we just focus below on the three effects for which positive evidence has been claimed. These come from the heating of the Galactic disk, features of UFDGs, the cusp/core problem for dwarf galaxies, recent observations by JWST of galaxies at higher redshift than allowed by standard galaxy formation models, and the triggering of white-dwarf explosions.

There are also many constraints and possible signatures for PBHs resulting from their accretion luminosity. These are associated with point source observations for individual PBHs~\cite{2017JCAP...10..034I, 2017PhRvL.118x1101G, 2019JCAP...06..026M} and with the generation of cosmic background radiation~\cite{1980Natur.284..326C} and effects on the thermal history of the Universe for a population of PBHs~\cite{10.1093/mnras/194.3.639}. The signatures all depend upon astrophysical assumptions and on the black hole environment (gas density and temperature), so there is inevitably some uncertainty. If the PBHs reside in galactic nuclei or halos, they will accrete local gas and stars. If they reside outside galaxies, they will still accrete intergalactic gas, the consequences of which depend on the (somewhat uncertain) state of the intergalactic medium. There are also {\it indirect} constraints associated with the $\mu$-distortions in the CMB spectrum associated with the dissipation by Silk damping of the density fluctuations invoked to generate PBHs~\cite{Chluba:2012we, Kohri:2014lza, 2018PhRvD..97d3525N, 2019JCAP...06..028B}. Although all these effects are usually presented as constraints, clearly they are also potential signatures. The positive accretion evidence comes from the existence of SMBHs in galactic nuclei, the spatial coherence of the X-ray and infrared source-subtracted backgrounds, this requiring the existence of bound baryon clouds earlier than expected in the standard scenario but as predicted by the Poisson effect, and a radio background. Finally PBHs may accrete dark matter in some period, this being relevant to the considerations of Section~\ref{sec:Mixed-Dark-Matter}.

Some of these points have also been made by Silk, who argues that intermediate-mass PBHs could be ubiquitous in early dwarf galaxies, being mostly passive today but active in their gas-rich past~\cite{Silk:2017yai}. This would be allowed by current active galactic nuclei observations~\cite{2013ARAA..51..511K, 2016ApJ...831..203P, Baldassare:2016cox} and early feedback from intermediate-mass PBHs could provide a unified explanation for many dwarf galaxy anomalies~\cite{2018PDU....22..137C}. Besides providing a phase of early galaxy formation and seeds for SMBHs at high redshift, they could:
	(1) suppress the number of luminous dwarfs;
	(2) generate cores in dwarfs by dynamical heating;
	(3) resolve the ``too big to fail'' problem;
	(4) create bulgeless disks;
	(5) form UFDGs and ultra-diffuse galaxies;
	(6) reduce the baryon fraction in Milky-Way-type 
		galaxies;
	(7) explain ultra-luminous X-ray sources in the 
		outskirts of galaxies;
	(8) trigger star formation in dwarfs via active 
		galactic nuclei.
They would also induce microlensing of extended radio sources~\cite{Inoue:2003hq, Inoue:2013ey}.

\subsection{Dynamical Heating of the Milky Way Disk}
\label{sec:Dynamical-Heating-of-the-Milky-Way-Disk}

As halo objects traverse the Galactic disk, they will impart energy to the stars there. This will lead to a gradual puffing up of the disk, with older stars being heated more than younger ones. This problem was first studied by Lacey~\cite{1984ASIC..117..351L} in the context of a very general analysis of disk-heating and then by Lacey \& Ostriker~\cite{1985ApJ...299..633L}, who argued that SMBHs could generate the observed disk-puffing if they provide the dark matter ($\fPBH \sim 1$). In particular, they claimed that this could explain:
	(1)	why the velocity dispersion of disc stars 
		$\sigma$ scales with age as $t^{1/2}$; 
	(2)	the relative velocity dispersions in the 
		radial, azimuthal and vertical directions; and 
	(3)	the existence of a high-energy tail of stars 
		with large velocity (\cf~Ipser \& 
		Semenzato~\cite{1985A&A...149..408I}).
In order to normalise the $\sigma( t )$ relationship correctly, the number density $n$ of the holes must satisfy $n\.M^{2} \approx 2 \times 10^{4}\.\Msun^{2}\,{\rm pc}^{-3}$. Combining this with the local halo density, $\rho_{\Hrm} = n\.M \approx 0.01\.\Msun\,{\rm pc}^{-3}$, then gives $M \approx 2 \times 10^{6}\.\Msun$. Wielen \& Fuchs~\cite{1988LNP...306..100W} claimed that black hole heating could also explain the dependence of the velocity dispersion upon Galactocentric distance. If some other mechanism explains disc heating, these arguments still give an upper limit on the fraction of dark matter in black holes:
\begin{equation}
	\fPBH( M )
		<
				( M / 10^{6}\.\Msun )^{-1}
				\, .
\end{equation}
For example, this would apply if the heating were due to giant molecular clouds, which is the more mainstream explanation.

Although disc heating can no longer be attributed to such large black holes, because this is now excluded by other constraints, Carr \& Lacey pointed out that it may still provide evidence for clusters of much smaller PBHs~\cite{1987ApJ...316...23C}. Indeed, since PBH clusters with mass close to the value of $10^{6}\.\Msun$ required for disc heating are {\it expected} to form (see Section~\ref{sec:Clustering-of-PBHs}), this might be regarded as a prediction if PBHs provide the dark matter. The stellar velocity change in close encounters is smaller in the cluster case because of the finite size of the clusters but most of the disc heating comes from distant encounters anyway. However, scattering off the clusters can only produce the high-velocity disc stars with $v > 100\,$km\,s$^{-1}$ if $r_{\crm} < 0.5\,( M_{\crm} / 10^{6}\.\Msun )\,{\rm pc}$, and the associated constraint is shown by the red region in Fig.~\ref{fig:rcvsmc}.

\subsection{Dynamical Heating of Ultra-Faint Dwarf Galaxies}
\label{sec:Dynamical-Heating-of-Ultra--Faint-Dwarf-Galaxies}

UFDGs have been used to constrain the PBH abundance~\cite{2020MNRAS.492.5247S}. This requires $\fPBH < 1$ for $M \gtrsim 10\.\Msun$, with the precise limit depending on model assumptions and the possible effect of an intermediate-mass black hole (IMBH) at their centre. It was suggested in Ref.~\cite{2018PDU....22..137C} that the argument could be reversed, with the observed critical radius and mass of UFDGs indicating that a large fraction of dark matter comprises stellar-mass compact objects. Although Ref.~\cite{2020MNRAS.492.5247S} claims that this is excluded, the presence of Poisson-induced clusters was not allowed for.

Figure~\ref{fig:rcvsmc} shows the most likely half-light radius and mass of recently observed UFDGs, using the tables of Ref.~\cite{2019ARA&A..57..375S} and the virial theorem to obtain the mass from the stellar velocity dispersion. Comparison with the predicted relation between between $r_{\crm}$ and $M_{\crm}$ supports the model providing $\fPBH\.M$ does not deviate much from our favoured value of $3\.\Msun$; the $0.3\.\Msun$ and $30\.\Msun$ cases are represented in the figure. In addition, the critical radius and mass match the ones obtained from the dynamical heating effect. This was first noted in Ref.~\cite{2018PDU....22..137C}, which also pointed out that the large mass-to-light ratios observed in UFDGs could be explained by a rapid PBH accretion phase after the formation of the clusters. Furthermore, the Galactocentric distances of the UFDGs are sufficiently large for them to evade the limits from collisional disruption, the Galactic tidal field and disc heating.

\subsection{Cusp/Core Problem}
\label{sec:Cusp/Core-Problem}

$N$-body simulations in the standard $\Lambda$CDM scenario yield galactic halos with central cusps~\cite{Navarro:1996gj}, contradicting observations of dwarf galaxies with central cores~\cite{deBlok:2009sp}. One resolution of this problem is to invoke stellar feedback{\,---\,}since this redistributes gas clouds, generates bulk motions and galactic winds{\,---\,}or reheating by dynamical friction of massive clumps~\cite{Boldrini:2019isx}. Another is to include dark matter self-interaction~\cite{Vogelsberger:2015gpr, Cyr-Racine:2015ihg}, this leading to cored halo profiles for sufficiently large cross-sections.

Boldrini {\it et al.}~\cite{Boldrini:2019isx} have used an $N$-body code to explore an alternative explanation, which involves a galactic halo comprising both particles and PBHs. As a specific example, they consider $10^{7}\.\Msun$ dwarf galaxies with PBHs in the mass window $25\,\text{--}\,100\.\Msun$. They find that cores can form as a result of dynamical heating of the CDM through PBH infall and two-body processes, providing PBHs provide at least $1\%$ of the non-baryonic halo. The halo shape depends upon the PBH mass distribution, this being subject to mass segregation, PBH mergers, possible slingshotting of light PBHs, and PBH interactions with the central black hole. This may resolve the cusp/core problem.

In the context of the Milky Way, the observed rotation curve near the centre indicates the presence of a supermassive central black hole of $4 \times 10^{6}\.\Msun$ and a ring of dark matter at a distance of about $100\,$pc with the usual $r^{-2}$ fall-off at larger distances. These features are predicted by detailed numerical simulations of PBHs in the central region of an all-black-hole halo~\cite{2021Univ....7...18T}. Because the PBHs scatter off each other and the central SMBH, they accumulate in a ring around the centre~\cite{Garcia-Bellido:2022fgh} and the density profile has a hill at a distance of about $100\,$pc for a galaxy like our own and $10\,$pc for dwarf galaxies. This configuration could be transitory but it currently seems universal for all sizes of galaxies and halos. The profile looks like a core at large distances, so one needs more precise observations of rotation curves in the inner parts of galaxies to test the model~\cite{Calcino:2018mwh}. Note that several IMBHs have been observed in the central molecular zone around the Galactic Centre~\cite{2020ApJ...890..167T}, which could could be interpreted as a subpopulation of massive PBH that have migrated to the centre of our Milky Way due to dynamical friction.
\newpage

\subsection{Primordial Black Holes as Seeds for Quasars and Galaxies}
\label{sec:Primordial-Black-Holes-as-Seeds-for-Quasars-and-Galaxies}

\begin{figure}[t]
	\centering
	\vs{-10mm}
	\includegraphics[width=0.85\textwidth]{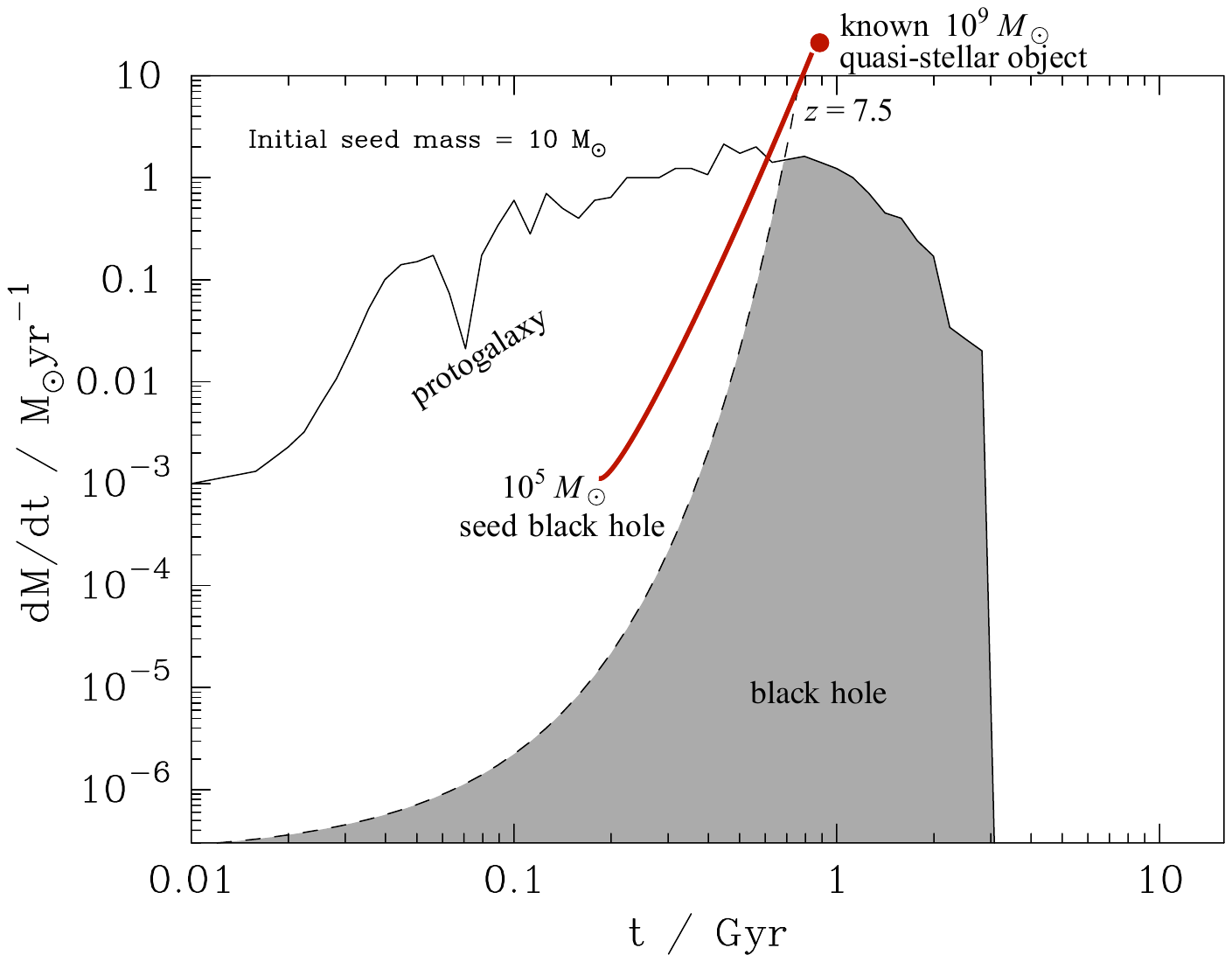}
	\vs{-8mm}
	\caption{
		Black hole growth rate as a function of time 
		for an initial PBH seed mass of $10\.\Msun$ 
		(black dashed line) and the rate of gas 
		deposition into the innermost $100\,$pc for the 
		galaxy model used in 
		Ref.~\cite{Archibald:2001gi} (black solid 
		line). Also shown (red line) is the growth 
		rate for a black hole which 
		evolves from a seed mass of $10^{5}\.\Msun$ to 
		the $10^{9}\.\Msun$ required for a 
		quasi-stellar object [extracted from 
		Ref.~\cite{2007ApJ...665..187L} by Hasinger 
		(private communication)]. The shaded area 
		corresponds to Eddington-limited accretion, 
		persisting until $0.5\,\text{--}\,0.7\,$Gyr, 
		with the accretion being sub-Eddington over the 
		subsequent $1\,$Gyr. From 
		Ref.~\cite{Archibald:2001gi}.
		\vs{1.5mm}
		}
	\label{fig:figY}
\end{figure}

The mainstream view is that the $10^{6}\,\text{--}\,10^{10}\.\Msun$ black holes in galactic nuclei form from dynamical processes {\it after} galaxies, but this proposal is becoming increasingly challenged by the high mass and redshift of some SMBHs. This has led to the suggestion that the SMBHs could form {\it before} galaxies, in which case they may have been seeded by PBHs. In this subsection the mass of the PBH is denoted by $m$, so for a monochromatic PBH mass function with $\fPBH < m / M$ (\ie~only one PBH per bound region of mass $M$), Eq.~\eqref{eq:initial} and the growth law $\delta \propto ( 1 + z )^{-1}$ imply that the mass bound by a seed could exceed the seed mass by a factor of $10^{3}$ at the redshift $z \sim 4$ when SMBHs are observed. For example, the softening of the pressure through $e^{+}e^{-}$ annihilation at $10\,$s naturally produces $10^{6}\.\Msun$ PBHs and the seed effect could then bind a region of $10^{9}\.\Msun$. Of course, one cannot infer that all the bound mass is swallowed by the black hole since that depends on the efficiency of accretion. Also the limit would not be attained anyway for $\fPBH > 10^{-3}$, since the filling factor of the bound regions reaches $1$ when $M \sim m/\fPBH$, so there will be competition between the seeds thereafter. If larger galaxies are produced by the mergers of smaller galaxies, with the original central black holes themselves possibly merging to form a single central SMBH, this would naturally explain the observed proportionality between the masses of the galaxy and the central black hole. However, one cannot be sure that the original SMBHs will all drift to the centre of the larger galaxy and merge. The accretion rate according to a recent calculation is indicated in Fig.~\ref{fig:figY}.

There is also the possibility that either the SMBHs or the PBHs which seeded them could help to generate the galaxies themselves. Hoyle \& Narlikar~\cite{1966RSPSA.290..177H} first suggested the seed picture of galaxy formation in the context of the steady state theory, arguing that a typical galactic mass $M_{\grm} \sim 10^{11}\.\Msun$ would require a seed of mass $m \sim 10^{9}\.\Msun$. Their model was not constrained by the existence of a radiation-dominated era but Ryan~\cite{1972ApJ...177L..79R} argued that SMBHs could also seed galaxies in the context of the big bang theory. Using a spherically-symmetric Newtonian cosmology, he showed that the hydrodynamic equations permit a solution in which the density contrast has a particular form in the radiation-dominated and matter-dominated eras. This gave expressions for the galactic mass $M_{\grm} \propto m^{2/5}$ and radius $R_{\grm} \propto m^{1/3}$. For the Milky Way, Ryan obtained $m \approx (1\,\text{--}\,9) \times 10^{6}\.\Msun$, which encompasses the now established mass of $4 \times 10^{6}\.\Msun$~\cite{1998ApJ...509..678G}. Gunn \& Gott~\cite{1972ApJ...176....1G} pointed out an interesting prediction of the seed theory. If each shell of gas virialises after it has stopped expanding and virialised, then one would expect the resultant galaxy to have a density profile $\rho( r ) \propto r^{-9/4}$. This does not agree with the standard Navarro--Frenk--White (NFW) profile~\cite{1996ApJ...462..563N}, but one would not expect the latter to apply within the radius of gravitational influence of the central black hole anyway.

Of course, one could not expect {\it all} galaxies to be produced by PBH seeds since the standard CDM model works well on large enough scales. However, one might want to invoke the PBH effect for early galaxies (\ie~below the mass $M_{\rm CDM}$ discussed in Section~\ref{sec:Seed-and-Poisson-Effects}). For a monochromatic PBH mass function, all the galaxies binding at a given redshift will have the same mass until the CDM fluctuations take over. For an extended mass function, the PBH seeds will naturally produce a range of galactic masses at a given redshift. However, there are two distinct situations. In the first, the PBHs are sufficiently rare that there is only one per galaxy. The galactic mass is then proportional to the seed mass and one would expect the galactic mass function to be the same as the PBH mass function, although only a small fraction of the Universe goes into these galaxies. In the second situation, the filling factor of the bound regions reaches $1$, so that the competition between seeds becomes important and one can no longer assume $M \propto m$. However, there should still be a simple relation between the mass spectrum of the holes, $\d n / \d m \propto m^{-\alpha}$, and that of the resulting galaxies, as we now demonstrate.

If $M_{\grm} \propto m^{\gamma}$, we expect the number of galaxies with mass in the range $( M,\.M +\mspace{1mu}\d M )$ to be $\d N_{\grm}( M )$ where $\d N_{\grm} / \d M \propto M^{( 1 - \gamma - \alpha ) / \gamma}$. For $\alpha < 3$, the considerations of Section~\ref{sec:Seed-and-Poisson-Effects} imply that one expects an $M$-dependent seed mass for $M < m_{\rm max}/ f_{\rm max}$, where $m_{\rm max}$ is the upper cut-off in the PBH mass function and $f_{\rm max}$ is the dark matter fraction on that scale. In this mass range, Eq.~\eqref{eq:seed} predicts $\gamma = \alpha - 1$, so one expects~\cite{Carr:2018rid}
\begin{equation}
	\frac{ \d N_{\grm} }{ \d M }
		\propto
				M^{-2}
		\quad
	( M < M_{\rm max}
		\equiv
				m_{\rm max} / f_{\rm max} )
				\, .
				\label{eq:gal}
\end{equation}
This prediction depends on the concept of an $M$-dependent seed mass, although the simple analytic treatment of this concept in Section~\ref{sec:Clustering-of-PBHs}. A needs to be backed up by numerical calculations. The form of the galactic mass function above $M_{\rm max}$ corresponds to the Poisson effect and one expects an exponential decline. For comparison, the Press--Schechter mass function is~\cite{Press:1973iz}
\begin{equation}
	\frac{ \d N_{\grm} }{ \d M }
		\propto
				M^{- 2 + \frac{ 3 + n }{ 6 }}\.
				\exp\!
				\Big[
					- ( M / M_{*} )
					^{\frac{ 3 + n }{ 3 }}
				\Big]
				\, ,
\end{equation} 
where $n$ is the power-spectrum index, $P( k ) \propto k^{n}$, which ranges from $1$ at large scales to $-\mspace{1.5mu}3$ at small scales in the $\Lambda$CDM model. Coincidentally, this corresponds to form~\eqref{eq:gal} at small scales. It is supposed to explain the Schechter galaxy luminosity function at large scales~\cite{1976ApJ...203..297S}:
\begin{equation}
	\frac{ \d N_{\grm} }{ \d L }
		\propto
				L^{\alpha} \exp( - L / L_{*} )
				\, ,
\end{equation} 
where $\alpha = - 1.07 \pm 0.07$ and $L_{*} = 1.2 \times 10^{10}\,L_{\odot}\.h^{-2}$ in the blue band. However, it also matches the prediction of Eq.~\eqref{eq:gal}, providing $m_{\rm max} / f_{\rm max} \approx 10^{12}\.\Msun$.

The possible r{\^o}le of PBHs as seeds for the SMBHs in galactic nuclei and early structures has been highlighted by recent results from the JWST, as reviewed by Dolgov~\cite{Dolgov:2023ijt}. These show that such structures form much earlier than expected in the standard $\Lambda$CDM scenario. For example, there is a population of red galaxies with stellar mass above $10^{10}\.\Msun$ at $7.4 < z < 9.1$ with central SMBHs~\cite{Labb__2023}. The {\it Atacama Large Millimeter/submillimeter Array} (ALMA) has observed an AGN at $z \sim 7$ whose large luminosity suggests a $10^{9}\.\Msun$ black hole~\cite{Endsley_2023}. The detection of X-rays from a $4 \times 10^{7}\.\Msun$ SMBH in a gravitationally lensed galaxy is reported at $z = 10.3$ by Bogdan {\it et al.}~\cite{Bogdan:2023ilu}, although they invoke seeds formed from the collapse of $10^{4}\,\text{--}\,10^{5}\.\Msun$ gas clouds rather than PBHs.

It has been claimed that such observations preclude the standard model but Dolgov~\cite{Dolgov:2023ijt} has emphasised that it is entirely consistent if one invokes a population of PBHs. Liu \& Bromm~\cite{Liu:2022bvr} have also argued that unusually massive galaxies at $z > 10$ could be evidence that structure formation is accelerated by $10^{9}\.\Msun$ PBHs making up $10^{-6}\,\text{--}\,10^{-3}$ of the dark matter. Goulding {\it et al.}~\cite{Goulding:2023gqa} also argue for a heavy seeding channel for the formation of SMBHs within the first billion years of cosmic evolution. An interesting feature of these observations is that the central black hole to stellar mass ratio is several orders of magnitude larger than observed locally, as expected if the black hole is acting as a seed. Seeding of dwarf galaxies by IMBHs has also been proposed and two examples of this have been found~\cite{Micic:2022mws, Yang:2023bnv}.
\newpage

Dolgov also links this proposal to the LVK observations, arguing that these are consistent with his earlier proposal of a lognormal mass function~\cite{PhysRevD.47.4244} centred around $10\.\Msun$. This would be compatible with a low QCD temperature of $70\,$MeV for a non-zero chemical potential. However, others argue that only some fraction of the LVK black holes could be primordial~\cite{2022arXiv220905959F}. He also argues that the SMBH binaries invoked to explain the NANOGrav data~\cite{NANOGrav:2023hfp} could be primordial.

\subsection{Exploding White Dwarfs}
\label{sec:Exploding-White-Dwarfs}

PBHs can trigger explosions of white dwarfs, leading to potentially observable signatures~\cite{2006PhRvL..96u1302F, 2015PhRvL.115n1301B, 2019PhRvD.100d3020A, 2021ApJ...914..138C, 2021ApJ...914..138C, 2018PhRvD..98k5027G, 2022PhRvD.105b3012A, 2022arXiv221100013S}). It is therefore interesting that some recently observed supernov{\ae}, the so-called {\it calcium-rich transients}~\cite{2010Natur.465..322P, 2017ApJ...836...60L}, do not trace the stellar density but are located off-centre from the host galaxies. Furthermore, they appear to originate from white dwarfs with masses of around $0.6\.\Msun$~\cite{2015A&A...573A..57M}, well below the Chandrasekhar limit, and they predominantly occur in old systems.

\begin{figure}[t!]
	\centering
	\vs{-1mm}
	\includegraphics[width=0.88\columnwidth]{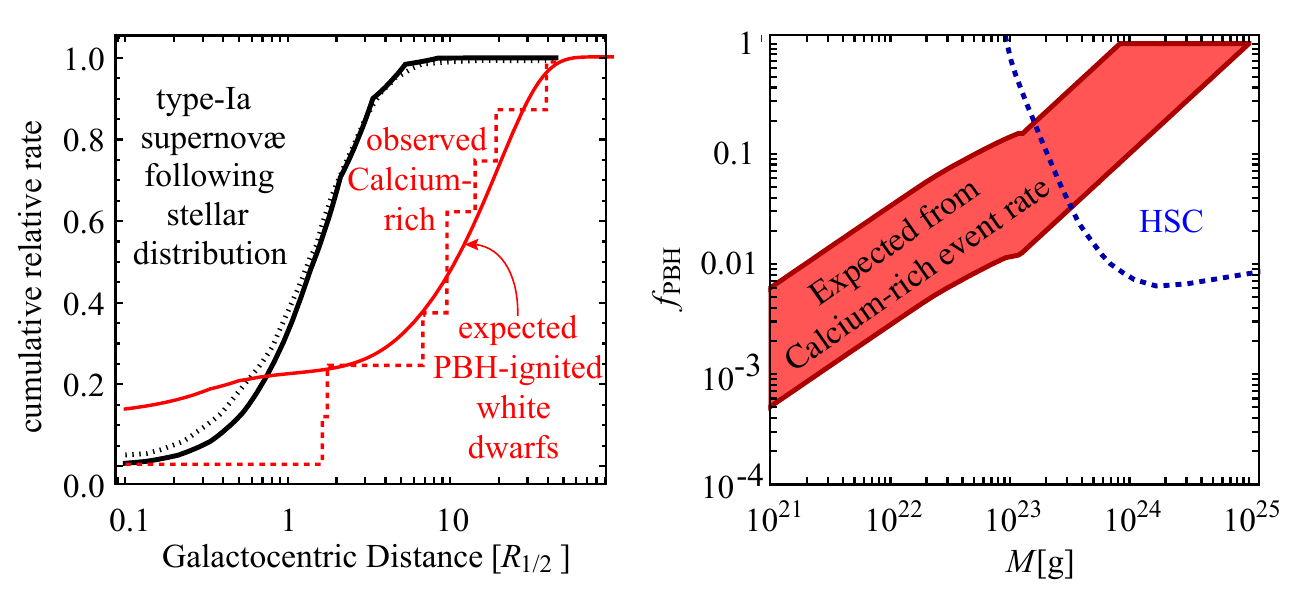}
	\caption{
		{\it Left panel:}
			Radial distributions of standard type-Ia 
			supernov{\ae} (black dashed) and 
			calcium-rich transients (red dashed) within 
			their host galaxies	compared to the 
			distribution of galactic stars (black 
			solid) and the expected PBH/white dwarf 
			interactions (red solid).
		{\it Right panel:}
			PBH dark matter fraction as a function of 
			mass required to produce the observed 
			supernova rate. Also shown (blue) is the 
			$95\mspace{1mu}\%$ CL exclusion region from 
			stellar microlensing in Andromeda by the 
			Subaru HSC~\cite{2019NatAs...3..524N, 2020PhRvD.101f3005S}.
			Figures adapted from 
			Ref.~\cite{2022arXiv221100013S}.
		}
		\label{fig:datacompare}
\end{figure}

It has been argued~\cite{2022arXiv221100013S} that these transient events could have been triggered by collisions with asteroidal-mass PBHs, with the associated event rate being within reach of current or near-future microlensing surveys~\cite{2019NatAs...3..524N, 2020PhRvD.101f3005S}. The left panel of Fig.~\ref{fig:datacompare} shows the cumulative type-Ia supernova event rate, normalised to the half-light radii of the host galaxies, as determined in Ref.~\cite{2017ApJ...836...60L}. This is different from standard type-Ia supernov{\ae}, whose locations closely follow the stellar distribution. However, the distribution of these events with galactocentric distance follows that expected from dark matter/white dwarf interactions. This, together with the unique chemical properties and atypical progenitors of calcium-rich transients, strongly supports the hypothesis that these events originate from such interactions.

The modelling of Ref.~\cite{2022arXiv221100013S} suggests that PBHs with $10^{21}\.\grm < M < 10^{23}\.\grm$ and $10^{-3} < \fPBH < 0.1$ are the most plausible triggers for these events (see right panel of Fig.~\ref{fig:datacompare}). Future observations will further explore this possibility. Interestingly, the {\it Zwicky Transient Facility} (ZTF), which is the largest ongoing systematic survey, classified 8 new events in its first $16$ months of operation~\cite{2020ApJ...905...58D}. In the year 2024, the {\it Large Synoptic Survey Telescope} (LSST){\,---\,}as part of the {\it Vera Rubin Observatory} (VRO){\,---\,}is expected to increase the search volume by at least two orders of magnitude. Furthermore, observations with the JWST should be able to resolve dwarf spheroidal galaxies for the current sample of calcium-rich transients~\cite{2020ApJ...905...58D}, these two populations being closely connected.

\subsection{PBH Accretion: Pregalactic Era}
\label{sec:PBH-Accretion:-Pregalactic-Era}

Here we focus on the accretion of gas by PBHs before the formation of galaxies or other cosmic structures. This problem was first studied in Ref.~\cite{10.1093/mnras/194.3.639}, although this work was later superseded by more detailed numerical investigations. It is usually associated with PBH constraints but it may also lead to positive evidence since PBH accretion in this period could ionise the Universe or generate to X-ray background~\cite{1980Natur.284..326C}. Reference~\cite{10.1093/mnras/194.3.639} assumes the Bondi accretion formula and covers epochs between matter-radiation equality at $t_{\rm eq} = 2.4 \times 10^{12}\,$s and galaxy formation. Before $t_{\rm eq}$, the sound-speed is $c_{\srm} = c /\!\sqrt{3}$ and one can show that there is very little accretion~\cite{Carr:1974nx}. After $t_{\rm eq}$, the accretion radius is increased, so the accretion rate is larger. However, the problem is complicated because the black hole luminosity will boost the matter temperature of the background Universe above the standard Friedmann value even if the PBH density is small, thereby reducing the accretion. Thus there are two distinct but related PBH constraints: one associated with the effects on the Universe's thermal history and the other with the generation of background radiation.

We assume that each PBH accretes at the Bondi rate, 
\begin{equation}
	\dot{M}
		=
				\frac{4\pi\.G^{2}\.n\.m_{\prm}\.
				M^{2}}
				{( k_{\Brm} T/m_{\prm} + 
				V^{2} )^{3/2}}
		\approx
				10^{11}
				\left(
					\frac{ M }{\.\Msun }
				\right)^{\!2}\.
				\left(
					\frac{ n }{ {\rm cm}^{-3} }
				\right)\!
				\left(
					\frac{ T }{ 10^{4}\.\Krm} 
				\right)^{\!-3/2}
				\grm\;\srm^{-1}
				\, ,
	\label{eq:Bondi}
\end{equation}
where a dot indicates a derivative with respect to cosmic time, $k_{\Brm}$ is the Boltzmann constant, $m_{\prm}$ is the proton mass, $V$ is the black hole velocity (neglected in the second expression since it is small before the formation of cosmic structures) and $n$ and $T$ are the number density and temperature of the gas at the black hole accretion radius,
\begin{equation}
	R_{\arm}
		\approx
				10^{14}
				\left(
					\frac{ M }{\.\Msun }
				\right)\!
				\left(
					\frac{ T }{ 10^{4}\.\Krm }
				\right)^{\!-1}
				{\rm\,cm}
				\, .
\end{equation}
Each PBH will initially be surrounded by an ionised (HII) region of radius $R_{\srm}$. If $R_{\rm a} > R_{\srm}$ or if the whole Universe is ionised (so that the individual ionised regions merge), the appropriate values of $n$ and $T$ are those in the background Universe ($\bar{n}$ and $\bar{T}$). In this case, after decoupling, $\dot{M}$ is epoch-independent so long as $\bar{T}$ has its usual Friedmann behaviour ($\bar{T} \propto z^{2}$). However, $\dot{M}$ decreases with time if $\bar{T}$ is boosted above the Friedmann value. If the individual ionised regions have not merged and $R_{\rm a} < R_{\srm}$, the values for $n$ and $T$ are those which pertain within the ionised region. In this case, $T$ is usually close to $10^{4}\,$K and pressure balance at the edge of the region implies $n \sim \bar{n}\.( \bar{T} / 10^{4}\.\Krm )$. This implies $\dot{M} \propto z^{5}$ if $\bar{T}$ is unaffected, so the accretion rate rapidly decreases in this phase.

We assume that accreted mass is converted into outgoing radiation with constant efficiency $\epsilon$, so that the associated luminosity is
\vs{-1mm}
\begin{equation}
	L
		=
				\epsilon\.\dot{M}c^{2}
				\,
				,
\end{equation}
until this reaches the Eddington limit,
\begin{equation}
	L_{\rm ED}
		 = 
				\frac{ 4\pi\.G\.M\mspace{1mu}
				c\.n\.m_{\prm} }
				{ \sigma_{\Trm} }
		\approx
				10^{38}\!
				\left(
					\frac{ M }{\.\Msun }
				\right)
				{\rm erg}\;\srm^{-1}
				\, ,
	\label{eq:Eddington}
\end{equation}
at which the Thomson drag of the outgoing radiation balances the gravitational attraction of the hole. (Here $\sigma_{\rm T}$ is the Thomson cross-section.) The assumption that $\epsilon$ is constant is simplistic and more sophisticated models allow it to be $\dot{M}$-dependent. We also assume that the spectrum of emergent radiation is constant, extending up to an energy $E_{\rm max} \equiv 10\,\eta\,$keV, with Ref.~\cite{10.1093/mnras/194.3.639} considering models with $\eta = 0.01$, $1$ and $100$. We must distinguish between the {\it local} effect of a PBH at distances sufficiently small that it dominates the effects of the others and the combined effect of all the PBHs on the mean conditions of the background Universe. Both effects are very dependent on the spectrum of the accretion-generated radiation and on the process by which this heats the background. The dominant process is Compton heating until a redshift~\cite{10.1093/mnras/194.3.639}
\begin{equation}
	\label{eq:zstar-def}
	z_{*}
		\approx
				10^{2}\,\Omega_{\mrm}^{1/3}\.
				(
					\eta\,\Omega_{\grm}
				)^{-2/3}
				\, ,
\end{equation}
where $\Omega_{\mrm} \approx 0.2$ and $\Omega_{\grm} \approx 0.05$ are the density parameters of the matter and gas, respectively.
\newpage

\begin{figure}[t]
	\centering
	\includegraphics[width=0.70\textwidth]{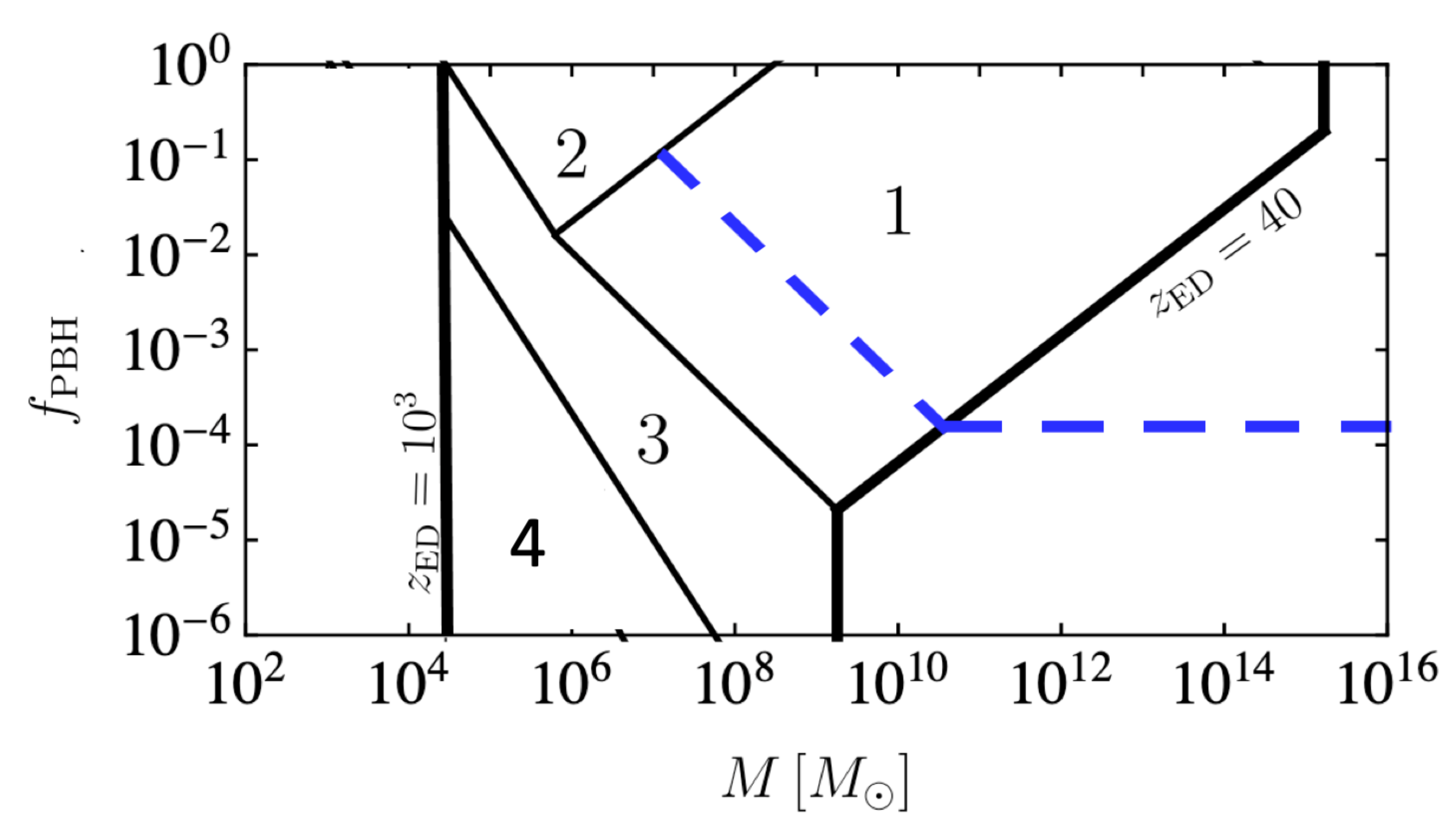}
	\caption{
		This shows how the evolution of 
		the background matter temperature depends 
		on the PBH mass and density, 
		adapted from Ref.~\cite{Carr:2020erq}.
		We assume $\epsilon = 0.1$, 
		$\Omega_{\grm} = 0.05$ and 
		$E_{\rm max} = 10\,$keV. The accretion rate 
		exceeds the Eddington limit for some period 
		after decoupling to the right of the line 
		$z_{\rm ED} = 10^{3}$ and the mass cannot be 
		regarded as constant the right of the line 
		$z_{\rm ED} = 40$. In domains (1) and (2), $T$ 
		is boosted above $10^{4}\,$K by Compton 
		heating; $t_{\rm ED}$ exceeds $t_{*}$ in domain 
		(1) but it is less than it in domain (2). In 
		domain (3), $T$ is boosted to $10^{4}\,$K but 
		not above it. PBHs reionise the Universe, with 
		no neutral phase after decoupling, throughout 
		these regions and they could generate the X-ray 
		background on the blue broken line.
		}
	\label{fig:extra}
\end{figure}

Providing the Bondi formula applies, the analysis of Ref.~\cite{10.1093/mnras/194.3.639} shows that a PBH will accrete at the Eddington limit for some period after decoupling if
\begin{equation}
	M
		>
				M_{\rm ED}
		\approx
				10^{3}\.
				\epsilon^{-1}\.
				\Omega_{\grm}^{-1}\,
				\Msun
				\, ,
\end{equation}
so this condition only applies for supermassive PBHs. This phase will persist until a redshift $z_{\rm ED}$ which depends upon $M$, $\Omega_{\rm PBH}$, $\Omega_{\grm}$ and $\epsilon$. The overall effect on the thermal history of the Universe is illustrated in Fig.~\ref{fig:extra}, which is adapted from Ref.~\cite{Carr:2020erq} and indicates the ($\Omega_{\rm PBH},\,M$) domain for which the Universe is reionised. $\bar{T}$ is boosted above $10^{4}\.\Krm$ in regions (1) and (2) with $z_{\rm ED} > z_{*}$ and $z_{\rm ED} < z_{*}$, respectively; it reaches $10^{4}\.\Krm$ but never exceeds it in region (3); it never reaches $10^{4}\.\Krm$ (so the Universe remains unionised) in region (4). Associated constraints from the background-radiation density in various wavebands are discussed in Ref.~\cite{10.1093/mnras/189.1.123} and the limit on the PBH density parameter in domain (1) is 
\begin{equation}
	\label{eq:accretion}
	\Omega_{\rm PBH}
		<
				( 10\.\epsilon )^{-5/6}\.
				\big(
					M / 2 \times 10^{5}\.\Msun
				\big)^{-5/6}\.
				\eta^{5/4}\.
					\big(
					\Omega_{\grm} / 0.05
				\big)^{-5/6}
				\, .
\end{equation}
This is shown by the blue line in Fig.~\ref{fig:extra}.

Two complications modify this analysis. First, accretion increases the PBH mass, although this has been assumed constant in the above analysis. During the Eddington phase, each PBH doubles its mass on the Salpeter timescale, $t_{\Srm} \approx 4 \times 10^{8}\,\epsilon\,$yr~\cite{Salpeter:1964kb}, so $M$ can only be regarded as constant if $z_{\rm ED} > z_{\Srm} \approx 10\,\epsilon^{-2/3}$ and corresponds to the condition~\cite{10.1093/mnras/194.3.639}
\vs{-1mm}
\begin{equation}
	\label{eq:Mconst}
	M
		<
				\begin{cases}
					10^{19}\,
					\Omega_{\grm}^{-1}\.
					\epsilon^{7/2}\.
					( \Omega_{\rm PBH}\.
					\eta )^{3/2}\,
					\Msun
						& (\text{region $1$})
					\\[1.5mm]
					10^{9}\,
					\Omega_{\grm}^{-1}\.
					\epsilon\,
					\Msun
						& (\text{region $3$})
					\, .
				\end{cases}
\end{equation}
This is indicated by the bold line on the right of Fig.~\ref{fig:extra}, where $z_{\rm ED} \approx 40$ for $\epsilon = 0.1$. Beyond this line, the PBH mass increases by the factor,
\begin{equation}
	\label{eq:salpetergrowth}
	\exp( t_{\rm ED} / t_{\Srm} )
		\approx
				\exp\!
				\big[
					( 0.1/ \epsilon )
					( 40 / z_{\rm ED} )^{3/2}
				\big]
				\, .
\end{equation}
The above analysis must be modified in this regime but, since most of the final black hole mass generates radiation with efficiency $\epsilon$, the current energy of the radiation produced is $E( M ) \approx \epsilon\.M c^{2} / z( M )$, where $z( M )$ is the redshift at which most of the radiation is emitted. The current background-radiation density is $\Omega_{\Rrm} \approx \epsilon\,\Omega_{\rm PBH}\.z( M )^{-1}$, so the constraint becomes
\begin{equation}
	\label{eq:accretion2}
	\Omega_{\rm PBH}
		<
				\epsilon^{-1}\.
				\Omega_{\Rrm}^{\rm max}\.
				z_{\Srm}
		\approx
				10^{-5}\.( 10\.\epsilon )^{-5/3}
				\, ,
\end{equation}
where $\Omega_{\Rrm}^{\rm max} \approx 10^{-8}$ in the X-ray band. This is shown by the flat part of the blue line in Fig.~\ref{fig:extra} and is equivalent to the so-called Soltan constraint~\cite{Soltan:1982vf}. This argument has been used to suggest that PBHs might generate the cosmic X-ray background~\cite{1980Natur.284..326C}.

The second complication is that the steady-state assumption fails if the Bondi accretion timescale, 
\begin{equation}
	\label{eq:steady}
	t_{\Brm}
		\approx
				10^{12}
				\left(
					\frac{ M }{ 10^{4}\.\Msun }
				\right)^{-1}\!
				\left(
					\frac{ T }{ 10^{4}\.\Krm }
				\right)^{\!3/2}
				\,\srm
				\, ,
\end{equation}
exceeds the cosmic expansion time (\ie~the Bondi formula is inapplicable at times earlier than $t_{\Brm}$). This arises if the mass within the accretion radius exceeds $M$, in which case the accretion radius is reduced to the value for which the gas mass contained is comparable to $M$. However, the Bondi formula becomes applicable by $t_{\rm ED}$ providing $M < 10^{4}\.\Msun$ and this applies for most of the expected PBH mass range~\cite{Carr:2020erq}.

Later, an improved numerical analysis of pregalactic PBH accretion was provided by Ricotti {\it et al.}~\cite{Ricotti:2007au}. They used a more realistic model for the efficiency parameter $\epsilon$, allowed for the increased density in the dark halo expected to form around each PBH and included the effect of the velocity dispersion of the PBHs [\ie~the $V$-term in Eq.~\eqref{eq:Bondi}] on the accretion in the period after cosmic structures start to form. They found much stronger accretion limits by considering the effects of the emitted radiation on the spectrum and anisotropies of the cosmic microwave background (CMB) rather than the background radiation itself and excluded $\fPBH = 1$ above $10\.\Msun$. However, this problem has subsequently been reconsidered by several groups, who argue that the limits are weaker than indicated in Ref.~\cite{Ricotti:2007au}. Ali-Ha{\"i}moud \& Kamionkowski~\cite{Ali-Haimoud:2016mbv} calculate the accretion on the assumption that it is suppressed by Compton drag and Compton cooling from CMB photons and allowing for the PBH velocity relative to the background gas. The point here is that accreting gas will have an inward velocity relative to the expanding background of CMB photons and the Thomson drag of these photons will inhibit accretion at sufficiently early times. They find the spectral distortions are too small to be detected, while the anisotropy constraints only exclude $\fPBH = 1$ above $10^{2}\.\Msun$. Horowitz~\cite{Horowitz:2016lib} performs a similar analysis and gets an upper limit of $30\.\Msun$. Serpico {\it et al.}~\cite{Serpico:2020ehh} argue that the spherical accretion approximation probably breaks down, with an accretion disc forming instead, and this affects the statistical properties of the CMB anisotropies. Provided the disks form early, these constraints exclude a monochromatic PBH distribution above $2\.\Msun$ as the dark matter.
\newpage

\subsection{PBH Accretion: X-Ray/Infrared/Radio Backgrounds}
\label{sec:PBH-Accretion:-X--Ray/Infrared/Radio-Backgrounds}

\begin{figure}[t]
	\centering
	\includegraphics[width=0.82\textwidth]{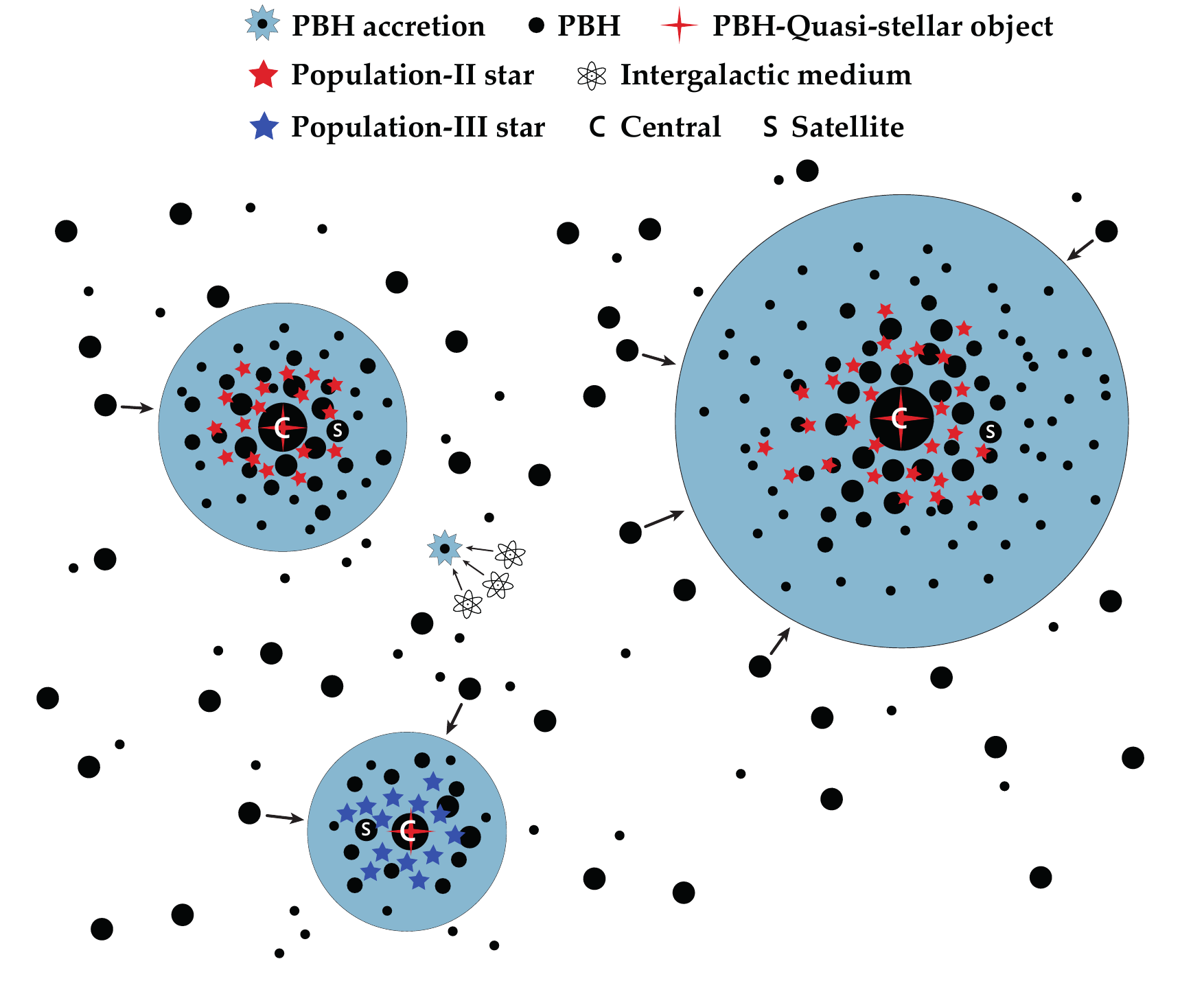}
	\caption{
		Illustration of PBH clustering at redshifts 
		$10\,\text{--}\,15$. Initially, PBHs 
		(black dots) capture baryons while accreting, 
		thereby contributing to the cosmic X-ray 
		background. Lighter PBHs later form halos 
		around more massive ones and initiate star 
		formation; the lowest mass halos first form 
		Population III stars, which generate a faint 
		cosmic infrared background, and the	higher mass 
		ones then yield Population II stars. The most 
		massive (central) supermassive PBH continues to 
		accrete and merge with other PBHs. It appears 
		as the central source in the infrared and X-ray 
		emission, with the smaller PBHs and stars 
		filling the halo as satellites.	Figure adapted 
		from Ref.~\cite{Cappelluti:2021usg}.}
	\label{fig:figZ2}
\end{figure}

\begin{figure}[h]
	\centering
	\vs{-5mm}
	\q\includegraphics[width=0.48\textwidth]{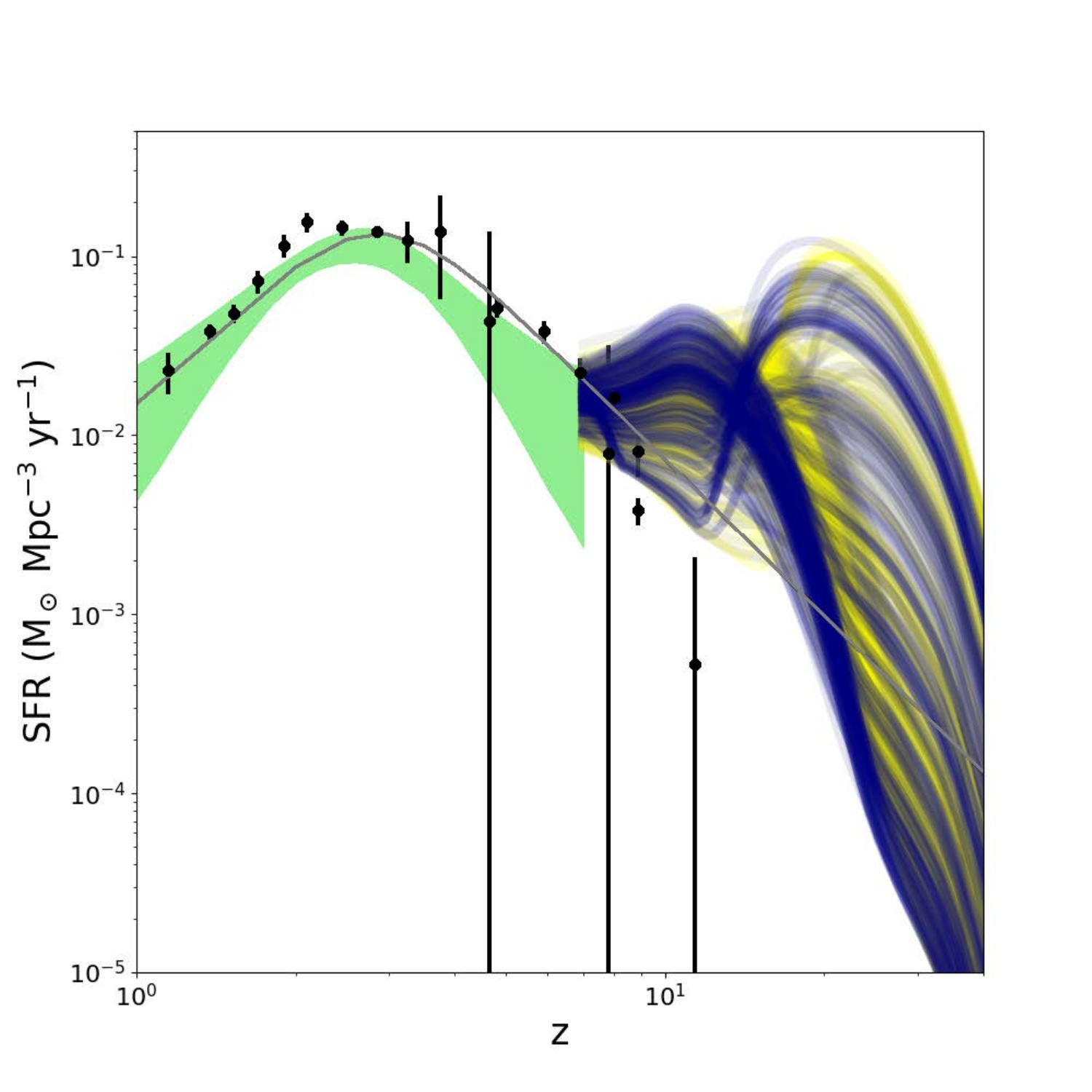}
	\includegraphics[width=0.48\textwidth]{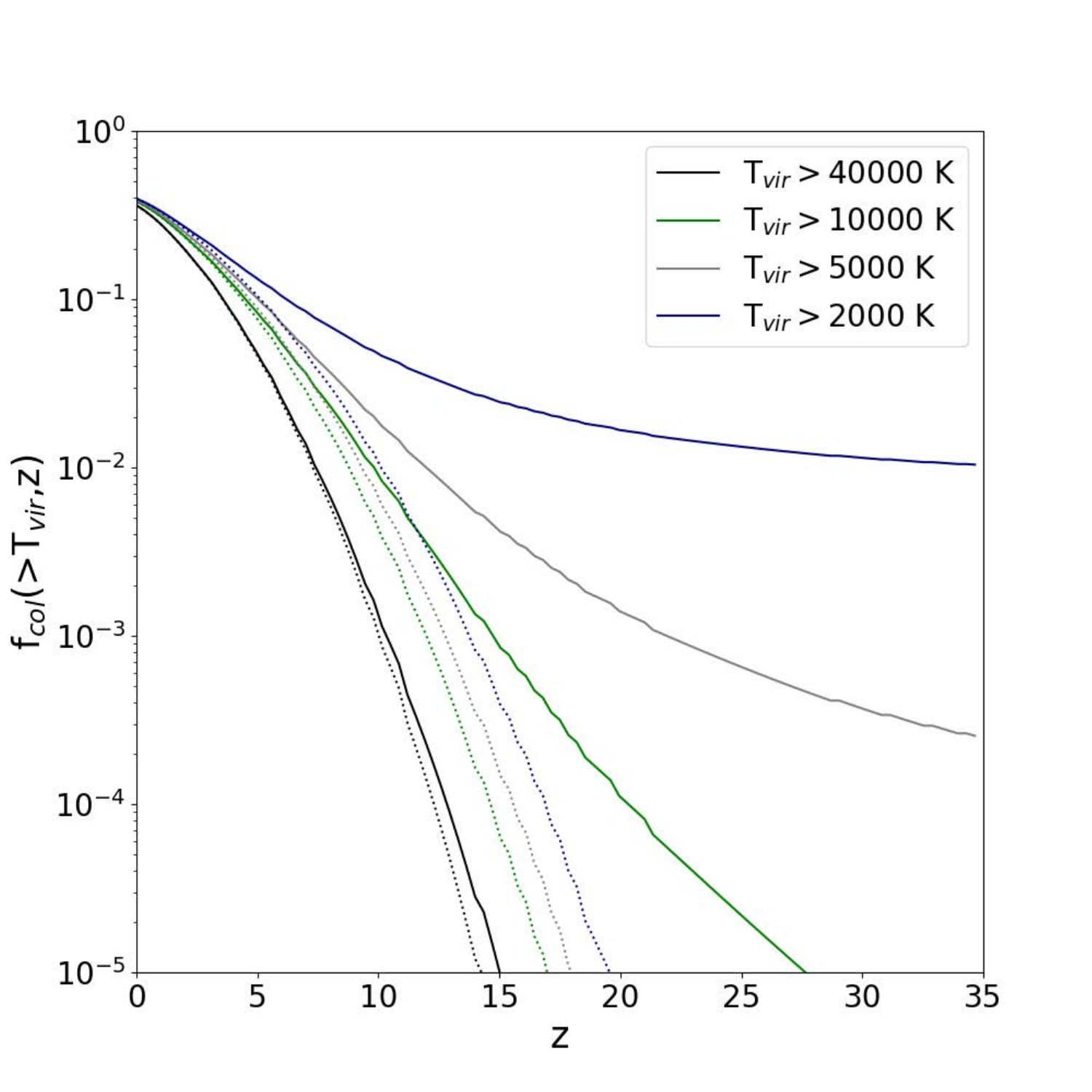}\,
	\caption{
		{\it Left panel:}
			Star formation rate (SFR) density versus 
			redshift for the thermal-history model (see 
			Section~
			\ref{sec:Thermal--History-Scenario}).
			The green band indicates compatibility with 
			measurements from extragalactic background 
			light and high-redshift surveys. The grey 
			continuous line indicates the best fit to 
			the data used.
		{\it Right panel:}
			Fraction of collapsed halos with mass 
			exceeding $M$ as a function of redshift for 
			various virilisation temperatures 
			$T_{\rm vir}$. Results are for both 
			particle dark matter (dotted lines) and PBH 
			dark matter (solid lines).
			From Ref.~\cite{Cappelluti:2021usg}.
			\vs{4mm}
		}
	\label{fig:figZ3}
\end{figure}

\begin{figure}[h]
	\centering
	\includegraphics[width=0.55\linewidth]{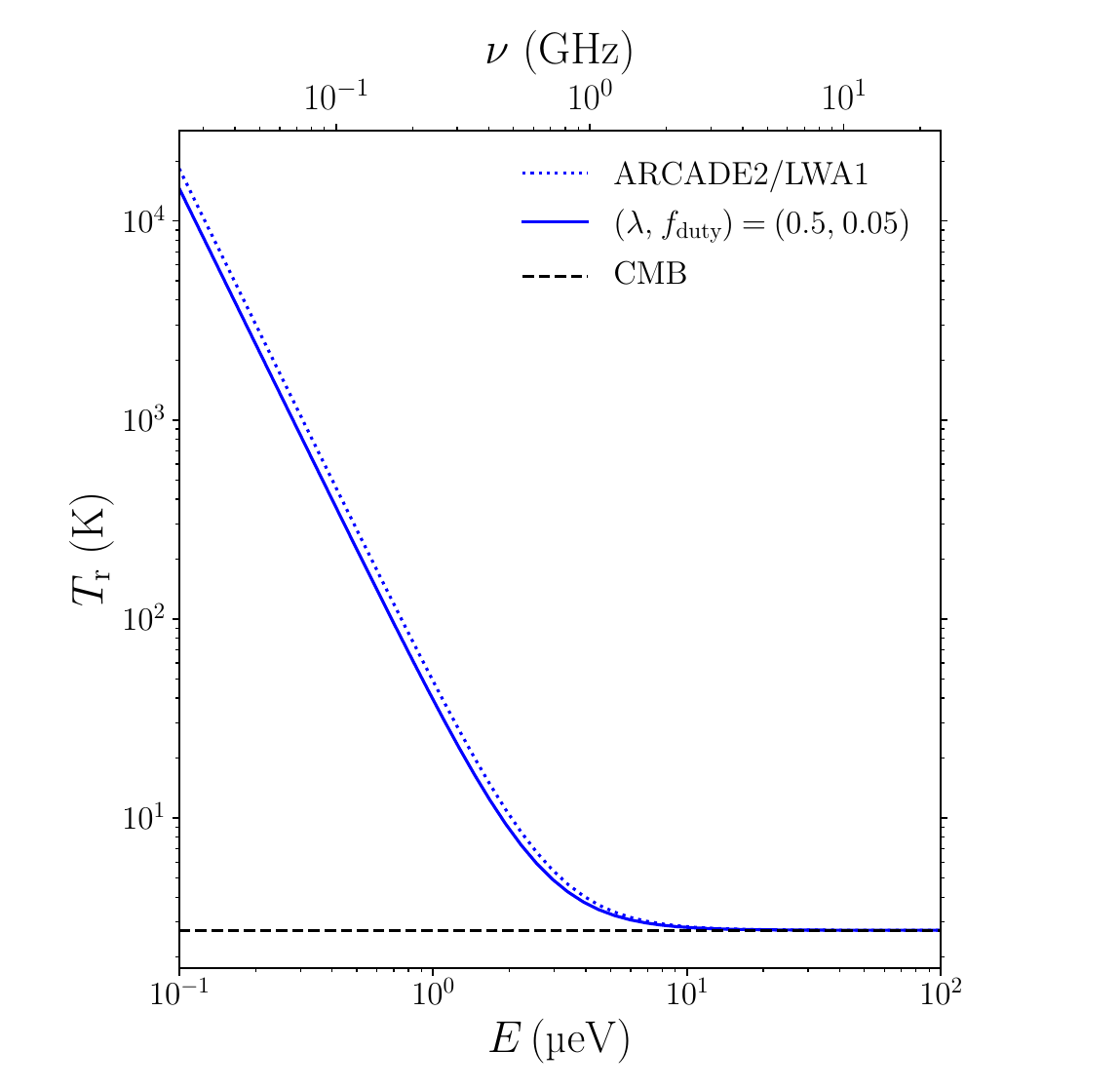}
	\caption{
		Net radio background temperature $T_{r}$ for 
		photon of energy $E$, due to accretion onto 
		supermassive PBHs. The CMB temperature (black 
		dashed line) is also shown. Figure adapted from 
		Ref.~\cite{Mittal:2021dpe}.
		}
	\label{fig:acc}
\end{figure}

The most compelling evidence for PBHs is associated with the X-ray and infrared source-subtracted cosmological backgrounds (requiring stellar PBHs) and the radio background (requiring supermassive PBHs). As shown by Kashlinsky and his collaborators~\cite{2013ApJ...769...68C, 2005Natur.438...45K, 2018RvMP...90b5006K, Kashlinsky:2016sdv}, the spatial coherence of the X-ray and infrared source-subtracted cosmological backgrounds suggests that black holes are required. Although these need not be primordial, the level of the infrared background suggests an overabundance of high-redshift halos and this could be explained by the Poisson effect discussed above if a significant fraction of the CDM comprises solar-mass PBHs. In these halos, a few stars form and emit infrared radiation, while PBHs emit X-rays due to accretion. It is challenging to find other scenarios that naturally produce such features. PBHs naturally explain both the amplitude and angular spectrum of the source-subtracted infrared anisotropies (as found by Spitzer). The infrared/X-ray coherence is a consistency argument but the infrared connection would apply even if there was no coherence (\ie~even if PBHs did not efficiently accrete). Kashlinksy also invokes cosmological advection at $z > 100$ to speed up the formation of early compact objects~\cite{2021PhRvL.126a1101K}.

Hasinger~\cite{2020JCAP...07..022H} has estimated the contribution from PBH baryon accretion and stars to the cosmic X-ray and infrared backgrounds, and analysed their cross-correlation using deep {\it Chandra} and {\it Spitzer} survey data~\cite{2013ApJ...769...68C}. Assuming Bondi capture and advection-dominated disc accretion, he finds that a population of $10^{-8}\,\text{--}\,10^{10}\.\Msun$ PBHs is consistent with the residual X-ray fluctuation signal, peaking at redshifts $z \approx 17\,\text{--}\,30$. Furthermore, he argues that PBHs could have an important bearing on a number of other phenomena:
	(1) amplifying primordial magnetic fields;
	(2)	modifying the reionisation history of the 
		Universe (while being consistent with recent 
		{\it Planck} 
		measurements~\cite{2020A&A...641A...6P});\;
	(3)	impacting on X-ray heating, thereby providing 
		a contribution to the entropy floor 
		observed in groups of 
		galaxies~\cite{1999Natur.397..135P};
	(4)	certain $21$-${\rm cm}$ absorption-line 
		features~\cite{2018Natur.555...67B} which could 
		be connected to radio emission from PBHs.

In a recent related study, Cappelutti {\it et al.}~\cite{Cappelluti:2021usg} have explored the high-redshift properties of PBH dark matter with an extended mass spectrum induced by the thermal history of the Universe (see Ref.~\cite{Carr:2019kxo} and Section~\ref{sec:Thermal--History-Scenario}). Their main results are summarised in Figs.~\ref{fig:figZ2} and~\ref{fig:figZ3}. Further findings are:
	(1)	a secondary peak of star formation at 
		$z \sim 15\,\text{--}\,20$ (beyond the 
		well-established observed peak at $z \sim 3$), 
		being driven by mini halos, which are likely to 
		host the first episode of Population III star 
		formation;
	(2)	a significant enhancement of the X-ray 
		background fluctuations and the unresolved 
		cosmic X-ray and infrared-background 
		cross-power spectrum, with only a minor effect 
		on the cosmic infrared-background fluctuations;
	(3)	an explanation of the integrated 
		$1\.{\rm nW}\.\mrm^{-2}\.{\rm sr}^{-1}$
		measured signal~\cite{2016MNRAS.455..282H, 
		2018RvMP...90b5006K}, which cannot be fully 
		accounted for by non-PBH cosmologies, with 
		$M > 1\.\Msun$ PBHs;
	(4)	the X-ray spectral-energy distribution of the 
		cosmic X-ray and infrared-background 
		cross-correlation signal also contains 
		information about their production mechanism 
		(\cf~Ref.~\cite{2018ApJ...864..141L}).
\newpage

Mittal \& Kulkarni~\cite{Mittal:2021dpe} discuss the generation of a radio background by accreting supermassive PBHs and claim that this could explain the excess observed by the second generation {\it Absolute Radiometer for Cosmology, Astrophysics and Diffuse Emission} (ARCADE2)~\cite{2011ApJ...734....5F} and the {\it Long Wavelength Array} (LWA1)~\cite{Dowell:2018mdb}. They take the comoving radio emissivity due to accreting PBHs to be
\begin{align}
	\epsilon_{\rm acc}( E )
		&=
				5.65 \times 10^{19}\.f_{\rm duty}\.
				( f_{\Xrm}\.\lambda )^{0.86}
				\left(
					\frac{ \fPBH\.\rho_{\rm DM} }
					{ 1\.{\rm kg}\.\mrm^{-3} }
				\right)
				\left(
					\frac{ E }{ 5.79\,{\mu\.{\rm eV}} }
				\right)^{\!-0.6}
				\srm^{-1}\mrm^{-3}
				\, .
				\label{eq:rem}
\end{align}
where $f_{\rm duty}$ is the duty cycle parameter (\ie~the probability that a black hole is active at a particular time), $\lambda$ is the luminosity in units of the Eddington value and $f_{\Xrm}$ is the ratio of the total X-ray luminosity to the bolometric luminosity. The specific intensity due to this emissivity is then
\begin{align}
	J_{{\rm acc}}( E, z )
		=
				\frac{ c }{ 4\pi }\.( 1 + z )^{3}
				\int_{z}^{1000}
					\frac{ \d z' }{ H( z' ) }\;
					\frac{ \epsilon_{\rm acc}
					\big[ E\.( 1 + z' ) / ( 1 + z ) 
					\big] }{ 1 + z' }
				\, .
				\label{eq:jacc}
\end{align}
This is shown in Fig.~\ref{fig:acc} for typical values for SMBHs, $f_{\Xrm} = 0.1$, $f_{\rm duty} = 5 \times 10^{-2}$, $\lambda = 0.5$, which shows that the observed excess (dotted blue line) and can be explained (solid blue line).

D'Agostino {\it et al.}~\cite{DAgostino:2022ckg} have examined the impact of PBH dark matter around a central SMBH on the accretion-disc luminosity of spiral galaxies and kinematic quantities. They find that suitable PBH masses are $10^{6} - 10^{12}\.\Msun$ for PBH fractions of $10^{-3} - 1$, with the best-fit configuration having $\fPBH = 1$ and $M = 10^{6}\.\Msun$.
\newpage

\begin{figure}[t]
	\centering
	\includegraphics[width=0.96\textwidth]{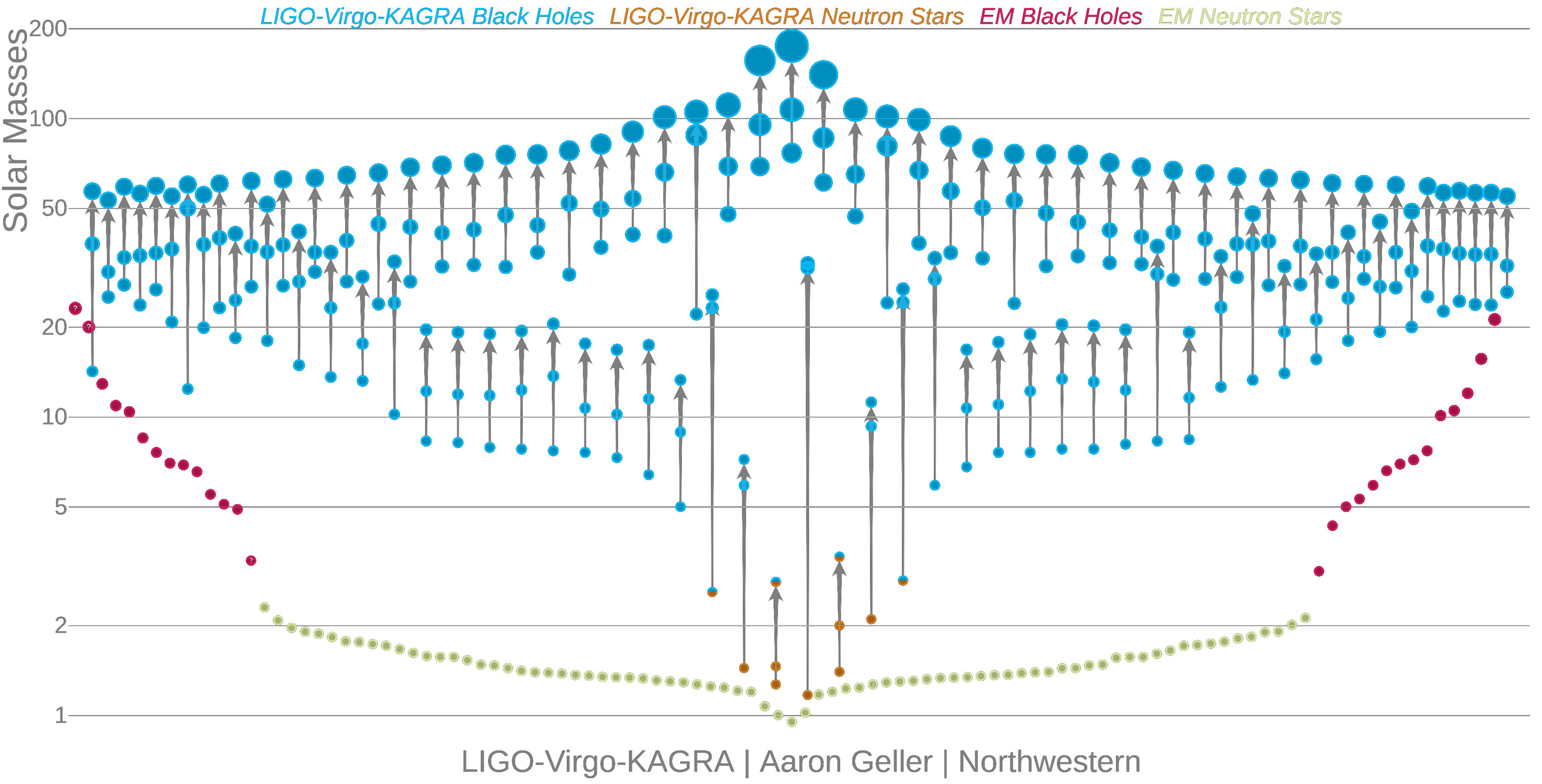}
	\vs{2mm}
	\caption{
		Summary of compact binary coalescences observed 
		by LIGO/Virgo, from 
		GWTC-3~\cite{2021arXiv211103606T} 
		(with all events having an astrophysical 
		probability greater than $0.5$), together with 
		black holes and neutron stars previously 
		constrained through electromagnetic 
		observations. Figure credit: LIGO-Virgo/Aaron 
		Geller/Northwestern University.
		\vs{5mm}
		}
	\label{fig:figLIGO}
\end{figure}

\section{Gravitational-Wave Evidence}
\label{sec:Gravitational--Wave-Evidence}

\noindent This section reviews the evidence for PBHs from gravitational-wave (GW) observations. The first detections of such waves were in 2015~\cite{Abbott:2016blz} and came from the merging of black holes which were unexpectedly large (around $30\.\Msun$) compared to those in X-ray binaries. This triggered renewed interest in PBHs, not only as a source of the mergers but also as a possible solution to the dark matter problem. Eight years later there are about $90$ detections in the third {\it Gravitational-Wave Transient Catalog} (GWTC-3)~\cite{2021arXiv211103606T}, as shown in Fig.~\ref{fig:figLIGO}, and the properties of the GW sources are still intriguing. A unifying PBH scenario may explain the merging rates, the black hole mass distribution, the existence of events in the high- and low-mass gaps for stellar remnants, the mergers with highly asymmetric masses, the reported low spins and some subsolar black hole candidates. PBHs in the stellar-mass range can also lead to a GW background at nanohertz frequency, with contributions from both the PBH binaries themselves and the waves generated at second order in perturbation theory by the density fluctuations which produce the PBHs. Such a background may have already been observed by {\it Pulsar Timing Arrays} (PTAs).

We note that connecting GW observations and the PBH abundance requires an understanding of the PBH clustering since this strongly impacts the PBH merger rates. It also requires an understanding of the PBH formation scenario, since this affects the amplitude of the scalar-induced GW background through the statistics of curvature fluctuations and the PBH mass distribution. In this section the mass of the PBH is denoted by $m$ and $f( m )$ denotes the PBH mass function.
\newpage

\subsection{Merger Rate of Compact Binaries}
\label{sec:Merger-Rate-of-Compact-Binaries}

One month after the first GW detection, Bird {\it et al.}~\cite{Bird:2016dcv} and Clesse \& Garc{\'\i}a-Bellido~\cite{Clesse:2016vqa} claimed that the expected PBH merging rates of binaries formed at late times in compact halos is compatible with the LVK analysis if they comprise all the dark matter. This was later supported by other studies~\cite{Blinnikov:2016bxu}. On the other hand, Sasaki {\it et al.}~\cite{Sasaki:2016jop} argued that the mergers of PBH binaries formed before matter-radiation equality, when neighbouring PBHs are sufficiently close to form pairs, are much more frequent and this implies that the fraction of dark matter in $30\.\Msun$ PBHs can be at most $1\%$. However, Raidal {\it et al.}~\cite{Raidal:2017mfl, Vaskonen:2019jpv} studied the evolution and merging of early PBH binaries in more detail with $N$-body simulations. They found that the rates are highly suppressed if PBHs significantly contribute to the dark matter, due to binary disruption by nearby PBHs, early-forming PBH clusters (see Section~\ref{sec:Clustering-of-PBHs}) and matter inhomogeneities. Other groups obtained similar results with both analytical and numerical methods~\cite{Ali-Haimoud:2017rtz} but without including all these effects.

The most recent rate analyses~\cite{2022arXiv220905959F, Escriva:2022bwe, Clesse:2020ghq, 2020JCAP...09..022J, 2021PhRvL.126e1302J} seem to exclude all the dark matter being in $\sim 30\.\Msun$ PBHs but it is still debated if a significant fraction of the dark matter could comprise solar-mass PBHs for the wide mass distribution expected if PBH formation is boosted at the QCD transition (see Section~\ref{sec:Thermal--History-Scenario}). Given the uncertainties related to the clustering of PBHs, it remains unclear which binary formation channel (early or late) is dominant in the solar-mass range. Another complication is that the perturbation of early binaries~\cite{2020PhRvD.101d3015V} and their formation by three-body capture~\cite{Franciolini:2022ewd} may significantly contribute to the merger rate. We now review some recent calculations of the merger rates for each production channel and show that they could be compatible with LVK observations if solar-mass PBHs dominate the dark matter.

{\bf Early Binaries\;}
A PBH binary can form when two PBHs are produced sufficiently close to each other, which happens regularly due to their Poisson distribution. The gravitational influence of one or more nearby PBHs prevents them from merging directly and allows them to form a binary which takes of order the age of the Universe to merge. For an arbitrary mass function $f( m )$, normalised to $1$, the differential merging rate $R^{\rm early} \equiv \drm R / ( \drm \ln m_{1}\,\drm \ln m_{2} )$ is given by~\cite{Raidal:2018bbj, Gow:2019pok, Kocsis:2017yty}
\bea 
\begin{split}
	R^{\rm early}(m_{1},\.m_{2})
		&=&
				1.6 \times 10^{6} \.
				f_{\rm sup}( m_{1},\,m_{2},\,z )\.
				\fPBH^{53/37}\.
				f( m_{1} )\.f( m_{2} )\mspace{-2mu}
				\left[
					\frac{ t( z ) }{ t_{0} }
				\right]^{-34/37} 
				\\[2.5mm] 
		&&
				\times\!
				\left(
					\frac{\,m_{1} +\,m_{2} }{\.\Msun }
				\right)^{\!-32/37}
				\left[
					\frac{\,m_{1}\.m_{2} }
					{ ( m_{1} +\,m_{2} )^{2} }
				\right]^{-34/37}
				{ \rm Gpc^{-3}\,yr^{-1} }
				\, .
				\label{eq:cosmomerg}
\end{split}
\eea
Assuming $m_{1} > m_{2}$ mass ordering, an additional factor of two must be included in these rates. For a monochromatic mass function, this reduces to the Sasaki {\it et al.}~\cite{Sasaki:2018dmp} expression, except for the suppression factor $f_{\rm sup}$. Analytical prescriptions derived from $N$-body simulations in Refs.~\cite{Raidal:2018bbj, Hutsi:2020sol} determine this suppression factor. It has a complex dependence on redshift, $\fPBH$ and the mass distribution $f( m )$. Nevertheless, if one assumes that the mass distribution has a sharp peak and that PBHs contribute significantly to the dark matter ($0.1 \lesssim \fPBH \leq 1$), one gets a useful low-redshift approximation~\cite{Clesse:2020ghq},
\vs{-1mm}
\begin{align}
	f_{\rm sup}
		&\approx
				2.3 \times 10^{-3} \fPBH^{-0.65}
				\, ,
				\\
				\notag
\end{align}
that is independent of the mass distribution. The more complete results of Refs.~\cite{Raidal:2018bbj, Hutsi:2020sol} have been tested against $N$-body simulations but only for monochromatic or lognormal distributions, which makes their validity uncertain for broad mass functions. Moreover, in this case, the value of $f_{\rm sup}$ depends on the low-mass cut-off~\cite{Escriva:2022bwe}. Binary disruption by nearby black holes has not been explored for very asymmetric binaries, which provides another source of uncertainty.

If one assumes $f( m \approx 2\.\Msun ) \approx 1$ and $\fPBH = 1$, one gets merger rates of order $10^{3}\;{\rm yr^{-1}\.Gpc^{-3}}$, which is in tension with the upper limits inferred from LIGO/Virgo observations. But one can have $\fPBH\gtrsim 0.1$ to explain merger rates. This could be obtained with a very wide mass distribution, implying that the PBH fraction in the solar-mass range is somewhat below $1$. However, the case of an extended mass distribution is more complex and the current prescription is not valid for arbitrary mass distributions. One must be cautious when applying them to a broad mass function, even if it exhibits a sharp solar-mass peak, because the influence of small mass black holes on binary disruption tends to be overestimated, even when they do not contribute significantly to the total PBH density. There are also subtleties related to the exact definition of the PBH mass distribution, which has led some studies~\cite{Raidal:2018bbj, Hutsi:2020sol} to use inconsistent prescriptions, as pointed out in Ref.~\cite{Escriva:2022bwe}. This can induce significant differences when the PBH density is inferred from GW observations.

Finally, large uncertainties remain in the formation rate of early binaries, the suppression factor for binaries with very low mass ratios, and the contribution to the merging rates of disrupted binaries~\cite{Raidal:2018bbj} that can dominate when $\fPBH \gtrsim 0.1$, the shape of the mass function or the clustering after matter-radiation equality. Strong claims relying on these merging rates are therefore probably premature. Nevertheless, Eq.~\eqref{eq:cosmomerg} gives a good estimate of the merging rates, at least in some cases.

{\bf Late Binaries\;}
PBH binaries can also form by capture in clusters at late times, in which case the merging rate 
becomes~\cite{Clesse:2016vqa, Clesse:2016ajp, Clesse:2020ghq, Bagui:2021dqi}
\vs{-1mm}
\bea 
	\mspace{-20mu}R^{\rm late}( m_{1},\,m_{2} )
		&=&
				R_{\rm clust}\.
				\fPBH^{2}\.f( m_{1} )\.f( m_{2} )\;
				\frac{ ( m_{1} +\,m_{2} )^{10/7} }
				{ (m_{1}\.m_{2})^{5/7}}\;
				{\rm yr^{-1}\,Gpc^{-3} }
				\, , 
				\label{eq:ratescatpure2}
\eea
where $R_{\rm clust}$ is a scaling factor that depends on the typical halo mass, size and virial velocity. Halo mass functions compatible with the standard $\Lambda$CDM scenario typically lead to $R_{\rm clust} \approx 1\,\text{--}\,10\;\rm yr^{-1}\.Gpc^{-3}$, as assumed by Bird {\it et al.}~\cite{Bird:2016dcv}. This is too low to explain the rates observed at the solar-mass scale, and a model with $f( m \approx 30\.\Msun ) \approx 1$ would be inconsistent with the increased rates expected for Poisson-induced clusters (see Section~\ref{sec:Clustering-of-PBHs}).

Poisson fluctuations induce an additional clustering compared to that expected in the $\Lambda$CDM scenario and in this case more realistic values of $R_{\rm clust}$ are between $100$ and $1000\,{\rm yr}^{-1}\,{\rm Gpc}^{-3}$. A value of around $400\,{\rm yr}^{-1}\,{\rm Gpc}^{-3}$ is needed to explain exceptional events like GW190425, GW190814 and GW19052 with the wide mass distribution expected in the 'thermal-history' scenario~\cite{Clesse:2020ghq} (see Section~\ref{sec:Thermal--History-Scenario}). The exact value of $R_{\rm clust}$, however, is still uncertain and model-dependent. It could have a more or less complex dependence on the component masses, $\fPBH$ and redshift.

From the PBH cluster properties identified in Section~\ref{sec:Clustering-of-PBHs}, one can estimate $R_{\rm clust}$ as a function of the PBH velocity and the enhanced local density contrast $\delta^{\rm local}$ compared to the cosmological dark matter density, which gives
\vs{-1mm}
\begin{align}
	R_{\rm clust}
		&=
				\frac{ 2 \pi\.\delta^{\rm local}\.
					\Omega_{\Mrm}^{2}\.\rho_{\crm}\.G }
					{ c }
				\left(
					\frac{ 85\.\pi }{ 6 \sqrt{2} }
				\right)^{\!2 / 7}\!
				\left(
					\frac{ c }{ \sqrt{2}\.v_{\rm vir} }
				\right)^{\!11 / 7}\!
				{\rm yr^{-1}\,Gpc^{-3} }
				\, .
\end{align} 
With realistic values of $\delta^{\rm local} = 3\,M_{\rm halo} / ( 4 \pi\.r_{\rm halo}^{3}\.\rho_{\rm DM}^{0} )$, assuming clusters with $M_{\rm halo} = 10^{6}\.\Msun$ and $r_{\rm halo} \approx 20$\,pc and the corresponding virial velocity, this gives $R_{\rm clust} \approx 100 \, \rm yr^{-1}\,Gpc^{-3}$, while for $M_{\rm halo} = 10^{7}\.\Msun$ and $r_{\rm halo} \approx 10$\,pc, one gets $R_{\rm clust} \approx 750$. Merger rates that are obtained with these values are consistent with those inferred from observations at the solar-mass scale.

Using the most recent merging rate calculations and the latest limits from GWTC-3~\cite{KAGRA:2021duu}, one can conclude that PBHs around $30\.\Msun$ can explain at most $0.1\,\text{--}\,1\.$\% of the dark matter, despite Poisson clustering reducing the merging rate of early binaries and increasing the rate of late ones. However, it is still possible that PBHs in the solar-mass range explain at least a significant fraction of the dark matter. In this case, the rate of late and early binaries are comparable. In order to distinguish these two binary formation mechanisms, or to distinguish late PBH binaries from standard astrophysical ones, it will be possible to use the next generation of GW detectors to measure the orbital eccentricities of inspiraling late binaries that are initially highly eccentric~\cite{Cholis:2016kqi}. At the solar-mass scale, the PBH merger rate also coincides with the neutron star merger rate inferred from GW observations. Apart from GW170817 (which had an electromagnetic counterpart), these could be PBHs at the QCD peak. PBH scenarios with extended mass functions generically lead to many asymmetric mergers (involving PBHs of order $10$ and $1\.\Msun$) and their rate is of order $10\;{\rm yr^{-1}\.Gpc^{-3}}$. On the other hand, the observed merging rates of asymmetric binaries like GW190814 challenges all current astrophysical models, as mentioned in the abstract of the LIGO/Virgo discovery paper of GW190814~\cite{2020ApJ...896L..44A}. In the next section, we expand the discussion of the evidence for PBHs from the mass distribution of the GW events.

\subsection{Mass Distribution}
\label{sec:Mass-Distribution}

\begin{figure}[t]
	\centering 
	\includegraphics[width = 0.75 \textwidth]{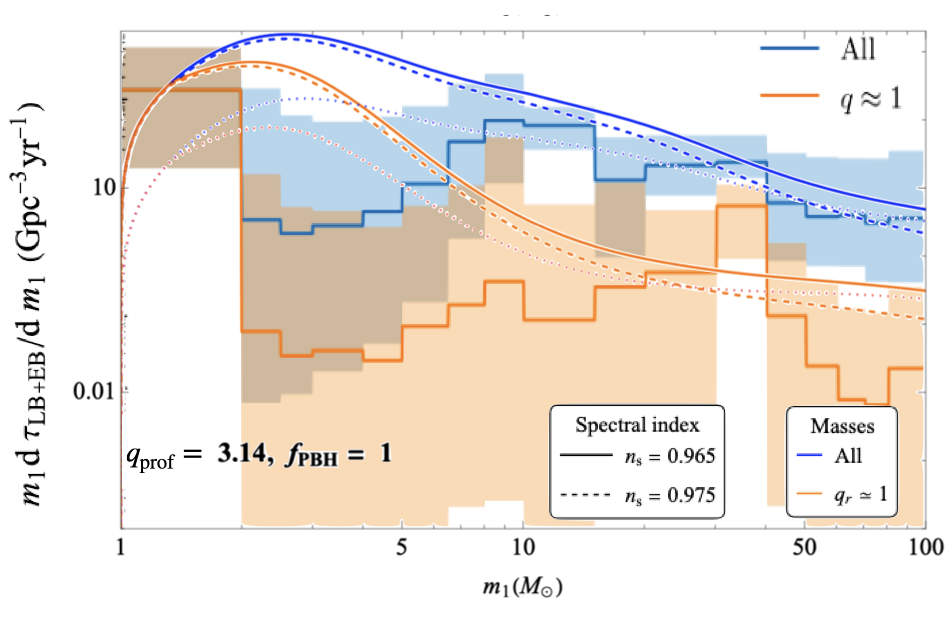}
	\caption{
		Reconstructed rate distribution and their
		90\% confidence intervals as a function of 
		$m_{1}$ for equal-mass (orange band) and 
		non-equal mass (blue band) mergers, using 
		GWTC-3 data, for the binned Gaussian 
		process model with uniform 
		$(\log m_{1},\log m_{2})$ bins.
		Superimposed are typical mass 
		distributions predicted by the simulations 
		of Ref.~\cite{Escriva:2022bwe}. These 
		include QCD effects and assume Gaussian 
		fluctuations for a nearly 
		scale-invariant power spectrum with 
		spectral index $n_{\rm s}$ and 
		$\fPBH = 1$ for the combined merger 
		rates of early and late binaries. The 
		parameter $q_{\rm prof}$ describes the 
		density profile. From 
		Ref.~\cite{KAGRA:2021duu}.
		}
		\vs{2mm}
	\label{fig:1DRate-LIGO}
\end{figure}

While merging rates indicate the PBH abundance, the mass distribution of the coalescences indicates the PBH mass function and therefore provides an important probe of the scenario. It was argued in Refs.~\cite{Clesse:2016vqa, Kovetz:2016kpi, Kovetz:2017rvv} that the PBH mass distribution could be reconstructed from the distribution of LVK merger events anticipated in the next observing run. It was also proposed that a quantity related to the merger-rate distribution, 
\begin{equation}
	\alpha
		\equiv
				- ( m_{1} + m_{2} )^{2}\.
				\frac{ \partial^{2} \ln R}
				{\partial m_{1}\.\partial m_{2}}
				\, ,
\end{equation}
could be used to identify a primordial component of black hole mergers: one should have $0.97 \lesssim \alpha \lesssim 1.05$ for early binaries and $\alpha \simeq 10/7$ for late binaries, independent of the PBH mass function~\cite{Kocsis:2017yty}. According to the most recent population analysis for GWTC-3~\cite{KAGRA:2021duu}, the large number of merger events with component masses around $30\.\Msun$ indicates a statistically preferred model with a peak at this mass scale. If these black holes are primordial, one might associate this excess with the QCD-induced feature expected at this scale. However, one should be cautious about this because an event excess need not imply a rate or black hole mass function excess. For example, the PBH mass distributions obtained with some of the most recent numerical simulations, taking into account the effects of the equation-of-state variations at the QCD epoch on the dynamics of the spherical inhomogeneities, do not lead to a significant peak in the rate distribution around $30\.\Msun$~\cite{2022arXiv220906196E}. Therefore, it is not straightforward to assess whether such models are favoured or disfavoured compared to stellar-black-hole models. This is illustrated in Fig.~\ref{fig:1DRate-LIGO}.
\newpage

Because the sensitivity of the detectors depends on frequency, they are more able to detect mergers with large chirp masses, even if these are less frequent than solar-mass mergers. In order to predict the observed mass distribution, one must therefore multiply the merging rates by the volume probed, which is fixed by the detector range:
\begin{equation}
	R_{\rm det}
		\approx
				\frac{ \sqrt{5} }{ 24 }\.
				\frac{ ( G\mspace{1mu}\Mcal\.
					c^{3} )^{5/6} }
				{ \pi^{2/3} }\,
				\frac{ 1 }{ 2.26 }
				\left[
					\int_{f_{\rm min}}^{f_{\rm max}}\.
					\frac{ f^{-\alpha } }{ S_{h}( f ) }
				\right]^{1/2}
				\, ,
\end{equation}
where $\Mcal$ is the binary chirp mass, $f_{\rm min} = 50\,$Hz is the minimum frequency used in the search, $f_{\rm max}$ is the smaller of the maximal search frequency ($\sim 4000\,$Hz) and the merger frequency [estimated as twice the innermost stable circular orbit (ISCO) frequency $f_{\rm ISCO} = 4400\.\Msun / ( m_{1} + m_{2} )\,{\rm Hz}$], and $S_{h}( f )$ is the spectral density of the detector noise. The exponent $\alpha$ is $7/3$ in the post-Newtonian limit up to the ISCO frequency and $2 / 3$ during the merging phase. For distant sources, one includes the redshift of the GW signal by replacing the source-frame chirp mass $\Mcal$ with $\Mcal\,( 1 + z )$ and adapting the probed frequency range. Another, more precise, method used in Ref.~\cite{Escriva:2022bwe} is to multiply the merger rates by the volume-time sensitivity $\langle VT \rangle$, computed by the LVK collaboration by injecting sets of waveform into the data, marginalised over all possible inclinations, sky locations, luminosity distances, spins etc.

\begin{figure}[t]
	\centering 
	\includegraphics[width = 0.7 \textwidth]{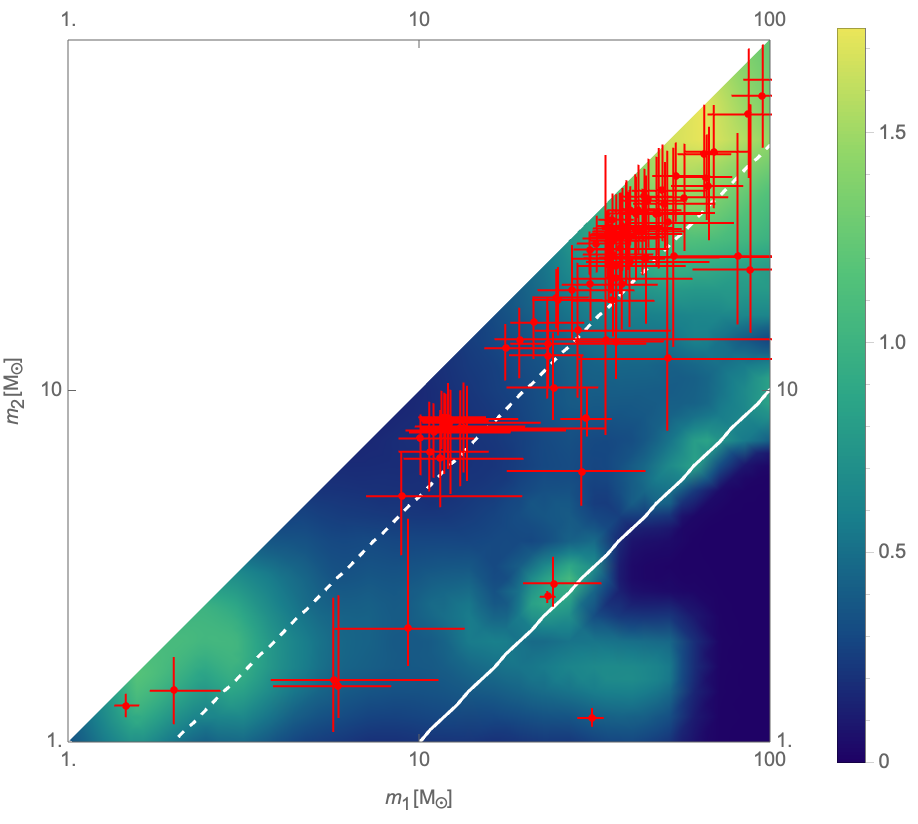}
	\caption{
		Expected probability distribution of PBH 
		mergers with masses $m_{1}$ and $m_{2}$ for a 
		mass function with $n_{\srm} = 0.97$ and LIGO 
		sensitivity for O3 run. Solid and dashed white 
		lines correspond to mass ratios 
		$q \equiv\,m_{2} / m_{1}$ of $0.1$ and $0.5$, 
		respectively. From Ref.~\cite{Escriva:2022bwe}.
		}
		\vs{5mm}
	\label{fig:Rate-LIGO-events}
\end{figure}

The resulting distribution of detections for the combination of early and late binaries in the unified model of Ref.~\cite{Carr:2019kxo}, but improved by using simulations of PBH formation at the QCD epoch~\cite{Escriva:2022bwe}, is shown in Fig.~\ref{fig:Rate-LIGO-events}. GW events are most likely observed between $30\.\Msun$ and $60\.\Msun$, which is where most observations lie. The second most likely region is the solar-mass scale, as expected for QCD peak. One coalescence has been observed in this region without an electromagnetic counterpart (GW190425) and attributed to a neutron star merger~\cite{Abbott:2020uma}. For early binaries, the event distribution would have a dominant peak at the solar-mass scale and so it is less likely that all the black holes between $20\.\Msun$ and $100\.\Msun$ were in early binaries. The distribution also exhibits a region with less expected events, between $5\.\Msun$ and $15\.\Msun$, where a few observations were made. This seems to conflict with the favoured PBH scenario but the importance of this gap depends on the model details and the statistical significance of this problem should be studied with Bayesian methods, as in Ref.~\cite{2022arXiv220905959F}.

Finally, some regions are not occupied by stellar black hole mergers but are a distinctive prediction of PBH models. These regions correspond to the pair-instability mass gap (above $60\.\Msun$), the subsolar region, the low-mass gap (between approximately $2.5$ and $5\.\Msun$) and binaries with low mass ratios. There are multiple observations of compact objects in these regions and GWTC-3 indicates that these regions exhibit only smooth transitions in merger rates~\cite{KAGRA:2021duu}. This is hard to explain with a stellar model but natural for PBHs. We now discuss each case in more detail.
\newpage

\begin{table*}
\vs{-4.5mm}
\begin{ruledtabular}
	\begin{tabular}{c c c c }
		Event & Mass 1 $[\.\Msun]$ & Mass 2 $[\.\Msun]$ 
		& Rate $\Rcal [\rm{yr^{-1} Gpc^{-3}} ]$\\
		\hline\\[-3.6mm]
		GW190403\_{0}51519	& $85^{+27.8}_{-33.0}$		& $20.0^{+26.3}_{-8.4}$		& \\ 
		GW190426\_{1}90642	& $105.5^{+45.3}_{-24.1}$	& $76.0^{+26.2}_{-36.5}$	& \\
		GW190519\_{1}53544	& $66.0^{+10.7}_{-12.0}$	& $40.5^{+11.0}_{-11.1}$	& \\
		GW190521 			& $95.3^{+28.7}_{-18.9}$	& $69.0^{+22.7}_{-23.1}$	
		& $[0.02\,\text{--}\,0.43]$\\
		GW190602\_{1}75927	& $69.1^{+15.7}_{-13.0}$	& $47.8^{+14.3}_{-17.4}$	& \\
		GW190706\_{2}22641	& $67.0^{+14.6}_{-16.2}$	& $38.2^{+14.6}_{-13.3}$	& \\
		GW190929\_{0}12149	& $80.8^{+33.0}_{-33.2}$	& $24.1^{+19.3}_{-10.6 }$	& \\
		GW191109\_{0}10717	& $65^{+11}_{-11}$			& $47^{+15}_{-13}$			& \\
		GW200220\_{0}61928	& $87^{+40}_{-23}$			& $61^{+26}_{-25}$			& \\[1mm]
		\hline\\[-3.8mm]
		GW190814			& $23.2^{+1.1}_{-1.0}$		& $2.59^{+0.08}_{-0.09}$ 	
		& $[1\,\text{--}\,23]$\\
		GW190917\_{1}14630	& $9.7^{+3.4}_{-3.9}$		& $2.1^{+1.1}_{-0.4}$		& \\
		GW190924\_{0}21846	& $8.9^{+7.0}_{-2.0}$		& $5.0^{+1.4}_{-1.9 }$		& \\
		GW191219\_{1}63120	& $31.1^{+2.2}_{-2.8}$		& $1.17^{+0.07}_{-0.06}$	& \\
		GW200105\_{1}62426	& $9.0^{+1.7}_{-1.7}$		& $1.91^{+0.33}_{-0.24}$	& \\
		GW200210\_{0}92254	& $24.1^{+7.5}_{-4.6}$		& $2.83^{+0.47}_{-0.42 }$	& \\ 
	\end{tabular}
\end{ruledtabular}
\label{tab:exceptional}
\vs{-1mm}
\caption{
	GW events reported in the catalogs
	GWTC-1~\cite{LIGOScientific:2018mvr}, 
	GWTC-2~\cite{LIGOScientific:2020ibl}, 
	GWTC-2.1~\cite{LIGOScientific:2021usb} and 
	GWTC-3~\cite{2021arXiv211103606T}, with at least 
	one component in the pair-instability mass gap 
	(above $60\.\Msun$) or low-mass gap (between 
	$2.5\.\Msun$ and $5\.\Msun$), or with low mass 
	ratios ($q \sim 0.1$), together with the mass 
	uncertainties. When available, the associated 
	merger rates are given.
	}
\end{table*}

\subsection{Black Holes in the Pair-Instability Mass Gap}
\label{sec:Black-Holes-in-the-Pair--Instability-Mass-Gap}

The most intriguing and unexpected GW observations involve black hole mergers with component masses in the pair-instability mass gap. Above a mass of around $60\.\Msun$, the temperature in the core of stars becomes so high during oxygen-burning that electron-positron pair production leads to a reduction of the pressure and core collapse. Below a mass of around $150\.\Msun$, the star explodes as a supernova without any remnant. As a result, stars are not expected to directly form black holes with mass between about $60\.\Msun$ and $150\.\Msun$. This pair-instability mass gap is well established theoretically, even if the exact range can be debated.

With component masses of $66\.\Msun$ and $85\.\Msun$, GW190521 is the first observation of at least one (and probably two) black holes in this mass gap. Reference~\cite{Nitz:2020mga} argues that the error bars on the masses might allow the two objects to be just at the limit of the mass gap. However, GWTC-3 contains several additional candidates 
(GW190403{\textunderscore}051519, 
GW190426{\textunderscore}190642, 
GW190519{\textunderscore}153544, 
GW190602{\textunderscore}175927, 
GW190706{\textunderscore}222661, 
GW190929{\textunderscore}012149, 
GW191109{\textunderscore}010717, 
GW200220{\textunderscore}061928) 
with components in the range from $61\.\Msun$ to $107\.\Msun$ (shown in {Tab.~\ref{tab:exceptional}} together with corresponding uncertainties), so this becomes unlikely. A deeper analysis of the merger rates provides statistical evidence for a smooth transition across the mass gap, with no evidence for a break~\cite{KAGRA:2021duu}. Also, there is some evidence for the dynamical formation of the progenitor binary~\cite{Gamba:2021gap}, with some Bayesian evidence in favour of a hyperbolic merger. PBHs offer an interesting alternative explanation for the events, still in the context of a dynamical capture, since their mass function should vary smoothly in the $10\,\text{--}\,100\.\Msun$ range. However, if they form from primordial density fluctuations, there should be a bump at around $30\,\text{--}\,100\.\Msun$ coming from the epoch of the pion formation (see Section~\ref{sec:Thermal--History-Scenario}).

In principle, black holes in the pair-instability mass gap could result from previous mergers (\eg~of two $30\,\text{--}\,50\.\Msun$ black holes). However, the observed merger rates require binary formation in dense black hole environments, such as globular clusters or active galactic nuclei disks. There are several difficulties with this scenario. First, it is intriguing that GW190521, GW190426\_{1}90642 and GW200220\_{0}61928 probably have two components in the mass gap; one would expect the secondary black holes to mainly merge with primary black holes between $20\.\Msun$ and $50\.\Msun$. Second, the velocity kick acquired in a previous merger typically exceeds the escape velocity of globular clusters. Third, the black hole mass and rate distribution above $20\.\Msun$ change continuously across the pair-instability mass gap. Even if one cannot rule it out, the previous-merger scenario may conflict with the latest observations.

\subsection{Black Holes in the Low-Mass Gap}
\label{sec:Black-Holes-in-the-Low--Mass-Gap}

Observations of X-ray binaries provide strong statistical evidence for a gap between approximately $2.5\.\Msun$ and $5\.\Msun$ in the stellar black hole mass distribution~\cite{Bailyn:1997xt, Farr:2010tu}. This can be explained by numerical simulations of supernova explosions due to carbon deflagration. A gap in the mass distribution of binary coalescences from GW observations was therefore expected. However, the latest catalogs~\cite{LIGOScientific:2020ibl, LIGOScientific:2021usb, 2021arXiv211103606T} reveal two mergers with at least one progenitor in the expected gap (GW190814 and GW200210\_{0}92254) and two others lying near the boundary of the gap (GW190924\_{0}21844, GW200115\_{0}42309). Overall, observations do not support the existence of this gap, even if the statistical significance is not yet very high and the mass uncertainties are large. The existence of black holes in the mass gap is also supported by microlensing observations of OGLE/{\it Gaia} towards the Galactic centre (see Section~\ref{sec:Galactic-Bulge}). If confirmed, this discrepancy between X-ray and GW observations cannot easily be explained with stellar models. However the existence of PBHs in the low-mass gap is natural since they would come from the large-mass tail of the QCD peak.

\subsection{Binaries with Asymmetric Masses}
\label{sec:A-Binary-with-Asymmetric-Masses}

The event GW190814~\cite{2020ApJ...896L..44A} is exceptional, not only because its secondary component lies in the lower mass gap, but also because of the observed very low mass ratio, $q \equiv\,m_{2} / m_{1} \approx 0.1$. The existence of such asymmetric binaries is not excluded for astrophysical black hole populations, but it is difficult to explain why their associated merger rate is only slightly lower than for binaries with similar masses. Asymmetric binaries are also interesting because one can reconstruct the individual spin of the primary component $\chi_{1}$, which is very low for GW190814: $\chi_{1} < 0.07$ at $90\%$ CL (see next subsection). This has led the LVK collaboration to claim that {\it ``the combination of mass ratio, component masses, and the inferred merger rate for this event challenges all current models of the formation and mass distribution of compact-object binaries."} On the other hand, such binaries are expected for PBH models~\cite{Clesse:2020ghq}, especially when one includes QCD-induced features in their mass function: a dominant bump at $m_{1} \approx 2.5\.\Msun$ and a smaller bump at $m_{1} \approx 30\,\text{--}\,50\.\Msun$. GW190814 could therefore involve PBHs from both bumps, as shown in Fig.~\ref{fig:Rate-LIGO-events}. We note that Eqs.~(\ref{eq:cosmomerg}, \ref{eq:ratescatpure2}) naturally give rates of order $R \sim 10\,{\rm yr}^{-1}\,{\rm Gpc}^{-3}$ with $f( m_{1} ) \sim 1$ and $f( m_{2} ) \sim 0.02$.
\newpage

GWTC-3 contains information about a few additional coalescences with mass ratios below $0.25$: GW200210\_{0}92254 (similar to GW190814), GW200115\_{0}42309 ($5.4\.\Msun$ and $1.4\.\Msun$) as well as GW191219\_{1}63121. For the latter, the secondary mass is around $1.2\.\Msun$ and $q \simeq 0.04$, which would make it the lowest mass neutron star ever observed. The mass properties of these GW events are indicated in Tab.~\ref{tab:candidate}. As shown in Fig.~\ref{fig:Rate-LIGO-events}, one can obtain a broad range of low-mass-ratio mergers for a wide PBH mass function. Future observations may soon provide additional information on the abundance of such asymmetric binaries and their merger rates. For PBHs, one cannot exclude LVK observing even lower mass ratios (down to $q \simeq 0.001$) using continuous-wave methods~\cite{Miller:2021knj, Miller:2020kmv} but specific waveforms must be assumed for such searches. For broad PBH mass distributions, asymmetric binaries also boost the stochastic GW background to a level observable with upcoming runs~\cite{Bagui:2021dqi}.

\subsection{Subsolar Candidates}
\label{sec:Subsolar-Candidates}

\begin{figure}[t]
	\centering
	\includegraphics[width=0.8\textwidth]{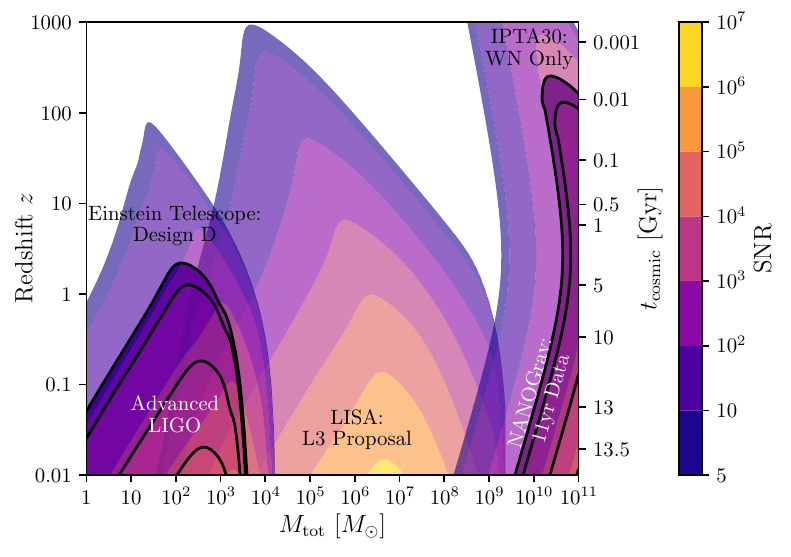}
	\caption{
		Signal-to-Noise Ratio (SNR) as a function of 
		source redshift and total source mass for 
		various GW detectors: 
			{\it Einstein Telescope} (design model D), 
			advanced LIGO, 
			LISA (L3 proposal), 
			IPTA30 (WN only) and 
			NANOGrav (11yr).
		From Ref.~\cite{Kaiser:2020tlg}.}
	\label{fig:horizon}
\end{figure}

Given the broad variety of PBH and stellar black hole scenarios, such models could accommodate very different mass, rate and spin distributions. It is therefore crucial to identify a clear signature of PBHs that cannot be explained by black holes of stellar origin. There are two observations that would point towards a PBH origin. The first is detecting a black hole merger from a redshift before the first stars formed (viz.~$z > 20$)~\cite{2020JCAP...08..039C}. For chirp masses above or around $30\.\Msun$, such redshifts will become accessible to the third generation of ground-based GW detectors, such as the {\it Einstein Telescope}~\cite{2020JCAP...03..050M} and {\it Cosmic Explorer}~\cite{Evans:2021gyd}. Their expected redshift range for different values of the signal-to-noise ratios is shown in Fig.~\ref{fig:horizon}, as a function of the total binary black hole mass. As indicated there, intermediate-mass or supermassive black holes will be probed by space interferometers like LISA~\cite{LISACosmologyWorkingGroup:2022jok} and pulsar timing arrays. The second signature would be the detection of a black hole smaller than $1\.\Msun$. Detecting a subsolar-mass black hole can already be achieved with the current generation of GW detectors. The first subsolar LIGO/Virgo searches were unsuccessful~\cite{Abbott:2018oah, Authors:2019qbw} but only focused on binaries with a primary component smaller than $2\.\Msun$. Similar searches were performed independently~\cite{Nitz:2022ltl, Nitz:2021vqh} and rate limits obtained for asymmetric binaries in Refs.~\cite{Nitz:2021mzz, Nitz:2020bdb}.

A recent re-analysis of data from the second Advanced LIGO run, including binaries with a $2\.\Msun < m_{1} < 10\.\Msun$ primary component and a subsolar secondary component, has revealed four subsolar black hole candidates with a signal-to-noise larger than $8$ and a false-alarm-rate smaller than $2$ per year~\cite{2021arXiv210511449P}. These threshold are the usual ones for LVK detections. Two candidates are seen by only a single detector (one in Livingston and the other in Hanford), so are more likely attributed to noise but the others are seen in several detectors.

The properties of these candidates, inferred from the template bank, are shown in Tab.~\ref{tab:candidate}. Some parameter estimation is ongoing to check their validity and confirm that the secondary component is indeed subsolar. Although noise might mimic such a signal, we tentatively consider these four candidates as possible PBHs. Recently, three additional subsolar triggers have been reported by the second half of the third LIGO/Virgo run~\cite{LVK:O3bSSM} based on the same threshold. Their properties are shown in Tab.~\ref{tab:candidate} but they were only detected by one of the pipelines used in the search. If more clear-cut candidates are observed in future observing runs, this would clearly provide evidence for PBHs. Alternative explanation exist, such as black holes formed from accretion of dark matter particles~\cite{Garani:2021gvc, Kouvaris:2018wnh}, by dissipative dark matter~\cite{Gurian:2022nbx, Shandera:2018xkn}, or neutron stars with non-standard equation of state (a subsolar neutron star may have already been observed~\cite{2022NatAs.tmp..224D}). However, PBHs would be the favoured interpretation of subsolar-mass compact objects.

\begin{table*}
\label{tab:candidate}
\vs{1.5mm}
\begin{ruledtabular}
\begin{tabular}{c c c c c c c c c c}
	Date 			& FAR [yr$^{-1}$]	& $m_{1} [\Msun]$	& $m_{2} [\Msun]$	& spin-1-$z$
	& spin-2-$z$	& H SNR				& L SNR				& V SNR				& Network SNR \\
	\hline
	2017-04-01	& 0.41 & 4.90 & 0.78 & $-0.05$ 
	& $-0.05$ & 6.32 & 5.94 & -- & 8.67 \\
	2017-03-08	& 1.21 & 2.26 & 0.70 & $-0.04$ 
	& $-0.04$ & 6.32 & 5.74 & -- & 8.54 \\
	\hline
	2020-03-08	& 0.20 & 0.78 & 0.23 
	& $\phantom{-}$0.57 & $\phantom{-}$0.02 & 6.31 
	& 6.28 & -- & 8.90\\
	2019-11-30	& 1.37 & 0.40 & 0.24 
	& $\phantom{-}$0.10 & $-0.05$ & 6.57 & 5.31 
	& 5.81 & 10.25\\
	2020-02-03	& 1.56 & 1.52 & 0.37 
	& $\phantom{-}$0.49 & $\phantom{-}$0.10 & 6.74 
	& 6.10 & -- & 9.10\\
\end{tabular}
\caption{Reported properties (date, component masses 
	and spins) of the subsolar triggers in 
	Refs.~\cite{2021arXiv210511449P, LVK:O3bSSM}, 
	with a network signal-to-noise ratio (SNR) $> 8$ 
	and false-alarm rate (FAR) $< 2\,{\rm yr}^{-1}$ in 
	the second and third observing runs of LIGO/Virgo.
	\vs{7mm}}
\label{tab:candidate}
\end{ruledtabular}
\end{table*}

\begin{figure}[t]
	\centering 
	\includegraphics[width = 0.65 \textwidth]{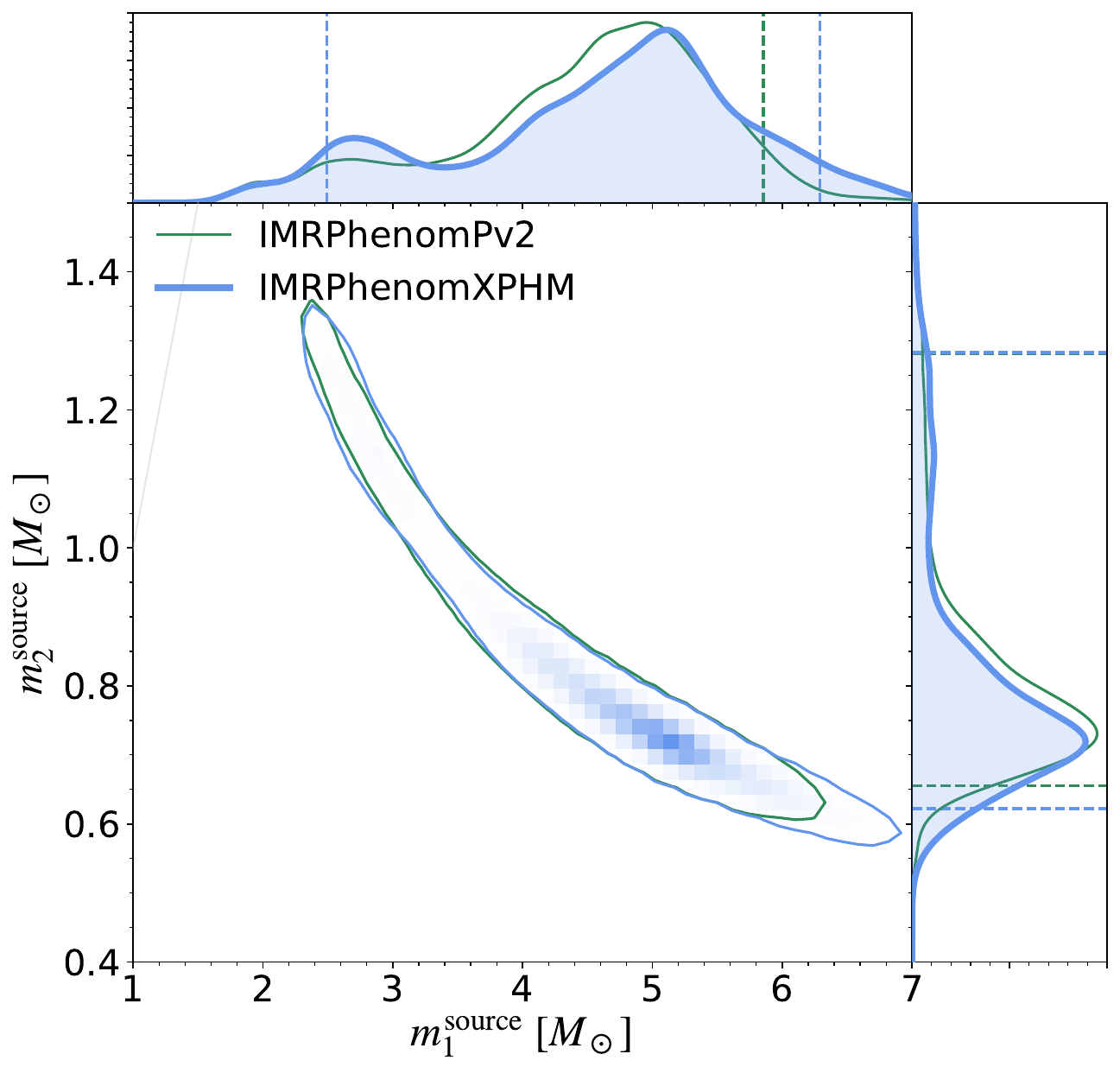}
	\caption{
		Posterior probability distributions of the 
		primary and secondary component masses $m_{1}$ 
		and $m_{2}$ of SSM170401, observed during
		the second observing run of Advanced LIGO, 
		assuming that the signal comes from a real GW 
		event. A subsolar-mass $m_{2}$ is preferred at 
		$84\%$ CL but one cannot rule out that the 
		signal is due to noise or a neutron star/black 
		hole merger. From Ref.~\cite{Morras:2023jvb}.
		}
	\label{fig:SSM-events}
\end{figure}

For one of these subsolar-mass triggers, SSM170401, a parameter estimation has been performed~\cite{Morras:2023jvb}, assuming that the signal comes from a real GW event. The results are consistent with expectation for a real signal. However, after removing a glitch, even if one considers more precise waveforms and a lower minimum frequency of $20\,$Hz, the signal-to-noise ratio does not increase and is even slightly reduced from $8.6$ to $7.94$ if one includes marginalisation over the analysed parameters. Nevertheless, the posterior probability distribution for the secondary mass supports a subsolar-mass at $84$\% CL. The posterior distributions are shown in Fig.~\ref{fig:SSM-events}. These results are not sufficiently statistically significant to claim a firm observation of a subsolar object, so the signal could be due to noise, but it remains an interesting possibility. In order to better assess this possibility, one would have to run similar parameter estimations for the other triggers reported in the third observing run. One also has to understand the origin of the different ranking statistics obtained with the various pipelines. But it is plausible that SSM170401 is the first subsolar-mass PBH observed by GW detectors. We emphasise that observing a subsolar-mass black hole is the only way to prove the existence of PBHs with the current generation of detectors, so this research line should be pursued in a cautious way, given the far-reaching consequence of such a discovery.

\begin{figure}[t]
	\centering 
	\includegraphics[width = 0.75 \textwidth]{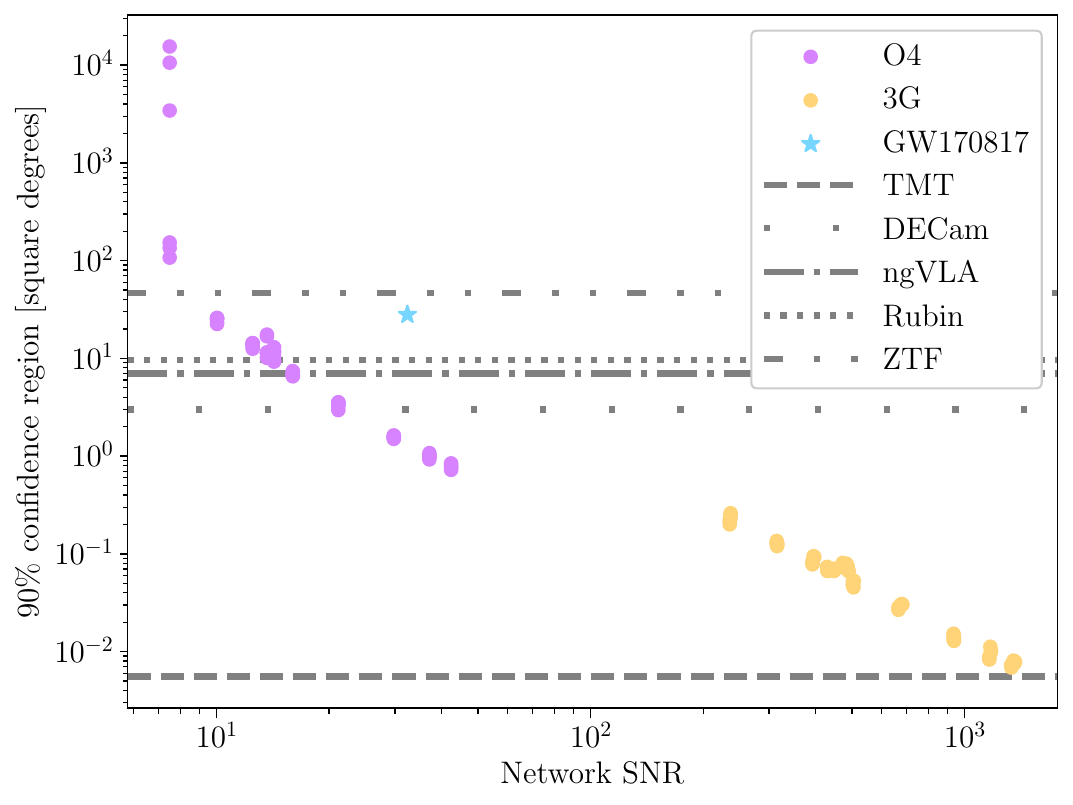}
	\caption{
		Localisation areas ($90\%$ CL) 
		as function of network SNR for all 
		subsolar-mass merger signals studied in 
		Ref.~\cite{Wolfe:2023yuu}. Coloured dots 
		indicate injected signals into the LVK 
		O4-design sensitivity network (purple) and a 
		network of {\it Cosmic Explorer} and 
		{\it Einstein Telescope} (yellow), 
		respectively. The light-blue star indicates the 
		near equal-mass binary neutron star GW170817.
		Also shown as gray lines are the fields of view 
		for the electromagnetic follow-up observations 
		with
			ZTF (dash-dot-dotted), 
			{\it Vera Rubin Observatory} (dotted), 
			{\it next-generation Very Large Array} 
				(ngVLA; dash-dotted), 
			{\it Dark Energy Camera} 
				(DECam; wide-spaced dots), 
		and 
			the {\it Thirty-Meter Telescope} 
			(TMT; dashed).
		From Ref.~\cite{Wolfe:2023yuu}.
		}
	\label{fig:sky-localisation}
\end{figure}

In a recent study, Wolfe {\it et al.}~\cite{Wolfe:2023yuu} have presented a detailed analysis of the measurability of the mass, spin and sky location for binary black hole mergers with least one subsolar-mass component. They find that next-generation detectors, such as {\it Cosmic Explorer} and the {\it Einstein Telescope}, will enable measurements of the source-frame component masses down to $\Ocal( 10^{-5} )\.\Msun$. Furthermore, due to the long duration of these signals, it should be possible to confidently identify sub-solar components at the threshold of detectability even during the fourth observing run of LVK, with these events being well sky-localised and possibly revealing characteristic spin information. This is indicated in Fig.~\ref{fig:sky-localisation}. However, for a clear attribution of these signals to black holes, future work on the development of waveforms{\,---\,}calibrated for subsolar-mass compact-object mergers and including tidal deformability{\,---\,}is necessary.
\newpage

\subsection{Black Hole Spins}
\label{sec:Black-Hole-Spins}

Another unexpected property of the first detected coalescences is their low effective spins, often compatible with zero. The effective spin $\chi_{\rm eff}$ of a binary is defined as
\begin{equation}
	\chi_{\rm eff}
		\equiv
			\frac{ 1 }{\,m_{1} +\,m_{2} }\.
			\Big[
				m_{1} S_{1} \cos( \theta_{\rm LS_{1}} )
				+
				m_{2} S_{2} \cos( \theta_{\rm LS_{2}} )
				\Big]
				\, ,
\end{equation}
where $m_{i}$ is the mass, $S_{i}$ the spin and $\theta_{\rm LS_{i}}$ the angle between the spin and the orbital momentum for each component ($i = 1,\mspace{1mu}2$). If the two masses are almost equal, a low effective spin may indicate that the individual spins are similar and anti-aligned, as expected for stellar black hole binaries. But the fact that $\chi_{\rm eff}$ remains low and compatible with zero for the vast majority of binaries observed so far~\cite{2021arXiv211103606T}, even for different progenitor masses, suggests that the individual spins are also low. This is further supported by the individual spin of the primary component of GW190814, $\chi_{1} \equiv S_{1} \cos \theta_{\rm LS_{1}} < 0.07$ at $90\%$ CL, this being well determined due to the mass asymmetry. Low spins are also most likely for the primary components of other asymmetric binaries, which disfavours a spin orientation that is orthogonal to the orbital plane for GW190814 by chance. Low spins for black hole binaries are not expected for popular stellar models, such as the Geneva model, since these should produce a wide effective spin distribution with a peak around $\chi_{\rm eff} \approx 0.8$.
\newpage

In some astrophysical scenarios, it is possible to produce such low spins (see \eg~Refs.~\cite{Belczynski:2017gds, Bavera:2020inc}), so an astrophysical origin cannot be excluded on this basis alone. But low spin is a generic outcome for PBHs since these must have almost vanishing spin at formation~\cite{DeLuca:2019buf}. This is because they are close to the Hubble horizon size at formation and therefore do not collapse much. Also the rare large initial fluctuations from which they form are expected to be nearly spherical~\cite{1986ApJ...304...15B}. While chaotic accretion may reduce the PBH spin, other factors{\,---\,}such as coherent accretion, previous mergers and hyperbolic encounters in dense PBH environments{\,---\,}are expected to broaden the spin distribution~\cite{2021PDU....3400882J}. In particular, for coalescences in the pair-instability mass gap, this would result in a distribution of $\chi_{\rm eff}$ centred on zero but with some non-zero values. There is some evidence for non-zero $\chi_{\rm eff}$ in the most recent catalogue. In particular, GW190521 has at least one component in the pair-instability gap. PBHs therefore provide a natural explanation for the low binary spins inferred from GW observations and this complements the other evidence.

In order to quantify the weight of this evidence, several studies~\cite{Fernandez:2019kyb, Garcia-Bellido:2020pwq, Wong:2020yig, DeLuca:2021wjr} have compared the Bayes factor for PBH and astrophysical black hole models. All of them favour either PBHs or a mixed population. However, more data and more precise PBH and astrophysical models are needed to better distinguish them with a Bayesian approach~\cite{Garcia-Bellido:2020pwq}.

Finally, some events in GWTC-3 seem to have spins compatible with a random spin orientation, which suggests a dynamical formation channel in a dense black hole environments~\cite{KAGRA:2021duu}. Active galactic nuclei and globular clusters are a plausible context for this but PBHs are also expected to be clustered due to Poisson fluctuations (Section~\ref{sec:Clustering-of-PBHs}). An important difference between the two scenarios is that black holes from secondary mergers should have all acquired large spins, whereas smaller spins are expected for PBHs. A statistical analysis of the spins of black holes in the pair-instability mass gap could soon discriminate between these explanations.

\subsection{Pulsar-Timing Arrays}
\label{sec:Pulsar--Timing-Arrays}

The NANOGrav collaboration have recently reported a possible detection of a GW background at nano-Hertz frequency~\cite{NANOGrav:2020bcs} using pulsar timing arrays (PTA). This has been confirmed by the {\it European Pulsar Timing Array} (EPTA)~\cite{Chen:2021rqp}, Parkes~\cite{Goncharov:2021oub} and IPTA~\cite{Antoniadis:2022pcn} collaborations, although the GW origin still had to be demonstrated. This claim was confirmed in June 2023, based on updated PTA data from the {\it North American Nanohertz Observatory for Gravitational Waves} (NANOGrav)~\cite{NANOGrav:2020bcs}, the {\it European Pulsar Timing Array} (EPTA)~\cite{Chen:2021rqp}, the {\it Parkes}~\cite{Goncharov:2021oub} and {\it International Pulsar Timing Array} (IPTA)~\cite{Antoniadis:2022pcn} collaborations. Strong evidence was also found for the Hellings--Downs correlations~\cite{1983ApJ...265L..39H} predicted by the quadrupolar nature of GWs.

\begin{figure}[t]
	\centering
	\vs{-3mm}
	\includegraphics[width=0.65\textwidth]{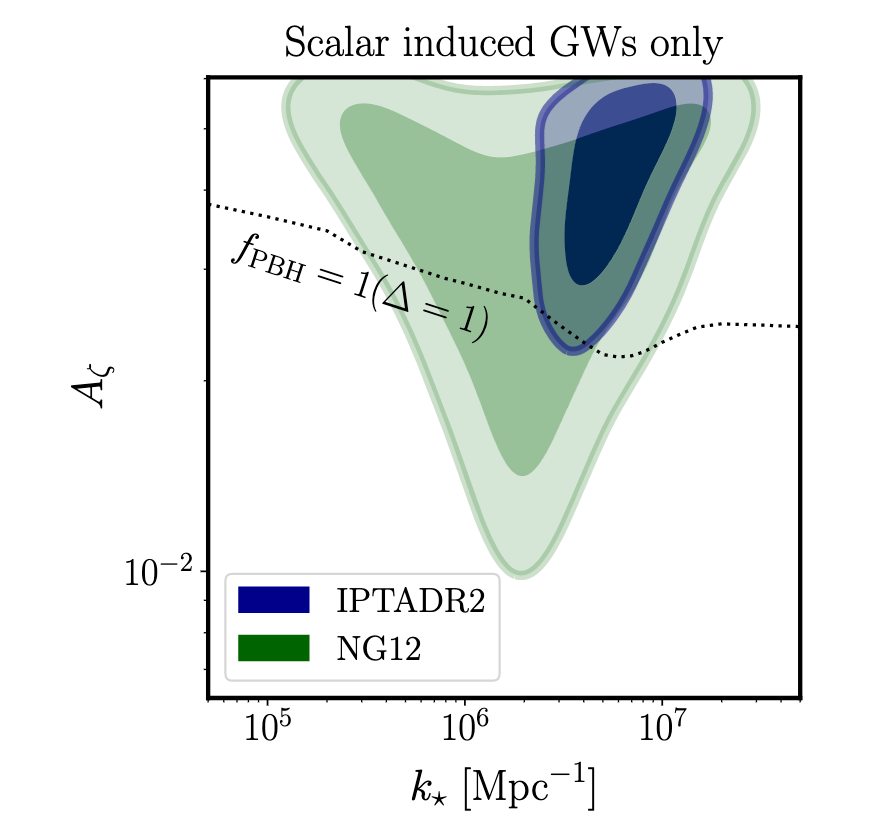}
	\caption{
		Two-dimensional posterior distributions 
		($1\sigma$ and $2\sigma$ contours) of the 
		primordial power spectrum amplitude $A_\zeta$ 
		at peak scale $k_{*}$, assuming a log-normal 
		shape of width $\Delta = 1$, from the NANOGrav 
		$12.5$-year (green) and second IPTA data 
		release (blue) observations of PTAs, if the 
		observed signal comes from scalar-induced GWs.
		The dotted line represents the expected 
		amplitude leading to $\fPBH = 1$ on which one 
		can identify the bumpy feature due to the 
		QCD-epoch leading to solar-mass PBHs. From 
		Ref.~\cite{Dandoy:2023jot}.
		}
	\label{fig:PTA}
\end{figure}

Such a GW background could have been sourced by the second-order perturbations associated with the large density fluctuations needed to form PBHs. Several groups have suggested this and the signal frequency would be associated with PBHs in the stellar~\cite{DeLuca:2020agl, Vaskonen:2020lbd, Kohri:2020qqd, Yi:2021lxc, Dandoy:2023jot, Inomata:2020xad} or even planetary~\cite{Domenech:2020ers} mass range. There were also claims that the amplitude of the PTA signal excludes PBHs as the source of the signal because it would require a PBH abundance larger than that associated with the dark matter~\cite{Inomata:2020xad, Dandoy:2023jot, Ellis:2023oxs}. However, the uncertain shape of the primordial power spectrum, the possibility of non-Gaussianity~\cite{Wang:2023ost, Franciolini:2023pbf} and non-Gaussian tails~\cite{Ferrante:2023bgz} in the fluctuation statistics~\cite{Inomata:2020xad}, the model dependence of the critical overdensity threshold~\cite{DeLuca:2023tun}, together with other uncertainties, make it difficult to estimate the PBH dark matter fraction required to explain such a signal. Nevertheless, the range $10^{-3} < \fPBH < 1$ could be accommodated. Given its shape, and assuming Gaussian fluctuations, it is still possible to set limits on and identify the most likely amplitude of the primordial power spectrum on scales that are relevant for the formation of stellar-mass PBHs. The results obtained using the most recent IPTA data and a log-normal primordial power spectrum are shown in Fig.~\ref{fig:PTA}, where the required amplitude to get $\fPBH = 1$ is also estimated~\cite{Dandoy:2023jot}.

Even if some papers claim that there is an inconsistency between the value of $\fPBH$ required for this model and the merger rates inferred from LVK observations, given the large uncertainties and model-dependence of the critical threshold~\cite{DeLuca:2023tun}, mass distributions, merger rates and possible non-Gaussianity, we regard the order-of-magnitude coincidence between the value of $\fPBH$ required for the NANOGrav signal and the LIGO/Virgo merger rates as favouring PBHs. Even if the NANOGrav detection has a GW origin, there are alternative sources, such as SMBH binaries or cosmic strings, and the recent analysis of Ref.~\cite{Ellis:2023oxs} compares possible astrophysical and cosmological explanations. Another PBH-related explanation, given that the signal for extended mass distribution is boosted, could be late binaries in clusters~\cite{Bagui:2021dqi} or SMBH binaries of primordial origin~\cite{Depta:2023qst}. It has also been proposed that such a GW background is produced by the mergers of stupendously large PBHs~\cite{Atal:2020yic}, or during a first-order phase transition at the origin of stellar-mass or heavier PBHs~\cite{Ashoorioon:2022raz}.

\newpage

\section{Potential Solutions to Cosmic Problems}
\label{sec:Potential-Solutions-to-Cosmic-Problems}

\noindent There are a number of other unresolved problems in cosmology which PBHs might help to solve. We discuss them separately from the problems discussed above, since they are less studied and the arguments 
are less compelling.

\subsection{Fast Radio Bursts and Transmuted Stars and Stellar Remnants}
\label{sec:Fast-Radio-Bursts-and-Transmuted-Stars-and-Stellar-Remnants}

Fast radio bursts (FRBs) are bright radio transient events with millisecond pulse widths and GHz frequencies; their nature is still unknown~\cite{Nicastro:2021cxs}. All of them are extragalactic and most are non-repeating. Many explanations have been proposed, including cataclysmic events involving merging neutron stars and stellar black holes, neutron star seismic activity, black hole accretion and active galactic nuclei (see Refs.~\cite{Katz:2018xiu, Petroff:2019tty, Xiao:2021omr} for recent reviews). FRBs may also be explained by collisions of neutron stars with planetary-mass PBHs. For example, Abramowicz {\it et al.}~\cite{Abramowicz:2017zbp} argue that this could apply if PBHs of around $10^{21}\,$g constitute around $1\%$ of the dark matter. This mechanism can also lead to the production of some $r$-process elements, as argued in Ref.~\cite{Fuller:2017uyd}. This corresponds to the only known mechanism to explain the surprisingly large amount of gold in the Universe, which neutron star mergers appear to underproduce by a large factor~\cite{Kobayashi:2020jes}. Interestingly, this suggestion would be consistent with the `thermal-history' model~\cite{Carr:2019kxo}, which is discussed in Section~\ref{sec:Thermal--History-Scenario}.

As pointed out by Kainulainen {\it et al.}~\cite{Kainulainen:2021rbg}, the Bondi formula implies that the characteristic time for accretion of neutron star material onto the PBH is 
\begin{equation}
	t_{\Brm}
		\approx
				2.5\,\text{yr}\,
				\bigg(
					\frac{ c_{\srm} }{ 0.6 }
				\bigg)^{\!3}
				\bigg(
					\frac{ 10^{15}\,\grm\,
					{\rm cm}^{-3} }
					{ \rho_{\crm} }
				\bigg)
				\bigg(
					\frac{ 10^{-12}\.\Msun }{ M }
				\bigg)
				\, .
\end{equation}
Using $c_{\srm} = 0.6$ as a characteristic sound speed in the core of a neutron star, this is longer than $10\,$yr for PBHs smaller than $2 \times 10^{-13}\.\Msun$. Reference~\cite{Kainulainen:2021rbg} suggests that one needs accretion on this timescale to explain the irregular bursts of the repeating FRBs, such as FRB 121102 (\cf~Ref.~\cite{Houben:2019xla}), through intermittent reconnection of the magnetic field-line bundles. Hence planetary-mass PBHs could be the source of repeating FRBs. Indeed, a PBH mass function with $\fPBH \sim 0.1$ in the range $10^{-14}\,\text{--}\,10^{-11}\.\Msun$ ($10^{19}\,\text{--}\,10^{22}\,$g) might explain all the FRBs.

\begin{figure}[tb]
	\includegraphics[width=0.82\linewidth]{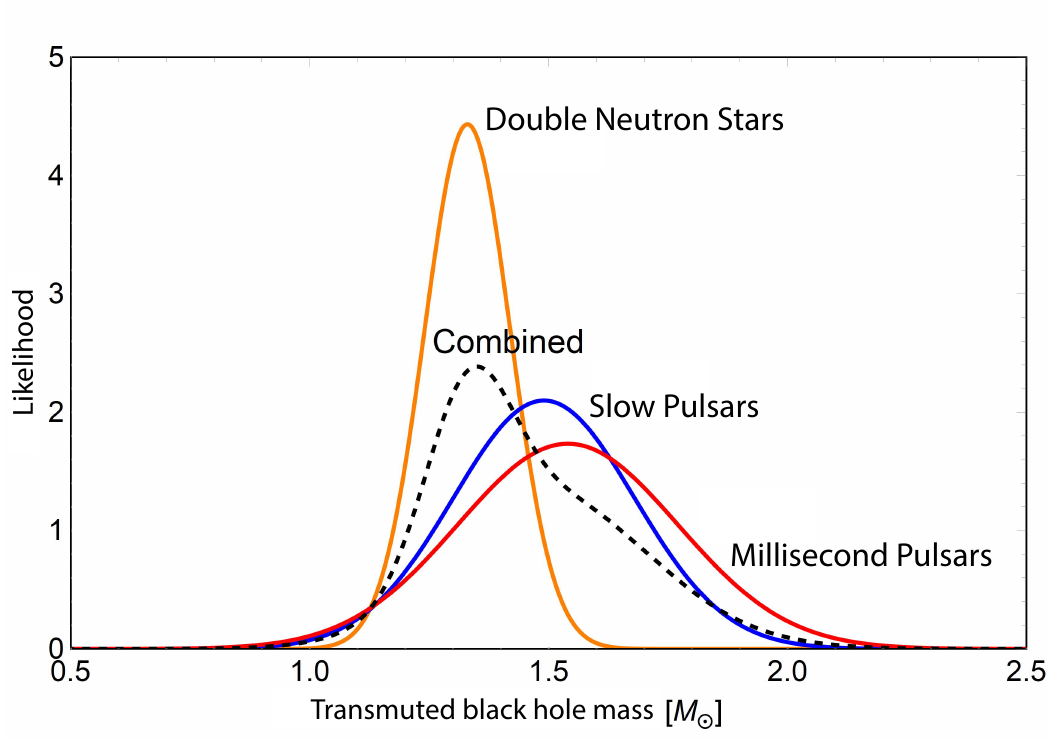}
	\caption{
		Expected mass distribution of transmuted 
		solar-mass black holes, assuming they
		track their neutron star progenitors, for the 
		subpopulations originating from slow pulsars 
		(blue), recycled fast-rotating millisecond 
		pulsars (red) and double neutron stars 
		(orange). Also shown is the combined 
		distribution (black dashed). Input parameters 
		for the Gaussian distributions of these 
		populations have been taken from 
		Ref.~\cite{2013ApJ...778...66K, 2016ARA&A..54..401O}. Adapted from 
		Ref.~\cite{2021PhRvL.126g1101T}.
		}
	\label{fig:nsmass}
\end{figure}

In a related proposal, Takhistov {\it et al.}~\cite{2021PhRvL.126g1101T} have suggested that small (\ie~subsolar) PBHs can be captured by neutron stars and then consume them, thereby converting them into black holes of $1\,\text{--}\,2\.\Msun$ (see Refs.~\cite{2013PhRvD..87l3524C, 2017PhRvL.119f1101F, 2018PhLB..782...77T} for earlier works on PBH-induced implosions of neutron stars). The resulting mass distribution is depicted in Fig.~\ref{fig:nsmass}, broken down into its subcomponents. Its shape is very different from that expected in standard astrophysical scenarios, so they conclude that the observation of a population of $1\,\text{--}\,2\.\Msun$ black holes provide support for the original PBHs.

The recently discovered {\it G objects} in the Galactic Centre~\cite{s41586-019-1883-y} may be clouds of gas bound by the gravitational field of stellar-mass black holes. As pointed out by Flores {\it et al.}~\cite{Flores:2023lll}, these could have primordial seeds. Specifically, they argue that light PBHs of mass in the range $10^{-16}\,\text{--}\,10^{-10}\.\Msun$ could have transmuted neutron stars into a population of $1\,\text{--}\,2\.\Msun$ black holes which subsequently develop gaseous atmospheres. This may explain the properties of the {\it G} objects, such as their avoidance of tidal disruption from the SMBH at the Galactic Centre, while generating the observed emission.

PBHs in the cores of ordinary stars (as well as neutron stars) could have observable consequences, as first suggested by Hawking~\cite{Hawking:1971ei}. For a range of stellar masses and metallicities, Bellinger {\it et al.}~\cite{Bellinger:2023wou} have delineated the evolution of stars hosting a PBH in their core. The lightest PBHs leave no evolutionary imprint but the larger ones can partially consume the star, with striking consequences for its spectrum. For example, a solar-type star could develop a convective core, which could not be produced via ordinary stellar evolution and might be asteroseismologically detectable. Also a giant star would exhibit pure-pressure-mode pulsations.

\subsection{Missing-Pulsar Problem}
\label{sec:Missing--Pulsar-Problem}

\begin{figure}[t]
	\centering 
	\includegraphics[width = 0.48 \textwidth]{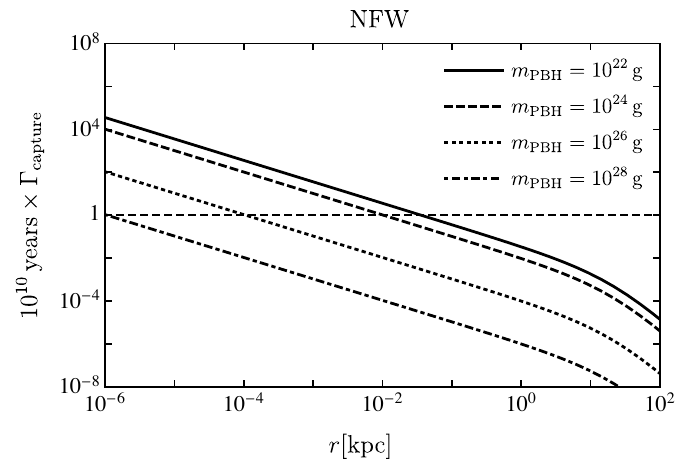}\\
	\includegraphics[width = 0.48 \textwidth]{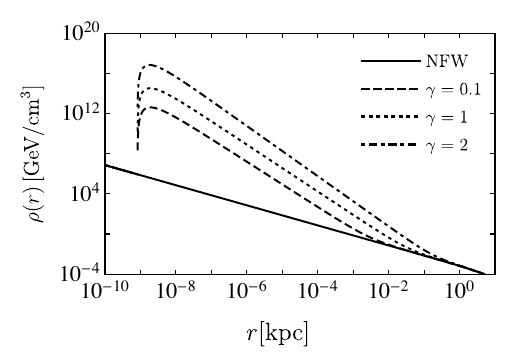}
	\hs{3mm}
	\includegraphics[width = 0.48 \textwidth]{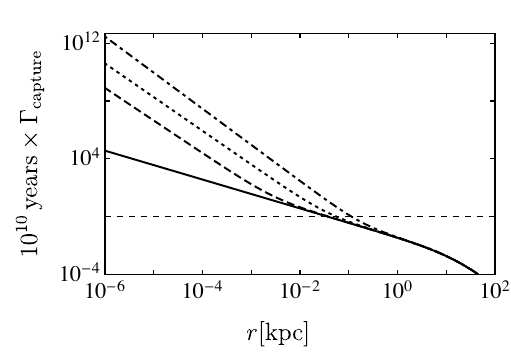}
	\caption{
		{\it Upper panel$\mspace{1mu}$}: Capture rate 
			\eqref{eq:Pcapture} as a function of 
			Galactocentric distance assuming an NFW 
			profile, $\fPBH = 1$, 
			$\sigma_{\vrm} = 7\,$km/s 
			and
				$\MPBH = 10^{22}\,$g	(solid),
				$10^{24}\,$g			(dashed),
				$10^{26}\,$g			(dotted),
				$10^{28}\,$g			(dot-dashed).
		{\it Lower-left panel$\mspace{1mu}$}: 
			Spike halo density profile \eqref{eq:rho}
			with
				$\gamma = 0.1$			(dashed),
				$\gamma = 1$			(dotted),
				$\gamma = 2$			(dot-dashed)
			as a function of Galactocentric distance, 
			compared to NFW profile (solid).
		{\it Lower-right panel$\mspace{1mu}$}: 
			Capture rate \eqref{eq:Pcapture} for 
			$\MPBH = 10^{22}\,$g as a function of 
			Galactocentric distance using the halo 
			density profiles of the left panel.
		}
	\label{fig:NS-Capture-Probability-Spike}
\end{figure}

One expects a large population of pulsars in the Galactic centre but none have been detected within the innermost $20\,$pc and there is a lack of old pulsars even at much larger distances~\cite{Bower:2018mta}. This may be accounted for if the Galactic halo comprises PBHs because they would sink into the centres of the pulsars due to dynamical friction and then consume them. Assuming a Maxwellian PBH velocity distribution, the capture rate for a PBH by a neutron star is~\cite{Capela:2013yf}
\begin{equation}
	\Gamma_{\rm cap}
		=
				\fPBH\.\sqrt{6\pi}\;
				\frac{ \rho_{\rm DM} }{ M }\.
				\frac{ 2\.G M_{\rm NS}\.R_{\rm NS} }
				{ \sigma_{\vrm}\.( 1 - 2\.G 
				M_{\rm NS} / R_{\rm NS}) }\!
				\left[
					1
					-
					\exp\!
					\left(
						-
						\frac{ 3\.E_{\rm loss} }
						{ M \sigma_{\vrm}^{2} }
					\right)
				\right]
				.
				\label{eq:Pcapture}
\end{equation}
Here, $\rho_{\rm DM}$ is the local dark matter density, $\sigma_{\vrm}$ is the velocity dispersion of the PBHs, $E_{\rm loss}$ is the energy lost in the process, and $R_{\rm NS}$ and $M_{\rm NS}$ are radius and mass of the neutron star, respectively. The upper panel of Fig.~\ref{fig:NS-Capture-Probability-Spike} shows the number of captures over $10^{10}\,$yr as a function of Galactocentric distance, assuming a NFW profile~\cite{Navarro:1995iw}. For PBHs smaller than about $10^{24}\,\grm$, no neutron star older than $10^{10}\,$yr survives within the innermost $10\.\fPBH\,$pc and this increases to $100\.\fPBH\,$pc for $M \sim 10^{20}\,$g.

This effect is significantly enhanced if the dark matter density near the Galactic centre is increased relative to an NFW profile due to adiabatic accretion of dark matter, as expected in the spike models of Gondolo and Silk~\cite{Gondolo:1999ef}. Sadeghian {\it et~al.}~\cite{Sadeghian:2013laa} suggest the following analytic approximation for the spike density profile, which we depict in the lower-left panel of Fig.~\ref{fig:NS-Capture-Probability-Spike} (see also Refs.~\cite{1980ApJ...242.1232Y, 1986ApJ...301...27B}):
\begin{align}
	\rho_{\rm sp}( r )
		&\propto
				\left(
					\frac{R_{\rm{sp}}}{r_{0} }
				\right)^{\!- \gamma}\!
				\left(
				 	1
					-
					\frac{ 2\.R_{\Srm} }{ r }
				\right)^{\!3}\!
				\left(
					\frac{ R_{\rm sp} }{ r }
				\right)^{\!- ( 9 - 2\.\gamma ) / 
				( 4 - \gamma )}
				\quad
				( 2\.R_{\Srm} < r < R_{\rm sp} )
				\, .
				\label{eq:rho}
\end{align}
Here, $R_{\Srm}$ is the Schwarzschild radius of Sgr\,${\rm A}^{\!*}$ and $R_{\rm sp}$ is the spike radius (see Refs.~\cite{Gondolo:1999ef, Nishikawa:2017chy} for details). The lower-right panel of Fig.~\ref{fig:NS-Capture-Probability-Spike} shows the capture rate for the profile \eqref{eq:rho} and indicates that a steeper profile (\ie~larger $\gamma$) results in more efficient removal. For $\gamma = 2$ and $\MPBH = 10^{22}\,$g, no neutron stars older than $10^{7}\fPBH\,$yr survive inside $10\,$pc. $N$-body simulations using more refined PBH velocity distributions and mass profiles will be needed to better assess this model.

\subsection{Hubble Tension}
\label{sec:Hubble-Tension}

The value of the Hubble constant reported by the Planck collaboration~\cite{Planck:2018vyg}, $H_{0} = 67.37 \pm 0.54\,$km/s/Mpc, is less than the value measured by Cepheids and Type Ia supernov{\ae}~\cite{Riess:2018byc}, $H_{0} = 73.52 \pm 1.62\,$km/s/Mpc. No standard explanation has been found so far. Some ways to weaken or resolve this tension have been discussed, for instance due to systematic uncertainties in the Cepheid colour-luminosity calibration~\cite{Mortsell:2021tcx}, or due to the fact that $30\%$ of the CMB dipole is not caused by the movement of the Sun, implying an error in Planck's dipole subtraction~\cite{Horstmann:2021jjg}. However, the current mainstream view is that this tension is real. Future research is clearly needed to clarify the situation.

If the Hubble tension prevails, PBHs could help to alleviate it. Eroshenko~\cite{Eroshenko:2021oeq} studied a scenario in which PBHs have 
	(1) a mass distribution peaking in the range 
		$10^{20}\,\text{--}\,10^{24}$\,g, and 
	(2) a weakly-relativistic velocity dispersion.
He points out that if these PBHs have been in dense clusters with initial masses $18\,\text{--}\,560\.\Msun$, their further dynamical evolution might have lead to a vastly increased merger rate, transforming about $10\%$ of their initial mass into gravitational waves. In turn, the Hubble constant might have increased by $\sim 5\%$ as compared to its value without these merger processes. This would imply a shortening of the first acoustic-peak scale, such that the present Hubble constant $H_{0}$ would be less than that deduced from local measurements. Papanikolaou~\cite{Papanikolaou:2023oxq} has recently argued that the Hubble tension could also be alleviated through the evaporation of PBHs with mass less than $10^{9}\,$g via dark radiation degrees of freedom having feeble couplings to the Standard Model. These would have been injected into the primordial plasma, increasing the effective number of neutrino species and in turn the value of the Hubble parameter. Papanaikolaou also argues that such PBHs can generate primordial magnetic fields~\cite{Papanikolaou:2023nkx}.
\newpage

\subsection{Falling Rotation Curves at Large Distances}
\label{sec:Falling-Rotation-Curves-at-Large-Distances}

There is increasing evidence for non-flat rotation curves of galaxies at $z \sim \Ocal( 1 )$~\cite{Genzel:2017jgd}, with the curves falling beyond twice the half-light radius (\cf~Fig.~\ref{fig:fig1} in Section~\ref{sec:Lensing-Evidence}). This could be accounted for if the dark matter were subdominant in the outer regions of galactic disks at those redshifts. This is expected if the dark matter comprises PBHs in dense clusters since{\,---\,}as they merge with larger halos{\,---\,}they would tend to fall to the centre. Moreover, in the standard $\Lambda$CDM scenario, galaxy halos are formed by successive mergers of smaller halos and galaxies move along cosmic web filaments, where most of the baryonic gas is located. For PBH dark matter, the galaxy halos have edges, so the rotation curves of galaxies should fall off at sufficiently large distances. This works for PBHs but not particle dark matter since only the former are affected by dynamical friction. The galactocentric distance at which this fall-off occurs depends on the type of galaxy and its merger history. In the case of the Milky Way, this happens at around $30\,$kpc, which is closer than the Large and Small Magellanic Clouds~\cite{Calcino:2018mwh}. This significantly weakens the microlensing constraints from the Large Magellanic Cloud, as discussed in Section~\ref{sec:Lensing-Evidence}. The estimate in Ref.~\cite{Calcino:2018mwh} suggests $\fPBH \simeq 0.4$ at $M \simeq 1\.\Msun$, which agrees with the MACHO microlensing result and the thermal-history scenario.

\subsection{Hyper-Velocity Stars}
\label{sec:Hyper--Velocity-Stars}

High-velocity stars have been recently detected in the Milky Way halo and observations suggest a Galactic bulge and LMC origin~\cite{2017MNRAS.469.2151B}. They could have resulted from slingshot interactions with massive PBHs~\cite{Garcia-Bellido:2017fdg} or massive black hole binaries~\cite{Yu:2003hj}. The existence of the former in the Galactic centre is supported by a recent observation~\cite{2017NatAs...1..709O}. High-velocity stars are also detected in the core of globular clusters~\cite{Lutzgendorf:2012tn}, which might also indicate a population of massive PBHs. A recent analysis~\cite{Montanari:2019rvu} has found a few candidates for high-velocity stars in the second data-release catalog of the {\it Gaia} mission, which seem to come from the centre of the Sagittarius dwarf galaxy, a highly compact tidal stream with a high concentration of black holes at the centre. These are probably primordial, given the stellar evolution in the tidally stripped dwarf galaxy. The core of the Sagittarius dwarf galaxy could be a cluster of a thousand PBHs with mass peaked around a solar mass~\cite{Montanari:2019rvu, Montanari:2020gcr}.

\subsection{Properties of Low-Mass X-Ray Binaries}
\label{sec:Properties-of-Low--Mass-X--Ray-Binaries}

The majority of so-called low-mass X-ray binaries have been detected in globular clusters or towards the Galactic centre, but not in the Galactic disk~\cite{Bahramian:2022msw}. This favours formation through a process of tidal capture. If stellar-mass PBHs provide a significant fraction of the dark matter, they should reside in both globular clusters and the Galactic bulge and their contribution may dominate that from black holes of stellar origin and neutron stars. This suggests that the compact objects in low-mass X-ray binaries could have a mass below the Chandrasekhar mass and are PBHs rather than neutron stars~\cite{2018PDU....22..137C}. Because the GW amplitude depends on the black hole mass, while the X-ray luminosity from a low-mass X-ray binaries is detectable whatever the PBH mass, a broad PBH mass distribution centred on a few solar masses would also explain why most low-mass X-ray binaries have a mass in that range, whereas GW events involve heavier black holes due to the detector optimal sensitivity at higher masses.

\subsection{Evaporating Black Holes}
\label{sec:Evaporating-Black-Holes}

There are numerous constraints on evaporating PBHs~\cite{2010PhRvD..81j4019C}. For example, the strongest limit on the collapse fraction $\beta( M )$ is associated with the ones with mass $M_{*} \approx 5 \times 10^{14}\,$g which are completing their evaporation at the present epoch. This comes from the extragalactic $100\,$MeV background~\cite{MacGibbon:1991vc}, and the general conclusion is that this could not have been generated by PBHs~\cite{2021RPPh...84k6902C}. Nevertheless, evaporating PBHs have been invoked to explain various other cosmological features. While none of the arguments may be compelling, they should certainly be regarded as potential signatures and will therefore be summarised here.

Several authors have suggested that the $511\,$keV annihilation-line radiation from the Galactic centre could result from the positrons emitted by the PBHs there~\cite{Bugaev:2008gw, Bambi:2008kx}, as first suggested by Okele and Rees~\cite{1980A&A....81..263O}. It has also been suggested that the pregalactic medium could have been reionised by PBH evaporations~\cite{Belotsky:2014twa}. Overproduction of ${}^{7}{\rm Li}$ by a factor $\sim 3$ is the most serious issue with the standard cosmological nucleosynthesis model~\cite{Coc:2017rpv} and astrophysical solutions (such as invoking particle decays to reduce the lithium abundance via neutron injection) have failed so far. However, PBHs evaporating at $10^{2}\,\text{--}\,10^{4}\,$s might provide a plausible alternative. For example, PBHs with an extended mass function around $10^{12}\,$g could provide an early injection of neutrons and then soft $\gamma$-rays. This combination could destroy some ${}^{7}{\rm Be}$ at $\sim 10^{2}\,\srm$, avoiding overproduction of ${}^{7}{\rm Li}$, and subsequently destroy excess Deuterium at $\sim 10^{3}\,\text{--}\,10^{4}\,\srm$.

Historically, there has been much interest in whether the explosions of PBHs of mass $M_{*}$ could be detected as $\gamma$-ray bursts. The extragalactic $\gamma$-ray background limit implies that the PBH explosion rate $\Rcal$ could be at most $10^{-6}\,{\rm pc}^{-3}\,{\rm yr}^{-1}$ if the PBHs are uniformly distributed or $10\,{\rm pc}^{-3}\,{\rm yr}^{-1}$ if they are clustered inside galactic halos~\cite{Page:1976wx}. The direct observational constraints on the explosion rate come both from $\gamma$-ray bursts and high-energy cosmic-ray showers. In the Standard Model, where the number of elementary particle species never exceeds around $100$, the likelihood of detecting such explosions is very low. However, the physics of the QCD phase transition is still uncertain and the prospects of detecting PBH explosions would be improved in less conventional models. In particular, Cline and colleagues have argued that the formation of a fireball at the QCD temperature could explain some of the short-period $\gamma$-ray bursts (\ie~those with duration less than $100\,$ms~\cite{Cline:1992ps}). In Ref.~\cite{Cline:2001tq} they claim to find $42$ BATSE ({\it Burst And Transient Source Experiment}) candidates of this kind and the fact that their distribution matches the spiral arms suggests that they are Galactic. In Ref.~\cite{Cline:2005xb} they report a class of short-period {\it Konus} bursts which have a harder spectrum than usual and identify these with exploding PBHs. In Ref.~\cite{Cline:2006nv} they find a further $8$ candidates in the {\it Swift} data. Overall they claim that the BATSE, {\it Konus} and {\it Swift} data correspond to a $4.5\sigma$ effect and that several events exhibit the time structure expected of PBH evaporations~\cite{Cline:2009ni}.

If PBHs of mass $M_{*}$ are clustered inside our own Milky Way halo, as expected, then there should also be a Galactic $\gamma$-ray background and, since this would be anisotropic, it should be separable from the extragalactic background. The ratio of the anisotropic to isotropic intensity depends on the Galactic longitude and latitude, the Galactic core radius and the halo flattening. Some time ago Wright~\cite{1996ApJ...459..487W} claimed that such a halo background had been detected in EGRET ({\it Energetic Gamma Ray Experiment Telescope}) observations between $30\,$MeV and $120\,$MeV and attributed this to PBHs. His detailed fit to the data, subtracting various other known components, required the PBH clustering factor to be comparable to that expected and the local explosion rate to be $\Rcal = 0.07\,\text{--}\,0.42\,{\rm pc}^{-3}\,{\rm yr}^{-1}$. A later analysis of EGRET data between $70\,$MeV and $150\,$MeV, assuming a variety of distributions for the PBHs, was given by Lehoucq {\it et al.}~\cite{Lehoucq:2009ge}. Subsequently, Ref.~\cite{Carr:2016hva} took into account several other physical factors, but this work only claimed to provide a constraint rather than positive evidence. Note that these analyses do not constrain PBHs of {\it initial} mass $M_{*}$ because these no longer exist. Rather, they constrain PBHs of {\it current} mass $M_{*}$, and this corresponds to an initial mass of $1.26\,M_{*}$\,. This contrasts to the situation with the extragalactic background, where the strongest constraint on $\beta( M )$ comes from the time-integrated contribution of the $M_{*}$ black holes.

It has sometimes been suggested that the tiny amount of antimatter in cosmic rays could be generated by PBH evaporations since these should emit particles and antiparticles in equal numbers~\cite{Barrau:1999sk}. Galactic antiprotons can constrain PBHs only in a very narrow range around $M \approx M_{*}$ since they diffusively propagate to the Earth on a timescale much shorter than the cosmic age. However, the evaporation of PBHs with $M > 10^{17}\,$g is expected to inject sub-GeV electrons and positrons into the Milky Way. These particles are shielded by the solar magnetic field for Earth-bound detectors, but not for {\it Voyager} 1, which is now beyond the heliopause. Boudaud \& Cirelli~\cite{Boudaud:2018hqb} use its data to constrain the number of PBHs in the Milky Way. However, these are only claimed as constraints.

If PBH evaporations leave stable Planck-mass relics, these might also contribute to the dark matter. This was first pointed out by MacGibbon~\cite{MacGibbon:1987my}, and has subsequently been explored in the context of inflationary scenarios by numerous authors~\cite{Barrow:1992hq, Carr:1994ar, Green:1997sz, Alexeyev:2002tg, Chen:2002tu, Barrau:2003xp, Chen:2004ft, Nozari:2005ah}. The problem is that such remnants would be challenging to detect. However, Lehmann {\it et al.}~\cite{Lehmann:2019zgt} point out that they may carry electric charge, making them visible to terrestrial detectors. They evaluate constraints and detection prospects in detail and show that this scenario can be explored within the next decade using planned experiments.
\newpage

\section{Evidence from Dark Matter and Unified Model}
\label{sec:Evidence-from-Dark-Matter-and-Unified-Model}

\noindent Of all the observational evidence discussed in this review, that pertaining to dark matter is most firmly established and this surely points towards new physics. The dark matter has about $26\%$ of the critical density and evidence for it has been found on many different scales{\;--\;}galactic halos, clusters of galaxies and intergalactic space. The usual assumption is that it comprises some form of cold dark matter (CDM) since, in this case, as progressively larger scales of structure bind from the initial fluctuations, the dark matter will inevitably be found on all these scales. It must be:
	(1)	dark enough to explain the lack of observations 
		across the whole electromagnetic spectrum;
	(2)	``cold'' in order to agree with the observed 
		large-scale structure and CMB anisotropies;
	(3)	non-interacting or weakly interacting to be 
		consistent with the dark matter distribution in 
		dwarf galaxies;
	(4)	of non-astrophysical origin since CMB 
		observations show that it was present in the 
		early Universe.
Only some new type of particle or a PBH can satisfy all these criteria.
\vs{2mm}

\subsection{PBHs versus Particle Dark Matter}
\label{sec:PBHs-versus-Particle-Dark-Matter}

Until recently, the most popular CDM candidates were WIMPs (weakly interacting massive particles) since they arise naturally in supersymmetric extensions of the standard model. However, neither direct searches (using accelerators and underground detectors) nor indirect searches (using astronomical observations) have provided any evidence for such particles and this has ruled out most versions of the scenario. This has led to the search for other particle candidates (sterile neutrinos~\cite{2005PhLB..620...17A, 2013PhRvD..87i3006C}, axions and axion-like particles~\cite{1978PhRvL..40..223w, 1978PhRvL..40..279W, 1977PhRvL..38.1440P}, ultralight bosons~\cite{2000PhRvL..85.1158H}) or even proposed modifications to the law of gravity (such as MOND~\cite{Milgrom:1983ca} or TeVeS~\cite{Bekenstein:2004ne}).

Black holes might also be regarded as a form of CDM. However, while black holes are certainly dark, it should be stressed that those which form at late times (\eg~as stellar remnants or in galactic nuclei) could not provide {\it all} the dark matter because they derive from baryons and are therefore subject to the well-known big bang nucleosynthesis constraint that baryons can have at most $5\%$ of the critical density~\cite{Cyburt:2003fe}. By contrast, PBHs formed in the radiation-dominated era before big bang nucleosynthesis and avoid this constraint. They should therefore be classified as non-baryonic and behave like any other form of CDM. There are certainly some dark baryons, including those in stellar remnants, but most of them are probably in the form of intergalactic gas. For recent reviews of PBHs as dark matter, see Refs.~\cite{Carr:2020xqk, 2021JPhG...48d3001G}.

It is sometimes argued that PBHs are more natural dark matter candidates than WIMPs or other particles since black holes holes definitely exist, so that one does not need to invoke new physics. It might be counter-argued that most models of PBH formation depend on processes in the early Universe which involve new physics. However, there exists at least one scenario, Critical Higgs Inflation, which only needs to include a non-minimal coupling of the Higgs field to gravity in the Lagrangian of the Standard Model~\cite{Garcia-Bellido:2017mdw}. This is very different from the particle dark matter scenario, which requires the addition of at least a new field beyond the Standard Model since the known particles cannot account for CDM. Such theoretical arguments cannot be decisive but we believe the observational evidence discussed in this review boosts the odds in favour of PBHs. Finally, we stress that PBHs could still play an important cosmological r{\^o}le even if they only contribute a small part of the dark matter. This resembles the situation with neutrinos: these definitely exist and are enormously important even though their cosmological density is low. However, as discussed below, PBHs and particles cannot both make significant contributions to the dark matter.

The PBH parameter space{\,---\,}just like the parameter space for the particle model{\,---\,}has been progressively constrained by various observations. In particular, many papers have constrained the function $\fPBH( M )$, usually for a monochromatic mass distribution. However, the comparison is not entirely valid since:
	(1) $f_{\rm WIMP}$ is not considered as a free 
		parameter in the WIMP case;
	(2) no theoretical model leads to monochromatic PBH 
		mass function; 
	(3) most limits on particle dark matter are rather 
		precise, whereas the PBH limits depend on many 
		astrophysical and theoretical uncertainties.

It is worth elaborating on the last point. All the limits below $10^{17}\,$g rely on the theory of black hole evaporation but this has never been proved experimentally and evaporation might be suppressed below some mass-scale in alternative models. Between $10^{-10}\.\Msun$ and $1\.\Msun$ (\ie~ten decades of mass) the limits come from microlensing surveys but these depend on assumptions about the dark halo (its density profile), the PBHs (their velocity and mass distributions and clustering) and the efficiency of ML detections. The claimed limits on $\fPBH$ are not very strong anyway, ranging from $10^{-2}$ to $1$. Most of the limits above $10\.\Msun$ rely on black hole accretion. However, this is not well understood either observationally or theoretically and could be more complex than usually assumed. In particular, accretion depends on the black hole velocity relative to gas and this depends on the exact model. A modest change in the PBH velocity distribution could change the abundance limits by several orders of magnitude. The limits above $10^{4}\.\Msun$ depend on the primordial power spectrum and could be evaded if PBHs formed from highly non-Gaussian fluctuations, as applies in a variety of theoretical models.

Given these uncertainties, PBHs or the relics of their evaporation could provide a significant fraction of the dark matter for the hundred decades of mass between $\sim 1\,$g (for PBHs formed around the grand unified theory scale) and $10^{6}\.\Msun$. They may therefore provide all the dark matter even if $\fPBH( M ) \sim 10^{-2}$ for any particular mass, thereby obviating many of the constraints. See Refs.~\cite{Carr:2016drx, Green:2016xgy, Kuhnel:2017pwq, Carr:2017jsz} for detailed analyses of extended PBH mass functions. In Section C below, we will place emphasis on a particular scenario in which most of the dark matter is in PBHs of around $1\.\Msun$ formed at the QCD epoch. While this is our preferred scenario, it should be stressed that other PBH advocates favour a scenario in which most of the dark matter comprises PBHs in the planetary mass range and there are certainly formation mechanisms which would allow this. Indeed, it has been argued that the PBH mass function could have its main peak in the planetary mass range and a second peak in the LIGO mass range (\eg~Ref.~\cite{2022arXiv220905959F}). This requires even more fine-tuning than usual (since there are two peaks) but is not excluded. Other alternatives have been proposed by Boehm {\it et al.}~\cite{Boehm:2020jwd}, who advocate dark matter in $30\,\text{--}\,100\.\Msun$ PBHs since they might escape microlensing and CMB constraints, and Coriano \& Frampton~\cite{Coriano:2020uny}, who advocate $10^{2}\,\text{--}\,10^{5}\.\Msun$ PBHs. Our reason for underemphasising these alternative scenarios in this review is that there is currently no positive evidence for them.
\newpage

\subsection{Mixed Dark Matter}
\label{sec:Mixed-Dark-Matter}

If most of the dark matter is in the form of elementary particles, these will be accreted around any small admixture of PBHs. In the case of WIMPs, this can even happen during the radiation-dominated era, since Eroshenko~\cite{Eroshenko:2016yve} has shown that a low-velocity subset will accumulate around PBHs as density spikes shortly after the WIMPs kinetically decouple from the background plasma. Their annihilation will give rise to bright $\gamma$-ray sources and comparison of the expected signal with Fermi-LAT data then severely constrains $\Omega_{\rm PBH}$ for $M > 10^{-8}\.\Msun$. These constraints are several orders of magnitude more stringent than other ones if one assumes a WIMP mass of $m_{\chi} \sim \Ocal( 100 )\.\GeV$ and the standard value of $\langle \sigma v \rangle^{} = 3 \times 10^{-26}\.{\rm cm}\.\srm^{-1}$ for the velocity-averaged annihilation cross-section. Boucenna {\it et al.}~\cite{Boucenna:2017ghj} have investigated this scenario for a larger range of values for $\langle \sigma v \rangle$ and $m_{\chi}$ and reach similar conclusions. Although this might be regarded as a PBH constraint, it also represents potential signature of even a small density of PBHs.

Besides the early formation of spikes around light PBHs, WIMP accretion can also occur by secondary infall around heavier PBHs~\cite{Bertschinger:1985pd}. This leads to a different halo profile and WIMP annihilations then yield a constraint $\fPBH \lesssim \Ocal( 10^{-9} )$ for the same values of $\langle \sigma v \rangle$ and $m_{\chi}$. This result was obtained by Adamek {\it et al.}~\cite{Adamek:2019gns} for solar-mass PBHs and the argument was extended by Carr {\it et al.}~\cite{Carr:2020mqm} to the mass range from $10^{-18}\.\Msun$ to $10^{15}\.\Msun$; this even includes stupendously large black holes (SLABs) larger than $10^{12}\.\Msun$~\cite{Carr:2020erq}. Recently, Gin{\'e}s {\it et al.}~\cite{Gines:2022qzy} used the same dark matter halo profile as Ref.~\cite{Carr:2020mqm} and their extended analysis of annihilation signals found similar results.

\begin{figure}[t]
	\centering
	\includegraphics[width = 0.48\linewidth]{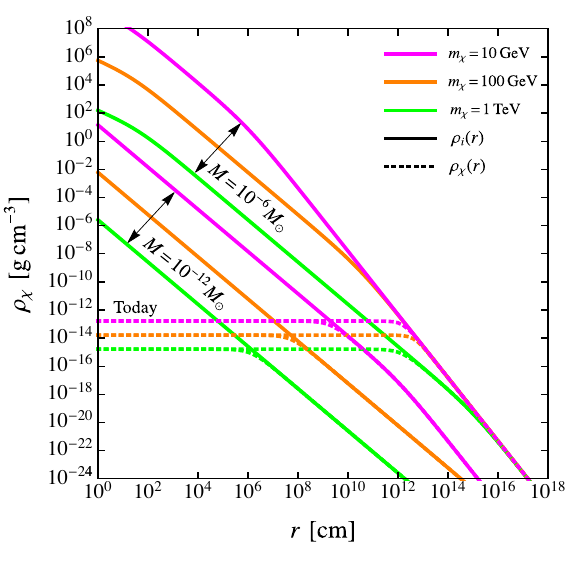}\;
	\includegraphics[width = 0.48\linewidth]{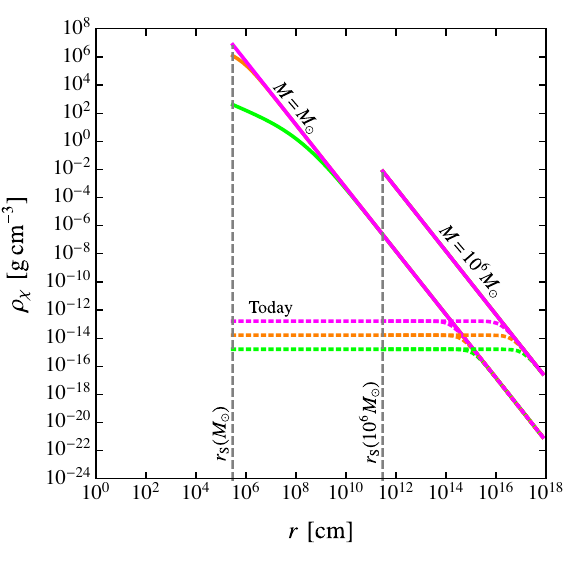} 
	\caption{
		Density profile before ($\rho_{i}$) and after 
		($\rho_{\chi}$) annihilations of WIMPs bound to 
		a PBH of mass 
			$10^{-12}\.\Msun$ or 
			$10^{-6}\.\Msun$ (left panel) 
		and 
			$1\.\Msun$ or 
			$10^{6}\.\Msun$ (right panel) 
		for $f_{\chi} \simeq 1$. We set 
			$m_{\chi}$ to be $10\.$GeV (magenta), 
			$100\.$GeV (orange) 
		and 
			$1\.$TeV (green).
		From Ref.~\cite{Carr:2020mqm}.
		}
	\label{fig:rhochi}
\end{figure}

The basis for all those constraints is the derivation of the density profile of the WIMP halos around the PBHs, which was given in Ref.~\cite{Carr:2020mqm}. Their result is depicted in Fig.~\ref{fig:rhochi}, which exhibits three initial scaling regimes,
\begin{equation}
	\label{eq:densityprofile2}
	\rho_{\rm \chi,\,spike}( r )
		\propto
				\begin{cases}
					f_{\chi}\.\rho_{\rm KD}\,
					r^{-3/4}
						& ( r < r_{\Crm} )
					\\[1mm]
					f_{\chi}\.\rho_{\rm eq}\,
					M^{3/2}\,r^{-3/2}
						& ( r_{\Crm} < r < r_{\Krm} )
					\\[1mm]
					f_{\chi}\.\rho_{\rm eq}\.
					M^{3/4}\,r^{-9/4}
						& ( r > r_{\Krm} )
					\, . on the relevant mass scale
				\end{cases}
\end{equation}
Here, $f_{\chi}$ is the dark matter fraction in the WIMPs, $\rho_{\rm KD}$ is the cosmological density when they kinetically decouple, $\rho_{\rm eq}$ is the density at matter-radiation equality, and $r_{\Crm}$ and $r_{\Krm}$ depend on $M$ and $m_{\chi}$. However, the dynamical evolution of the halo needs to be taken into account, since WIMP annihilations significantly change the profile from its initial form and produce a flat core. The derivation of these results and further details can be found in Ref.~\cite{Carr:2020mqm}. Using $N$-body simulations, Boudaud {\it et al.}~\cite{Boudaud:2021irr} obtained exactly the same density profile, confirming the three-fold behaviour of Eq.~\eqref{eq:densityprofile2}.

The most stringent constraints come from extragalactic $\gamma$-ray background observations. The differential flux of $\gamma$-rays is produced by the {\it collective} annihilations of WIMPs around PBHs at all redshifts~\cite{Ullio:2002pj},
\begin{equation}
	\label{eq:flux}
	\frac{ \drm\Phi_{\gamma} }{ \drm E\.\drm\Omega }
	\bigg|_{\rm eg}
		 =
		 		\int\limits_{0}^{\,\infty}\drm z\;
				\frac{ e^{-\tau^{}_{\Erm}( z,\.E )} }
				{ 8\pi H( z ) }
				\frac{ \drm N_{\gamma} }{ \drm E }
				\int\!\drm M\;
				\Gamma( z )\.
				\frac{ \drm n_{\rm PBH}( M ) }
				{ \drm M }
				\, ,
\end{equation}
where $H( z )$ is the Hubble rate at redshift $z$, ``eg'' indicates extragalactic and $n_{\rm PBH}$ is the PBH number density. Also $\Gamma( z ) \equiv \Gamma_{0}\,h( z )^{2/3}$, where $\Gamma_{0} = \Upsilon\.f_{\chi}^{1.7}\,M /\.\Msun$ is the WIMP annihilation rate around each PBH, with $\Upsilon \equiv 1.2 \times 10^{34}\,\srm^{-1}\,( \TeV / m_{\chi})^{4/3}$, $h( z ) = H( z )/100$ and $\tau^{}_{\Erm}$ is the optical depth at redshift $z$ resulting from
	(1)	photon-matter pair production,
	(2)	photon-photon scattering, and
	(3)	photon-photon pair production~\cite{Cirelli:2009dv, Slatyer:2009yq}. The numerical expressions for both the energy spectrum $\drm N_{\gamma} / \drm E$ and the optical depth are taken from Ref.~\cite{Cirelli:2010xx}. Integrating over the energy and solid angle leads to a flux
\begin{equation}
	\label{eq:flux-normalised}
	\Phi_{\gamma,\,{\rm eg}}
		 = 
				\frac{ \fPBH\,
				\rho_{\rm DM} }{ 2 H_{0}\.\Msun }\,
				\Upsilon\.f_{\chi}^{1.7}
				\tilde{N}_{\gamma}( m_{\chi} )
				\, ,
\end{equation}
where $\rho_{\rm DM}$ is the present dark matter density and $\tilde{N}_{\gamma}$ is the number of photons produced: 
\begin{equation}
	\label{eq:tildeNgamma}
	\tilde{N}_{\gamma}( m_{\chi} )
		\equiv
				\int_{z_{\star}}^{\infty}\!\drm z\;
				\int_{E_{\rm th}}^{m_{\chi}}\!\drm E\;
				\frac{ \drm N_{\gamma} }{ \drm E }
				\frac{ e^{-\tau^{}_{\Erm}( z,\.E )} }
				{ h( z )^{1/3} }
				\, .
\end{equation}
Here, the lower limit in the redshift integral corresponds to the epoch of galaxy formation, assumed to be $z_{\star} \sim 10$. The analysis becomes more complicated after $z_{\star}$. The lower limit in the energy integral is the threshold for detection.

\begin{figure}[t]
	\centering
	\vs{-1mm}
	\includegraphics[width = 0.43\linewidth]{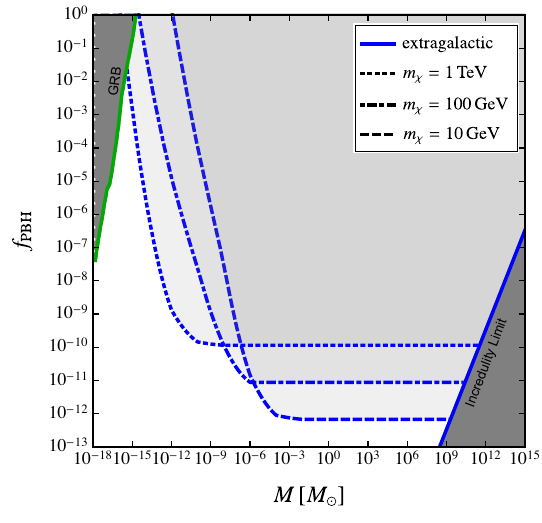} 
	\includegraphics[width = 0.52\linewidth]{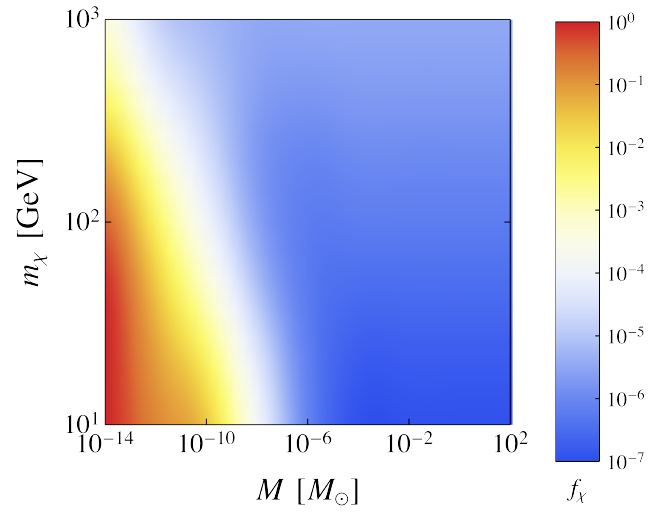}
	\caption{
		{\it Left panel$\mspace{1mu}$}: constraints on 
		$\fPBH$ as a function of PBH mass from 
		extragalactic $\gamma$-ray background for 
			$m_{\chi} = 10\.{\rm GeV}$ (dashed line), 
			$100\.{\rm GeV}$ (dot-dashed line) 
		and 
			$1\.{\rm TeV}$ (dotted line), 
		setting $\sv = 3 \times 10^{-26}\.
		$cm$^{3}$s$^{-1}$, and $\fPBH + f_{\chi} = 1$.
		Also shown is the incredulity limit 
		(one black hole within the particle horizon).
		{\it Right panel$\mspace{1mu}$}: the colour 
		shows the fraction of WIMPs $f_{\chi}$ as a 
		function of the PBH and WIMP masses. From 
		Ref.~\cite{Carr:2020mqm}.
		}
		\vs{-1mm}
	\label{fig:fPBHWIMP}
\end{figure}

Comparing the integrated flux with the Fermi sensitivity $\Phi_{\rm res}$ yields
\begin{align}
	\label{eq:egbound}
	\fPBH
		&\lesssim
				\frac{ 2 M\,H_{0}\,\Phi_{\rm res} }
				{ \rho_{\rm DM}\,
				\Gamma_{0}\,\tilde{N}_{\gamma}
				( m_{\chi} ) }
		\approx
				\begin{cases}
					2 \times 10^{-9}\,
					( m_{\chi} / {\rm TeV} )^{1.1}
						& \hbox{($M \gtrsim M_{*}$)}
					\\[1.5mm]
					\!1.1\times 10^{-12}
					\left(
						m_{\chi} / {\rm TeV}
					\right)^{-5.0}
					\left(
						M / 10^{-10}\.\Msun
					\right)^{\!-2}
						& \hbox{($M \lesssim M_{*}$)}
					\, ,
				\end{cases}
\end{align}
where $M_{*}$ is given by
\begin{equation}
	\label{eq:masschange}
	M_{*}
		\approx
				2 \times 10^{-12}\.
				( m_{\chi} / {\rm TeV} )^{-3.0}\.\Msun
				\, .
\end{equation}
The full constraint is shown by the blue curves in the left panel of Fig.~\ref{fig:fPBHWIMP} for a WIMP mass of $10\.$GeV (dashed line), $100\.$GeV (dot-dashed line) and $1\.$TeV (dotted line). We note that the extragalactic bound intersects the cosmological incredulity limit (corresponding to one PBH within the particle horizon) at a mass
\begin{equation}
	\label{eq:MiL}
	M_{\rm eg}
		=
				\frac{ 2\.H_{0}\.\Msun\.\Phi_{\rm res}
				\.M_{\Erm} }{ \alpha_{\Erm}\.
				\rho_{\rm DM}\,\Upsilon\,
				\tilde{N}_{\gamma}( m_{\chi} ) }
		\approx
				5 \times 10^{12}\.\Msun\,
				( m_{\chi} / {\rm TeV} )^{1.1}
				\, ,
\end{equation}
where we have used our fit for $\tilde{N}_{\gamma}( m_{\chi} )$.

The above analysis can be extended to the case in which WIMPs do not provide most of the dark matter~\cite{Carr:2020mqm}. The right panel of Fig.~\ref{fig:fPBHWIMP} shows the results, with the values of $f_{\chi}$ being indicated by the coloured scale as a function of $M$ and $m_{\chi}$. This shows the maximum WIMP density if most of the dark matter comprises PBHs of a certain mass and complements the constraints on the PBH density if most of the dark matter comprises WIMPs with a certain mass and annihilation cross-section. The left panel of Fig.~\ref{fig:fPBHWIMP} can also be applied in the latter case, with all the constraints weakening as $f_{\chi}^{-1.7}$. The important point is that even a small value of $\fPBH$ may imply a strong upper limit on $f_{\chi}$.

\begin{figure}[t]
	\includegraphics[width = 0.48\linewidth]{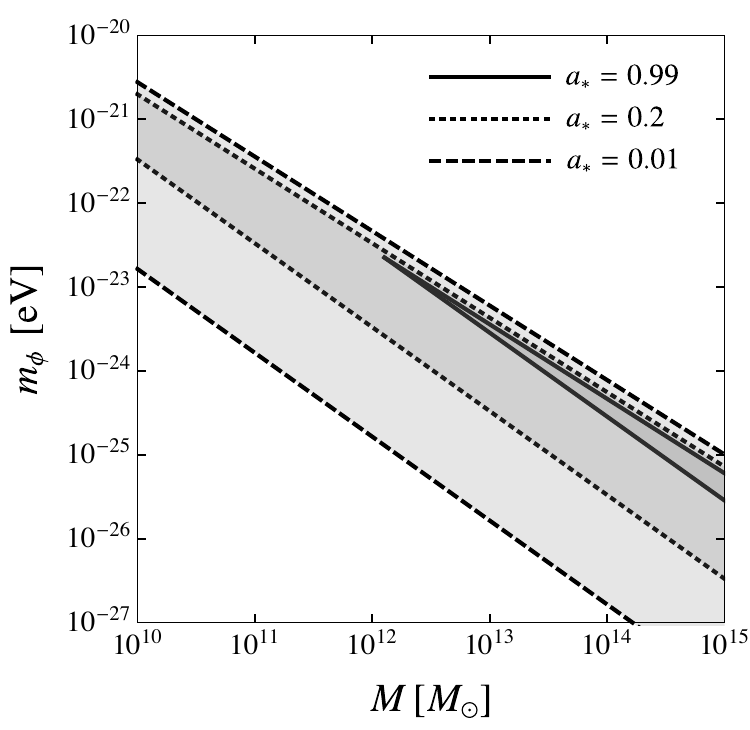} 
	\caption{
		Superradiance constraints on the mass 
		$m_{\phi}$ of a hypothetical boson as a 
		function of PBH mass $M$. Results are shown 
		for the {\it observed} black hole spin 
		parameter
			$a_{*} = 0.99$ (solid line), 
			$a_{*} = 0.2$ (dashed line) and 
			$a_{*} = 0.01$ (dotted line),
		defined in terms of the angular momentum 
		$J$ as $a_{*} \equiv J c / ( G M^{2} )$.
		From Ref.~\cite{Carr:2020erq}.
		}
	\label{fig:BHboson}
\end{figure}

If $M \gtrsim 10^{-11}\.\Msun$ and $m_{\chi} \gtrsim 100\.$GeV, both the WIMP and PBH fractions are at most $\Ocal( 10\% )$. Since neither of them can provide all the dark matter, this motivates a consideration of the situation with $\fPBH + f_{\chi} \ll 1$, which requires a {\it third} dark matter candidate. Particles which are not produced through the mechanisms discussed above or which avoid annihilation include axion-like particles~\cite{Abbott:1982af, Dine:1982ah, Preskill:1982cy}, sterile neutrinos~\cite{Dodelson:1993je, Shi:1998km}, ultra-light or ``fuzzy'' dark matter~\cite{Hu:2000ke, Schive:2014dra}. The latter is interesting because of its potential interplay with stupendously large black holes~\cite{Carr:2020erq}. If light bosonic fields exist in nature, they could accumulate around rotating SMBHs and form a condensate, leading to superradiant instabilities~\cite{Press:1972zz}. As shown in Fig.~\ref{fig:BHboson}, which is taken from Ref.~\cite{Carr:2020erq}, this leads to strong constraints on the mass $m_{\phi}$ of such a hypothetical boson.
\newpage

Recently, Kadota \& Hiroyuki~\cite{Kadota:2022cij} studied mixed dark matter scenarios consisting of PBHs and self-annihilating WIMPs through their synchrotron radiation at the radio frequency in the presence of galactic magnetic fields. This results in bounds on $\fPBH$ in the range $10^{-8}$\,\text{--}\,$10^{-5}$, depending on the WIMP annihilation channel and mass ($10$\,\text{--}\,$10^{3}$\,\GeV). These authors also investigated~\cite{Tashiro:2021xnj} the enhancement of heating and ionisation of the intergalactic medium due to WIMP annihilation, and found that the constraints on $\fPBH$ from CMB observations are comparable to or even tighter than those utilising $\gamma$-ray data. Using $21$-cm observations and assuming strong Lyman-$\alpha$ coupling, they find improved PBH limits for the canonical WIMP annihilation cross-section $\langle \sigma v \rangle = 3 \times 10^{-26}\,{\rm cm}^{3}\,\srm^{-1}$: $\fPBH < 4 \times 10^{-10}$ for $m_{\chi} = 100\,$GeV and $\fPBH < 2 \times 10^{-9}$ for $m_{\chi} = 1\,$TeV.

Choi {\it et al.}~\cite{GilChoi:2023qrz} have elaborated on a novel way to discriminate PBHs dressed by particle dark matter from PBHs without a halo. They show that microlensing abundance constraints on the former are significantly weakened and infer that dressed PBHs could provide all the dark matter in the range $0.1$ to $10^{2}\.\Msun$. They also demonstrate that diffractive lensing of chirping GWs from binary mergers can identify lenses consisting of dressed PBHs.

\subsection{Thermal-History Scenario}
\label{sec:Thermal--History-Scenario}
 
Here we explore the possibility of explaining all the observational evidence for PBHs with a single unified theoretical scenario, as first described in Ref.~\cite{Carr:2019kxo}. Because the evidence covers a wide range of masses, this requires that the PBH mass function also be extended. We will argue that this could be a natural consequence of the standard thermal history of the Universe, combined with the dependence of the critical threshold for PBH formation on the sound-speed. In particular, one expects a rather specific mass function in the planetary to intermediate mass range, which is independent of the origin of the density fluctuations. For all formation scenarios, PBH clustering is another key ingredient in explaining the evidence and the inevitable clustering induced by the Poisson fluctuations was discussed in Section~\ref{sec:Clustering-of-PBHs}. We must also consider the relationship between the primordial fluctuations on large and small cosmological scales (as probed by the CMB and PBHs, respectively), with particular emphasis on scenarios producing non-Gaussian tails in the fluctuation statistics~\cite{Ezquiaga:2019ftu}.

Reheating at the end of inflation fills the Universe with radiation. In the standard cosmological model, it remains dominated by relativistic particles until matter-radiation equality with an energy density decreasing as the fourth power of the temperature. As time increases, the number of relativistic degrees of freedom remains constant until around $200\,$GeV, when the temperature of the Universe falls to the mass thresholds of the Standard Model particles. The first particle to become non-relativistic is the top quark at $172\,$GeV, followed by the Higgs boson at $125\,$GeV, the $Z$ boson at $92\,$GeV and the $W$ boson at $81\,$GeV. At the QCD transition, occurring at a temperature of around $200\,$MeV, protons, neutrons and pions condense out of the free light quarks and gluons. A little later the pions become non-relativistic and then the muons, with $e^{+}e^{-}$ annihilation and neutrino decoupling occuring at around $1\,$MeV.

\begin{figure}[t]
	\centering
	\vs{-3mm}
	\includegraphics[width = 0.70\textwidth]{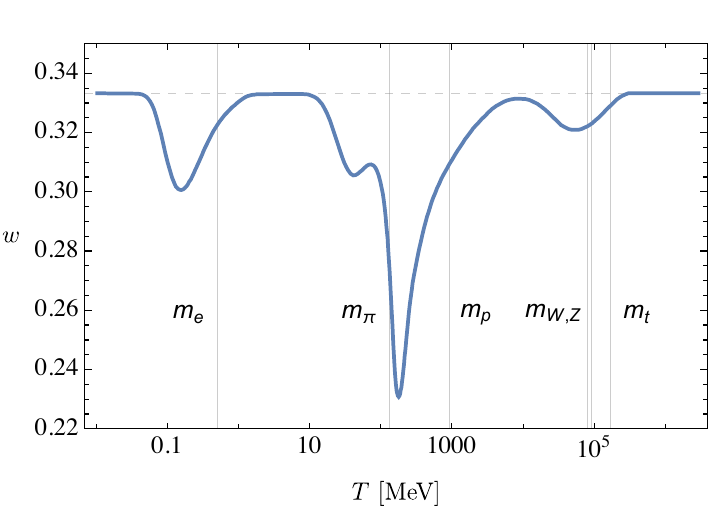}
	\caption{
		Equation-of-state parameter $w$ as a function 
		of temperature $T$, from 
		Ref.~\cite{Carr:2019kxo}. The grey vertical 
		lines correspond to the masses of the electron, 
		pion, proton/neutron, $W / Z$ bosons and top 
		quark. The grey dashed horizontal line 
		correspond to $w = 1 / 3$.
		}
	\label{fig:g-and-w-of-T}
\end{figure}

Whenever the number of relativistic degrees of freedom suddenly drops, it changes the effective equation-of-state parameter $w \equiv p/ \rho\mspace{1.5mu}c^{2}$. As shown in Fig.~\ref{fig:g-and-w-of-T}, there are thus four periods in the thermal history of the Universe when $w$ decreases. After each of these, $w$ resumes its relativistic value of $1 / 3$ but because the threshold overdensity required for PBH production $\delta_{\crm}$ is sensitive to the equation-of-state parameter $w( T )$, the sudden drop modifies the probability of gravitational collapse of any large curvature fluctuations. This results in pronounced features in the PBH mass function even if the primordial fluctuations have a uniform power spectrum. If the PBHs form from Gaussian inhomogeneities with root-mean-square amplitude $\delta_{\rm rms}$, the fraction of horizon patches undergoing collapse to PBHs on mass scale $M$ is~\cite{Carr:1975qj}
\vs{-1.5mm}
\begin{align}
	\beta( M )
		&\approx
				{\rm Erfc}\!
				\left[
					\frac{
						\delta_{\crm}
						\big(
							w[ T( M ) ]
						\big) }
					{ \sqrt{2}\,\delta_{\rm rms}( M )}
				\right]
				,
				\label{eq:beta-t}
\end{align}
where the mass and temperature are related by
\vs{-0.5mm}
\begin{align}
	T
		\approx
				200\,\sqrt{\.\Msun / M}\;\MeV
				\, .
\end{align}
Thus $\beta( M )$ is exponentially sensitive to $w( M )$ and the present CDM fraction for PBHs of mass $M$ is 
\vs{-1.5mm}
\begin{align}
	\fPBH( M ) 
		\equiv
				\frac{ 1 }{ \rho_{\rm CDM} }
				\frac{ \drm\.\rho_{\rm PBH}( M ) }
				{ \drm \ln M }
		\approx
				2.4\;\beta( M )\.
				\sqrt{\frac{ M_{\rm eq} }{ M }}
				\, ,
				\label{eq:fPBH}
\end{align}
where $M_{\rm eq} = 2.8 \times 10^{17}\.\Msun$ is the total horizon mass at matter-radiation equality and the numerical factor is $2\.( 1 + \Omega_{\Brm} / \Omega_{\rm CDM} )$ with $\Omega_{\rm CDM} = 0.245$ and $\Omega_{\Brm} = 0.0456$~\cite{Aghanim:2018eyx}.

\begin{figure}[t]
	\centering 
	\includegraphics[width = 0.65 \textwidth]{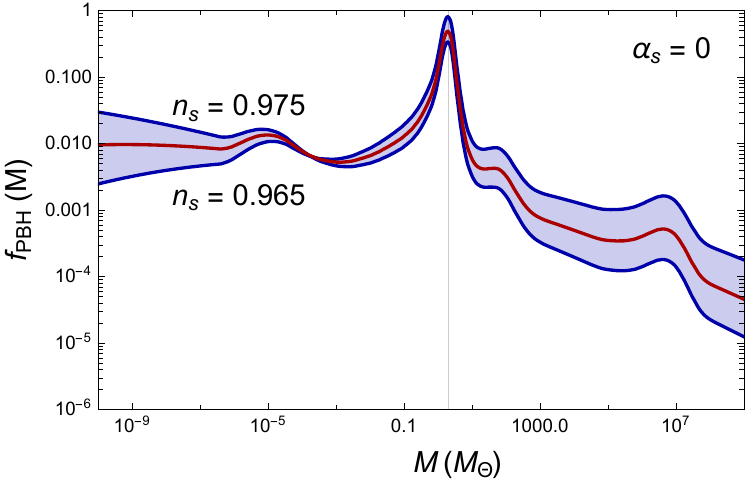}\\[7mm]
	\includegraphics[width = 0.65 \textwidth]{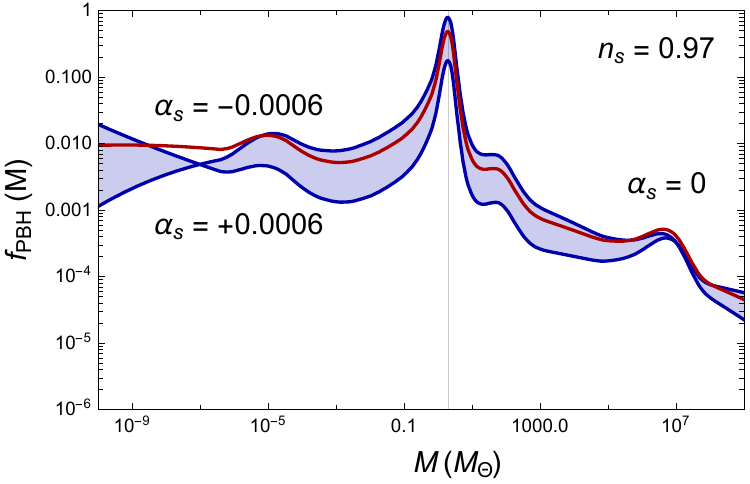}
	\caption{
		The effect of the spectral tilt (upper panel) 
		and its running (lower panel) on the thermal-
		history-induced mass function $\fPBH( M )$.
		From Ref.~\cite{Garcia-Bellido:2022fgh}
		}
\label{fig:effect-of-tilt-and-running}
\end{figure}

There are many inflationary models and they predict a variety of shapes for $\delta_{\rm rms}( M )$. Some models produce an extended plateau or dome-like feature in the power spectrum. For example, this applies for two-field models like hybrid inflation~\cite{Clesse:2015wea} and even some single-field models like Critical Higgs Inflation~\cite{Ezquiaga:2017fvi}. This results in a quasi-scale-invariant spectrum,
\vs{-0.5mm}
\begin{align}
	\delta_{\rm rms}( M )
		=
				A
				\left(
					\frac{ M }{\.\Msun }
				\right)^{\!(1 - n_{\srm}) / 
				4\.+\.\alpha_{\srm}/8 \ln( M/\Msun )}
				\; ,
				\label{eq:delta-power-law}
\end{align}
where the spectral index $n_{\srm}$, its running $\alpha_{\srm} \equiv \d \ln n_{\srm}( k ) / \d \ln k$ and the amplitude $A$ are treated as free phenomenological parameters. This could represent any spectrum with a broad peak, such as might be generically produced by a second phase of slow-roll inflation. We choose $A = 0.0218$ for $n_{\srm} = 0.986$ and $\alpha_{\srm} = - 0.0018$, in order to get an integrated abundance $\fPBH = 1$. Following Refs.~\cite{Carr:2019hud, Garcia-Bellido:2019vlf}, the ratio of the PBH mass and the horizon mass at re-entry is assumed to be $0.8$. The resulting mass functions are represented in Fig.~\ref{fig:effect-of-tilt-and-running}. There is a dominant peak at $M \simeq 2\.\Msun$ and three additional bumps at $10^{-5}\.\Msun$, $30\.\Msun$ and $10^{6}\.\Msun$. Earlier works~\cite{Jedamzik:1996mr, Byrnes:2018clq, Clesse:2020ghq, Jedamzik:2020omx} have also discussed the effect of the QCD transition on PBH formation but not that of the other transitions. All the effects were first considered in a unified way in Ref.~\cite{Carr:2019kxo}. Depending on the running and tilt, these bumps are more or less pronounced, as can be seen in Fig.~\ref{fig:effect-of-tilt-and-running}. This shows that small variations of the tilt and its running can dramatically change the normalisation and thus the height of the main peak.

Two recent numerical studies of PBH formation at the QCD epoch by Escriv{\`a} {\it et al.}~\cite{Escriva:2022bwe} and Musco {\it et al.}~\cite{Musco:2023dak} have treated the effects of the varying sound speed and equation of state during PBH formation more consistently. Both give a somewhat different PBH mass function, although the same qualitative feature of a peak at around $1\.\Msun$ is retained. This may be explained by subtle differences in the assumed curvature profiles, the interpolation method between data points of the equation of state as a function of the temperature and by the exact recipe for the computation of the PBH mass distribution. The precise mass distribution is therefore still uncertain and model-dependent. This can impact the analysis of PBH merger rates. Moreover, different fundamental physics around the QCD epoch, such as gluon condensation, could drastically change the shape of the mass distribution around $1\.\Msun$~\cite{Garcia-Bellido:2021zgu}. A detailed spectroscopic mass resolution around that peak could thus reveal the nature of the phase transition and possibly new fundamental physics.

\begin{figure}
	\centering
	\includegraphics[width = 0.42 \textwidth]{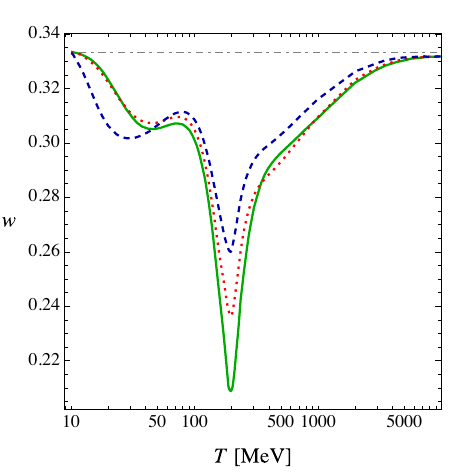}\q
	\includegraphics[width = 0.51 \textwidth]{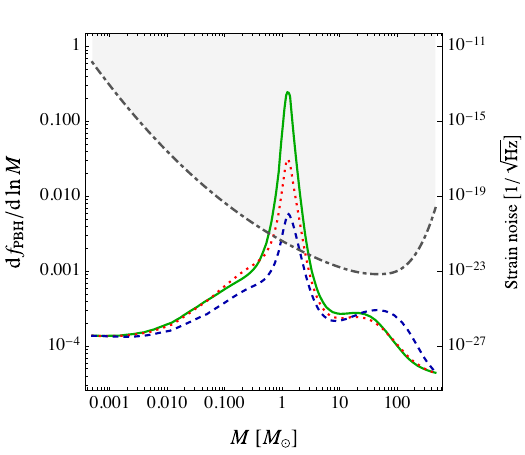}
	\caption{
			{\it Left panel:} Effective 
				equation-of-state parameter as a 
				function of temperature. Shown are the 
				three cases of different lepton flavour 
				asymmetry and the $1/3$ for a 
				radiation fluid. The standard scenario 
				[case (1)] is indicated by the green 
				solid line.
	 		{\it Right panel:} 
				Spectral density of the PBH dark matter 
				fraction as a function of PBH mass, for 
				the three cases. The green solid line 
				indicates the standard scenario. Also 
				shown is the LIGO sensitivity curve 
				(grey dot-dashed line) from 
				Ref.~\cite{LIGOScientific:2018mvr}, 
				for equal-mass mergers and using the 
				maximal GW frequency $f^{}_{\rm max} 
				\approx 4400\.\Msun / M$ for the 
				conversion from frequency to mass. From 
				Ref.~\cite{Bodeker:2020stj}.
			}
	\label{fig:fPBH}
\end{figure}

Another example of how fundamental physics can change the shape of the PBH mass function concerns lepton-flavour asymmetries. Observationally allowed lepton flavour asymmetries~\cite{Oldengott:2017tzj} could dramatically change the functional form of the bumps and even their location~\cite{Bodeker:2020stj}. Figure~\ref{fig:fPBH} illustrates this for the three examples analysed in Ref.~\cite{Bodeker:2020stj}:
\begin{itemize}

	\item[(1)] 
		$\ell_{e} = \ell_{\mu} = \ell_{\tau} = 
		-\,5.3 \times 10^{-11}$~[green, solid line];

	\item[(2)] 
	 	$\ell_{e} = 0$ and $\ell_{\mu} = -\,\ell_{\tau} 
		= 4 \times 10^{-2}$~[red, dotted line];

	\item[(3)]
	 	$\ell_{e} = -\,8 \times 10^{-2}$ and 
		$\ell_{\mu} = \ell_{\tau} = 4 \times 10^{-2}$
		~[blue, dashed line],

\end{itemize}
where the lepton flavour asymmetries are defined as 
\begin{equation}
	\ell_{\alpha}
		\equiv
				\frac{ n_{\alpha} - n_{\bar{\alpha}}
				+ n_{\nu_{\alpha}}
				- n_{\bar{\nu}_{\alpha}} }
				{ s }
				\; ,
				\quad
				\alpha\.\in\.\{ e,\,\mu,\,\tau \}
				\; .
				\label{eq:leptonasymmetry}
\end{equation}
Here, $n_{\alpha}$ and $n_{\bar{\alpha}}$ denote the number density of the particles and antiparticles, and $s$ is the entropy density. For adiabatic expansion, the $\ell_{\alpha}$ are conserved between the electroweak transition at $T \simeq 160\,$GeV and electron-positron annihilation at $T \simeq 10\,\MeV$. The baryon asymmetry of the Universe is constrained from CMB observations~\cite{Aghanim:2018eyx} and primordial element abundances~\cite{Pitrou:2018cgg} to be $\eta = 8.7 \times 10^{-11}$. Constraints on the lepton asymmetries are, however, many orders of magnitude weaker and allow for a total lepton asymmetry~\cite{Oldengott:2017tzj}
\begin{equation}
	| \ell_{e} + \ell_{\mu} + \ell_{\tau} |
		<
				1.2 \times 10^{-2}
				\, .
				\label{eq:ell-total}
\end{equation}
In particular, as pointed out in Refs.~\cite{Stuke:2011wz, Middeldorf-Wygas:2020glx, Vovchenko:2020crk}, scenarios with unequal lepton flavour asymmetries (before the onset of neutrino oscillations) are observationally almost unconstrained and therefore open up a whole new parameter space for the evolution of the Universe around the QCD transition.

The left panel of Fig.~\ref{fig:fPBH} shows how non-zero flavour asymmetries weaken the softening of $w$ during the transition, with even $\ell_{e} = 0$ yielding a pronounced effect. Two cases of unequal lepton flavour asymmetry are chosen for illustrative purposes. Note that a lepton flavour asymmetry always weakens the softening of the equation of state during the QCD transition, as it adds leptons to the Universe which do not interact strongly. This is different from the smaller effect at the pion/muon plateau, where lepton flavour asymmetries can lead to either stiffening or softening (\cf~the two cases of unequal flavour asymmetry) as pions and muons become non-relativistic. The corresponding result for the PBH dark matter fraction is depicted in the right panel of Fig.~\ref{fig:fPBH}. Due to the exponential enhancement of Eq.~\eqref{eq:beta-t}, the three cases differ significantly. The effect of lepton flavour asymmetries are currently being explored. As shown by Gao \& Oldengott~\cite{Gao:2021nwz}, the cosmic QCD transition might even become first order if the asymmetries are sufficiently large, with a dramatic impact on the PBH mass function.
\newpage

\subsection{Comparing Evidence with Thermal-History Model}
\label{sec:Comparing-Evidence-with-Thermal--History-Model}

\begin{figure}[t]
	\centering
	\vs{-7mm}
	\includegraphics[width=0.89\textwidth]{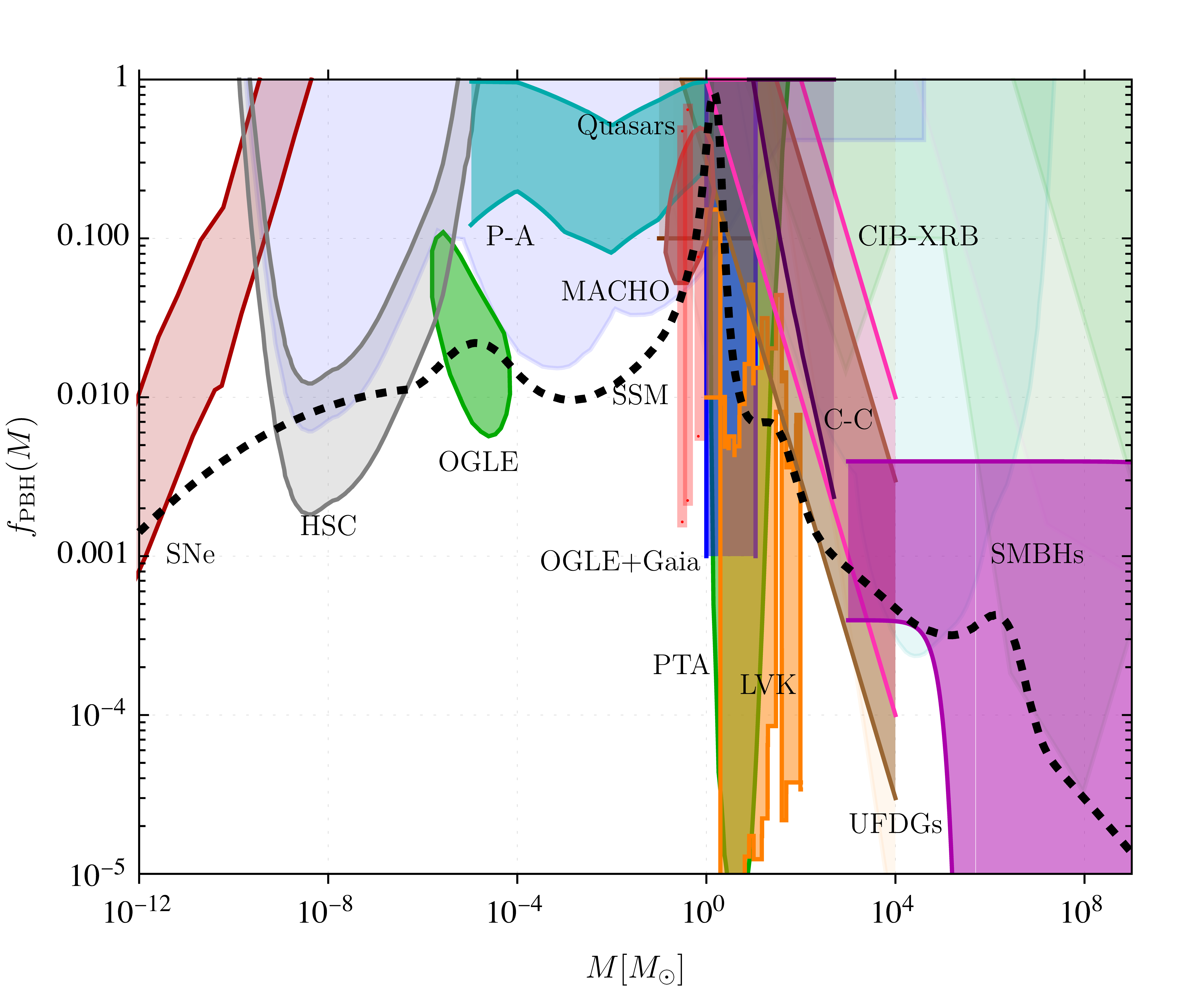}
	\caption{PBH mass function with peaks induced 
		by the thermal history of the Universe (thick, 
		dashed curve; \cf~Ref.~\cite{Carr:2019kxo}).
		This replicates the positive features of 
		Fig.~\ref{fig:PositiveEvidence} but also 
		includes various monochromatic constraints on 
		$\fPBH( M )$ (light-shaded regions), taken from 
		Ref.~\cite{carr2020primordial}.
		}
		\vs{-1.5mm}
	\label{fig:figZ1}
\end{figure}

In Figs.~\ref{fig:PositiveEvidence} and~\ref{fig:figZ1}, we have indicated the PBH mass and dark matter fraction required to explain the various type of observational evidence discussed in this review. We now explain the derivation of these regions in more detail, considering the lensing, dynamical and GW arguments in turn. However, just as for PBH constraints, all these estimates are based on various assumptions and subject to significant uncertainties. In particular PBH properties (such as mass function, clustering etc.) can modify the different regions. Unless indicated otherwise, we assume a monochromatic PBH mass function.

{\bf PBH dark matter fraction from lensing evidence.} We have estimated the PBH dark matter fraction for six types of lensing evidence in the following way:
\vs{-3mm}
\begin{itemize}

	\item {\it For HSC}, we have reinterpreted the 
		limits of Ref.~\cite{Niikura:2017zjd}. Instead 
		of assuming no detection, we have computed 
		the $2 \sigma$ confidence intervals for $\fPBH$ 
		assuming that one PBH microlensing event was 
		observed. The limit is identified with a band 
		using simple Poisson statistics. All the 
		assumptions are therefore identical to those of 
		Ref.~\cite{Niikura:2017zjd}.
	\vs{-1mm}

	\item {\it For OGLE}, we show the $2\sigma$ allowed 
		region provided in Fig.~8 of 
		Ref.~\cite{Niikura:2019kqi}, combining the OGLE 
		confidence region with the HSC exclusion 
		region.
	\vs{-1mm}

	\item {\it For POINT-AGAPE pixel-lensing (P-A)}, we
		provide a band of possible $\fPBH$ values based 
		on the Table 9 of 
		Ref.~\cite{CalchiNovati:2005cd}, where 
		$2\sigma$ confidence intervals were estimated 
		for different PBH masses.

	\item {\it For quasars}, there is not yet a robust 
		estimation of a confidence region despite the 
		fact that we consider them as a strong evidence 
		for a large fraction of dark matter made of 
		compact objects. We therefore consider a 
		relatively broad range of PBH masses, between 
		$10^{-2}\.\Msun$ and $10\.\Msun$, and a lower 
		limit of $\fPBH > 0.1$, but these are only 
		order-of-magnitude estimates.

	\item {\it For MACHO}, the $2\sigma$ region comes 
		from Refs.~\cite{1996ApJ...461...84A, 
		2007A&A...469..387T}. We have argued that the 
		MACHO microlensing events are real and 
		plausibly due to PBHs, despite EROS and OGLE 
		later claiming more stringent limits.
	\vs{-1mm}

	\item {\it For OGLE+Gaia}, we assume that all the 
		microlensing events are due to PBHs. The 
		indicated band comes from the lowest and 
		highest mean masses shown in 
		Fig.~\ref{fig:figX}, obtained by the analysis 
		of Ref.~\cite{2020A&A...636A..20W}, and there 
		is a somewhat arbitrary lower bound, 
		$\fPBH \approx 10^{-3}$, below which PBHs are 
		unlikely to explain so many microlensing 
		events.
	\vs{-1mm}

\end{itemize}
All these estimates are subject to large uncertainties. In particular, the size of PBH clusters could alter the PBH fraction inferred from microlensing of stars towards the Magellanic Clouds. Figure~\ref{fig:PositiveEvidence} represents the ideal situation, in which all the observations are due to PBHs. However, since alternative origins are not excluded, it is possible that only a some of them are.
\vs{3mm}

{\bf PBH dark matter fraction from dynamical and accretion evidence.} We have estimated the PBH dark matter fraction for the five types of dynamical and accretion evidence as follows:
\vs{-1mm}
\begin{itemize}

	\item {\it For SNe}, the band corresponds to the 
		one shown in Fig.~\ref{fig:datacompare}, which 
		is taken from Ref.~\cite{2022arXiv221100013S}.
	\vs{-1mm}

	\item {\it For UFDGs}, we assume 
		$0.3 < \fPBH\.M/\.\Msun < 30$, with a 
		maximum mass of $10^{4}\.\Msun$. These values 
		are somewhat arbitrary and do not result from a 
		robust statistical analysis, but the large 
		cosmological, astrophysical and observational 
		uncertainties would limit the validity of a 
		more refined analysis. The values shown 
		correspond to the two extreme cases displayed 
		in Fig.~\ref{fig:rcvsmc}. With our simple 
		modelling of PBH clustering, these values could 
		simultaneously explain 
			(1)	the minimum size of observed UFDGs, 
			(2) the relation between their radius and 
				mass, 
		and 
			(3) their very large mass-to-light ratios 
				due to PBH gas accretion.
	\vs{-1mm}

	\item {\it For C-C (core-cusp)}, the displayed 
		region is based on the lower limits from Fig.~7 
		of Ref.~\cite{Boldrini:2019isx}, extrapolated 
		up to a mass of $10^{3}\.\Msun$ since there is 
		no theoretical restriction on this, other than 
		there being enough PBHs for dynamical heating 
		to be efficient.
	\vs{-1mm}

	\item {\it For CIB-XRB correlations}, the band 
		corresponds to 
		$1 < \fPBH\.M/\.\Msun < 100$ up to a 
		mass of $10^{4}\.\Msun$. We have not used a 
		rigorous statistical analysis to get more 
		precise values but the proposed band agrees 
		with the order-of-magnitude estimate in 
		Ref.~\cite{Kashlinsky:2016sdv}.
	\vs{-1mm}

	\item {\it For SMBHs}, we have followed 
		Ref.~\cite{Carr:2019kxo} in assuming a linear 
		relation between the central IMBH or SMBH mass 
		and stellar mass, as suggested by the gray band 
		in Fig.~1 of Ref.~\cite{Kruijssen:2013cna}, as 
		well as a Press--Schechter halo mass function.
		The upper limit neglects the effects of 
		accretion and mergers. The origin of IMBHs and 
		SMBHs in clusters and galaxies is therefore 
		related to the PBH mass distribution. The lower 
		limit assumes that the PBH mass increased 
		during the pregalactic era at the Bondi rate, 
		given by Eq.~\eqref{eq:Bondi}, until it reached 
		the Eddington limit, given by 
		Eq.~\eqref{eq:Eddington}. However, this is 
		subject to the large uncertainties in the 
		accretion process. We note that 
		Fig.~\ref{fig:effect-of-tilt-and-running} 
		suggests $\fPBH( M ) \propto M^{-1/4}$ for 
		$M > 10\.\Msun$ in the thermal-history model, 
		which corresponds to a power-law function with 
		$\alpha \approx 9/4$.
\vs{-1mm}

\end{itemize}

The proposed regions depend on the very uncertain physical processes which underlie the dynamical evolution of PBH clusters and the accretion process throughout cosmic history. Most of the dynamical and accretion evidence relates to the mass range from one to a billion solar masses and therefore complements the microlensing and GW evidence. This clearly favours models with an extended PBH mass distribution. In this case, all the evidence depends on Poisson-induced clustering. Since this is determined by the product $\fPBH\,M$, PBHs with different masses would contribute to the effect in such a way that a distribution not crossing the proposed region can still provide the required effect. Although such extended mass distributions may conflict with the stringent constraints on the CMB distortions and anisotropies, these can be relaxed for realistic accretion models if PBH formation is associated with curvature fluctuations with non-Gaussian tails~\cite{2022arXiv221207969F}.
\vs{3mm}

{\bf PBH dark matter fraction from GW evidence.} We have estimated the PBH dark matter fraction for various types of GW evidence. For this purpose, we have considered the merger rates of early binaries for a monochromatic model, but obviously the range of component masses associated with the observed coalescences points to an extended mass distribution.
\vs{-1mm}
\begin{itemize}

	\item {\it For subsolar-mass candidates}, we have 
		used the three subsolar-mass triggers in the 
		LVK O3b observing run to derive credible 
		$2\sigma$ intervals for $\fPBH$ in the 
		monochromatic case. More precisely, we use the 
		chirp mass associated with each candidate to 
		compute an associated interval for $\fPBH$, 
		this mass being well reconstructed for GW 
		events. In each case, it is sufficiently below 
		$1\.\Msun$ to guarantee that at least one of 
		the components is a subsolar-mass compact 
		object for a real GW signal. We then assume 
		that the two components have the same mass and 
		use the volume-time sensitivity 
		$\langle VT \rangle$ obtained in 
		Ref.~\cite{LVK:O3bSSM} for each mass. When 
		combined with the expected merger rate of early 
		binaries, using simple Poisson statistics, one 
		can compute a $2\sigma$ interval for $\fPBH$ 
		for each subsolar-mass trigger.
	\vs{-1mm}

	\item {\it For LVK}, we have used the inferred 
		merger rates in intervals at different masses, 
		for the O1, O2 and O3 runs and the binned 
		Gaussian process model of 
		Ref.~\cite{KAGRA:2021duu}, considering only the 
		rates for equal-mass mergers shown in their 
		Fig.~4 (orange regions). For each bin, we have 
		then compared this with the expected merger 
		rates of early binaries for a monochromatic 
		model to infer the $90\%$ CL intervals for 
		$\fPBH$. This model does not allow $\fPBH$ to 
		reach $1$ but it could still exceed $0.1$ at 
		solar-mass scale, bearing in mind the 
		uncertainties associated with the rate model.

	\item {\it For PTAs}, we have used the 
		two-dimensional marginalised posterior 
		distributions of the primordial power spectrum 
		amplitude and pivot wavelength mode calculated 
		for IPTA observations in 
		Ref.~\cite{Dandoy:2023jot}, as shown in 
		Fig.~\ref{fig:PTA}; this assumes a log-normal 
		primordial power spectrum. In order to 
		translate these contours into $\fPBH$ 
		constraints, we have assumed a critical 
		overdensity threshold for PBH formation at the 
		QCD epoch and a corresponding PBH abundance in 
		agreement with the recent numerical simulations 
		of Ref.~\cite{Escriva:2022bwe}. The obtained 
		region is extremely sensitive to the threshold 
		values and the method used to compute the PBH 
		abundance, so it can vary by several orders of 
		magnitudes.

\end{itemize}
\newpage

The uncertainties and model dependence of the merger rates, the exact PBH formation mechanism, and the various possibilities for the statistics of the curvature fluctuations, clearly blur the calculated intervals for GW mergers and PTAs. Nevertheless, GW observations hint at a peak in $\fPBH$ at the solar-mass scale, as expected for PBHs forming at QCD epoch. These observations complement and agree well with the lensing, dynamical and accretion evidence. The fit with the data is consistent with all the constraints if one uses a running spectral index for the power spectrum at PBH scales, as suggested in Ref.~\cite{Hasinger:2020ptw}. In this case, we are considering a spectral tilt $n_{\srm} = 0.986$ and running $\alpha_{\srm} = - 0.0018$ at PBH scales. These values are remarkably close to those measured at CMB scales, which may hint that Critical Higgs Inflation generates the full matter power spectrum.

\subsection{PBHs, Baryogenesis and the Fine-Tuning Problem}
\label{sec:PBHs,-Baryogenesis-and-the-Fine--Tuning-Problem}

In this section we argue that the coincidence between the dark matter and baryonic densities suggests a common origin, which is linked to baryogenesis. There are several ways in which PBHs could have induced baryogenesis and, in this sense, the observed baryon asymmetry might be considered positive evidence for them. In particular, we emphasise that the density coincidence is naturally explained if the PBHs have around a solar mass and form at the QCD epoch. This is because each region collapsing to a PBH is surrounded by a hot expanding shell, in which all the Sakharov conditions for efficient low-temperature electroweak baryogenesis are satisfied. This provides a natural connection between the PBH collapse fraction and the cosmological baryon asymmetry.

The important point is that observations imply that only a tiny fraction of the early Universe could have collapsed into PBHs. The current density parameter $\Omega_{\rm PBH}$ of PBHs which form at a redshift $z$ is related to the initial collapse fraction $\beta$ by~\cite{Carr:1975qj} 
\begin{equation}
	\Omega_{\rm PBH}
		=
				\beta\;
				\Omega_{\Rrm}\.
				( 1 + z )
		\approx
				10^{6}\.
				\beta
				\left(
					\frac{ t }{ \srm }
				\right)^{\!-1/2}
		\approx
				10^{18}\.
				\beta
				\left(
					\frac{ M }{ 10^{15}\.\grm }
				\right)^{\!-1/2}
				,
\label{beta}
\end{equation}
where $\Omega_{\Rrm} \approx 5 \times 10^{-5 }$ is the density parameter of the CMB. The $( 1 + z )$ factor arises because the radiation density scales as $( 1 + z )^{4}$, whereas the PBH density scales as $( 1 + z )^{3}$. The CDM density parameter is $\Omega_{\rm CDM} \approx 0.25$, so $\beta$ must be tiny even if PBHs provide all of it. For example, PBHs of $10^{15}\,$g, the smallest ones avoiding evaporation, would have $\beta \sim 10^{-18}$, while those of $1\.\Msun$ would have $\beta \sim 10^{-9}$. More generally, any limit on $\Omega_{\rm PBH}$ places a constraint on $\beta( M )$. Requiring $\Omega_{\rm PBH} < \Omega_{\rm CDM}$ above $10^{15}\.$g is the most basic constraint but there are many stronger ones, as discussed in numerous papers (\eg~Ref.~\cite{Carr:1994ar}). In particular, interesting constraints are associated with PBHs smaller than $10^{15}\.$g since they would have evaporated by now. The strongest one is the $\gamma$-ray limit associated with PBHs evaporating at the present epoch.

The fine-tuning of the collapse fraction required to explain the dark matter is sometimes regarded as a criticism of the PBH dark matter proposal. Furthermore, it implies even more fine-tuning on the amplitude of the density fluctuations generating the PBHs, since Eq.~\eqref{eq:beta-t} implies that $\beta$ depends exponentially on this amplitude. However, the small value of $\beta$ implied by Eq.~\eqref{beta} itself derives from the high photon-to-baryon ratio associated with the small baryon asymmetry of the Universe (BAU) at the QCD epoch ($\eta \sim 10^{-9}$). The origin of this asymmetry and the PBH collapse fraction are thus linked. The standard assumption is that high-energy physics generates the baryon asymmetry everywhere simultaneously via out-of-equilibrium particle decays or a first-order phase transition well before PBH formation. However, there is no direct evidence for this and{\,---\,}even if the process occurs{\,---\,}it may not provide all the baryon asymmetry required.

\begin{figure}[t]
	\vs{-14mm}
	\centering 
	\includegraphics[width = 0.88 \textwidth]{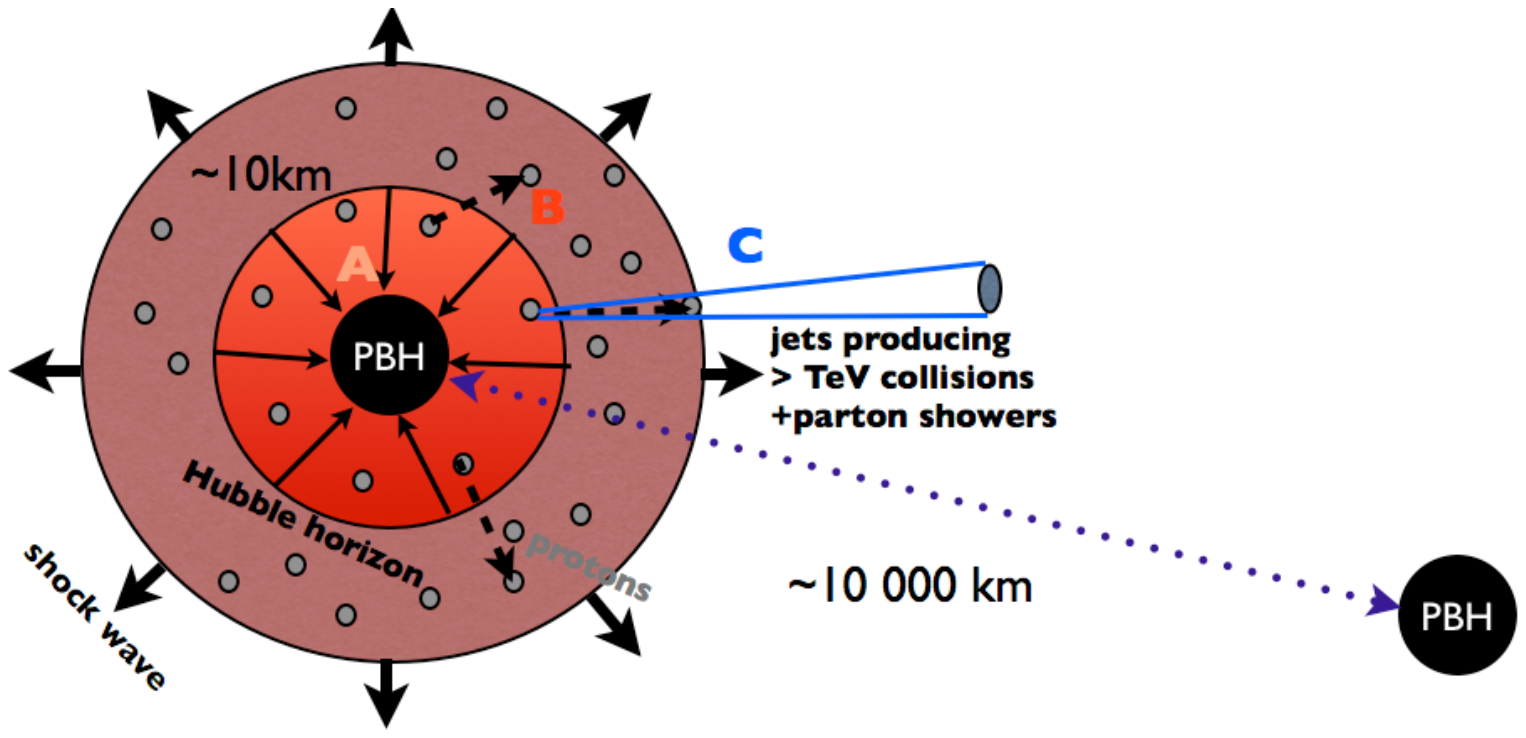}
	\vs{-15mm}
	\caption{
		Qualitative representation of the three steps 
		in the discussed scenario: gravitational 
		collapse of the curvature fluctuation at 
		horizon re-entry; sphaleron transition in hot 
		spot around the PBH, producing 
		$\eta \sim \Ocal( 1 )$ locally through 
		electroweak baryogenesis; propagation of 
		baryons to rest of Universe through jets, 
		resulting in the observed BAU with 
		$\eta \sim 10^{-9}$. From 
		Ref.~\cite{Garcia-Bellido:2019vlf}.
		}
	\label{fig:BAU}
\end{figure}

Garc{\'i}a-Bellido {\it et al.}~\cite{Garcia-Bellido:2019vlf} have proposed an alternative scenario in which the gravitational collapse to PBHs at the QCD epoch can resolve both these problems. The collapse is accompanied by the violent expulsion of surrounding material, which might be regarded as a sort of ``primordial supernova". Such high density {\it hot spots} provide the out-of-equilibrium conditions required to generate a baryon asymmetry~\cite{Sakharov:1967dj} through the well-known electroweak sphaleron transitions responsible for Higgs windings around the electroweak vacuum~\cite{Asaka:2003vt}. The charge-parity symmetry violation of the Standard Model then suffices to generate a local baryon-to-photon ratio of order one. The hot spots are separated by many horizon scales but the outgoing baryons propagate away from them at the speed of light and become homogeneously distributed well before BBN. The large initial local baryon asymmetry is thus diluted to the tiny observed global BAU. This naturally explains why the observed BAU is of order the PBH collapse fraction and why the baryons and dark matter have comparable densities. The situation is illustrated qualitatively in Fig.~\ref{fig:BAU}. Only at the QCD epoch are the conditions such that over-the-barrier sphalerons can occur via the heating of the surrounding plasma due to the violent shockwave at PBH formation. This is because the rapid quenching at QCD temperatures prevents baryon washout~\cite{Garcia-Bellido:2019vlf}.

The energy available for hot spot electroweak baryogenesis can be estimated as follows. Energy conservation implies that the change in kinetic energy due to the collapse of matter within the Hubble radius to the Schwarzschild radius of the PBH is
\begin{align}
	\Delta K
		\simeq
				\left(
					\frac{ 1 }{ \gamma }
					-
					1
				\right)\mspace{-1mu}
				M_{\Hrm}
		=
				\left(
					\frac{ 1 - \gamma }{ \gamma^{2} }
				\right)\mspace{-1mu}
				M
				\, ,
\end{align}
where $M_{\Hrm}$ is the Hubble mass and $\gamma$ is the size of the black hole compared to the Hubble horizon. The energy acquired per proton in the expanding shell is $E_{0} = \Delta K / ( n_{\prm}\,\Delta V)$, where $\Delta V = ( 1\.-\.\gamma^{3} )\.V_{\Hrm}$ is the difference between the Hubble and PBH volumes, so $E_{0}$ scales as $( \gamma\.+\.\gamma^{2}\.+\.\gamma^{3} )^{-1}$. For a PBH formed at $T \approx \Lambda_{\rm QCD} \approx 140\.\MeV$, the effective temperature is $T_{\rm eff} = 2\.E_{0} / 3 \approx 5\.$TeV, which is well above the sphaleron barrier and induces a charge-parity violation parameter $\dCP( T ) \sim 10^{-5}\.( T / 20\.\GeV )^{-12}$~\cite{Shaposhnikov:1998ewc}. The appropriate reheat temperature $T_{\rm rh}$ is the geometric mean of $T_{\rm eff}$ and the temperature of the thermal plasma, $T_{\rm pl} \sim 10\,$MeV, right after the collision of the shockwave with the plasma, since this corresponds to the centre-of-mass energy. This gives $T_{\rm rh} \sim 5\,$GeV, which implies $\dCP \sim 1$. The final baryon number comes from the Boltzmann equation, with all three temperatures playing a r{\^o}le: $T_{\rm eff}$ for baryon-number violation, $T_{\rm rh}$ for CP-violation and $T_{\rm pl}$ for baryon washout~\cite{Garcia-Bellido:2019vlf}. The production of baryons can therefore be very efficient, giving $\eta \sim 1$ locally. Note that PBH formation at a higher temperature transition would result in a smaller value of $\dCP$, which is why the QCD transition is optimal. Since the temperature at the QCD epoch is close to the proton mass $m_{\prm}$, the ratio of the photon and baryon densities, $n_{\gamma}\.T / n_{\brm}\mspace{1mu}m_{\prm}$, is just of order the photon-to-baryon ratio $S \sim \eta^{-1}$, so this proposal naturally links the BAU to the PBH collapse fraction ($\eta \sim \beta$). The observed ratio of the dark matter and baryon densities is also explained provided $\gamma \approx 0.8$.
\newpage

The spectator-field mechanism for PBH production from curvature fluctuations~\cite{2017PhRvD..96f3507C} also avoids the need for a fine-tuned peak in the power spectrum, which has long been considered a major drawback of PBH scenarios. One still needs fine-tuning of the mean field value to produce the observed values of $\eta$ and $\beta$, which are both around $10^{-9}$. However, the stochasticity of the field during inflation ensures that Hubble volumes exist with all possible field values and one can explain the fine-tuning by invoking a single anthropic constraint on the value of $\eta$. In this scenario, the curvature fluctuations at the origin of PBH formation are seeded after inflation when the field transiently dominates the energy density of the Universe. One attractive feature is that the primordial power spectrum could remain small on all scales, as in standard slow-roll inflation, because the curvature fluctuations are generated by non-Gaussian tails in the curvature fluctuation statistics. The scenario is discussed in Ref.~\cite{Carr:2019hud} and depends on the fact that only a small fraction of patches will have the PBH and baryon abundance required for galaxies to form. Here we can briefly review the possibilities proposed to solve the fine-tuning issue of the PBH abundance.

\newpage

\section{Conclusions}
\label{sec:Conclusions}

\noindent Previous literature has usually focussed on PBH {\it constraints}, identifying the mass ranges in which their density could be large enough to have interesting cosmological consequences but without providing positive evidence that this is the case. In this paper we have taken the opposite approach, identifying the observational data which specifically support the PBH proposal. In this sense, we are following the lead of the {\it positivist} philosophers of the early 20th century.

What is striking is the large number of strands of evidence for PBHs which we have described in this paper. We have mentioned numerous arguments. One may argue about the strength of any particular strand, be it on observational or theoretical grounds, but the collective impact seems very strong{\,---\,}like a rope whose strength far surpasses that of any individual fibre. We have also considered observations which could {\it potentially} be explained by PBHs, even though there are other explanations here and the evidence is less secure. Of course, the division between weak and strong evidence is not clear-cut and amounts to merely selecting a confidence level.

As regards the dark matter problem, it is interesting that much of the evidence point towards dark matter in PBHs with around a solar mass, this also corresponding to the peak of the mass function of ordinary stars, both being close to the Chandrasekhar mass. This coincidence is naturally explained if PBHs form at the QCD epoch and this is our preferred scenario. Other PBH advocates have claimed that this is precluded and favour a scenario in which the dark matter comprises PBHs in the planetary mass range. Although this alternative scenario is certainly not excluded, our justification for emphasising our own favoured scenario is that there is currently no {\it positive} evidence for it.

As regards the GW events, we have argued that the black holes which dominate the LIGO/Virgo/ KAGRA detections can only provide a small fraction of the dark matter but this is expected if PBHs have an extended mass function since the GW events inevitably peak at a higher mass than the density. The {\it natural} proximity of the PBH and stellar masses might be regarded as unfortunate from the perspective of distinguishing between astrophysical and primordial sources. However, other characteristics (such as spin) may remove this degeneracy. In any case, it seems most likely that one has a combination of the two, with evidence for at least some primordial component coming from observations of black holes in the two mass gaps in which stellar remnants are not expected.

As regards seeding the SMBHs in galactic nuclei our favoured scenario predicts that there should be at least some PBHs large enough to seed the SMBHs. However, the evidence for this is certainly not definitive, the primary uncertainty being the amount of growth of the seeds as a result of accretion. Since most of the final SMBH mass may derive from accretion at late times (\ie~from ordinary baryonic material), we emphasise that the SMBHs are merely seeded by PBHs. Although we have not discussed them much here, there could even be a population of stupendously large black holes (\ie~larger than $10^{12}\.\Msun$) of primordial origin, although again most of their mass might come from baryonic accretion.

An important aspect of our approach is that it is very broadly based, being {\it multi-production-mode} (invoking different PBH formation mechanisms), {\it multi-epoch} (the relevant data coming from many different redshifts), {\it multi-scale} (the predicted PBH mass function having bumps on several different scales) and {\it multi-probe} (invoking many types of observational evidence). The last point implies that several new or impending telescopes (such as the LSST, {\it Einstein Telescope}, SKA, LOFAR or JWST) may be brought to bear on the PBH proposal.
\newpage

\section*{Acknowledgements}
\noindent S.C.~acknowledges support from the Belgian Francqui Foundation through a start-up grant and the Belgian fund for research F.R.S.-FNRS through a {\it Mandat d'impulsion scientifique}. J.G.-B.~acknowledges support from the Spanish Research Project PID2021-123012NB-C43 [MICINN-FEDER] and the Centro de Excelencia Severo Ochoa Program CEX2020-001007-S at IFT. It is a pleasure to thank Anne Green, Jim Rich and Will Sutherland for elucidating remarks on the MACHO and EROS microlensing surveys, Gianfranco Bertone for an initial discussion on the missing-pulsar problem, Michael Brockamp for simulating remarks on cryogenic bulk acoustic wave cavity measurements, Dominik Schwarz for sharing his insights into the apparent Hubble tension, and G{\"u}nther Hasinger for multiple insights and discussions on the r{\^o}le of PBHs for early structure formation. We also acknowledge stimulating discussions with Earl Bellinger, Matthew Caplan and Selma de Mink on various aspects of PBHs and valuable insights from the referee. Finally, we would like to thank our collaborators in previous PBH studies since they have all indirectly contributed to this review: 
Nancy Aggarwal, 
Yacine Ali-Ha{\"i}moud, 
Richard Arendt, 
Eleni Bagui, 
Gregory Baltus, 
Sofiane Boucenna, 
Chris Byrnes, 
Francesca Calore, 
Nico Cappelluti, 
Sarah Caudill, 
Federico De Lillo, 
Valerio De Luca, 
Antoine Depasse, 
Kontantinos Dimopoulos, 
Gia Dvali, 
Albert Escriv{\`a}, 
Jose Ezquiaga, 
Ga{\'e}tan Facchinetti, 
Maxime Fays, 
Pierre Fleury, 
Heather Fong, 
Gabriele Franciolini, 
Andr{\'e} F{\"u}zfa, 
Katie Freese, 
Jonathan Gilbert, 
Ariel Goobar, 
Tomohiro Harada,
Nicolas Herman, 
Mark Hindmarsh, 
Rajeev Jain, 
Alexander Jenkins, 
Cristian Joana, 
Shasvath Kapadia, 
Christos Karathanasis, 
Alexander Kashlinsky, 
Kazunori Kohri, 
Sachiko Kuroyanagi, 
L{\'e}onard Lehoucq, 
Jim Lidsey, 
Matteo Lucca, 
Jane MacGibbon,
Kei-ichi Maeda, 
Ryan Magee, 
Mario Mart{\'i}nez, 
Katarina Martinovic, 
Alexis Men{\'e}ndez-V{\'a}zquez, 
Andrew Miller, 
Ester Ruiz Morales, 
Gonzalo Morr{\'a}s, 
Ilia Musco, 
Savvas Nesseris, 
Tommy Ohlsson, 
Charlotte Owen, 
Don Page,
Theodoros Papanikolaou, 
Marco Peloso, 
Khun Phukon, 
Sasha Polnarev, 
Vivian Poulin, 
Alvise Raccanelli, 
Martti Raidal, 
Sebastien Renaux-Petel, 
Antonio Riotto, 
Maria Sakellariadou, 
Ziad Sakr, 
Marit Sandstad, 
Marco Scalisi, 
Yuuiti Sendouda, 
Olga Sergijenko, 
Pasquale Serpico, 
Jos{\'e} Francisco 
Nu{\~n}o Siles, 
Joe Silk, 
Ioanna Stamou, 
Yuichiro Tada, 
Tomi Tenkanen, 
Jesus Torrado, 
Caner Unal, 
Ville Vaskonen, 
Hardi Veerm{\"a}e, 
Vincent Vennin, 
Luca Visinelli, 
David Wands, 
{\L}ukasz Wyrzykowski, 
Jun'ichi Yokoyama, 
Sam Young and
Michael Zantedeschi.
\newpage

\section*{Acronyms}
\vs{-6mm}
\begin{tabular}{l@{\hskip 4mm}l}
	\\[1.5mm]	
	ARCADE
		& {\bf A}bsolute {\bf R}adiometer for 
			{\bf C}osmology, {\bf A}strophysics and 
			{\bf D}iffuse {\bf E}mission
	\\[1.5mm]
	ALMA
		& {\bf A}tacama {\bf L}arge {\bf M}illimeter/submillimeter {\bf A}rray
	\\[1.5mm]
	BAU
 		& {\bf b}aryon {\bf a}symmetry of the 
			{\bf U}niverse
	\\[1.5mm]
	BATSE
 		& {\bf B}urst {\bf A}nd {\bf T}ransient 
			{\bf S}ource {\bf E}xperiment
	\\[1.5mm]
	BBN
 	 	& {\bf b}ig {\bf b}ang {\bf n}ucleosynthesis
	\\[1.5mm]
	CDM
	 	& {\bf c}old {\bf d}ark {\bf m}atter
	\\[1.5mm]
	CL
 		& {\bf c}onfidence {\bf l}evel
	\\[1.5mm]
	CMB
 		& {\bf c}osmic {\bf m}icrowave {\bf b}ackground
	\\[1.5mm]
	DESI
		& {\bf D}ark {\bf E}nergy {\bf S}pectroscopic 
			{\bf I}nstrument
	\\[1.5mm]
	EGRET
 		& {\bf E}nergetic {\bf G}amma {\bf R}ay 
			{\bf E}xperiment {\bf T}elescope
	\\[1.5mm]
	EPTA
 		& {\bf E}uropean {\bf P}ulsar {\bf T}iming 
			{\bf A}rray
	\\[1.5mm]
	EROS
 		& {\bf E}xp{\'e}rience pour la {\bf R}echerche 
			d'{\bf O}bjets {\bf S}ombres
	\\[1.5mm]
	FAR
 		& {\bf f}alse {\bf a}larm {\bf r}ate
	\\[1.5mm]
	FRB
 		& {\bf f}ast {\bf r}adio {\bf b}urst
	\\[1.5mm]
	GW
 		& {\bf g}ravitational {\bf w}ave
	\\[1.5mm]
	GWTC
 		& {\bf G}ravitational-{\bf W}ave 
			{\bf T}ransient {\bf C}atalog
	\\[1.5mm]
	HSC
 		& {\bf H}yper {\bf S}uprime-{\bf C}amera
	\\[1.5mm]
	HST
 		& {\bf H}ubble {\bf S}pace {\bf T}elescope
	\\[1.5mm]
	IMBH
	 	& {\bf i}ntermediate {\bf m}ass {\bf b}lack 
			{\bf h}ole
	\\[1.5mm]
	IPTA
 		& {\bf I}nternational {\bf P}ulsar {\bf T}iming 
			{\bf A}rray
	\\[1.5mm]
	ISCO
 		& {\bf i}nnermost {\bf s}table {\bf c}ircular 
			{\bf o}rbit
	\\[1.5mm]
	JWST
 		& {\bf J}ames {\bf W}ebb {\bf S}pace 
			{\bf T}elescope
	\\[1.5mm]
	KAGRA
 		& {\bf Ka}mioka {\bf Gra}vitational-Wave 
			Detector
	\\[1.5mm]
	LIGO
 		& {\bf L}aser {\bf I}nterferometer 
			{\bf G}ravitational-Wave {\bf O}bservatory
	\\[1.5mm]
	LISA
 		& {\bf L}aser {\bf I}nterferometer {\bf S}pace 
			{\bf A}ntenna
	\\[1.5mm]
	LOFAR
		& {\bf Lo}w {\bf F}requency {\bf Ar}ray
	\\[1.5mm]
	LVK
 		& {\bf L}IGO/{\bf V}irgo/{\bf K}AGRA
	\\[1.5mm]
	LSS
 		& {\bf l}arge-{\bf s}cale {\bf s}tructure
	\\[1.5mm]
	LSST
 		& {\bf L}arge {\bf S}ynoptic {\bf S}urvey 
			{\bf T}elescope
	\\[1.5mm]
	LWA1
		& {\bf L}ong {\bf W}avelength {\bf A}rray 
			({\bf 1}st station)

\end{tabular}
\newpage

\begin{tabular}{l@{\hskip 4mm}l}

	MACHO
 		& {\bf Ma}ssive {\bf C}ompact {\bf H}alo 
			{\bf O}bject
	\\[1.5mm]
	NANOGrav
 		& {\bf N}orth {\bf A}merican {\bf N}anohertz 
			{\bf O}bservatory for {\bf Grav}itational 
			Waves
	\\[1.5mm]
	NFW
 		& {\bf N}avarro-{\bf F}renk-{\bf W}hite
	\\[1.5mm]
	OGLE
 		& {\bf O}ptical {\bf G}ravitational 
			{\bf L}ensing {\bf E}xperiment
	\\[1.5mm]
	PBH
 		& {\bf p}rimordial {\bf b}lack {\bf h}ole
	\\[1.5mm]
	POINT-AGAPE
 		& {\bf P}ixel-lensing {\bf O}bservations with 
			the {\bf I}saac {\bf N}ewton 
			{\bf T}elescope
		 \\
 		& -- {\bf A}ndromeda {\bf G}alaxy 
			{\bf A}mplified {\bf P}ixels 
			{\bf E}xperiment
	\\[1.5mm]
	QCD
 		& {\bf q}uantum {\bf c}hromo {\bf d}ynamics
	\\[1.5mm]
	SKA
 		& {\bf S}quare {\bf K}ilometer {\bf A}rray
	\\[1.5mm]
	SMBH
 	 	& {\bf s}uper{\bf m}assive {\bf b}lack
			{\bf h}ole
	\\[1.5mm]
	SNR
 		& {\bf s}ignal-to-{\bf n}oise {\bf r}atio
	\\[1.5mm]
	SLAB
 		& {\bf s}tupendously {\bf l}arge 
			{\bf b}lack {\bf h}ole
	\\[1.5mm]
	SSM
 		& {\bf s}ub-{\bf s}olar {\bf m}ass
	\\[1.5mm]
	UFDG
 		& {\bf u}ltra-{\bf f}aint {\bf d}warf 
			{\bf g}alaxy
	\\[1.5mm]
	VRO
 		& {\bf V}era {\bf R}ubin {\bf O}bservatory
	\\[1.5mm]
	WIMP
 		& {\bf w}eakly {\bf i}nteracting 
			{\bf m}assive {\bf p}article
	\\[1.5mm]
	ZTF
 		& {\bf Z}wicky {\bf T}ransient {\bf F}acility

\end{tabular}
\newpage


\setlength{\bibsep}{7.5pt}
\bibliography{refs}

\end{document}